\documentclass[prd,aps,twocolumn,a4paper,showkeys,nofootinbib]{revtex4-1}

\usepackage{amsmath}
\usepackage{amsfonts}
\usepackage{amssymb}	
\usepackage{graphicx}
\usepackage{bm}
\usepackage{color}
\usepackage{commath}
\usepackage{accents}
\usepackage{float}
\allowdisplaybreaks

\usepackage{hyperref}
\usepackage{makecell}
\usepackage{multirow}

\newlength{\dhatheight}
\newcommand{\doublehat}[1]{%
	\settoheight{\dhatheight}{\ensuremath{\hat{#1}}}%
	\addtolength{\dhatheight}{-0.35ex}%
	\hat{\vphantom{\rule{1pt}{\dhatheight}}%
		\smash{\hat{#1}}}}

\def\e{{\rm e}}

\def\GMc2{G M_{\odot} c^{-2}}

\def\eps{\epsilon}

\def\O{\mathcal{O}}

\def\lm{{\ell m}}

\def\lm{{\ell m}}

\def\lm{{\ell m}}

\def\l{{\ell }}

\def\F{{\cal F}}

\def\O{{\cal O}}

\newcommand\be{\begin{equation}}
	\newcommand\ee{\end{equation}}

\def\ha{{\hat{a}}}

\def\Teukode{{\texttt{Teukode}}}
\def\htail_nc{{\hat{h}^{{\rm nc_{tail}}}_\lm}}

\def\TEOBResumS{\texttt{TEOBResumS}}

\usepackage{color}
\definecolor{cyan}{rgb}{0,0.9,0.9}
\definecolor{orange}{rgb}{0.9,0.5,0}
\definecolor{magenta}{rgb}{1,0,1}
\definecolor{purple}{rgb}{0.8,0.4,0.8}
\definecolor{gray}{rgb}{0.8242,0.8242,0.8242}
\definecolor{dodgerblue}{rgb}{0.12, 0.56, 1.0}
\definecolor{darkgrey}{rgb}{0.5,0.5,0.5}
\definecolor{darkgreen}{rgb}{0,0.65,0}
\definecolor{lilac}{rgb}{0.8, 0.4, 1}

\begin{document}
	
	\title{Exploiting Newton-factorized, 2PN-accurate, waveform multipoles in effective-one-body models for spin-aligned noncircularized binaries}
	
	\author{Andrea \surname{Placidi}${}^{1,2}$}
	\author{Simone \surname{Albanesi}${}^{3,4}$}
	\author{Alessandro \surname{Nagar}${}^{4,5}$}
	\author{Marta \surname{Orselli}${}^{1,2}$}
	\author{Sebastiano \surname{Bernuzzi}${}^{6}$}
	\author{Gianluca \surname{Grignani}${}^{1}$}
	\affiliation{${}^1$Dipartimento di Fisica e Geologia, Universit\`a di Perugia,
		INFN Sezione di Perugia, Via A. Pascoli, 06123 Perugia, Italia}
	\affiliation{${}^2$Niels Bohr Institute, Copenhagen University,  Blegdamsvej 17, DK-2100 Copenhagen \O{}, Denmark}
	\affiliation{${}^3$Dipartimento di Fisica, Universit\`a di Torino, via P. Giuria 1, 
		10125 Torino, Italy}
	\affiliation{${}^4$INFN Sezione di Torino, Via P. Giuria 1, 10125 Torino, Italy} 
	\affiliation{${}^5$Institut des Hautes Etudes Scientifiques, 91440 Bures-sur-Yvette, France}
	\affiliation{${}^6$Theoretisch-Physikalisches Institut, Friedrich-Schiller-Universit{\"a}t 
		Jena, 07743, Jena, Germany}  
	
	\date{\today}
	\begin{abstract}
		We present a new approach to factorize and resum the post-Newtonian (PN) waveform for generic equatorial motion
		to be used within effective-one-body (EOB) based waveform models. The new multipolar
		waveform factorization improves previous prescriptions in that: (i) the generic Newtonian contribution is factored out from each
		multipole; (ii) the circular part is factored out and resummed using standard EOB methods and (iii) the residual, 2PN-accurate,
		noncircular part, and in particular the tail contribution, is additionally resummed using Pad\'e approximants.
		The resulting waveform is validated in the extreme-mass-ratio limit by comparisons with nine (mostly nonspinning) 
		numerical waveforms either from eccentric inspirals, with eccentricities up to $e=0.9$, or dynamical captures . 
		The resummation of the noncircular tail contribution  is found essential to obtain excellent (${\lesssim}0.05$~rad at periastron for $e=0.9$) 
		analytical/numerical agreement and to considerably improve the  prescription with just the Newtonian prefactor.
		In the comparable mass case, the new 2PN waveform shows only a marginal improvement over the previous 
		Newtonian factorization, though yielding maximal unfaithfulness $\simeq 10^{-3}$ with the 28 publicly available numerical 
		relativity simulations with eccentricity up to $\sim 0.3$ (except for a single outlier that grazes $10^{-2}$). 
		We finally use test-particle data to validate the waveform factorization proposed by 
		Khalil et al.~[Phys.~Rev.~104 (2021) 2, 024046] and conclude that its amplitude can be considered 
		reliable (though less accurate, $\sim 6\%$ fractional 
		difference versus $1.5\%$ of our method) only up to eccentricities $\sim 0.3$.
	\end{abstract}
	\maketitle
	%\tableofcontents
	%
	%==========================================================
	%
	
	\section{Introduction}
	\label{Sec:Int}
	
	One of the greatest achievements of recent years is the detection of gravitational waves (GW) from coalescing binary systems by the LIGO-Virgo-Kagra (LVK) collaboration~\cite{LIGOScientific:2021djp}. This opened up a new exciting way of investigating the properties of spacetime in regime of strong gravitational field.
	Moreover, the scientific community is already working hard to plan the construction of 
	ground based detectors of third generation~\cite{Reitze:2021gzo, Couvares:2021ajn, Punturo:2021ryo, Katsanevas:2021fzj, Kalogera:2021bya, McClelland:2021wqy}, 
	such as Einstein Telescope~\cite{Maggiore:2019uih} and Cosmic Explorer~\cite{Evans:2021gyd}, 
	with extremely high sensitivity. In addition, 
	the space based detector LISA~\cite{LISA:2017pwj} will 
	open a window in the low frequency band allowing the potential detection of
	new kinds of sources, such as  Intermediate and Extreme-Mass-Ratio 
	Inspirals~\cite{Amaro-Seoane:2007osp, Amaro-Seoane:2018gbb, Babak:2017tow}.
	This gives rise to the compelling need of improving the 
	analytical waveform models used to interpret the detected GW signals~\cite{Purrer:2019jcp}.
	In particular, the question of how to incorporate eccentric effects in the waveform models for coalescing binaries, 
	both on the analytical and the numerical side, is under intense 
	study~\cite{Hinder:2017sxy,Hinderer:2017jcs, Chiaramello:2020ehz, Islam:2021mha, Albanesi:2021rby, Liu:2021pkr,Yun:2021jnh, Tucker:2021mvo, Setyawati:2021gom, Nagar:2021gss, Nagar:2021xnh, Cho:2021oai, Khalil:2021txt}.
	More specifically, Ref.~\cite{Chiaramello:2020ehz} introduced an efficient and accurate, yet simple, approach 
	to generalize the effective-one-body (EOB) quasi-circular waveform and radiation-reaction of the \TEOBResumS{} 
	model to the eccentric case, basically consisting in replacing the standard quasi-circular Newtonian prefactors in 
	these functions with the general expression that follows from taking the time-derivatives of the Newtonian mass 
	and current multipoles. 
	% hyperbolic generalization GW190521
	The same idea has been used to describe dynamical captures of black 
	holes~\cite{Nagar:2020xsk}, leading to an EOB model able to describe
	binaries with generic orbits~\cite{Nagar:2021gss}.
	This breakthrough in the description of noncircular systems even allowed a new 
	analysis of GW190521~\cite{Gamba:2021gap}, showing that the dynamical capture scenario 
	is favorite against the quasi-circular highly-precessing scenario proposed
	by the LVK Collaboration~\cite{LIGOScientific:2020iuh}.
	Moreover, a recent work~\cite{Nagar:2021xnh} showed that the inclusion of higher-order post Newtonian (PN)
	terms in the EOB potentials allows to construct a model that is both faithful in the quasi-circular case, similarly to
	the native quasi-circular model, as well as for eccentric inspirals.
	
	%In practice, thus
	Nonetheless, high-PN, noncircular, analytical information computed in several 
	works~\cite{Mishra:2015bqa,Khalil:2021txt} has not been incorporated yet 
	within~\TEOBResumS{}, although it is currently present in other models~\cite{Khalil:2021txt,Liu:2021pkr,RamosBuades:2021}.
	In particular, the 2PN-accurate waveform of Ref.~\cite{Khalil:2021txt} has been included
	in a new EOB model for eccentric coalescence, {\tt SEOBNRv4HME}~\cite{RamosBuades:2021}
	that builds upon the quasi-circular model {\tt SEOBNRv4HM}~\cite{Bohe:2016gbl,Cotesta:2018fcv},
	that is rather different from \TEOBResumS{} either in structure and performance~\cite{Rettegno:2019tzh,Albertini:2021tbt}.
	The purpose of this work is to exploit currently known analytical results up to 2PN so to include 
	them in \TEOBResumS{}-based eccentric models~\cite{Nagar:2021gss,Nagar:2021xnh}.
	In particular, we exploit here the analytical 2PN waveform information present in its original 
	form in both Refs.~\cite{Mishra:2015bqa,Khalil:2021txt}\footnote{Note that, although Ref.~\cite{Khalil:2021txt}
		also computed new spin-dependent eccentric  waveform corrections up to 2PN, 
		we will not use them here, but postpone their inclusion in a suitable factorized waveform to future work.}. 
	In practice, each multipole is factorized using the general Newtonian prefactor,
	as proposed in Ref.~\cite{Chiaramello:2020ehz}, that is then multiplied by
	a resummed quasicircular correction and by the residual noncircular correction at 2PN.
	This latter contribution is then additionally resummed. This procedure is extensively
	tested against different types of numerical waveform data both in the  test-mass limit and for comparable-mass binaries.
	
	The paper is organized as follows. In Sec.~\ref{Sec:PNExpWavef} we review available analytical results up to 2PN accuracy 
	and calculate the genuine noncircular contribution at 2PN accuracy obtained by factoring out, multipole by multipole, 
	both the generic Newtonian prefactor and the circular part. This analytic approach is validated in Sec.~\ref{sec:testmass}
	by comparison with numerical waveforms emitted by a test-particle orbiting a Kerr black hole along eccentric orbits. 
	This gives rise to the need of implementing additional resummation strategies, notably on the noncircular tail 
	contribution, so to obtain analytical waveforms that are reliable {\it also} for large eccentricities, $e\sim 0.9$.
	In Sec.~\ref{sec:testmass_hyp} we perform similar analyses for a few illustrative hyperbolic encounters and
	dynamical capture scenario in the test-mass limit. Section~\ref{sec:comparable_masses} focuses on
	comparable-mass binaries and provides direct phasing comparisons between Numerical Relativity (NR) 
	waveforms and the EOB eccentric model of Ref.~\cite{Nagar:2021xnh} updated with the new 2PN-accurate
	factorized and resummed waveform.
	Finally, in Sec.~\ref{sec:qc}, for the test-mass limit case, we provide analogous waveform comparisons with
	the factorization scheme proposed in Ref.~\cite{Khalil:2021txt} (and used in Ref.~\cite{RamosBuades:2021}), 
	whose distinctive feature is that {\it only} the circular Newtonian prefactor, instead of the generic one, 
	is factorized for each multipole.
	Concluding remarks are collected in Sec.~\ref{sec:end}. The paper is finally completed by a few technical
	Appendixes. If not otherwise specified, we use geometrical units with $G=c=1$.

	\section{Newton-factorized EOB waveform}
	\label{Sec:PNExpWavef}
	
	The 2PN waveform in EOB coordinates was recently obtained by Khalil et al.~\cite{Khalil:2021txt}.
	For the instantaneous contribution, they obtained the expressions for the modes in EOB coordinates at 2PN accuracy by translating to EOB coordinates the results of Refs.~\cite{Mishra:2015bqa,Boetzel:2019nfw} which are originally expressed in harmonic coordinates. The tail contributions are derived starting from the results of Ref.~\cite{Hinderer:2017jcs} which are subsequently extended in Ref.~\cite{Khalil:2021txt} to include higher order corrections in the eccentricity and also to higher modes. The final result in Ref.~\cite{Khalil:2021txt} is then written using the standard factorization of the circular part, adding the noncircular contribution as a correction. 
	Here we exploit
	the PN-expanded results of Ref.~\cite{Khalil:2021txt} (neglecting the spin part) but then we employ a rather different factorization scheme. Before doing so,
	it is pedagogically useful to recall here all the steps, in order to fix the notation and to keep the discussion as self-contained as possible.
	In this section we mostly focus on the dominant mode with $\ell=m=2$. The results for the subdominant modes are presented in Appendix~\ref{App:HM}. Note also that in Sec.~\ref{Sec:Alt_PN_corrections} we discuss some issues related to the modes with $m=0$.
	
	\subsection{The 2PN waveform in EOB coordinates}
	\label{sec:2PN}
	To start with, let us recall that we work with the following multipolar decomposition of 
	the strain waveform
	\begin{equation}
		h_+ - i h_\times = D_L^{-1} \sum_{\ell=2}^{\ell_{\text{max}}} \sum_{m=-\ell}^{\ell} h_{\ell m} \, {}_{-2}Y_{\ell m},
	\end{equation}
	where $D_L$ is the source distance in radiative coordinates and ${}_{-2}Y_{\ell m}$ 
	are the spin-weight -2 spherical harmonics. Each waveform mode is additionally
	factorized as~\cite{Mishra:2015bqa}
	\begin{equation}
		\label{basicStruct}
		%	h_\lm=\frac{4GM\nu}{c^4}\sqrt{\dfrac{\pi}{5}} e^{-im\varphi} \hat{H}_{\ell m}, 
		h_\lm=4\nu\sqrt{\dfrac{\pi}{5}} e^{-im\varphi} \hat{H}_{\ell m}, 
	\end{equation}
	where each $\hat{H}_\lm$ mode is decomposed in the sum of instantaneous 
	and hereditary (or tail) terms 
	\be
	\hat{H}_\lm= \hat{H}^{\text{inst}}_\lm + \hat{H}^{\text{tail}}_\lm ,
	\ee
	that enter at different PN orders. 
	
	\subsubsection{Instantaneous contributions}
	\label{SubSec:inst_hlm}
	The instantaneous contributions to the modes for
	nonspinning binaries were derived in Ref.~\cite{Will:1996zj,Gopakumar:2001dy} up to 2PN order and 
	in Ref.~\cite{Mishra:2015bqa} up to 3PN order, where they are expressed in harmonic coordinates. 
	The 2PN-accurate $\ell=m=2$ mode in harmonic coordinates  explicitly reads
	\begin{widetext}
		\begin{align}
			\label{instH22_indians}
			\big(\hat{H}^{\text{inst}}_{22}\big)_{h} &= \frac{1}{r_h}+r_h^2 \dot{\varphi}_h^2+2 i r \dot{r}_h \dot{\varphi}_h-\dot{r}_h^2 \nonumber\\
			&+ \frac{1}{c^2} \bigg\{\frac{1}{r_h^2} \bigg(\frac{\nu}{2}-5\bigg)+r_h \left[\dot{\varphi}_h^2 \left(\frac{78 \nu}{21} +\frac{11}{42}\right) - i  \dot{r}_h \dot{\varphi}_h(\dot{r}_h^2+r_h^2\dot{\varphi}_h^2)\left(\frac{27\nu}{7}  -\frac{9}{7}\right)\right]\cr
			&-\frac{\dot{r}_h^2}{r_h}\left(\frac{16\nu}{7}+\frac{15}{14}\right)+i  \dot{r}_h \dot{\varphi}_h \left(\frac{45 \nu}{7} +\frac{25}{21}\right) + (\dot{r}_h^4-r_h^4 \dot{\varphi}_h^4)\left(\frac{27\nu}{14}  -\frac{9}{14} \right) \bigg\} \cr
			&+ \frac{1}{c^4} \bigg\{\frac{1}{r_h^3}\left(\frac{79\nu^2}{126}+\frac{181\nu}{36}+\frac{757}{63}\right)+ \dot{\varphi}_h^2\left(\frac{ 13133 \nu ^2}{1512}-\frac{5225 \nu}{216} -\frac{11891}{1512}\right) \cr
			&+\frac{ \dot{r}_h^4}{r_h} \left(\frac{214 \nu ^2}{21}+\frac{83 \nu}{21} -\frac{557}{168}\right) +\frac{ i \dot{r}_h \dot{\varphi}_h}{r_h} \left(\frac{2852 \nu ^2}{189}-\frac{3767 \nu}{189} -\frac{773}{189}\right) \cr
			&-\frac{\dot{r}_h^2}{r_h^2}\left(\frac{467 \nu ^2}{126}+\frac{2789 \nu}{252} -\frac{619}{252}\right) +(\dot{r}_h^2 r_h^4 \dot{\varphi}_h^4-\dot{r}_h^6+r_h^6 \dot{\varphi}_h^6-r_h^2\dot{r}_h^4 \dot{\varphi}_h^2) \left(\frac{1111\nu ^2}{168} -\frac{589 \nu}{168} +\frac{83}{168}\right)  \cr
			&-  i r_h^2 \dot{r}_h \dot{\varphi}_h^3 \left(\frac{1703 \nu ^2}{84}-\frac{103 \nu}{12} -\frac{433}{84}\right) - i \dot{r}_h^3 \dot{\varphi}_h \left(\frac{211 \nu ^2}{9}+\frac{731 \nu}{63} -\frac{863}{126}\right) \cr
			&+r_h^3  \dot{\varphi}_h^4  \left(-\frac{2995 \nu ^2}{252}+\frac{19 \nu}{252} +\frac{835}{252}\right) + ir_h^3 \dot{r}_h^3 \dot{\varphi}_h^3 \left(\frac{1111 \nu ^2}{42}-\frac{589 \nu}{42} +\frac{83}{42}\right) \cr
			&-r_h  \dot{r}_h^2 \dot{\varphi}_h^2 \left(\frac{58 \nu ^2}{21}+\frac{169 \nu}{14} -\frac{11}{28}\right)+ i (r_h \dot{r}_h^5 \dot{\varphi}_h+\dot{r}_h r_h^5 \dot{\varphi}_h^5)\left(\frac{1111 \nu ^2}{84}-\frac{589 \nu}{84} +\frac{83}{84}\right) \bigg\}+ \mathcal{O}\left(\dfrac{1}{c^6}\right),
		\end{align}
	\end{widetext}
	where $\varphi_h$ is the harmonic orbital phase, $r_h$ is the mass-reduced radial harmonic 
	coordinate defined as $r_h\equiv R_h/M$, with $M=m_1+m_2$ where $m_1$ and $m_2$ are the individual  masses of the binary system.
	For completeness we explicitly report here the transformation from harmonic to EOB coordinates, following
	the same steps of Ref.~\cite{Khalil:2021txt} using the relations that can be derived from 
	%the Appendix of 
	Ref.~\cite{Bini:2012ji}.
	The EOB dynamics is expressed using mass-reduced phase-space 
	variables $(r,\varphi,p_r,p_\varphi)$, which are related to the physical ones by $r\equiv R/M$ (relative separation), 
	$p_r=P_R/\mu$ (radial momentum), $\varphi$ (orbital phase), $p_\varphi= P_\varphi/(\mu M)$ (angular momentum)
	and $t=T/M$ (time) where $\mu\equiv m_1 m_2/M$. As for the relative separation, we will henceforth use its inverse $u \equiv 1/r$.
	Moreover, to better keep track of the PN counting, we will make explicit the dependence on $1/c$,
	recalling that $u \sim p_r^2 \sim p_\varphi^2 u^2 \sim 1/c^2$. 
	%The dynamical variables we employ are the EOB orbital phase $\varphi$ and the rescaled quantities
	%\begin{equation}
		%	\label{eobVar}
		%	u\equiv \frac{G M}{r_e \, c^2}, \quad p_r \equiv \frac{(\hat{\mathbf{n}}_{e}\cdot \mathbf{p}_{e})}{\mu \, c}, \quad p_{\varphi} \equiv \frac{\lvert\hat{\mathbf{n}}_{e}\times \mathbf{p}_{e}\rvert \, c}{\mu},
		%\end{equation}
	%all three being adimensional and expressed in terms of the EOB coordinates $\mathbf{r}_{e}$=$\hat{\mathbf{n}}_{e} \, r_e$ and conjugate 
	%momenta $\mathbf{p}_{e}$ \cite{missing}. Here $G$ is Newton’s gravitational constant, $c$ the speed of light, $M = m_1 + m_2$ the total mass of the system 
	%and $\mu = m_1 m_2 / M$ its reduced mass. 
	%To better keep track of the PN counting we also introduce the dimensionless 
	%parameter $\eta\equiv 1/c^2$, bearing in mind that $u \sim p_r^2 \sim p_\varphi^2 u^2 \sim 1/c^2$. In this scheme a $n$PN term is proportional to $\eta^{2n}$.  
	Using the transformation laws given in 
	%Appendix~E2 of 
	Ref.~\cite{Bini:2012ji}, we obtain the following relations between
	the quantities $(r_h,\dot{r}_h,\varphi_h,\dot{\varphi}_h)$ and the EOB canonical variables $(u,\varphi,p_r,p_\varphi)$
	%Therefore, for our purpose, we can use as a starting point the transformation laws given in Appendix E of Ref.~\cite{Bini:2012ji} and adapt them to 
	%our case. In particular, while always remaining in the center of mass reference frame, the harmonic dynamical (or dynamical harmonic?) 
	%variables we need to transform are the relative separation $r_h$, its time derivative $\dot{r}_h$, the orbital phase $\varphi_h$ and the angular velocity $\dot{\varphi}_h$.
	%The resulting canonical transformations at 2PN are
	\begin{widetext}
		\begin{flalign}
			\label{rhTransf}
			r_h & =\frac{1}{u}+ \dfrac{1}{c^2} \left[\frac{1}{2} \nu  \left(\frac{3 p_r^2}{u}+p_\varphi^2 u-1\right)-1\right] \cr
			&-\frac{ \nu}{8 c^4 u} \bigg[(5-3 \nu ) p_r^4+2 p_r^2 u \bigg(28-3 (\nu -1) p_\varphi^2 u\bigg)+u^2 \bigg(2 (\nu -19)+(\nu +1) p_\varphi^4 u^2+p_\varphi^2 u(1-3 \nu)\bigg)\bigg] \cr
			&+\mathcal{O}\left(\frac{1}{c^6}\right),\\
			\label{dotrhTransf}
			\dot{r}_h & = p_r + \frac{1}{2c^2}  p_r \bigg[(2 \nu -1) p_r^2+u \bigg( (4 \nu -1) p_\varphi^2 u-6 -4 \nu \bigg)\bigg] \cr
			&+ \frac{1}{8c^4} \bigg[(3-8 \nu ) p_r^5-2 p_r^3 u \bigg(2 (8 \nu -5)+\left(2 \nu ^2+12 \nu -3\right) p_\varphi^2 u\bigg) \bigg. \cr
			&\bigg.+p_r u^2 \bigg(-10 \nu ^2+78 \nu +12 +\left(8 \nu ^2-16 \nu +3\right) p_\varphi^4 u^2+2 \left(\nu ^2-55 \nu +6\right) p_\varphi^2 u\bigg)\bigg] +\mathcal{O}\left(\frac{1}{c^6}\right),\\
			\label{phihTransf}
			\varphi_h & =\varphi + \dfrac{1}{c^2}  \nu p_r p_\varphi u-\frac{1}{4c^4}  \nu  p_r p_\varphi u \bigg((4 \nu +2) p_r^2+u \left(-3\nu +15+2 p_\varphi^2 u\right)\bigg)+\mathcal{O}\left(\frac{1}{c^6}\right),\\
			\label{dotphihTransf}
			\dot{\varphi}_h &= p_\varphi u^2 +\frac{1}{2c^2} p_\varphi u^2 \bigg[u \bigg((\nu -1) p_\varphi^2 u-2\bigg)-(3 \nu +1) p_r^2\bigg] \cr
			&-\frac{1}{8c^4} p_\varphi u^2 \bigg[-\left(15 \nu ^2+11 \nu +3\right) p_r^4+2 p_r^2 u \bigg(2 \left(\nu ^2-24 \nu -3\right)+3 \left(3 \nu ^2-\nu -1\right) p_\varphi^2 u\bigg) \bigg. \cr
			& \bigg.+u^2 \bigg(2 \left(\nu ^2-9 \nu +2\right)+\left(\nu ^2+5 \nu -3\right) p_\varphi^4 u^2-2 \left(3 \nu ^2-17 \nu +2\right) p_\varphi^2 u\bigg)\bigg]+\mathcal{O}\left(\frac{1}{c^6}\right).
		\end{flalign}
	\end{widetext}
	Replacing these equations into Eq.~\eqref{instH22_indians} above, yields
	\begin{widetext}
		\begin{align}
			\label{instH22}
			\hat{H}^{\text{inst}}_{22} &= u-p_r^2+2 i p_r p_\varphi u+p_\varphi^2 u^2 +\frac{1}{c^2}\bigg\{ i \left(\frac{\nu}{7} -\frac{5}{7}\right)p_r^3 p_\varphi u - i p_r p_\varphi u^2 \bigg[\left(\frac{ 4\nu}{7}+ \frac{185}{21}\right) - \left(\frac{\nu}{7} -\frac{5}{7}\right) p_\varphi^2 u\bigg] \bigg. \cr
			&\bigg.+ \left(\frac{\nu}{14} -\frac{5}{14}\right) (p_\varphi^4 u^4-p_r^4) +\left(\frac{3\nu}{14} +\frac{64}{14}\right)  p_r^2 u+u^2 \bigg[ (\nu -4)+\left(\frac{31\nu}{14} -\frac{157}{42}\right)p_\varphi^2 u\bigg]\bigg\} \cr
			&+\frac{1}{c^4} \bigg\{\left(\frac{17 \nu ^2}{168}+\frac{13 \nu}{168} -\frac{5}{24}\right)(p_r^6+p_r^4p_\varphi^2 u^2-p_\varphi^6 u^6) -i\left(\frac{17 \nu ^2}{84}+\frac{13 \nu}{84}-\frac{5}{12}\right)p_r p_\varphi u(p_r^4+ p_\varphi^4 u^4)  \bigg]\bigg. \cr
			&\bigg.- p_r^2 u^2 \bigg[ \left(\frac{13 \nu ^2}{63}+\frac{151 \nu}{18}+\frac{1055}{252}\right)  + \left(\frac{17 \nu ^2}{504}+\frac{13 \nu}{504}-\frac{5}{72}\right)p_\varphi^4 u^2+ \left(\frac{313 \nu ^2}{252}+\frac{85 \nu}{252} -\frac{101}{252}\right) p_\varphi^2 u\bigg]\bigg. \cr
			&\bigg.- \left(\frac{85 \nu ^2}{168}+\frac{55 \nu}{168} +\frac{425}{168}\right)p_r^4 u + i p_r^3 p_\varphi u^2 \bigg[\left(\frac{62 \nu ^2}{63}+\frac{11 \nu}{126}+\frac{695}{126}\right)-\left(\frac{17 \nu ^2}{42}+\frac{13 \nu}{42}-\frac{5}{6}\right) p_\varphi^2 u \bigg. \cr
			&\bigg.- i p_r p_\varphi u^3 \bigg[\left(\frac{523 \nu ^2}{189}+\frac{2452 \nu}{189}-\frac{193}{27}\right) +\left(\frac{8 \nu ^2}{21}-\frac{29 \nu}{14}-\frac{67}{28}\right)p_\varphi^2 u\bigg]+u^3 \bigg[\left(\frac{205 \nu ^2}{126}-\frac{49 \nu}{18}+\frac{190}{63}\right)\bigg.\bigg. \cr
			&\bigg.\bigg.-\left(\frac{671 \nu ^2}{504}+\frac{1375 \nu}{504}-\frac{481}{72}\right)p_\varphi^4 u^2+\left(\frac{127 \nu ^2}{27}-\frac{2710 \nu}{189}-\frac{5519}{1512}\right)p_\varphi^2 u\bigg]\bigg\}+\mathcal{O}\left(\dfrac{1}{c^6}\right).
		\end{align}
	\end{widetext}
	This result coincides with Eq.~(83) of Ref.~\cite{Khalil:2021txt} by replacing $p_\varphi^2 u^2=p^2-p_r^2 $. The canonical transformations 
	above have also been applied to the other multipoles up to $\ell=6$ and have been obtained independently
	of Ref.~\cite{Khalil:2021txt}, which appeared while the current paper was in preparation. Our results precisely coincide with those
	of Ref.~\cite{Khalil:2021txt}. However, as an additional, independent, validation of the transformations applied to the various 
	multipoles we have shown that the 2PN angular momentum flux given by Eq.~(3.70) of Ref.~\cite{Bini:2012ji} can be obtained
	combining together the various multipoles once written in EOB coordinates, as explained in Appendix~\ref{App:angularRRcheck}.

	\subsubsection{Hereditary contributions}
	\label{SubSec:HeredContr}
	
	The hereditary components $\hat{H}^{\text{tail}}_\lm$ are taken from Ref.~\cite{Khalil:2021txt}.
	These contributions were computed as an expansion in eccentricity and using the Keplerian 
	parametrization (KP), according to the method outlined in \cite{Hinderer:2017jcs}. 
	The resulting tail contributions are initially expressed in terms of the frequency parameter 
	$x \equiv (GM \Omega/c^3)^{2/3}$, the eccentricity $e$ and the phase variable $\chi$, 
	which together with the semilatus rectum $p$ parameterize the motion as 
	\begin{equation}
		r = \frac{p}{1+e \, \rm{cos} \, \chi }.
	\end{equation}
	In particular the parameters $p$ and $e$ used in Refs.~\cite{Khalil:2021txt,Hinderer:2017jcs} and in this paper are defined in analogy with Newtonian mechanics by the relations\footnote{We warn the reader that the eccentricity $e$ defined here is different from the time eccentricity $e_t$ that appears in the context of the quasi-Keplerian parametrization. }
	\begin{equation}
		\label{eq: e&p}
		e=\frac{r_+-r_-}{r_++r_-}, \qquad p= \frac{2 r_+ r_-}{r_++r_-},
	\end{equation}
	where $r_{\pm}$ are the turning point of the radial motion, respectively apastron and periastron.
	
	From Ref.~\cite{Khalil:2021txt}, the $\ell=m=2$ dominant tail component at 2PN accuracy, and up $\mathcal{O}(e^6)$ ,
	reads
	\begin{widetext}
		\begin{align}
			\label{tailH22KP}
			\big(\hat{H}^{\text{tail}}_{22}\big)_{KP} &= \frac{ 2\pi x^{5/2}}{c^3} \bigg[ 1 + e \bigg( \frac{11 \e^{- i \chi}}{8}+\frac{13 \e^{ i \chi} }{8}\bigg) + e^2 \bigg(\frac{5}{8} \e ^{-2i \chi} + \frac{7}{8} \e ^{2i \chi}+4\bigg)
			+ e ^3 \bigg(\frac{121 \e^{- i \chi}}{32}+\frac{143 \e^{i \chi}}{32}+\frac{3 \e^{-3 i \chi}}{32}+\frac{ \e^{3 i \chi}}{12}\bigg)\cr
			&+ e^4 \bigg(\frac{25 \e^{-2i \chi}}{16}+ \frac{203 \e^{2 i \chi }}{96} - \frac{5 \e ^{4 i \chi}}{96} + \frac{65}{8}\bigg) + e^5 \bigg(\frac{55 \e ^{-i\chi}}{8}+\frac{6233 \e ^{i\chi}}{768}+\frac{15 \e ^{-3i\chi}}{64}+\frac{281 \e ^{3i\chi}}{1536}+\frac{53 \e ^{5\chi}}{7680}\bigg)  \cr
			&+ e^6 \bigg(\frac{175 \e ^{-2i\chi}}{64}+\frac{1869 \e ^{2i\chi}}{512} - \frac{449 \e ^{4i\chi}}{3840}+\frac{31 \e ^{6i\chi}}{23040} + \frac{30247}{2304}\bigg) \bigg]. %+\mathcal{O}(e^7)+\mathcal{O}\left(\frac{1}{c^5}\right)
		\end{align}
	\end{widetext}
	Analogously to Ref.~\cite{Khalil:2021txt}, we want to recast the expressions above in terms of EOB phase-space variables.
	In particular, we want to use $(u, p_r, p_\varphi)$ only, avoiding the explicit appearance of the time-derivatives of the 
	momenta. This will eventually simplify the numerical implementation and differs from the choice made in Ref.~\cite{Khalil:2021txt}.
	Since we are working at 2PN accuracy, to this aim we can use the following 
	Newtonian relations\footnote{Corrections to the leading Newtonian order would enter at 2.5PN order in the waveform.}
	\begin{align}
		\label{LO:x}
		x & =\frac{1 - e^2}{p},\\
		\label{LO:u}
		u & = \frac{1+e \, \rm{cos} \chi }{p}, \\
		\label{LO:pphi}
		p_\varphi &= \sqrt{p},\\
		\label{LO:pr}
		p_r  &= \frac{e \, \rm{sin} \chi}{\sqrt{p}}.		 
	\end{align}
	Moreover, since Eq.~\eqref{tailH22KP}  is given as an expansion in the eccentricity $e$, it is important to identify the proper variables, 
	in terms of the new set $(u, p_r, p_\varphi)$, which are of the same order in the eccentricity. One can show that 
	Eq.~\eqref{tailH22KP} translates into an expansion in both $p_r$ and $\dot{p}_r$, with the latter related to $p_\varphi$ 
	through the Newtonian equation of motion
	\begin{equation}
		\label{prdot}
		\dot{p}_r = u^2 (p_\varphi^2 u -1).
	\end{equation} 
	Using Eqs.~\eqref{LO:x}-\eqref{LO:pr}, one can show that $p_r\sim e$ and $(p_\varphi^2 u -1)\sim e$. 
	In this way, from Eq.~\eqref{tailH22KP} one obtains an expression in terms of $(u, p_r, \dot{p}_r)$ 
	which contains also half integer powers of $u$, and that reads\footnote{Here the round brackets collect terms at the same order in eccentricity}
	\begin{widetext}
		\begin{align}
			\label{eq:H22u}
			\hat{H}_{22}^{\rm tail}& = \frac{2\pi u^{5/2}}{c^3}\bigg[1+\bigg(\frac{\dot{p}_r}{2 u^2}+\frac{i p_r}{4 \sqrt{u}}\bigg)-\frac{\dot{p}_r^2}{8 u^4}-\bigg(\frac{\dot{p}_r^3}{96 u^6}+\frac{7 i p_r \dot{p}_r^2}{32 u^{9/2}}-\frac{7 p_r^2 \dot{p}_r}{32 u^3}-\frac{7 i p_r^3}{96 u^{3/2}}\bigg)+\bigg(\frac{7 \dot{p}_r^4}{384 u^8}+\frac{i p_r \dot{p}_r^3}{12 u^{13/2}}\cr
			&-\frac{p_r^2 \dot{p}_r^2}{64 u^5}-\frac{i p_r^3 \dot{p}_r}{96 u^{7/2}}+\frac{p_r^4}{48 u^2}\bigg)-\bigg(\frac{13 \dot{p}_r^5}{1920 u^{10}}-\frac{i p_r \dot{p}_r^4}{768 u^{17/2}}+\frac{73 p_r^2 \dot{p}_r^3}{768 u^7}+\frac{49 i p_r^3 \dot{p}_r^2}{384 u^{11/2}}-\frac{35 p_r^4 \dot{p}_r}{384 u^4}-\frac{89 i p_r^5}{3840 u^{5/2}}\bigg)\cr
			&+\bigg(\frac{109 \dot{p}_r^6}{46080 u^{12}}-\frac{i p_r \dot{p}_r^5}{64 u^{21/2}}+\frac{137 p_r^2 \dot{p}_r^4}{1536 u^9}+\frac{137 i p_r^3 \dot{p}_r^3}{1152 u^{15/2}}-\frac{65 p_r^4 \dot{p}_r^2}{768 u^6}-\frac{23 i p_r^5 \dot{p}_r}{640 u^{9/2}}+\frac{p_r^6}{96 u^3}\bigg)\bigg].
		\end{align}
	\end{widetext}
	These can then be eliminated by
	using Eq.~\eqref{prdot}, which, after an expansion in $\dot{p}_r$ (i.e., in eccentricity) gives
	\begin{align}
		p_\varphi  &=  
		%\sqrt{\frac{\dot{p}_r+u^2}{u^3}} = 
		\frac{1}{\sqrt{u}}+\frac{\dot{p}_r}{2 u^{5/2}}-\frac{\dot{p}_r^2}{8 u^{9/2}}
		+\frac{\dot{p}_r^3}{16 u^{13/2}}-\frac{5 \dot{p}_r^4}{128 u^{17/2}}\nonumber\\
		&+\frac{7 \dot{p}_r^5}{256 u^{21/2}}
		-\frac{21 \dot{p}_r^6}{1024 u^{25/2}} + \mathcal{O}(\dot{p}_r^7),
	\end{align}
	which yields
	\begin{align}
		\frac{1}{\sqrt{u}}  &=  p_\varphi \bigg(1-\frac{\dot{p}_r}{2 u^2}+\frac{3 \dot{p}_r^2}{8 u^4}-\frac{5 \dot{p}_r^3}{16 u^6}\nonumber\\
		&+\frac{35 \dot{p}_r^4}{128 u^8}
		-\frac{63 \dot{p}_r^5}{256 u^{10}}+\frac{231 \dot{p}_r^6}{1024 u^{12}}\bigg) + \mathcal{O}(\dot{p}_r^7).
	\end{align}
	%	
	%The latter relation is thus used on every half-integer power of $u$ in the expression obtained at the previous step.
	Once this is inserted into Eq.~\eqref{eq:H22u}, one precisely obtains Eq.~(102) of \cite{Khalil:2021txt} which we 
	rewrite here for completeness in our notation
	\begin{widetext}
		\begin{align}
			\label{eq:102K}
			\hat{H}_{22}^{\rm tail} &= \frac{2 \pi}{c^3} \bigg[p_\varphi u^3+\frac{1}{4} i p_r u^2-\bigg(\frac{7 \dot{p}_r^3p_\varphi}{96 u^3}+\frac{7 i p_r \dot{p}_r^2}{32 u^2}-\frac{7}{32} p_r^2 \dot{p}_rp_\varphi-\frac{7}{96} i p_r^3 u\bigg)\cr
			&+\bigg(\frac{3 \dot{p}_r^4p_\varphi}{32 u^5}+\frac{i p_r \dot{p}_r^3}{12 u^4}-\frac{p_r^2 \dot{p}_r^2p_\varphi}{8 u^2}-\frac{i p_r^3 \dot{p}_r}{96 u}+\frac{1}{48} p_r^4p_\varphi u\bigg)-\bigg(\frac{173 \dot{p}_r^5p_\varphi}{1920 u^7}-\frac{i p_r \dot{p}_r^4}{768 u^6}+\frac{p_r^2 \dot{p}_r^3p_\varphi}{192 u^4}+\frac{49 i p_r^3 \dot{p}_r^2}{384 u^3}-\frac{31 p_r^4 \dot{p}_rp_\varphi}{384 u}-\frac{89 i p_r^5}{3840}\bigg)\cr
			&+\bigg(\frac{97 \dot{p}_r^6p_\varphi}{1152 u^9}-\frac{i p_r \dot{p}_r^5}{64 u^8}+\frac{p_r^2 \dot{p}_r^4p_\varphi}{16 u^6}+\frac{137 i p_r^3 \dot{p}_r^3}{1152 u^5}-\frac{47 p_r^4 \dot{p}_r^2p_\varphi}{384 u^3}-\frac{23 i p_r^5 \dot{p}_r}{640 u^2}+\frac{p_r^6p_\varphi}{96}\bigg)\bigg]
		\end{align}
	\end{widetext}
	Note that this expression, once interpreted within the EOB framework, is ambiguous, since here $\dot{p}_r$ actually only refers to
	the time derivative of the {\it Newtonian} radial momentum obtained from the Newtonian equations of motion. Although there are no 
	strong arguments that may prevent us to promote it to the derivative of the {\it relativistic} radial momentum as defined within the 
	resummed EOB dynamics, we prefer to simplify the logic and have an expression that {\it avoids} $\dot{p}_r$, and 
	only uses $(u,p_r,p_\varphi)$. We thus use again Eq.~\eqref{prdot} to transform Eq.~\eqref{eq:102K} as 
	%Instead of using $\dot{p}_r$, we prefer to express the final result in terms of $p_\varphi$ using again Eq.~\eqref{prdot}. This would in fact avoid possible issues when comparing our analytical results to NR simulations, due to the fact that here with $\dot{p}_r$ we only mean its Newtonian expression. {\mo {Please, can somebody write this more precisely?}}\anp{If we don't have any strong argument here we could just say that we want to express the tails in the same variables we have in the instantaneous part, where we use $p_\varphi$}.
	%Thus, the final expression we obtain in terms of the variables $(u, p_r, p_\varphi)$ reads
	%Eq.~\eqref{tailH22KP} becomes 
	%
	\begin{widetext}
		%	\begin{align}
			%	\label{tailH22}
			%\hat{H}^{\text{tail}}_{22} &= \frac{\pi}{c^3} \bigg[120 p_r^6 p_\varphi-3 i p_r^5 \left(138 p_\varphi^2 u-227\right)-30 p_r^4 p_\varphi u \left(47 p_\varphi^4 u^2-125 p_\varphi^2 u+70\right) \bigg. \cr
			%
			%&\bigg.+10 i p_r^3 u \left(137 p_\varphi^6 u^3-558 p_\varphi^4 u^2+693 p_\varphi^2 u-188\right)+60 p_r^2 p_\varphi u^2 \left(12 p_\varphi^8 u^4-49 p_\varphi^6 u^3+51 p_\varphi^4 u^2+39 p_\varphi^2 u-53\right) \bigg. \cr
			%
			%
			%&\bigg.-15 i p_r u^2 \left(12 p_\varphi^{10} u^5-61 p_\varphi^8 u^4+60 p_\varphi^6 u^3+234 p_\varphi^4 u^2-464 p_\varphi^2 u+27\right) \bigg. \cr
			%
			%
			%&\bigg.+ 2 p_\varphi u^3 \left(485 p_\varphi^{12} u^6-3429 p_\varphi^{10} u^5+10410 p_\varphi^8 u^4-17470 p_\varphi^6 u^3+16965 p_\varphi^4 u^2-8925 p_\varphi^2 u+7724\right) \bigg].
			%\cr
			% &+ \mathcal{O}(p_r^7)+\mathcal{O}\left(\frac{1}{c^5}\right).
			%
			%\end{align}
		\begin{align}
			\label{tailH22}
			\hat{H}^{\text{tail}}_{22} &= \frac{2 \pi}{c^3}\bigg[\bigg(\frac{1931}{1440}-\frac{595p_\varphi^2 u}{384}+\frac{377p_\varphi^4 u^2}{128}-\frac{1747p_\varphi^6 u^3}{576}+\frac{347p_\varphi^8 u^4}{192}-\frac{381p_\varphi^{10} u^5}{640}+\frac{97p_\varphi^{12} u^6}{1152}\bigg)\cr
			&-i p_r u^2 \bigg(\frac{9}{256}-\frac{29p_\varphi^2 u}{48}+\frac{39p_\varphi^4 u^2}{128}+\frac{5p_\varphi^6 u^3}{64}-\frac{61p_\varphi^8 u^4}{768}+\frac{p_\varphi^{10} u^5}{64}\bigg)\cr
			&-p_r^2 p_\varphi u^2\bigg(\frac{53}{192}-\frac{13p_\varphi^2 u}{64}-\frac{17p_\varphi^4 u^2}{64}+\frac{49p_\varphi^6 u^3}{192}-\frac{p_\varphi^8 u^4}{16}\bigg)-i p_r^3 u \bigg(\frac{47}{288}-\frac{77p_\varphi^2 u}{128}+\frac{31p_\varphi^4 u^2}{64}-\frac{137p_\varphi^6 u^3}{1152}\bigg)\cr
			&-p_r^4p_\varphi u\bigg(\frac{35}{192}-\frac{125p_\varphi^2 u}{384}+\frac{47p_\varphi^4 u^2}{384}\bigg)+i p_r^5 \bigg(\frac{227}{3840}-\frac{23p_\varphi^2 u}{640}\bigg)+\frac{p_r^6p_\varphi}{96}\bigg].
			%\cr
			% &+ \mathcal{O}(p_r^7)+\mathcal{O}\left(\frac{1}{c^5}\right).
			%
		\end{align}
		%Here the original expansion in eccentricity translates into an expansion in both $p_r$ and $\dot{p}_r$, with the latter related to $p_\varphi$ through the Newtonian equation of motion
		%\begin{equation}
			%\dot{p}_r = u^2 (p_\varphi^2 u -1).
			%\end{equation} 
		%
		
		%More details on the transition from Eq.~\eqref{tailH22KP} to Eq.~\eqref{tailH22} are given in Appendix~\ref{App:Tail_in_pr_u_pphi}. \\
	\end{widetext}
	In this way, the tail contribution is written with the same variables 
	that we use for the instantaneous part.
	Moreover, leaving $p_r$ and $\dot{p}_r$ in the expressions leads 
	to two issues: (i) the waveform becomes progressively unreliable at 
	merger since $p_r$ may become singular (although this is easily solved
	by replacing $p_r$ with $p_{r_*}$, the momentum conjugate to some, suitably defined,
	Regge-Wheeler tortoise coordinate~\cite{Damour:2007xr});
	(ii) an expression that includes $\dot{p}_r$ is evidently less efficient from the computational point of view.
	Finally, note that the absence of logarithmic terms in Eq.~\eqref{tailH22} is due 
	to a dedicated phase redefinition performed already at the level of Eq.~\eqref{tailH22KP} 
	that reabsorbs these terms along with the gauge parameter (see Sec.~III C 
	of \cite{Hinderer:2017jcs} for further details).

	%
	%==========================================================
	%
	\subsection{Newton-factorized modes and 2PN residual noncircular corrections ($m \neq 0 $)}
	\label{Sec:NewFactors}
	The PN expanded waveform needs to be factorized and resummed in order to improve its strong-field
	behavior and to incorporate its information within state-of-the-art EOB models. The factorization (and resummation)
	scheme for the circular waveform was introduced long ago~\cite{Damour:2008gu} and it has been progressively 
	improved with successive layers of sophistication~\cite{Nagar:2016ayt,Messina:2018ghh,Nagar:2019wrt,Nagar:2020pcj} 
	to achieve the best possible strong-field robustness and accuracy.
	To start with, Ref.~\cite{Damour:2008gu} proposed to factor out the PN-expanded multipolar waveform for 
	spin-aligned binaries as
	\be
	\label{eq:hlm_fact_l}
	h_\lm = h_\lm^{(N,\epsilon)}\hat{h}_\lm ,
	\ee
	where $h_\lm^{(N,\epsilon)}$ indicates the Newtonian contribution, obtained by taking the $\ell$-th time derivative 
	of the mass or current multipole moments~\cite{Thorne:1980ru}, while $\hat{h}_\lm$ indicates the post-Newtonian
	corrections. In Eq.~\eqref{eq:hlm_fact_l}, $\epsilon=0,1$ depending on whether $\ell+m$ is even or odd. 
	Since Ref.~\cite{Damour:2008gu} was dealing with circularized binaries, $h_\lm^{(N,\epsilon)}$ was simplified in the circular approximation, while the $\hat{h}_\lm$ was then additionally factorized and resummed in various
	ways, allowing to obtain a remarkable level of agreement with the numerical results concerning very different binaries
	on quasi-circular orbits (from large mass ratios to comparable mass black holes).
	Ref.s~\cite{Chiaramello:2020ehz,Nagar:2020xsk,Nagar:2021gss} pointed out that an easy, and surprisingly
	accurate, way of generalizing the EOB resummed quasi-circular waveform to eccentric inspirals is to replace the
	quasi-circular Newtonian prefactor (for each multipole) with its general counterpart valid along generic orbits, i.e.~without neglecting the time derivatives of $r$ and $\Omega$. The validity of this prescription has then been tested
	extensively (including higher modes) by Ref.~\cite{Albanesi:2021rby} in the test-particle limit 
	and in Refs.~\cite{Chiaramello:2020ehz, Nagar:2021gss,Nagar:2021xnh} for comparable mass binaries, via 
	detailed comparisons with the available NR simulations.
	To incorporate the 2PN results computed in this paper in such a factorization scheme, Eq.~\eqref{eq:hlm_fact_l}
	is generalized to explicitly separate circular and noncircular contributions, that is 
	\be
	\label{eq:hlm_fact}
	h_\lm = h_\lm^{(N,\epsilon)_{\rm c}}\hat{h}_\lm^{(N,\epsilon)_{\rm nc}}\hat{h}_\lm^{\rm c} \hat{h}^{{\rm nc}}_\lm,
	\ee
	where we have
	\begin{itemize}
		\item[(i)] $h_\lm^{(N,\epsilon)_{\rm c}}$: the Newtonian circular factor. 
		\item[(ii)]$\hat{h}_\lm^{(N,\epsilon)_{\rm nc}}$: the Newtonian {\it noncircular} correction, 
		which is simply obtained from the corresponding mass and current Newtonian multipole moments,
		and 
		%that is explicitly detailed 
		whose expression is given in detail in Ref.~\cite{Albanesi:2021rby}.
		\item[(iii)] $\hat{h}_\lm^{\rm c}$: the PN residual circular correction, which is then resummed using 
		several analytical prescriptions which we will briefly recall below. We use here the $\hat{h}_\lm$'s
		functions entering the last avatar of the \TEOBResumS{} waveform model~\cite{Nagar:2020pcj,Riemenschneider:2021ppj}.
		\item[(iv)] $\hat{h}^{{\rm nc}}_\lm$: the PN residual {\it noncircular} correction, which are computed here
		for the first time.
	\end{itemize}
	
	Ref.s~\cite{Chiaramello:2020ehz,Nagar:2020xsk,Nagar:2021gss,Albanesi:2021rby} adopted 
	the waveform factorized scheme of Eq.~\eqref{eq:hlm_fact} for the waveform, although imposing 
	$\hat{h}^{{\rm nc}}_\lm=1$ for simplicity\footnote{Notice that the waveform factorization of Ref.~\cite{Khalil:2021txt}
		is different from the one proposed here, since it only factorizes the circular parts $h_\lm^{(N,\epsilon)}\hat{h}_\lm$
		while the parts that belong to the Newtonian noncircular factor are kept in PN-expanded form.}.
	Our aim here is to explicitly determine  the $\hat{h}^{\rm nc}_\lm$ noncircular correction factors at 2PN 
	accuracy by factorizing the PN-expanded multipoles  obtained in the previous section.
	The procedure is rather straightforward, although it needs the 2PN-expanded EOB equations of motion
	in order to recast the Newton-normalized relativistic correction into a meaningful residual PN-expansion.
	The generic Newtonian prefactors are all listed in Ref.~\cite{Albanesi:2021rby}, so there is 
	no need to repeat here their derivation. Our factorization procedure consists in three steps: 
	(i) starting from $h_\lm$, we factor out the generic Newtonian prefactor and (ii) we replace 
	the derivatives with the 2PN expanded equations of motion and expand the residual at 2PN, 
	then (iii) we factor out the circular part in order to single out the residual noncircular correction $\hat{h}^{\rm nc}_\lm$. Indeed, for this prescription to work properly one has to be sure that no spurious poles are introduced by the factorization. Even though this is not the case for the majority of the spherical modes, all the modes with $m=0$ happen to show this kind of problematic behavior, since their Newtonian factor is entirely noncircular and thus goes to zero in the circular limit. We defer to the next section the discussion of a possible alternative prescription for these modes.
	
	Focusing here on the $(2,2)$ mode to exemplify the general factorization procedure, we have 
	\begin{align}
		\label{h22N}
		h^{(N,0)_{\rm c}}_{22} &=-8\sqrt{\dfrac{\pi}{5}} (r \Omega)^2 e^{-2i\varphi}, \\
		h^{(N,0)_{\text{nc}}}_{22} &= 1 - \frac{\Ddot{r}}{2 r \Omega^2} - \frac{\Dot{r}^2}{2 (r\Omega)^2} + i \bigg( \frac{2\Dot{r}}{r \Omega} + \frac{\Dot{\Omega}}{2 \Omega^2} \bigg). 
		\label{eq:ncN22}
	\end{align}
	The factorization of the general Newtonian factor $h_\lm^{(N,\epsilon)_{\rm c}}\hat{h}_\lm^{(N,\epsilon)_{\rm nc}}$
	yields
	%Newtonian factor in front, so that the resulting expression is equivalent to the original one once the 
	%time-derivatives of $r$ and $\Omega$ are replaced with the equations of motion at 2PN. 
	%We have thus $h_{\ell m} = h^{(N,\epsilon)}_\lm \hat{h}^{(\eps)}_\lm$ with the total PN factor defined as
	\begin{equation}
		\label{NNH}
		\hat{h}_\lm^{(\eps)} \equiv  T_{\rm 2PN} \left[\dfrac{h_\lm}{    \left( h_\lm^{(N,\epsilon)_{\rm c}}\hat{h}_\lm^{(N,\epsilon)_{\rm nc}}\right)_{\text{EOMs}}} \right] ,
	\end{equation}
	where the operator $T_{\rm 2PN}$ indicates that the expression within the square brackets 
	is expanded at 2PN order.
	The subscript ``EOMs" indicates that in the Newtonian term we replace the time-derivatives 
	with the corresponding EOB equations of 
	motion\footnote{The equations of motion for the first time-derivatives can be computed directly 
		from the EOB Hamiltonian; see Appendix B of Ref.~\cite{Bini:2012ji} for their explicit expression 
		at 2PN order. From there the computation of the higher order derivatives follows from 
		a straightforward iteration.}.
	At this point the expression \eqref{NNH} is a rational function of $(u,p_r,p_\varphi)$ 
	of the type  ``1 + PN terms"\footnote{ We remind the reader that, as explained above, 
		this factorization cannot be used for the $m=0$ modes since they do not have
		a circular part to factorize. We will discuss this particular case 
		in Sec.~\ref{Sec:Alt_PN_corrections}.}.  
	
	Now that the PN factor is singled out we can also factor out its circular component
	\begin{equation}
		\hat{h}^{(\eps)_{\text{c}}}_\lm \equiv \lim_{p_r \to 0}  \hat{h}^{(\eps)}_\lm,
	\end{equation}
	in order to obtain the total noncircular PN factor we are interested in, that is
	\begin{equation}
		\hat{h}^{\rm nc}_\lm \equiv T_{\rm 2PN}\left[\dfrac{ \hat{h}^{(\eps)}_\lm}{  \hat{h}^{(\eps)_{\text{c}}}_\lm} \right] ,
	\end{equation}
	where again $T_{\rm 2PN}$ indicates that we are retaining all combinations of powers of $(u, p_r,p_\varphi)$ up to 2PN accuracy.
	
	In this way the noncircular PN factor $ \hat{h}^{\text{nc}}_\lm$ 
	amounts to a collection of all the relativistic noncircular contributions not yet included into the model. 
	%Focusing on the latter, 
	Moreover, we can split 
	%it 
	$\hat{h}^{{\text{nc}}}_\lm $ into a tail factor and an instantaneous 
	factor\footnote{This splitting can be performed with ease since tail and instantaneous contributions are consistently well separated by the PN ordering}, namely
	\be
	\hat{h}^{{\text{nc}}}_\lm  = \hat{h}^{\text{nc}_{\rm tail}}_\lm \, \hat{h}^{\text{nc}_{\rm inst}}_\lm,
	\ee
	and eventually trade the radial momentum $p_r$ for $p_{r_*} \equiv (A/B)^{1/2} \, p_r$,
	with $(A/B)^{1/2}$ truncated at 2PN accuracy.
	In addition, to simplify the structure of the analytical expressions we are using, we expand 
	each of the new factors in $p_{r_*}$ up to the fourth order\footnote{We have verified that this choice gives an 
		excellent approximation to the full expressions for all cases considered.}.
	Focusing now on the $\ell=m=2$ dominant mode, the so obtained tail factor is
	\begin{widetext}
		\begin{align}
			\label{eq:tail22}
			\hat{h}_{22}^{\rm nc_{tail}}=1 + \dfrac{1}{c^3}\dfrac{\pi}{(p_\varphi^2 u + 1)^2}\bigg[-i \left(\dfrac{9p_{r_*} u }{64}  \hat{t}^{22}_{ p_{r_*} } +\dfrac{457}{456}\dfrac{p_{r_*}^3}{(p_\varphi^2 u + 1)^2} \hat{t}^{22}_{p_{r_*}^3}\right)
			+ \dfrac{5729}{1440}\dfrac{ p_{r_*}^2 p_\varphi u}{(p_\varphi^2 u + 1)} \hat{t}^{22}_{p_{r_*}^2 }  + \dfrac{133}{80}\dfrac{ p_{r_*}^4 p_\varphi}{(p_\varphi^2 u + 1)^3} 
			\hat{t}^{22}_{p_{r_*}^4 }\bigg]
		\end{align}
	\end{widetext}
	where $(\hat{t}^{22}_{p_{r_*}}, \hat{t}^{22}_{p^3_{r_*}},\hat{t}^{22}_{p_{r_*}^2},\hat{t}^{22}_{p_{r_*}^4})$ 
	are the following polynomials in $y\equiv p_\varphi^2 u$ (with alternate signs) 
	\begin{align}
		\label{eq:tp1}
		\hat{t}^{22}_{p_{r_*}} &= 1 + \dfrac{24341}{405}y-\dfrac{290}{3}y^2 + \dfrac{1606}{9}y^3 - \dfrac{13979}{81}y^4\nonumber\\
		& + 101 y^5 - \dfrac{1504}{45}y^6 + \dfrac{388}{81}y^7,\\
		\label{eq:tp2}
		\hat{t}^{22}_{p_{r_*}^3} &= 1+\dfrac{46372}{2285}y - \dfrac{134587}{2285}y^2+\dfrac{45492}{457}y^3\nonumber\\
		&-\dfrac{54397}{457}y^4+ \dfrac{43924}{457}y^5 - \dfrac{112697}{2285}y^6\nonumber\\
		&+\dfrac{32212}{2285}y^7-\dfrac{776}{457}y^8,\\
		\label{eq:tp3}
		\hat{t}^{22}_{p_{r_*}^2} & = 1-\dfrac{27552}{5729}y+\dfrac{2910}{337}y^2-\dfrac{70015}{5729}y^3 \nonumber\\
		& +\dfrac{61785}{5729}y^4 - \dfrac{34674}{5729}y^5 + \dfrac{10952}{5729}y^6 -\dfrac{1455}{5729}y^7,\\
		\label{eq:tp4}
		\hat{t}^{22}_{p_{r_*}^4} & = 1 - \dfrac{35260}{1197}y +\dfrac{78500}{1197}y^2-\dfrac{34825}{342}y^3 \nonumber\\
		& + \dfrac{265975}{2394}y^4 - \dfrac{96727}{1107}y^5 + \dfrac{6305}{171}y^6 \nonumber\\
		&- \dfrac{3245}{342}y^7 + \dfrac{2425}{2394}y^8 .
	\end{align}

	The instantaneous factor is conveniently separated in amplitude and phase 
	\be
	\label{hinstnc}
	h^{\rm nc_{inst}}_{22}=f_{22}^{\rm nc_{inst}}e^{\rm i \delta_{22}^{\rm nc_{inst}}},
	\ee
	which respectively are given by
	\begin{widetext}
		\begin{align}
			\label{APNnc22}
			f^{\rm nc_{inst}}_{22} & = 1 - \frac{p_{r_*}^2}{c^2  \left(p_\varphi^2 u+1\right)^3} \bigg[ \left(\frac{1}{14}-\frac{31 \nu}{14} \right)\hat{f}^{22}_{1^{\rm 1PN}}+\frac{p_{r_*}^2}{u \left(p_\varphi^2 u+1\right)^2}\left(\frac{5}{7}-\frac{8 \nu}{7} \right)\hat{f}^{22}_{p_{r_*}^{\rm 1PN}}
			\bigg]\nonumber\\
			&+\frac{p_{r_*}^2}{ c^4\left(p_\varphi^2 u+1\right)^4} \bigg[  u  \left(\frac{65}{252}+\frac{211 \nu}{126} +\frac{139 \nu ^2}{63}\right)\hat{f}^{22}_{u^{\rm 2PN}} +\frac{p_{r_*}^2}{\left(p_\varphi^2 u+1\right)^2} \left(\frac{1613}{504}-\frac{1567 \nu }{504}-\frac{71 \nu ^2}{72}\right)\hat{f}^{22}_{p_{r_*}^{\rm 1PN}}
			\bigg],\\
			\label{PhPNnc22}
			\delta_{22}^{\rm nc_{inst}}&= \dfrac{1}{c^2}\dfrac{p_{r_*}p_\varphi}{(p_\varphi^2 u +1)^2}\bigg[u\left(\dfrac{25}{21}-\dfrac{18}{7}\nu\right)\hat{\delta}^{22}_{u^{\rm 1PN}} + \dfrac{p_{r_*}^2}{(p_\varphi^2 u +1)^2}\left(\dfrac{55}{21}-\dfrac{34\nu}{7}\right)\hat{\delta}^{22}_{p_{r_*}^{\rm 1PN}}\bigg]\nonumber\\
			&+\dfrac{1}{c^4}\dfrac{p_{r_*}p_\varphi u}{(p_\varphi^2 u +1)^3}\bigg[u\left(\dfrac{7}{27}-\dfrac{416}{189}\nu-\dfrac{652}{189}\nu^2\right)\hat{\delta}^{22}_{u^{\rm 2PN}} + \dfrac{p_{r_*}^2}{(p_\varphi^2 u +1)^2}\left(\dfrac{20945}{2646}-\dfrac{17321}{1323}\nu+\dfrac{134}{1323}\nu^2\right)\hat{\delta}^{22}_{p_{r_*}^{\rm 2PN}}
			\bigg].
		\end{align}
	\end{widetext}
	Similarly to the tail factor, the functions $(\hat{f}_i,\hat{\delta}_i)$ are polynomials in $y$
	whose explicit expressions are
	\begin{align}
		\hat{f}^{22}_{1^{\rm 1PN}} = 1&-\frac{7 (1+3 \nu )}{1-31 \nu }y-\frac{(451-177 \nu )}{3-93 \nu }y^2 \nonumber \\ & -\frac{3  (3+5 \nu )}{1-31 \nu }y^3,  \\
		\hat{f}^{22}_{p_{r_*}^{\rm 1PN}}  = 1&-\frac{65-216 \nu }{3 (5-8 \nu )}y-\frac{5 (115-72 \nu ) }{3 (5-8 \nu)}y^2  \nonumber \\ 
		& +\frac{305-264 \nu }{3 (5-8 \nu )}y^3, \\
		\hat{f}^{22}_{u^{\rm 2PN}} =  1&-\frac{44767+28618 \nu -7276 \nu ^2}{42 \left(65+422 \nu +556 \nu ^2\right)}y \cr
		&+\frac{ 132507-87244 \nu -29672 \nu ^2}{14 \left(65+422 \nu +556 \nu ^2\right)}y^2\cr
		&-\frac{ 134789+27920 \nu +9472 \nu ^2}{42 \left(65+422 \nu +556 \nu ^2\right)}y^3\cr
		&-\frac{3  \left(637-1448 \nu +512 \nu ^2\right)}{2 \left(65+422 \nu +556 \nu ^2\right)}y^4 \cr
		&-\frac{3 \left(418+355 \nu +710 \nu ^2\right)}{7 \left(65+422 \nu +556 \nu ^2\right)} y^5, \\
		\hat{f}^{22}_{p_{r_*}^{\rm 2PN}}= 1&-\frac{ 16399-169738 \nu -9902 \nu ^2}{7 \left(1613-1567 \nu -497 \nu ^2\right)}y\cr
		&-\frac{256835-145513 \nu +7405 \nu ^2}{4839-4701 \nu -1491 \nu ^2}y^2  \cr
		&+\frac{4 \left(292018-194489 \nu -76057 \nu ^2\right)}{21 \left(1613-1567 \nu -497 \nu ^2\right)}y^3 \cr
		&-\frac {449937+29671 \nu -80119 \nu ^2}{7 \left(1613-1567 \nu -497 \nu ^2\right)}y^4\cr
		&-\frac{ 34361-33826 \nu -3518 \nu ^2}{7 \left(1613-1567 \nu -497 \nu ^2\right)}y^5\cr
		&+\frac{15  \left(5-13 \nu +\nu ^2\right)}{1613-1567 \nu -497 \nu ^2}y^6, \nonumber \\
		\hat{\delta}^{22}_{u^{\rm 1PN}} =  1 & + \frac{125-102\nu}{25-54 \nu} y,  \\
		\hat{\delta}^{22}_{p_{r_*}^{\rm 1PN}} = 1 & + \frac{42 \left( 5-2 \nu \right)}{55-102 \nu} y 
		- \frac{35 \left( 7-6 \nu \right)}{55-102 \nu} y^2, \\
		\hat{\delta}^{22}_{u^{\rm 2PN}}  =  1 & + \frac{39761-20950 \nu -21236 \nu ^2}{14 \left(49-416 \nu-652 \nu ^2 \right)} y\nonumber\\
		& -\frac{3 \left(3709-2556 \nu-408 \nu ^2\right)}{14 \left(49-416 \nu-652 \nu ^2 \right)}y^2\nonumber\\
		&-\frac{3 \left( 767-2551 \nu -1070 \nu ^2\right)}{7 \left(49-416 \nu -652 \nu ^2\right)}y^3 \ ,\\
		\hat{\delta}^{22}_{p_{r_*}^{\rm 2PN}} = 1 &+ \frac{66624-84120 \nu+300789 \nu ^2}{41890-69284 \nu +536 \nu ^2} y \nonumber\\
		& -\frac{ 292601-144464 \nu ^2-298528 \nu}{ 41890-69284 \nu+536 \nu ^2} y^2 \nonumber\\
		& +\frac{7 \left( 28217+17672 \nu-1664 \nu ^2\right)}{41890 -69284 \nu+536 \nu ^2 }y^3\nonumber\\
		& + \frac{3 \left(6473-23284 \nu+6856 \nu ^2\right)}{41890-69284 \nu+536 \nu ^2}y^4 \  \label{eq:hatdelta22_4}.
	\end{align}
	
	We anticipate here that in Sec.~\ref{sec:resum} we will argue that the polynomials 
	$\hat{t}^\lm_{p_{r_*}^n}(y)$ of the tail factor need a proper resummation in order for the 2PN corrections to 
	have a robust behavior in strong field.
	%==========================================
	% Fig.01: insplunge: expanded tail
	%==========================================
	\begin{figure*}
		\center
		\includegraphics[width=0.31\textwidth]{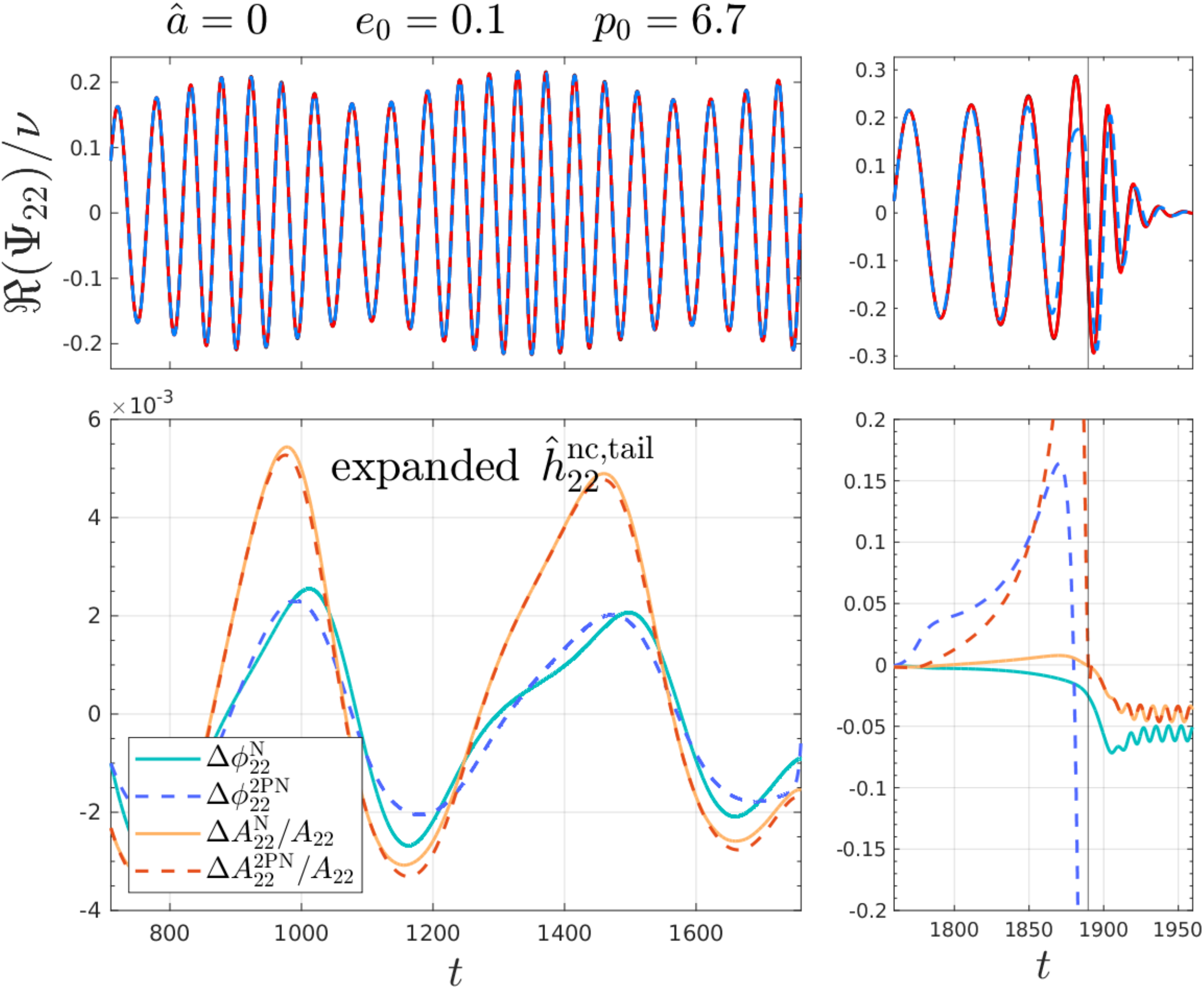}
		\hspace{0.2cm}
		\includegraphics[width=0.31\textwidth]{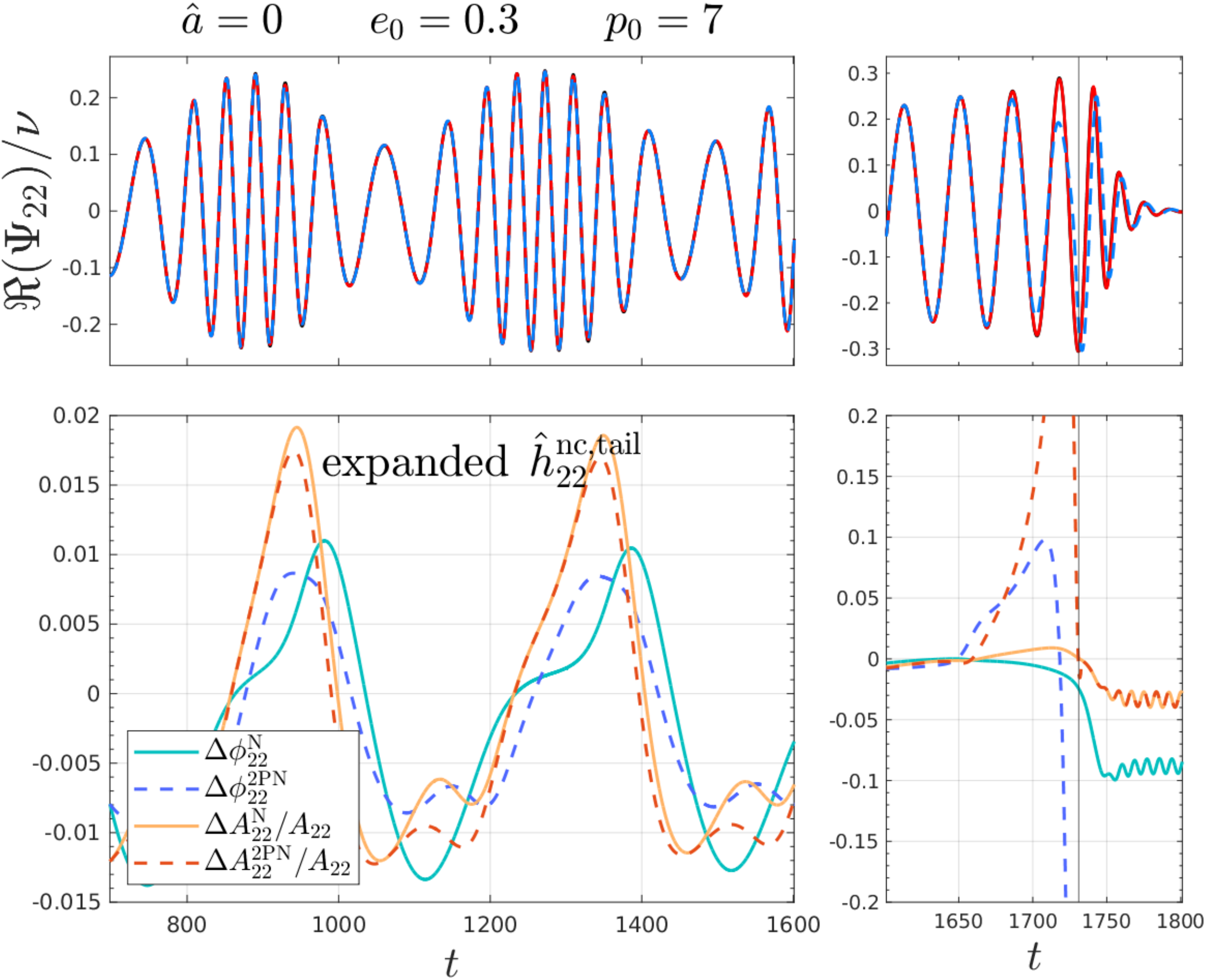}
		\hspace{0.2cm}
		\includegraphics[width=0.31\textwidth]{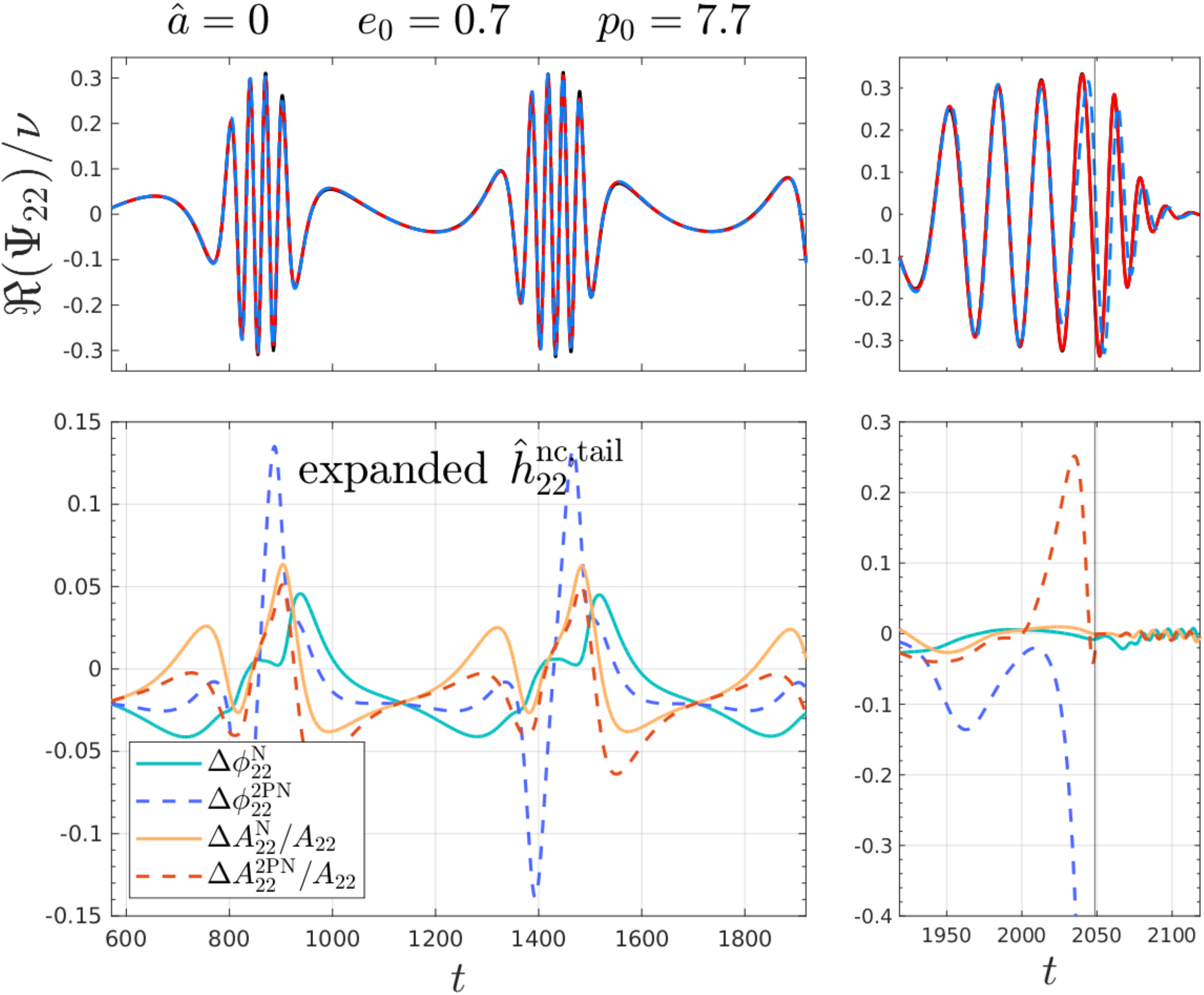} \\
		\vspace{0.3cm}
		\includegraphics[width=0.31\textwidth]{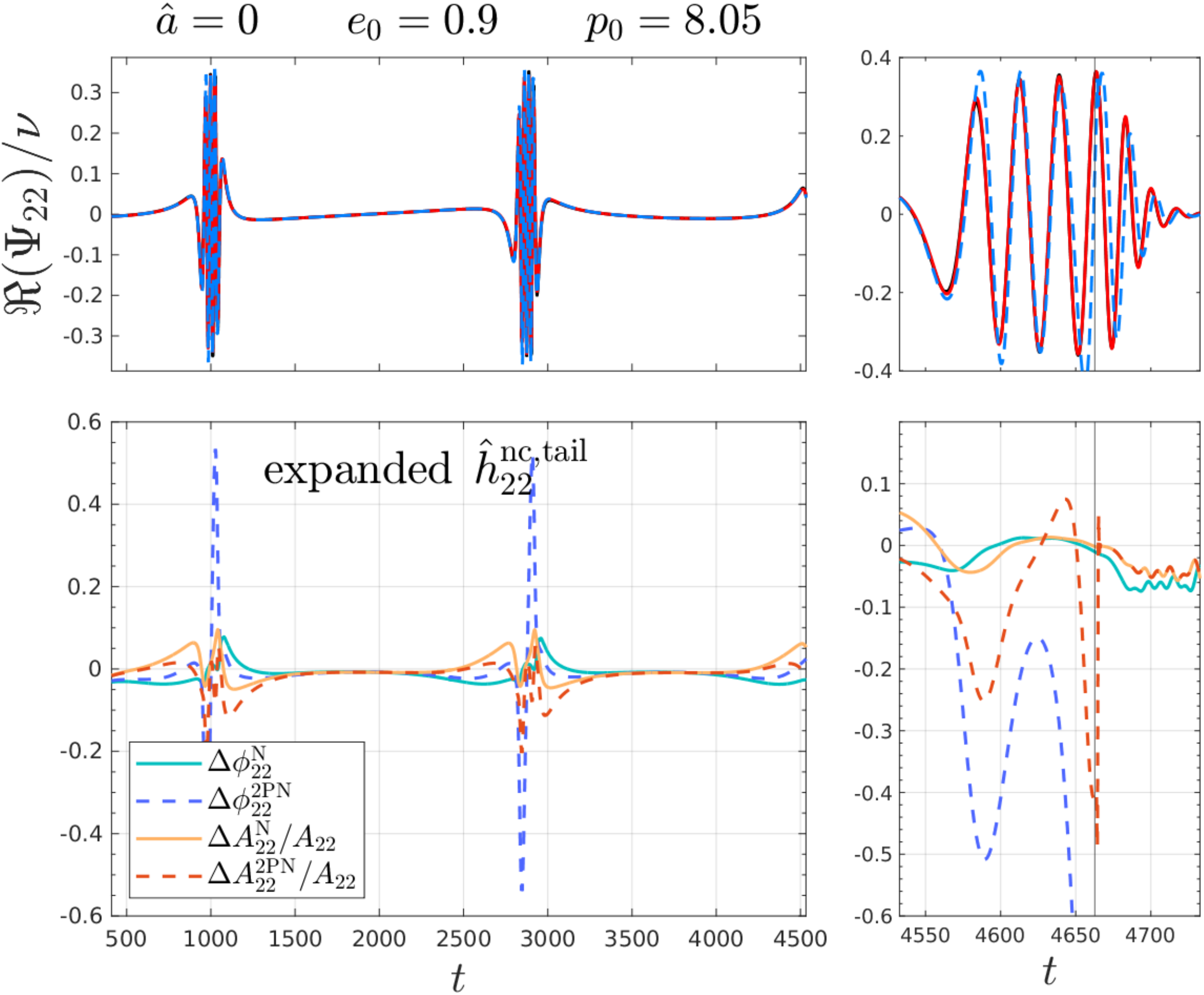}
		\hspace{0.2cm}
		\includegraphics[width=0.31\textwidth]{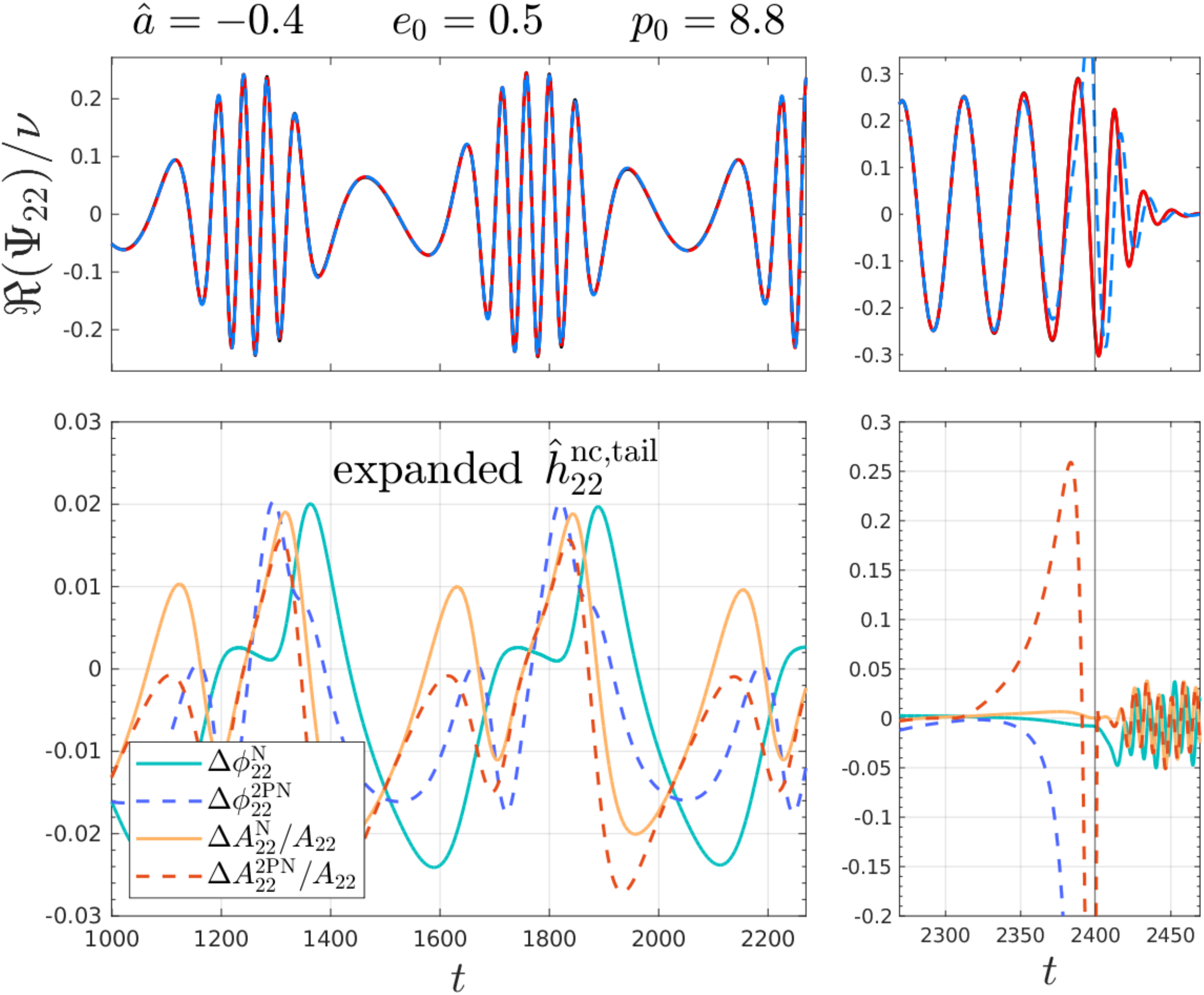}
		\hspace{0.2cm}
		\includegraphics[width=0.31\textwidth]{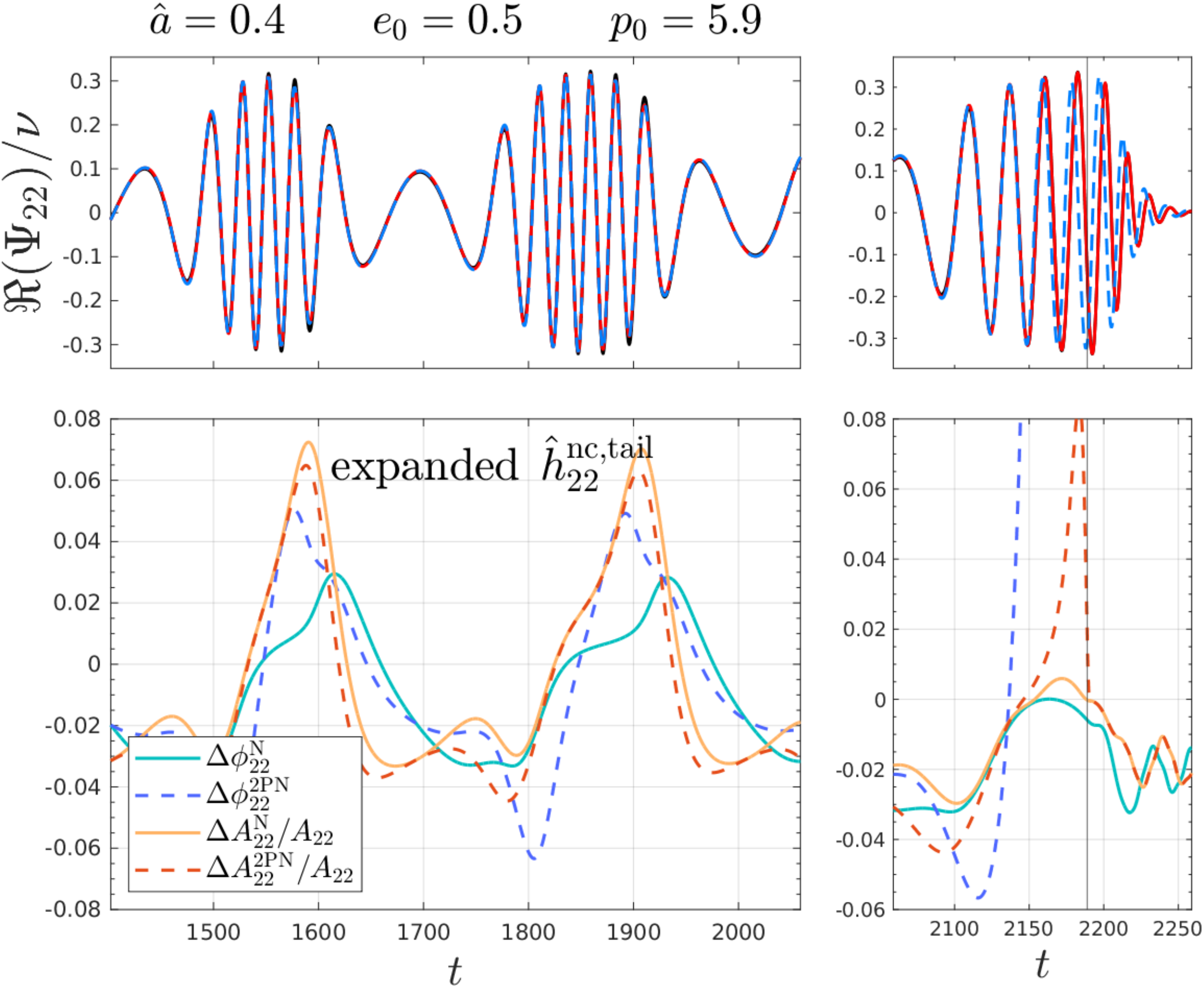}
		\caption{\label{fig:testmass_inspl_expanded_tail} 
			Comparisons with {\it nonresummed} noncircular tail factor: 
			comparing analytical and numerical $\ell=m=2$ waveforms for the transition from 
			inspiral to plunge of a test-particle 
			on a Kerr black hole with spin parameter $\hat{a}$. We consider different
			configurations with $\nu = 10^{-3}$.
			Each panel displays the numerical waveform (black, indistinguishable) and two EOB 
			waveforms: (i) the solid-red one with noncircular information only in the Newtonian prefactor,
			(ii) the dashed-blue one with noncircular 2PN corrections with the 
			{\it nonresummed} tail $\hat{h}^{{\rm nc_{tail}}}_{22}$ of Eq.~\eqref{eq:tail22}.
			The bottom panel shows both the phase differences and the relative amplitude 
			differences with respect to the numerical waveform. We use
			dashed lines for the differences corresponding to the wave with 2PN corrections.
			The vertical line marks the merger-time, corresponding to the peak of the numerical amplitude.}
	\end{figure*}
	%===========================
	%
	%==========================================================
	%
	\subsection{2PN noncircular corrections for the $m=0$ modes}
	\label{Sec:Alt_PN_corrections}
	As mentioned in the previous section, the factorization scheme presented therein 
	cannot be applied successfully to the spherical modes with $m=0$, because of the 
	vanishing of their Newtonian factor in the circular limit. Nevertheless, we can 
	still build a model for them that is well behaved and incorporates the 2PN 
	noncircular information of the starting PN-expanded waves. In particular the 
	alternative to Eq.~\eqref{eq:hlm_fact} we propose in this case is
	\begin{equation}
		h_{\ell0} = \hat{S}_{\rm eff} \left(h_{\ell0}^{(N,\epsilon)}  + \doublehat{h}_{\ell0}\right),
	\end{equation}
	where the PN correction, which indeed is fully noncircular, is given by
	\begin{equation}
		\doublehat{h}_{\ell0} = T_{\rm 2PN} \left[\left(\dfrac{h_{\ell0}-h_{\ell0}^{(N,\epsilon)}}{\hat{S}_{\rm eff}}\right)_{\rm EOMs}\right].
	\end{equation}
	Similarly to the other prescription, this quantity comes out naturally split 
	into an instantaneous and a tail part,
	\begin{equation}
		\doublehat{h}_{\ell0} = 	\doublehat{h}_{\ell0}^{\rm tail}+\doublehat{h}_{\ell0}^{\rm inst}. 
	\end{equation}
	Here we prefer to express the PN corrections using $(r,p_{r_*},\dot{p}_{r_*})$, without writing 
	$\dot{p}_{r_*}$ in terms of $(r, p_{r_*}, p_\varphi)$. 
	The reason is that in this case writing the corrections in $p_\varphi$ leads
	to terms that do not vanish in the circular limit since they are not proportional to powers
	of $p_{r_*}$\footnote{Specifically there are contributions that vanish in the circular limit only when $p_\varphi$ is replaced with its corresponding quasi-circular PN expansion in terms of $u$.} . The corresponding PN corrections for the mode $(2,0)$ read 
	\begin{widetext}
		\begin{align}
			\doublehat{h}_{20}^{\rm tail} &= - \frac{\pi }{960 u^{10} c^3} \sqrt{\frac{\dot{p}_{r_*}+u^2}{6 u}} \bigg[960 \dot{p}_{r_*} u^{10}+960 \dot{p}_{r_*}^2 u^8+240 \left(-3 \dot{p}_{r_*}^3 u^6+p_{r_*}^2 \dot{p}_{r_*} u^9\right)+80 u^4 \left(7 \dot{p}_{r_*}^4+2 p_{r_*}^4 u^6\right)\cr
			&-5 \left(95 \dot{p}_{r_*}^5 u^2+26 p_{r_*}^2 \dot{p}_{r_*}^3 u^5+11 p_{r_*}^4 \dot{p}_{r_*} u^8\right)+\left(417 \dot{p}_{r_*}^6+110 p_{r_*}^2 \dot{p}_{r_*}^4 u^3-45 p_{r_*}^4 \dot{p}_{r_*}^2 u^6+2 p_{r_*}^6 u^9\right) \bigg]\\
			\doublehat{h}_{20}^{\rm inst} &= \frac{1}{14 \sqrt{6}u^2 c^2}\bigg[\dot{p}_{r_*} u^2 (-19+\nu )+3 \left(\dot{p}_{r_*}^2-p_{r_*}^2 u^3\right) (3+5 \nu )+6 p_{r_*}^2 \dot{p}_{r_*} u (3+5 \nu )+3 p_{r_*}^4 u^2 (3+5 \nu )\bigg] \cr
			&+ \frac{1}{504 \sqrt{6} u^3 c^4} \bigg[\dot{p}_{r_*} u^4 \left(1052-2803 \nu -53 \nu ^2\right)+p_{r_*}^2 u^5 \left(-743+7009 \nu -571 \nu ^2\right) +3 \dot{p}_{r_*}^2 u^2 \left(545-430 \nu +28 \nu ^2\right)\cr
			&-3 \dot{p}_{r_*}^3 \left(79+25 \nu +5 \nu ^2\right)+18 p_{r_*}^2 \dot{p}_{r_*} u^3 \left(81+404 \nu +65 \nu ^2\right) -9 p_{r_*}^2 \dot{p}_{r_*}^2 u \left(115+121 \nu +65 \nu ^2\right)\cr
			&+6 p_{r_*}^4 u^4 \left(79+133 \nu +185 \nu ^2\right)-9 p_{r_*}^4 \dot{p}_{r_*} u^2 \left(151+217 \nu +125 \nu ^2\right)-3 p_{r_*}^6 u^3 \left(187+313 \nu +185 \nu ^2\right)\bigg].
		\end{align}
	\end{widetext}
	Inspired by what we did here, we also rewrote the 2PN noncircular corrections for the modes $m \neq 0$
	(both the instantaneous and the tail part) by replacing $p_\varphi$ by its expression in terms of 
	$(u,p_{r_*},\dot{p}_{r_*})$ using the 2PN-accurate equations of motion. 
	Since this approach lays outside the main logic of the paper, 
	we will discuss it separately in Appendix~\ref{App:2PN_prrdot}.
	
	\section{Waveform validation: eccentric inspirals in the large mass ratio limit}
	\label{sec:testmass}
	
	\begin{table}[t]
		\caption{\label{tab:Teukode_ecc} Numerical eccentric simulations considered in this work.
			We use $\nu=10^{-3}$ to drive the transition from inspiral to merger.  
			For each eccentric simulation we report the spin parameter $\ha$, 
			the initial/final values of eccentricity and semilatus rectum, 
			and the merger time $t_{\rm mrg}$.
			The final values of eccentricity and semilatus rectum are evaluated at $t_{\rm sep}$   
			since they are not defined for later times, 
			where $t_{\rm sep}$ 
			is the time when the semilatus rectum equals the separatrix and the radial turning
			points cease to exist (definitions of $e$, $p$ and separatrix written in terms of radial 
			turning points can be found in Sec.II A of Ref.~\cite{Albanesi:2021rby}).}
		\begin{center}
			\begin{ruledtabular}
				\begin{tabular}{c c c c c c c} 
					$\ha$ & $e_0$ & $p_0$ & $e_{\rm sep}$ & $p_{\rm sep}$ & $t_{\rm sep}$ & $t_{\rm mrg}$  \\
					\hline
					\hline
					$0.0$ & $0.1$ & $6.700$ & $0.107$ & $6.213$ & $1459$ & $1890$ \\ 
					$0.0$ & $0.3$ & $7.000$ & $0.305$ & $6.611$ & $1382$ & $1731$ \\ 
					$0.0$ & $0.7$ & $7.700$ & $0.694$ & $7.388$ & $1916$ & $2049$ \\ 
					$0.0$ & $0.9$ & $8.050$ & $0.891$ & $7.783$ & $4570$ & $4663$ \\ 
					$-0.4$ & $0.5$ & $8.800$ & $0.501$ & $8.426$ & $2182$ & $2387$ \\ 
					$0.4$ & $0.5$ & $5.900$ & $0.490$ & $5.415$ & $2092$ & $2192$ \\ 
				\end{tabular}
			\end{ruledtabular}
		\end{center}
	\end{table}
	
	%=========================================
	% Fig. 02: behavior of p_varphi2 u
	%=========================================
	\begin{figure}[t]
		\center
		\includegraphics[width=0.45\textwidth]{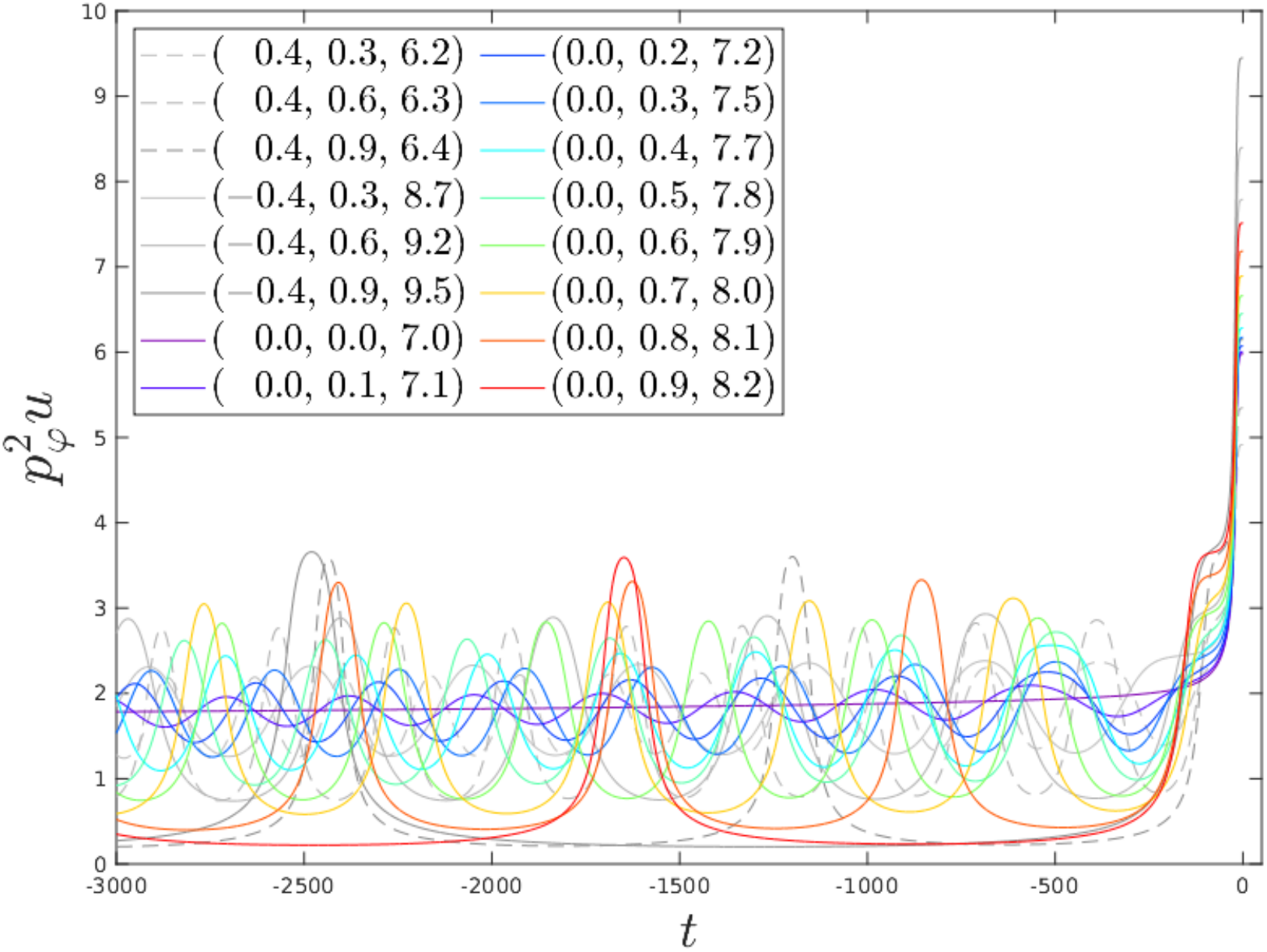}
		\caption{\label{fig:pphi2u} Last part of the time evolution of $p_\varphi^2 u$ 
			for different combinations of $(\hat{a},e_0,p_0)$. 
			Note that during the plunge $p_\varphi^2 u$ can grow up to $\sim 10$. 
			This growth is mostly responsible of the unacceptably large
			analytical/numerical phase disagreement during the plunge, at 2PN-accuracy, 
			seen in Fig.~\ref{fig:testmass_inspl_expanded_tail}.}
	\end{figure}
	%=========================================
	
	%=========================================
	% Fig. 03: resummed tail subseries
	%=========================================
	\begin{figure}[t]
		\center
		\includegraphics[width=0.45\textwidth]{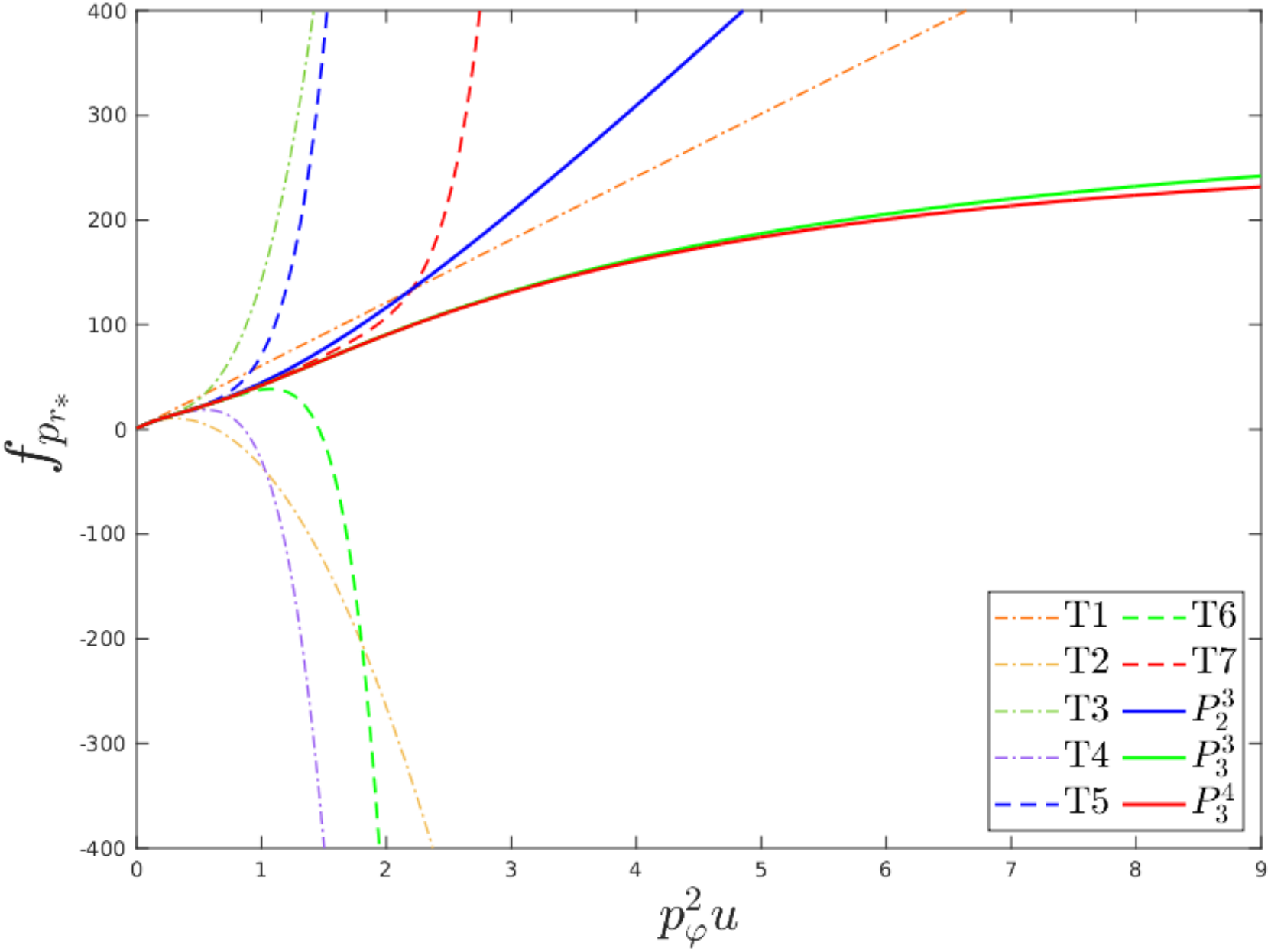}
		\caption{\label{fig:f_prstar_u}
			Behavior of various truncations of the $\hat{t}^{22}_{p_{r_*}}$ polynomial
			of Eq.~\eqref{eq:tp1}. The various truncations of $\hat{t}^{22}_{p_{r_*}}$ oscillate
			and become very large for values of $p^2_\varphi u$ of the order of those reached during the plunge,
			see Fig.~\ref{fig:pphi2u}. A straightforward diagonal Pad\'e approximant tapers the behavior of the polynomials
			in strong field and eventually improves the behavior of the waveform there.}
	\end{figure}
	
	Let us now move to explore the performance of our new factorized waveform,
	starting from the case of elliptic inspirals in the large mass ratio limit. We recall that 
	Ref.~\cite{Albanesi:2021rby} validated the simple Newton-factorized waveform, without 
	the 2PN noncircular correction, in the test-mass limit, i.e. considering 
	the motion of a test-particle around 
	a Kerr black hole. The validation was relying on comparisons between the analytic EOB waveform
	and the numerical solution of the Teukolsky equation, obtained using the 2+1 
	time-domain code {\Teukode}~\cite{Harms:2014dqa}.  
	It was considered either: (i) the geodesic
	motion along elliptic orbits and (ii) the full transition from the eccentric 
	inspiral to merger and ringdown.
	The outcome of that study was that, even without the 2PN correction, the analytic waveform delivers
	a rather accurate approximation of the exact waveform up to mild values of the initial eccentricity,
	both for amplitude and phase (see e.g. Fig.~13 of Ref.~\cite{Albanesi:2021rby}).
	To start with, we use precisely the expressions for ($\hat{h}_{22}^{\rm nc_{tail}},h_{22}^{\rm nc_{inst}})$ 
	given in Eqs.~\eqref{eq:tail22}, \eqref{hinstnc}
	and redo the comparison of Ref.~\cite{Albanesi:2021rby}, where the interested reader can find more 
	technical details regarding the dynamics.  
	The eccentric numerical waveforms used in this work are listed in Table~\ref{tab:Teukode_ecc}.
	We also recall that the quasi-circular part
	of the waveform we use is precisely the same of Ref.~\cite{Albanesi:2021rby}.
	In Fig.~\ref{fig:testmass_inspl_expanded_tail} we report different configurations
	aiming at comprehensively cover the parameter space. The first four panels from left to right refer
	to nonspinning binaries with increasing eccentricity $e_0=(0.1, 0.3, 0.7, 0.9)$, 
	while the last two panels refer to two spinning binaries with $\ha=\pm0.4$ and initial 
	eccentricity $e_0=0.5$. 
	For low eccentricity, up to $e\simeq 0.3$, the 2PN corrections improve the
	phase agreement during the inspiral, but for higher eccentricity the phase of
	the wave with only Newtonian corrections is more accurate.
	Moreover, in all the cases the analytical/numerical agreement visibly 
	deteriorates as one gets closer to plunge and merger, both at the level of the phase and of the amplitude.
	Indeed, careful analysis of the geodesic case highlights that the reliable behavior of 
	the waveform during the eccentric inspiral is related to {\it cancellations} between 
	the tail and instantaneous factors.
	By contrast, the inaccurate behavior of the analytical waveform during the plunge
	is related to the fact that the quantity $p_\varphi^2u$, which appears everywhere in Eq.~\eqref{eq:tail22}, becomes rather large during late plunge up to merger, as shown in Fig.~\ref{fig:pphi2u}. 
	The growth of $p_\varphi^2 u$ makes the eccentric corrections too large with respect to the instantaneous terms and 
	the cancellations mentioned above are no longer possible, leading to 
	the observed loss in accuracy.
	This issue is also responsible for the large phase disagreement near the 
	periastra of configurations with high eccentricity. To cure this behavior we need to 
	implement specific resummation strategies, as we will discuss in the next section.

	\subsection{Resummation of the hereditary residual noncircular factor}
	\label{sec:resum}
	%==========================================
	% Fig.04: insplunge: resummed tail
	%==========================================
	\begin{figure*}
		\center
		\includegraphics[width=0.31\textwidth]{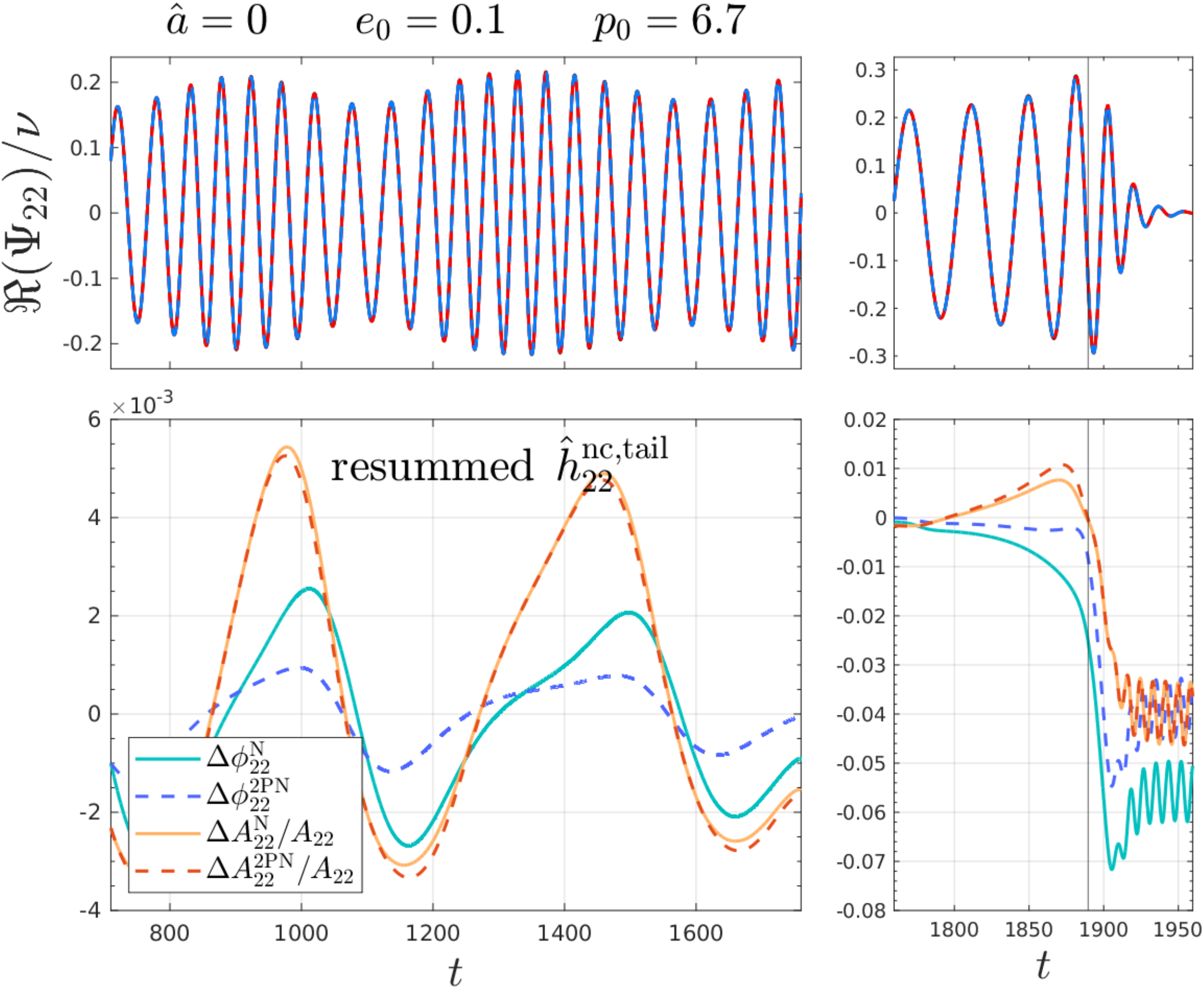}
		\hspace{0.2cm}
		\includegraphics[width=0.31\textwidth]{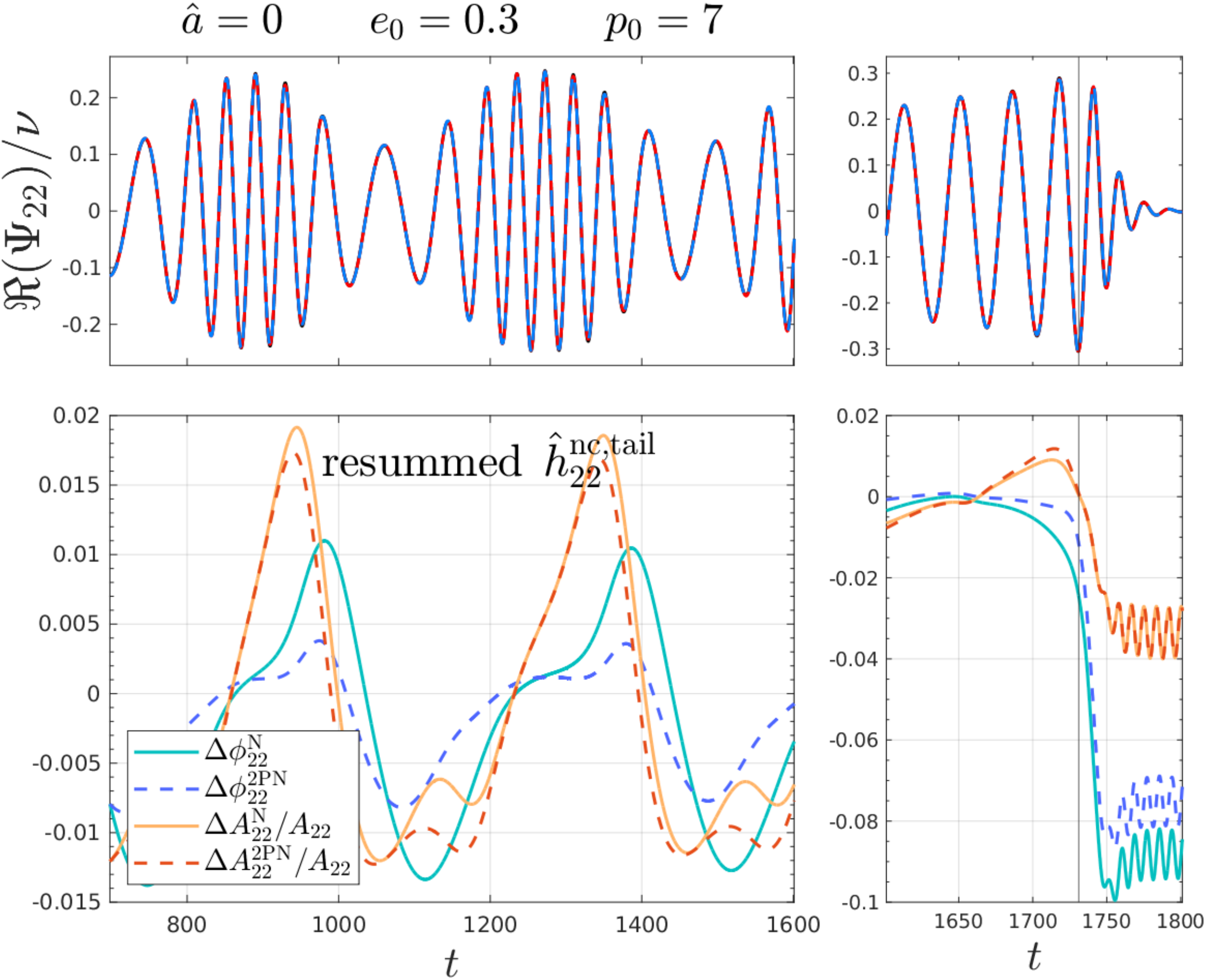}
		\hspace{0.2cm}
		\includegraphics[width=0.31\textwidth]{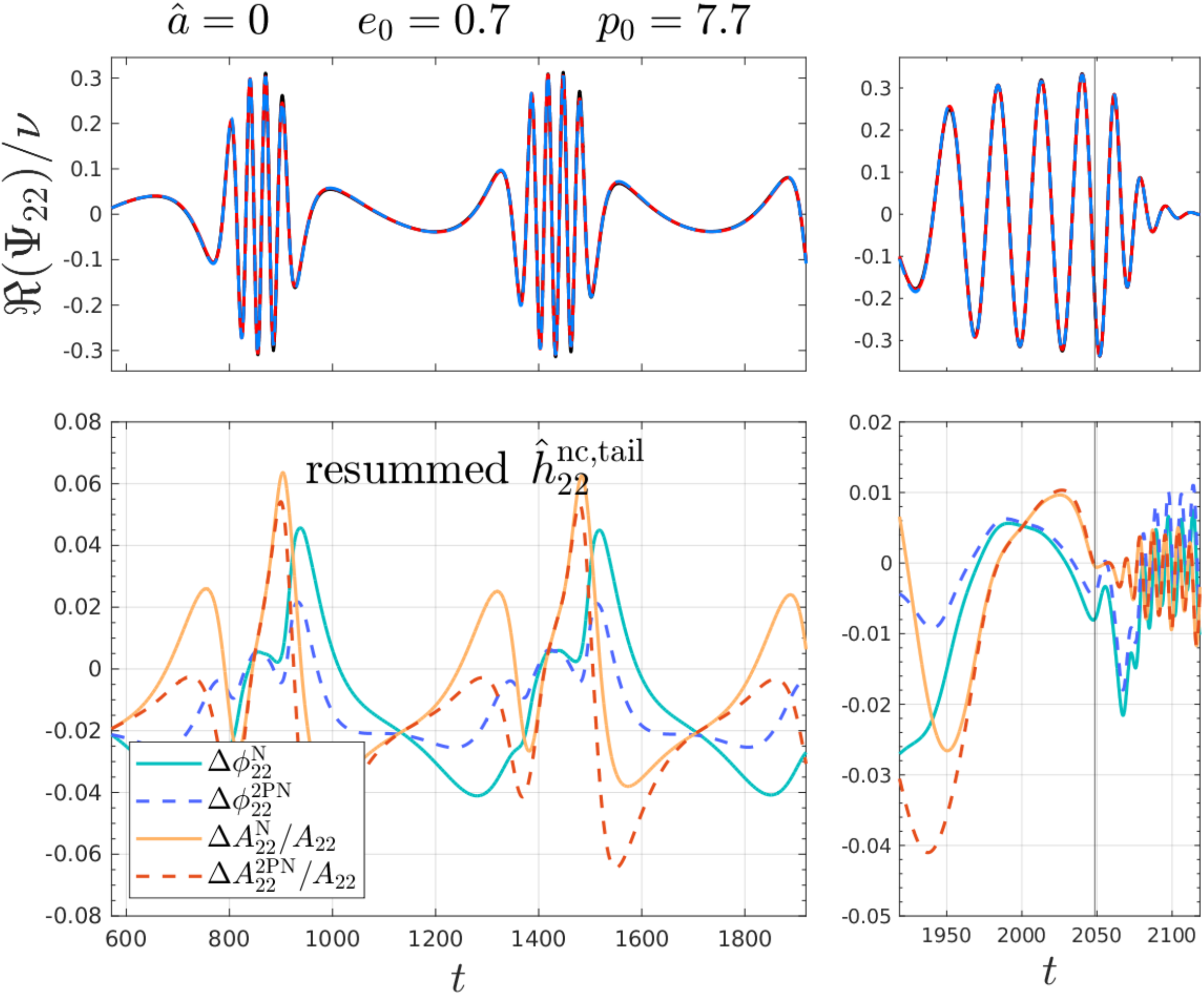}\\
		\vspace{0.3cm}
		\includegraphics[width=0.31\textwidth]{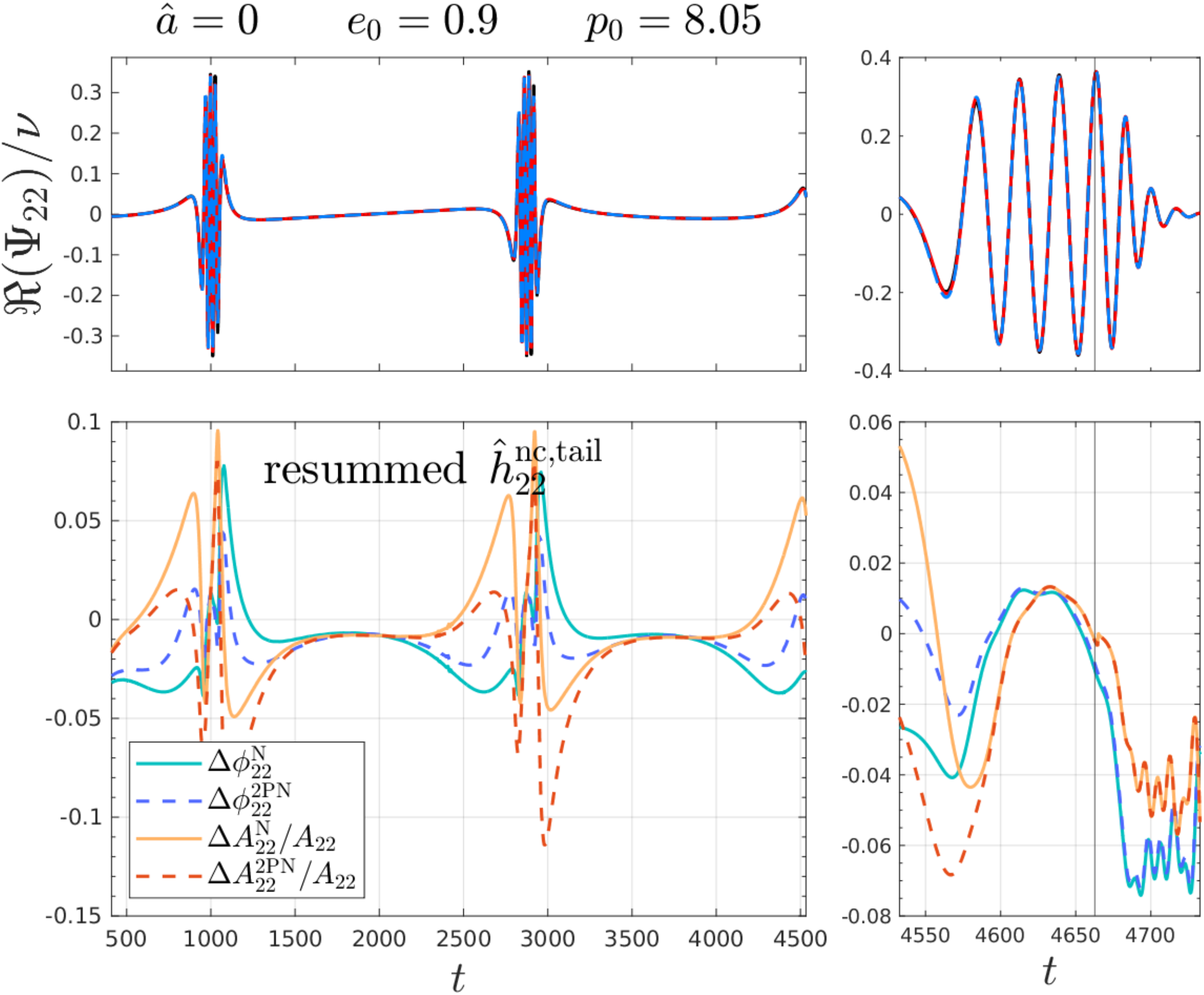}
		\hspace{0.2cm}
		\includegraphics[width=0.31\textwidth]{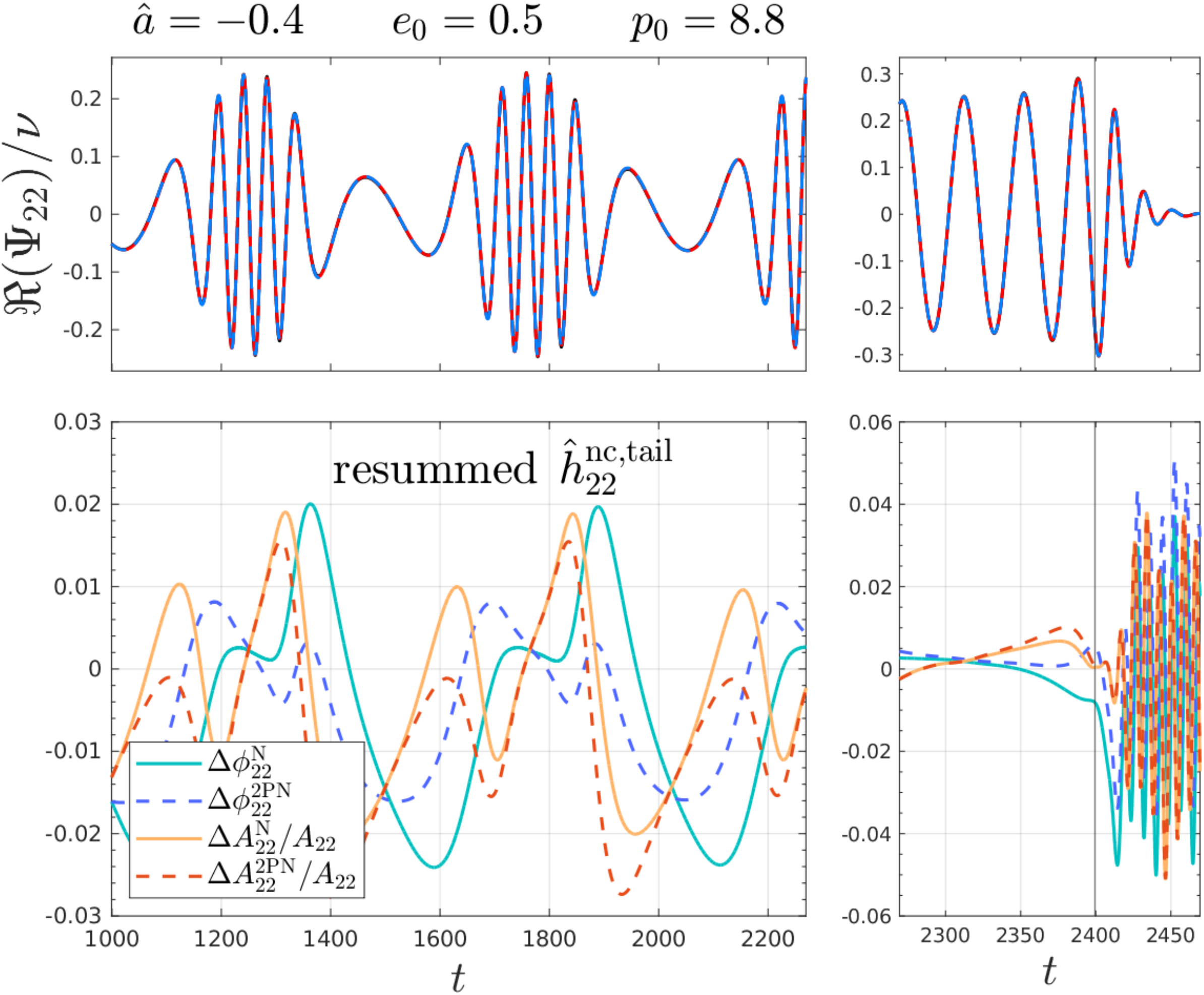}
		\hspace{0.2cm}
		\includegraphics[width=0.31\textwidth]{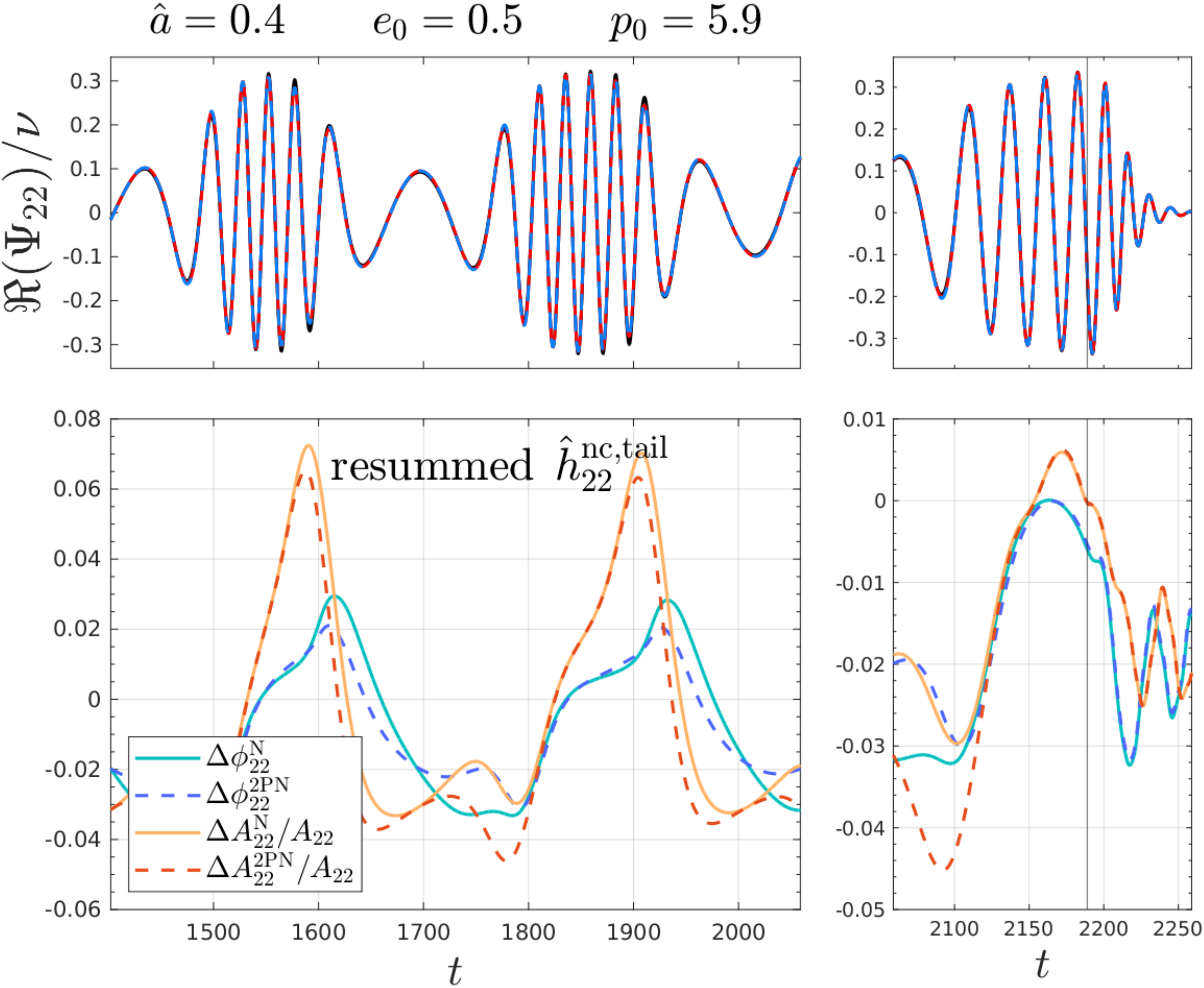}
		
		\caption{\label{fig:testmass_inspl_resum_tail}
			Comparisons with {\it resummed} noncircular tail factor:
			comparing analytical and numerical $\ell=m=2$ waveforms for the transition 
			from inspiral to plunge of a test-particle 
			on a Kerr black hole with spin parameter $\hat{a}$. We consider different configurations 
			with $\nu = 10^{-3}$.
			Each panel displays the numerical waveform (black, indistinguishable) and two EOB 
			waveforms: (i) the solid-red one with noncircular information only in the Newtonian prefactor,
			(ii) the dashed-blue one with noncircular 2PN corrections with the 
			tail $\hat{h}^{{\rm nc_{tail}}}_{22}$ of Eq.~\eqref{eq:tail22} 
			{\it resummed} following the procedure discussed in
			Sec.~\ref{sec:resum}.
			The bottom panel shows both the phase differences and the relative 
			amplitude differences with respect to the numerical waveform. We use
			dashed lines for the differences corresponding to the wave with 2PN corrections.
			The vertical line marks the merger-time, corresponding to the peak of the numerical amplitude.
			The resummation strongly improves the analytical/numerical agreement 
			with respect to Fig.~\ref{fig:testmass_inspl_expanded_tail}.}
	\end{figure*}
	%==========================================
	%
	%==========================================
	% Fig.05: insplunge: resummed delta-inst
	%==========================================
	\begin{figure*}
		\center
		\includegraphics[width=0.31\textwidth]{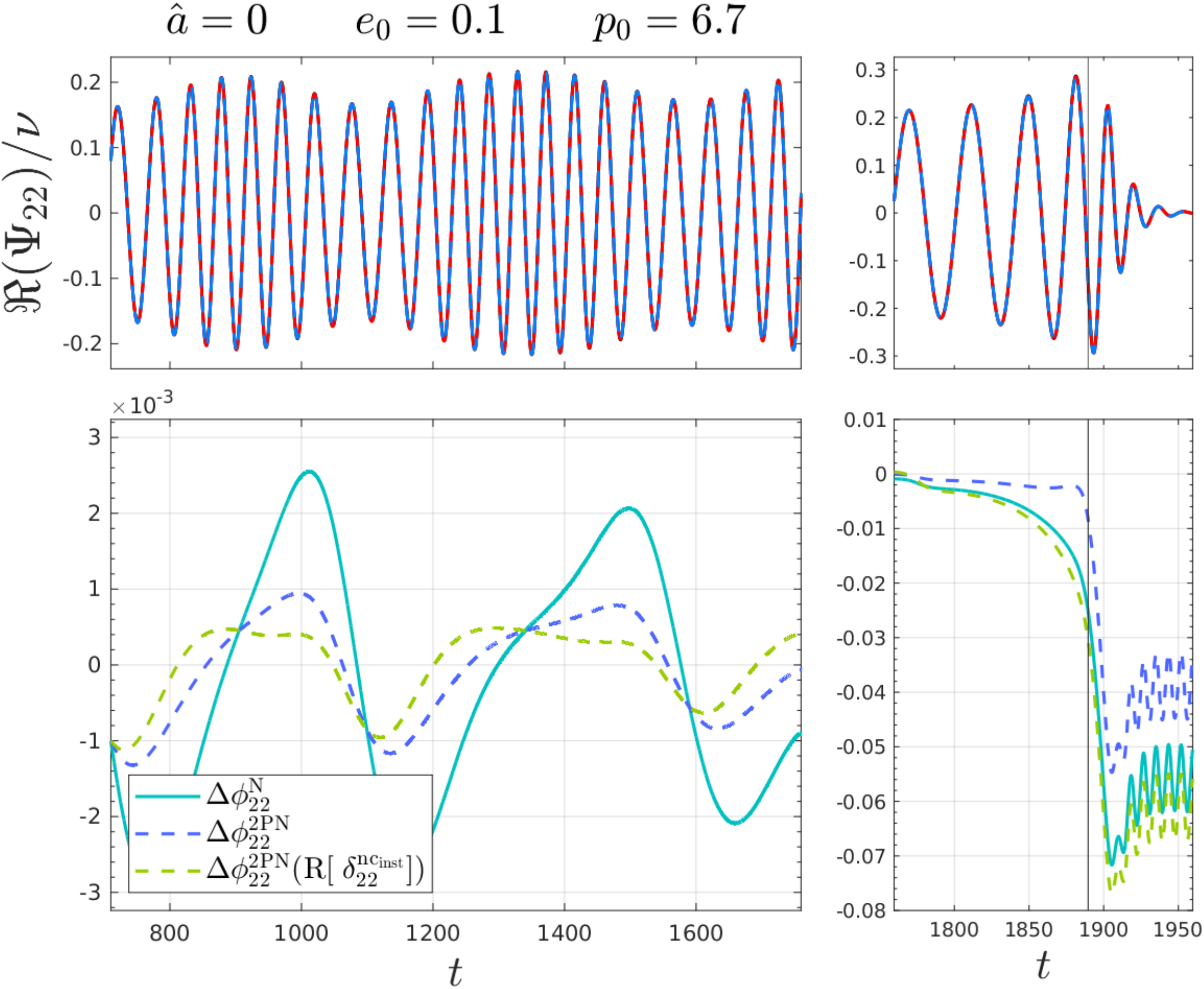}
		\hspace{0.2cm}
		\includegraphics[width=0.31\textwidth]{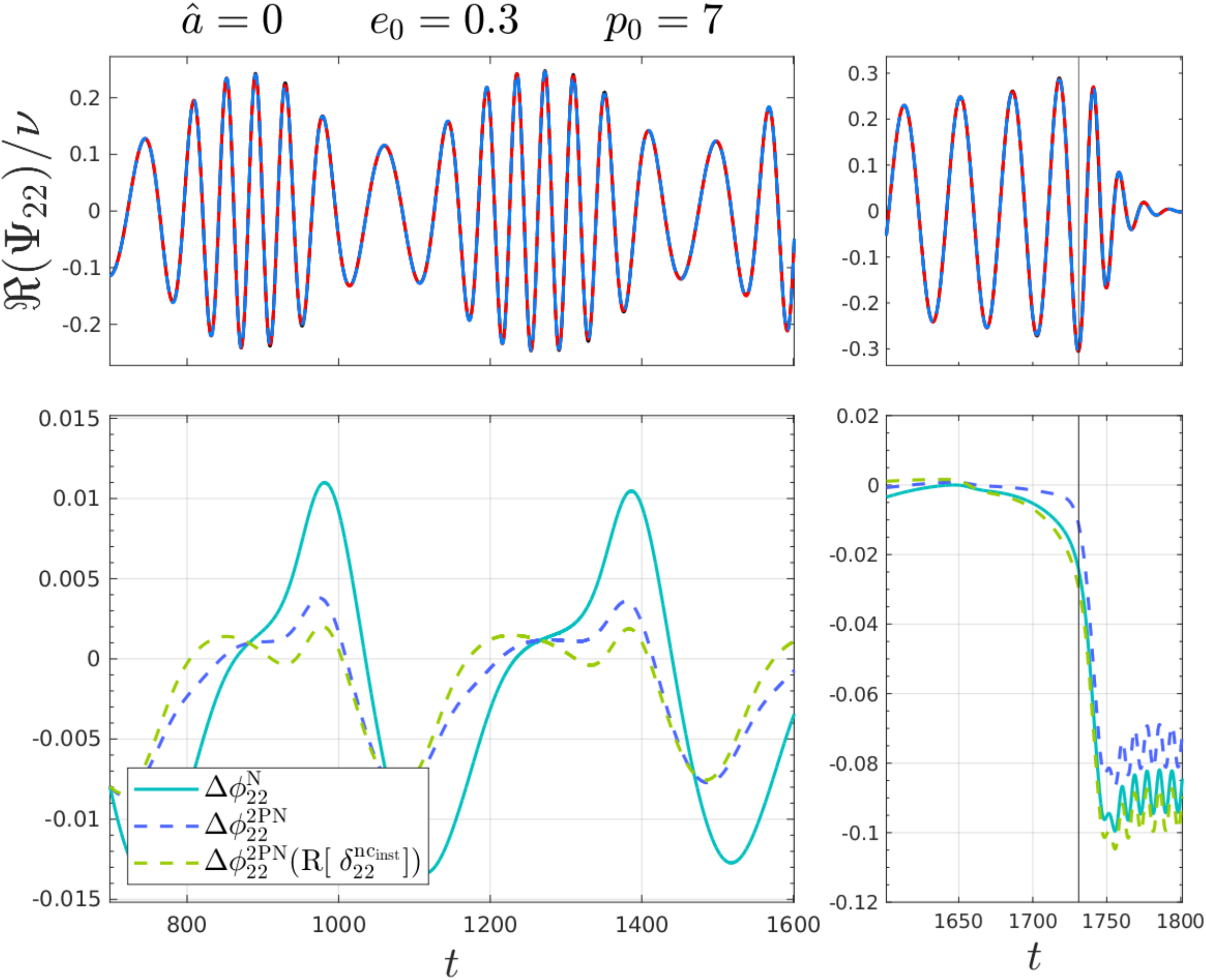}
		\hspace{0.2cm}
		\includegraphics[width=0.31\textwidth]{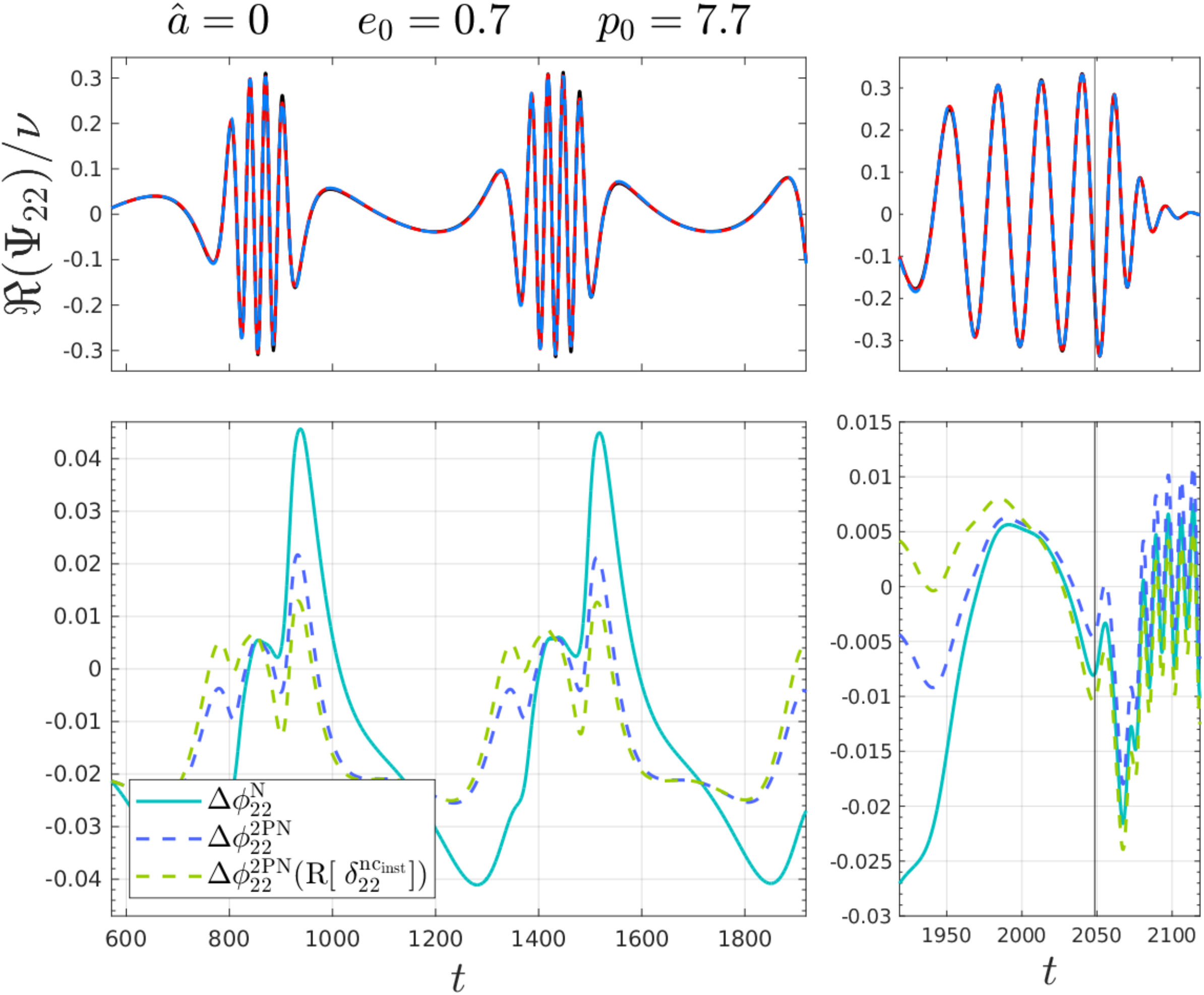}\\
		\vspace{0.3cm}
		\includegraphics[width=0.31\textwidth]{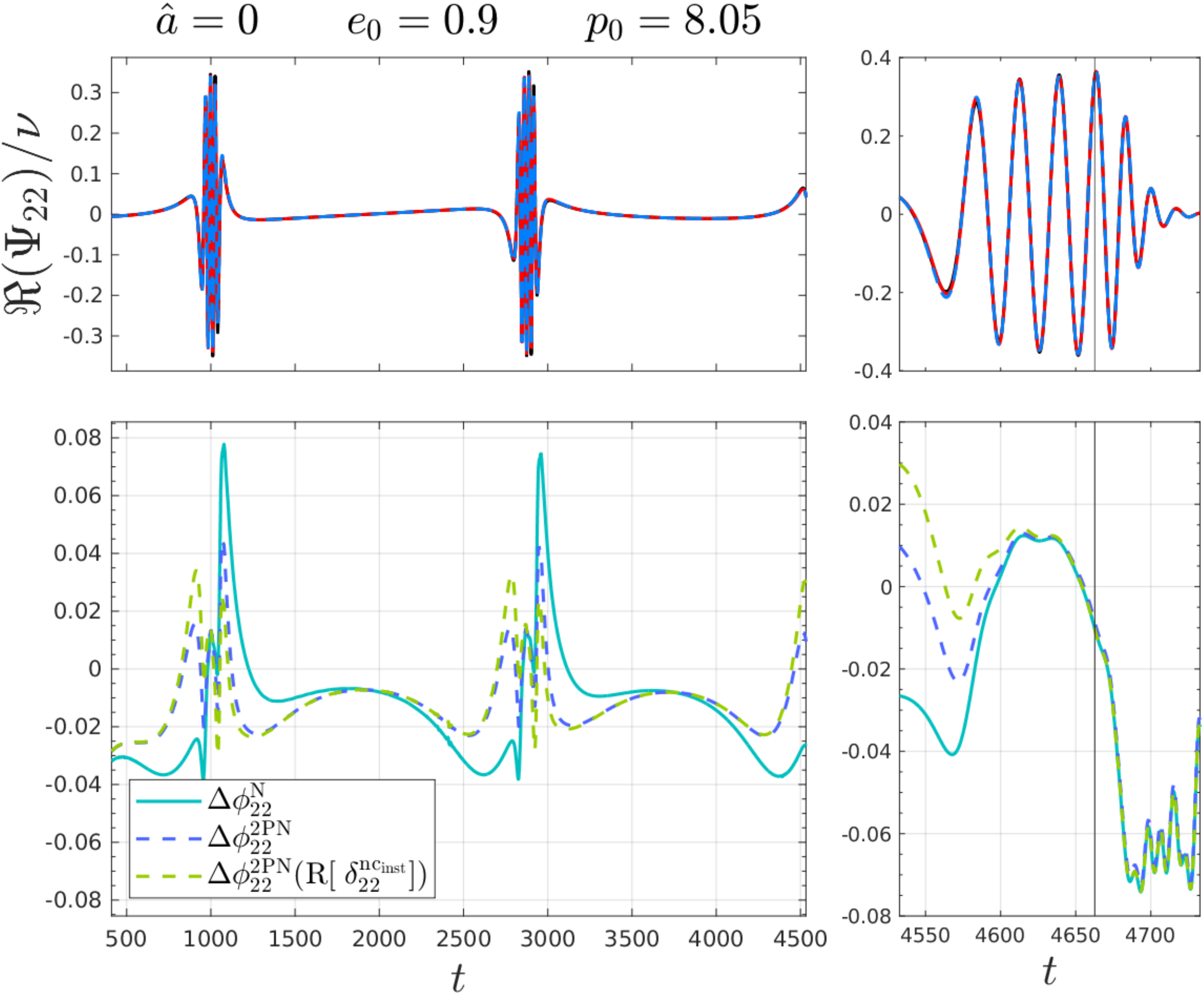}
		\hspace{0.2cm}
		\includegraphics[width=0.31\textwidth]{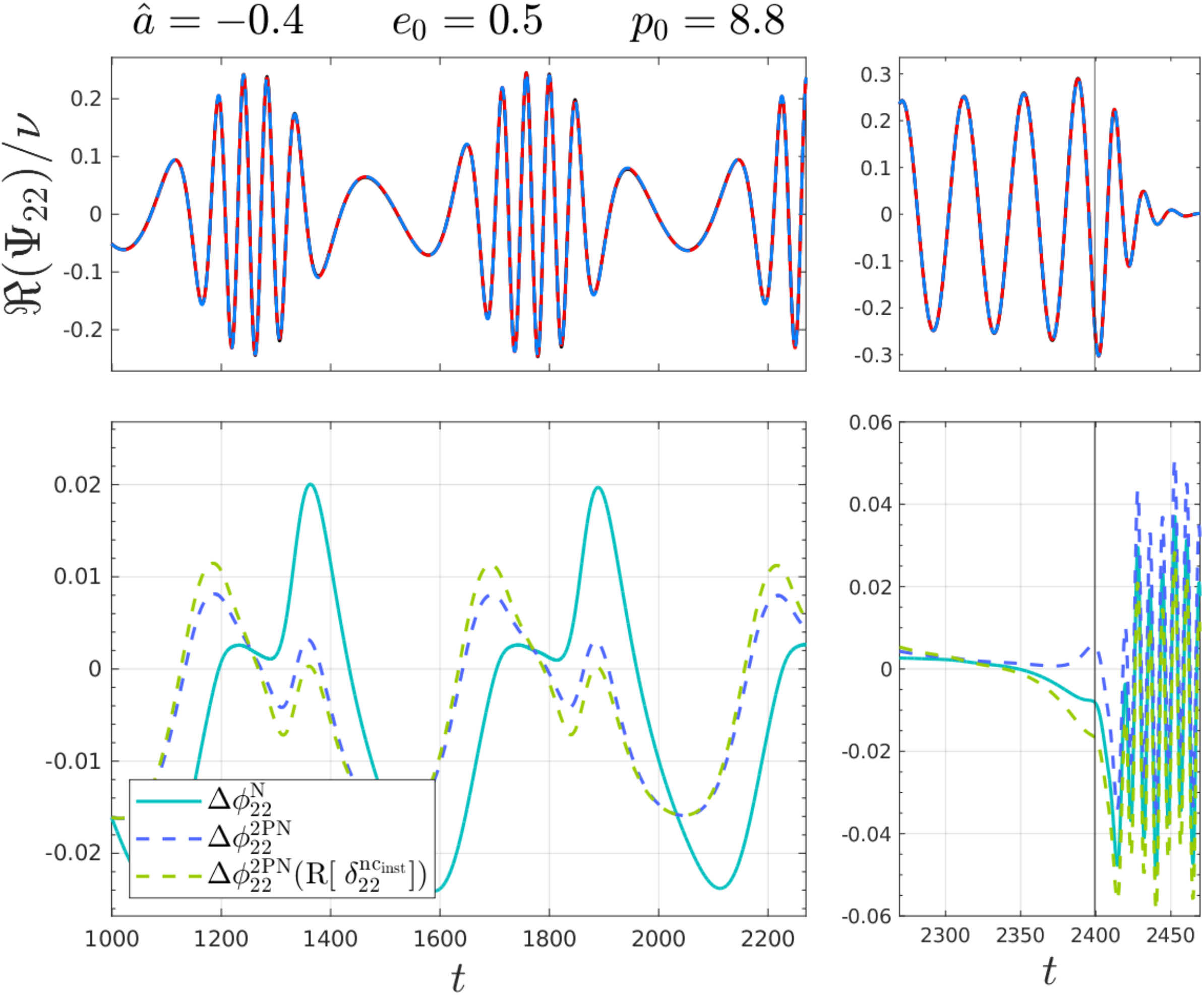}
		\hspace{0.2cm}
		\includegraphics[width=0.31\textwidth]{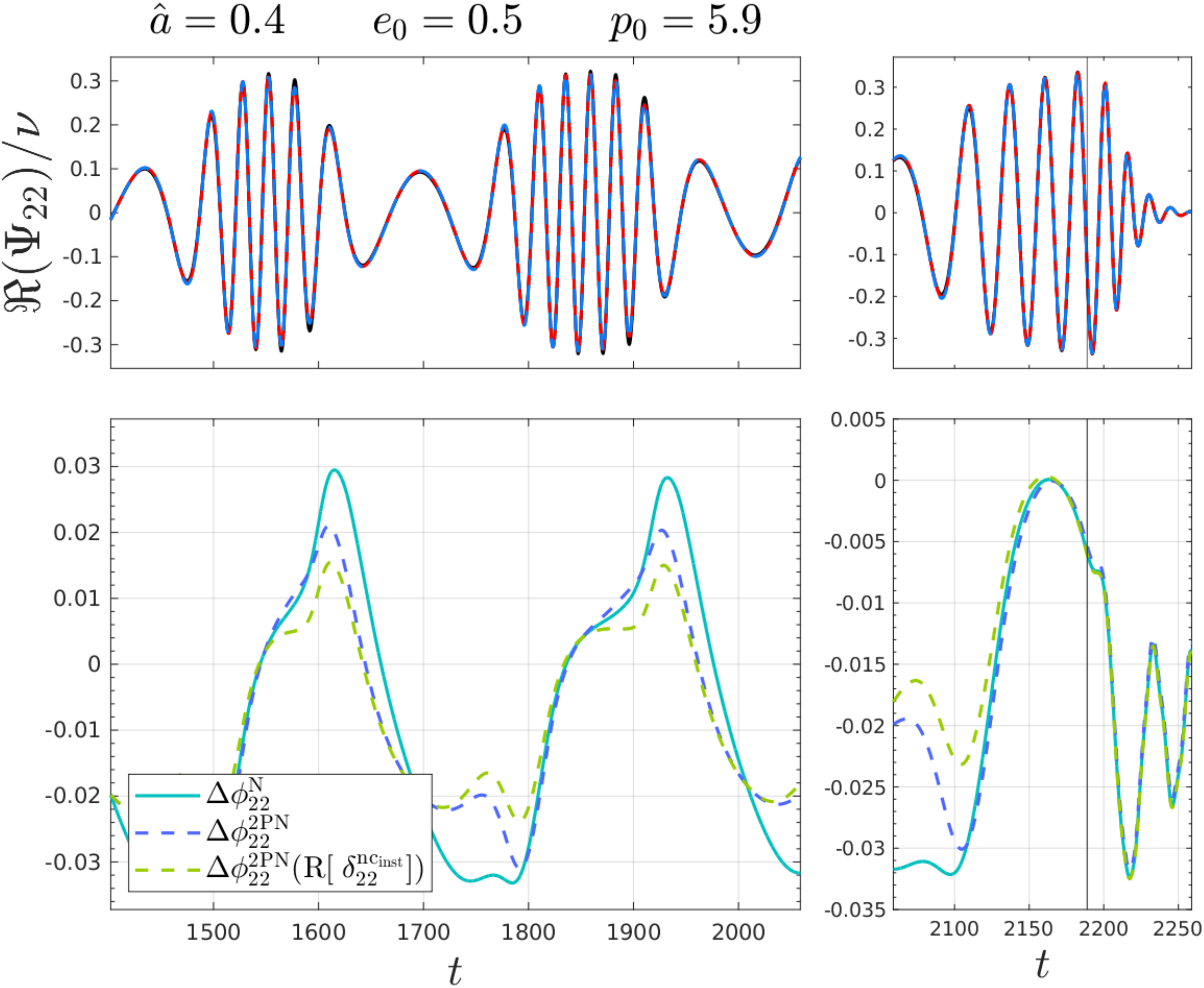}  
		\caption{\label{fig:testmass_inspl_resum_inst}
			Same configurations of Fig.~\ref{fig:testmass_inspl_resum_tail}, but here we focus on the phase
			and we show also the analytical/numerical agreement obtained considering 
			the resummed noncircular tail and the resummed instantaneous noncircular correction (dashed light green). 
			The color scheme of the other
			differences is the same of Fig.~\ref{fig:testmass_inspl_resum_tail}: solid light blue for  
			the wave with only the generic Newtonian prefactor, dashed blue for 
			the wave with 2PN corrections with resummed tail and Taylor expanded
			instantaneous corrections.}
	\end{figure*}
	Let us start this section by going back to the structure of the tail factor. 
	In its native form, it is a 1.5PN accurate term that  is expanded in eccentricity 
	up to $e^6$. We have seen in the sections above that this expansion
	in eccentricity, after the factorization of the Newtonian contribution, can be 
	recasted in a rational function of $(u,p_r,p_\varphi)$, Eq.~\eqref{eq:tail22}. 
	In particular, the expansion in the eccentricity $e$ can be rewritten as 
	an expansion in the radial momentum $p_r$ and $\dot{p}_r$ which can be subsequently 
	recasted in a form where one is left with several polynomials in $y=p_\varphi^2 u$
	that are all, formally, at {\it Newtonian order}. 
	Figure~\ref{fig:pphi2u} shows the behavior of $y$ versus time for different 
	eccentric configurations: $y$ {\it is not}
	a small quantity\footnote{Note that $y=p_\varphi^2u=1$ at Newtonian order for circular orbits, 
		where $p_\varphi = p_\varphi^{\rm N,circ}=1/\sqrt{u}$.}.
	For the nonspinning configurations considered in Fig.~\ref{fig:pphi2u}, 
	it oscillates between 0 and 4 during the eccentric inspiral and may reach values $\sim 7$ up to merger.
	We thus wonder whether an argument that can be so large
	may eventually generate some nonphysical behavior 
	for the functions
	$(\hat{t}^{22}_{p_{r_*}}, \hat{t}^{22}_{p^3_{r_*}},\hat{t}^{22}_{p_{r_*}^2},\hat{t}^{22}_{p_{r_*}^4})$,
	especially given the fact that they stem from an expansion in eccentricity
	within a PN expansion.
	%================================
	% Fig.06: 2PN phase compensation
	%================================
	\begin{figure}[t]
		\center
		\includegraphics[width=0.22\textwidth]{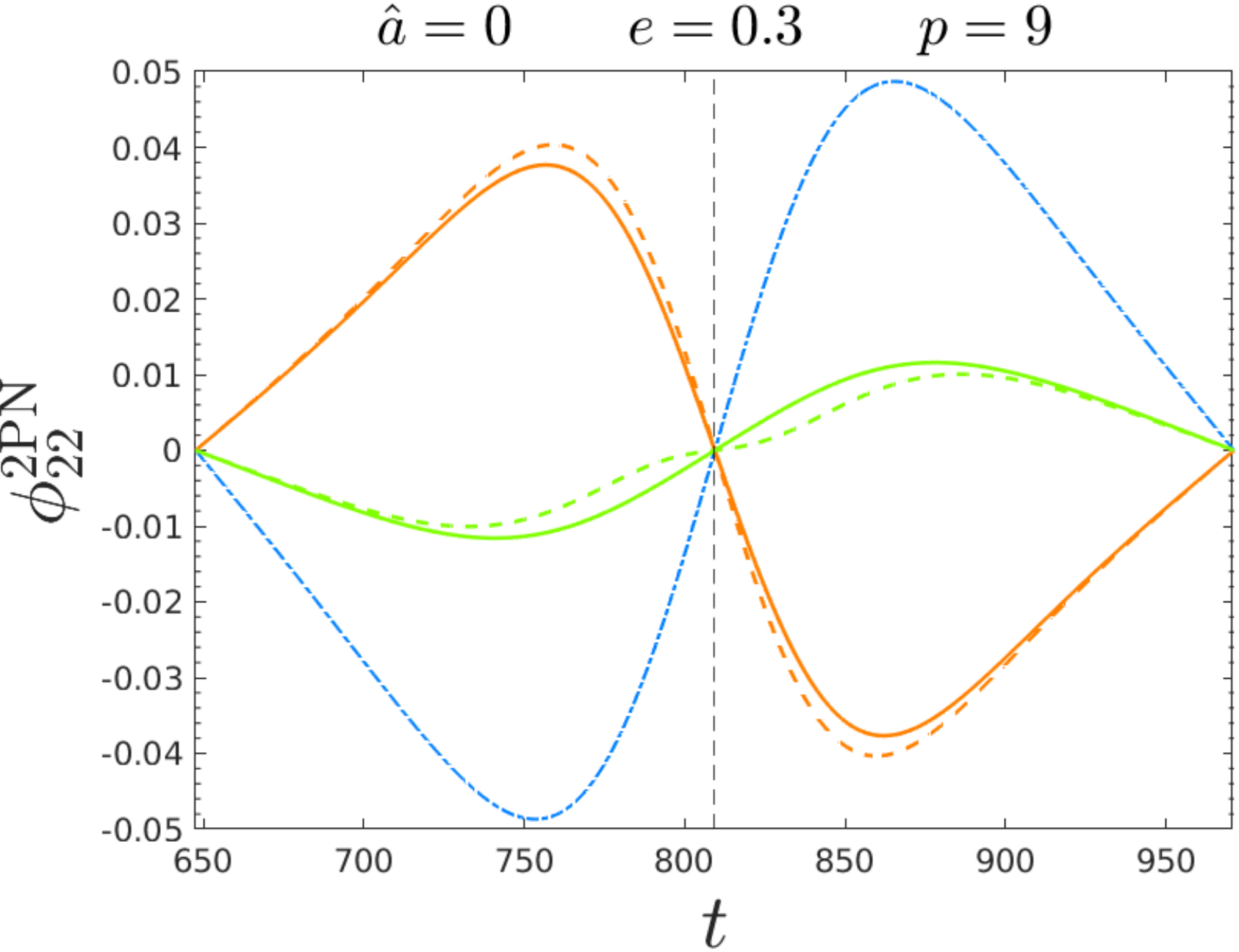}
		\includegraphics[width=0.22\textwidth]{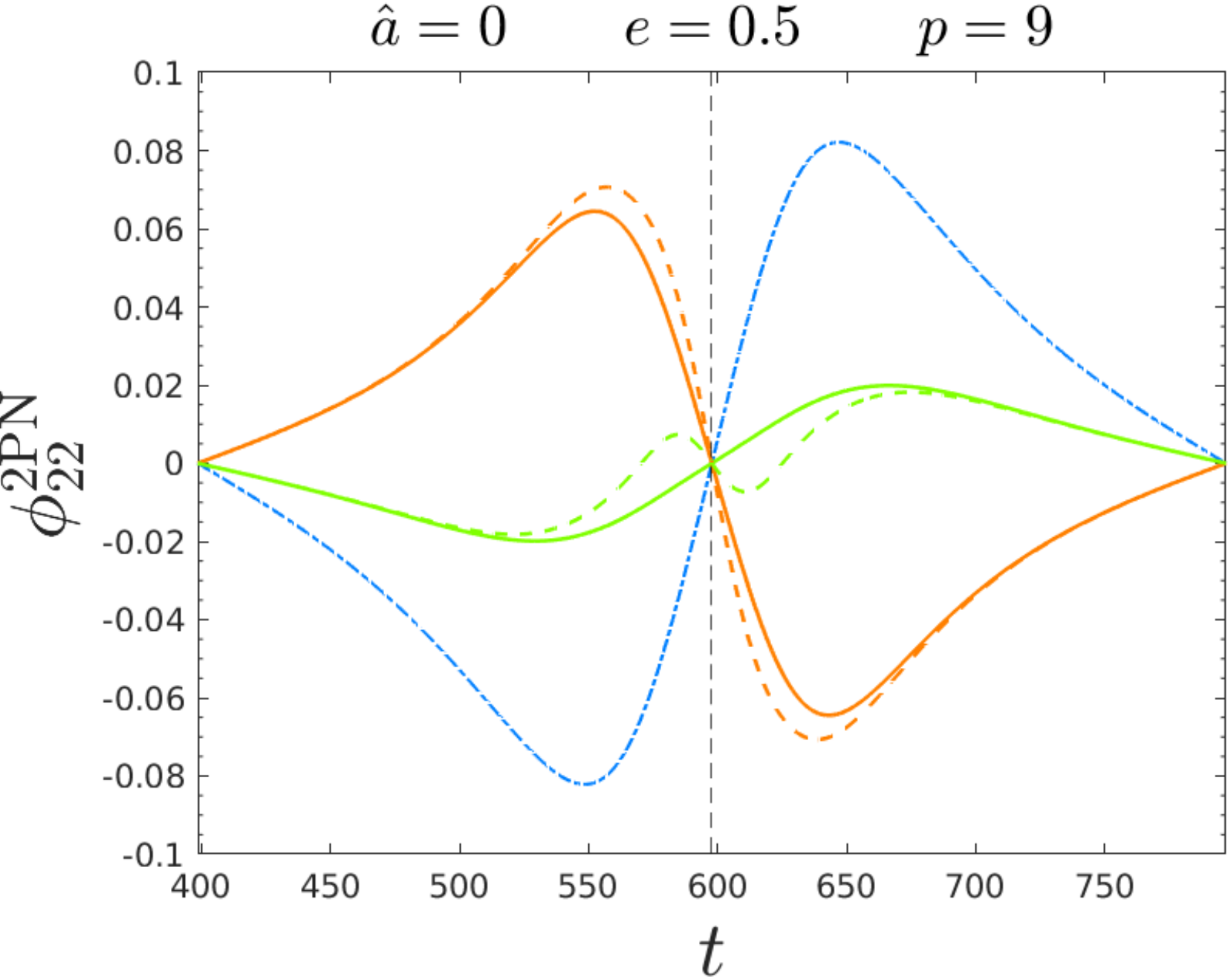} \\
		\vspace{0.3cm}
		\includegraphics[width=0.22\textwidth]{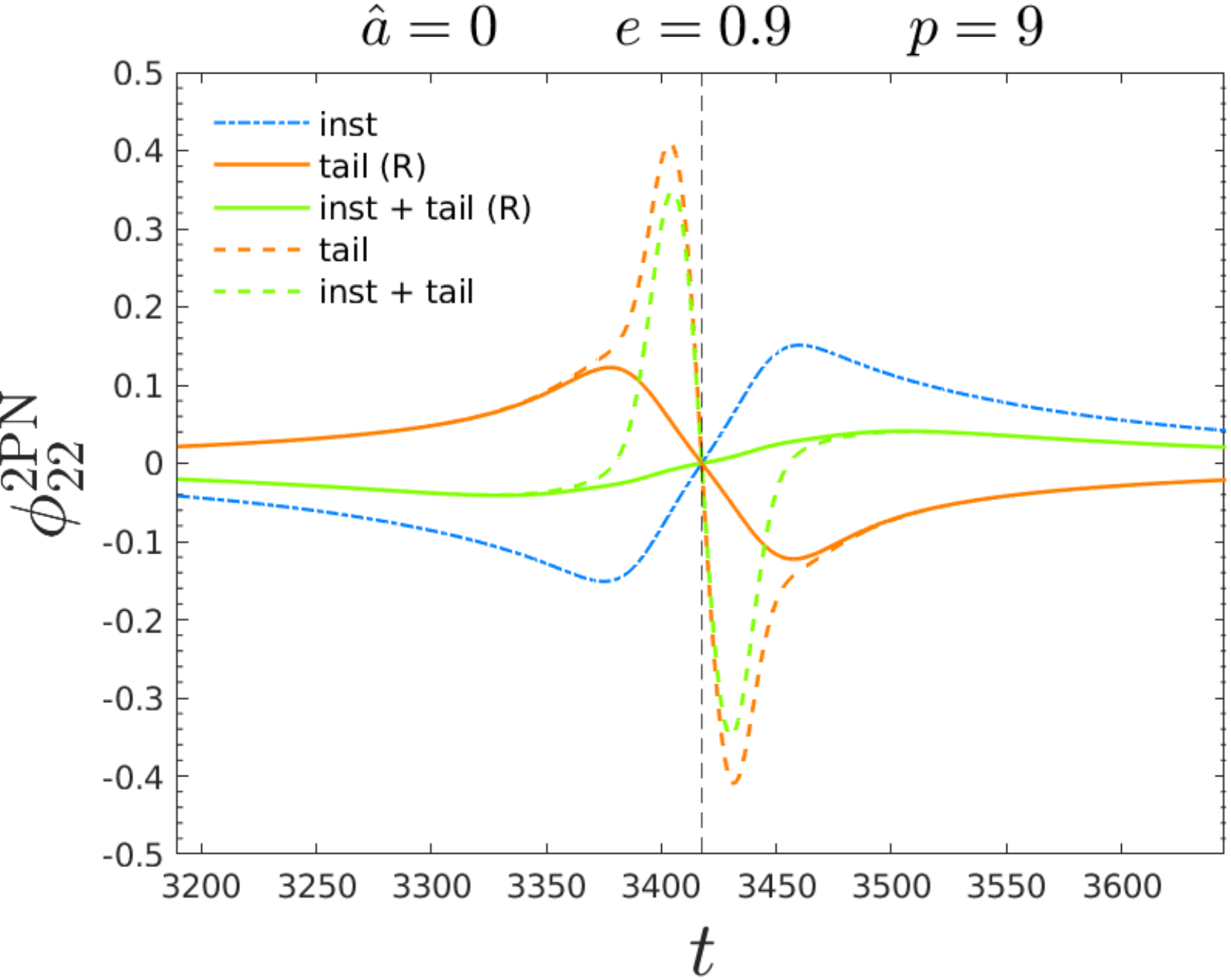}
		\includegraphics[width=0.22\textwidth]{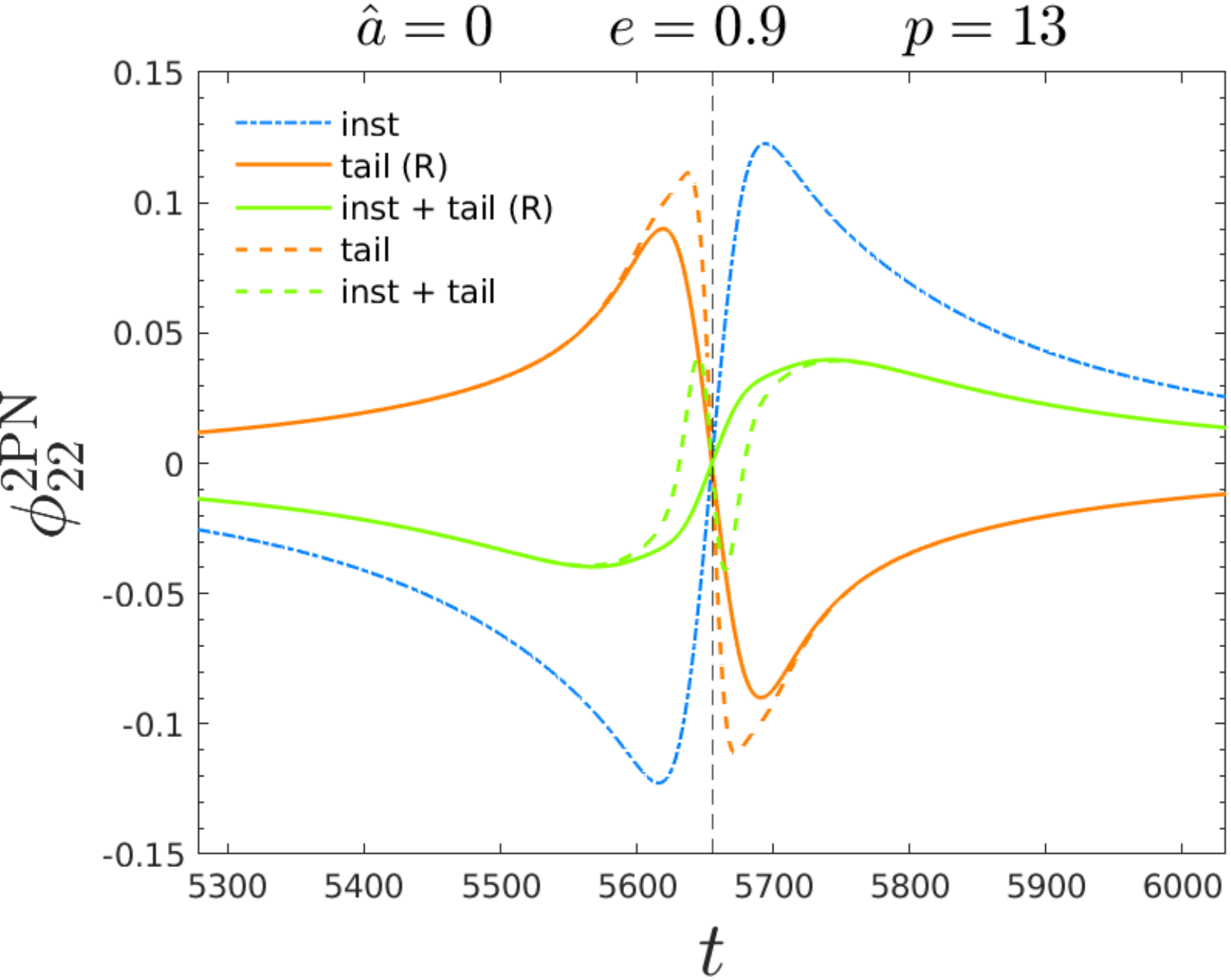}
		\caption{\label{fig:compensations_tail} Instantaneous and hereditary noncircular 2PN
			corrections to the quadrupolar phase for four nonspinning geodesic cases 
			$(e,p)=(0.3,9), (0.5,9), (0.9,9), (0.9,13)$. The instantaneous phase corrections
			are shown with dash-dotted blue lines, while 
			the orange lines are for the phase contributions of the resummed eccentric tail 
			(dashed for the expanded results and solid for the resummed ones).
			The corresponding sums between instantaneous and hereditary are shown in green with the same 
			style-scheme of the considered tail. The vertical dashed line marks the periastron passage.
			For $e=0.9$ we do not show the whole radial period in order to highlight the periastron.}
	\end{figure}
	%
	%=============================================
	% Fig.07: 2PN phase compensation - inst resum
	%=============================================
	\begin{figure}[t]
		\center
		\vspace{0.5cm}
		\includegraphics[width=0.22\textwidth]{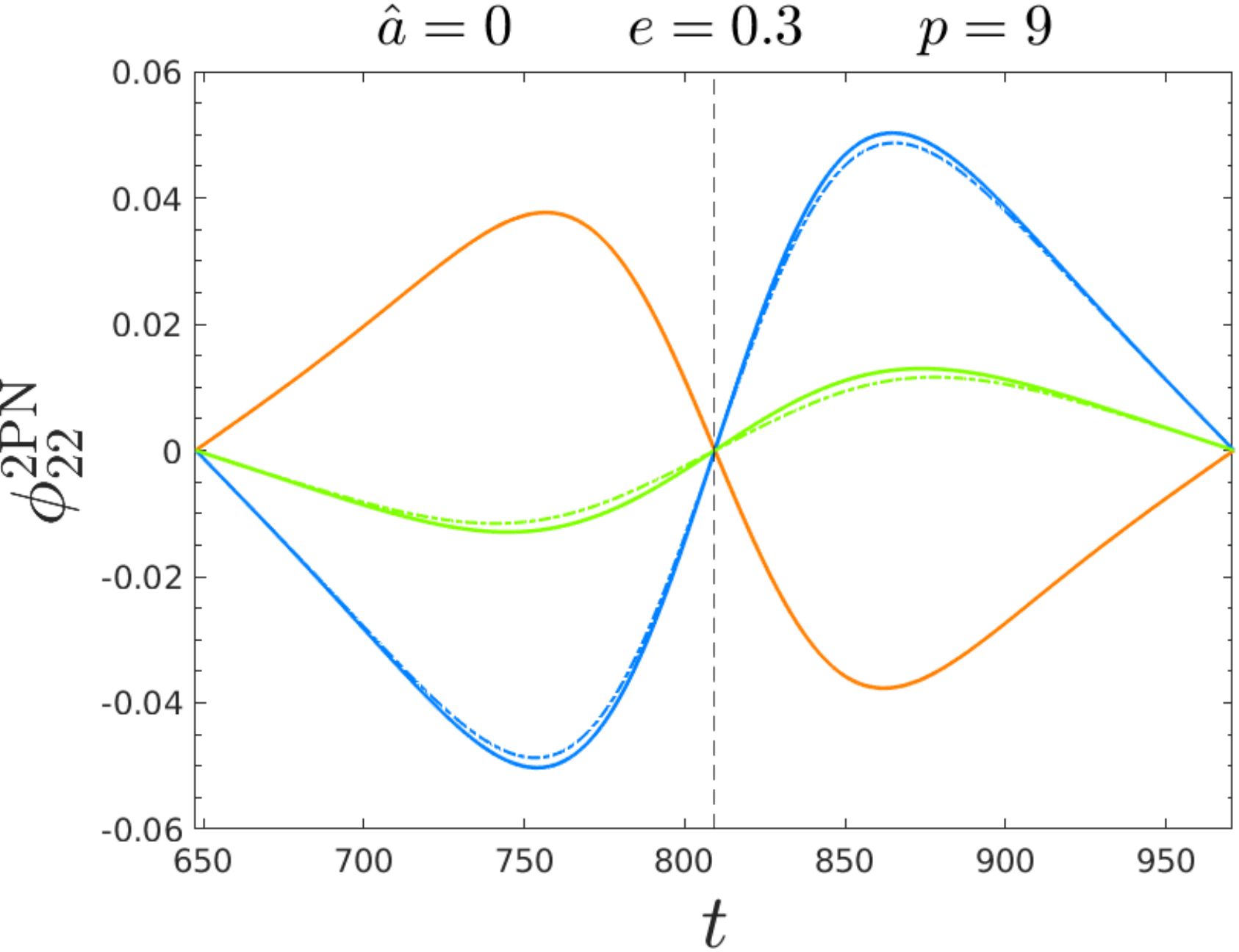}
		\includegraphics[width=0.22\textwidth]{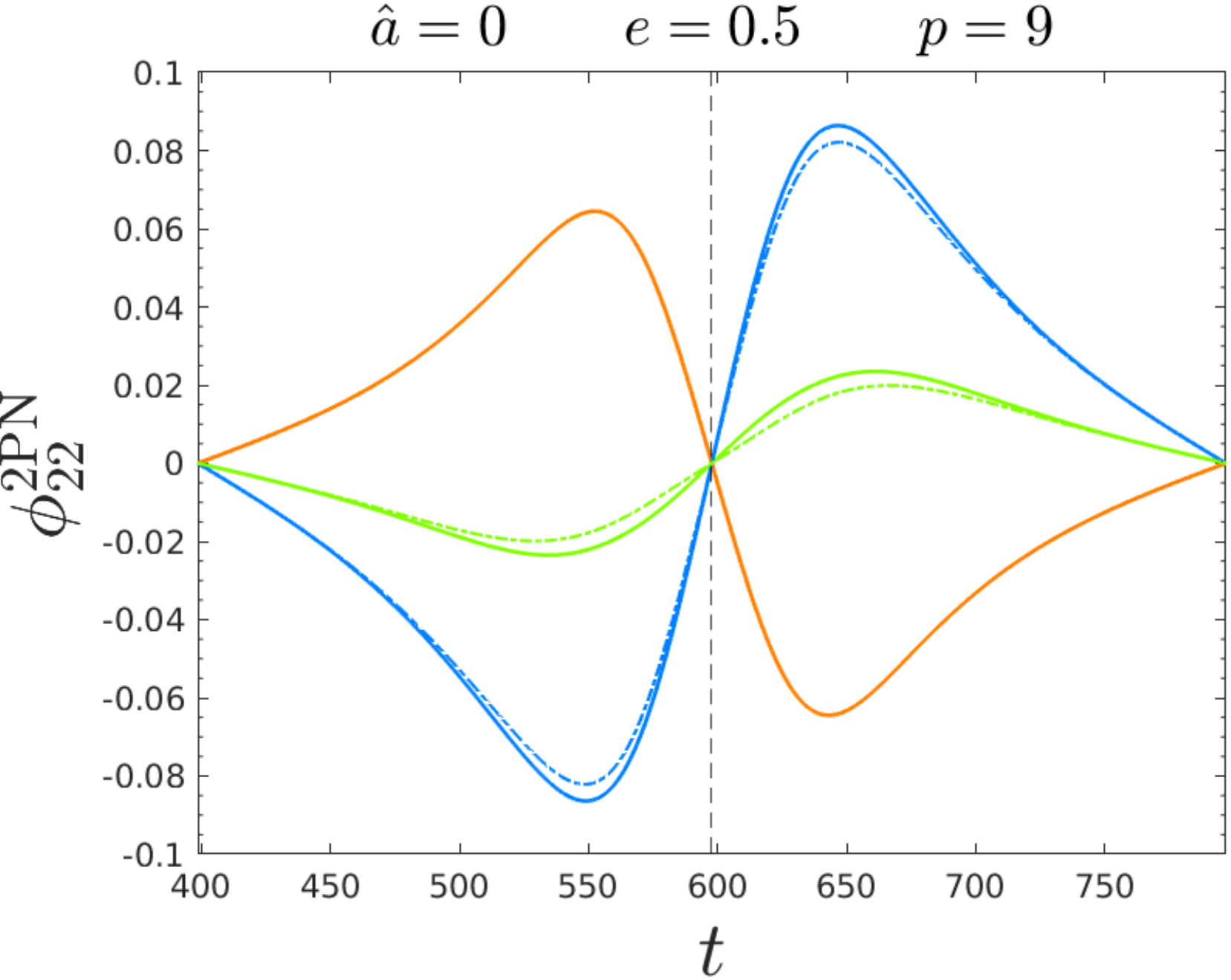} \\
		\vspace{0.3cm}
		\includegraphics[width=0.22\textwidth]{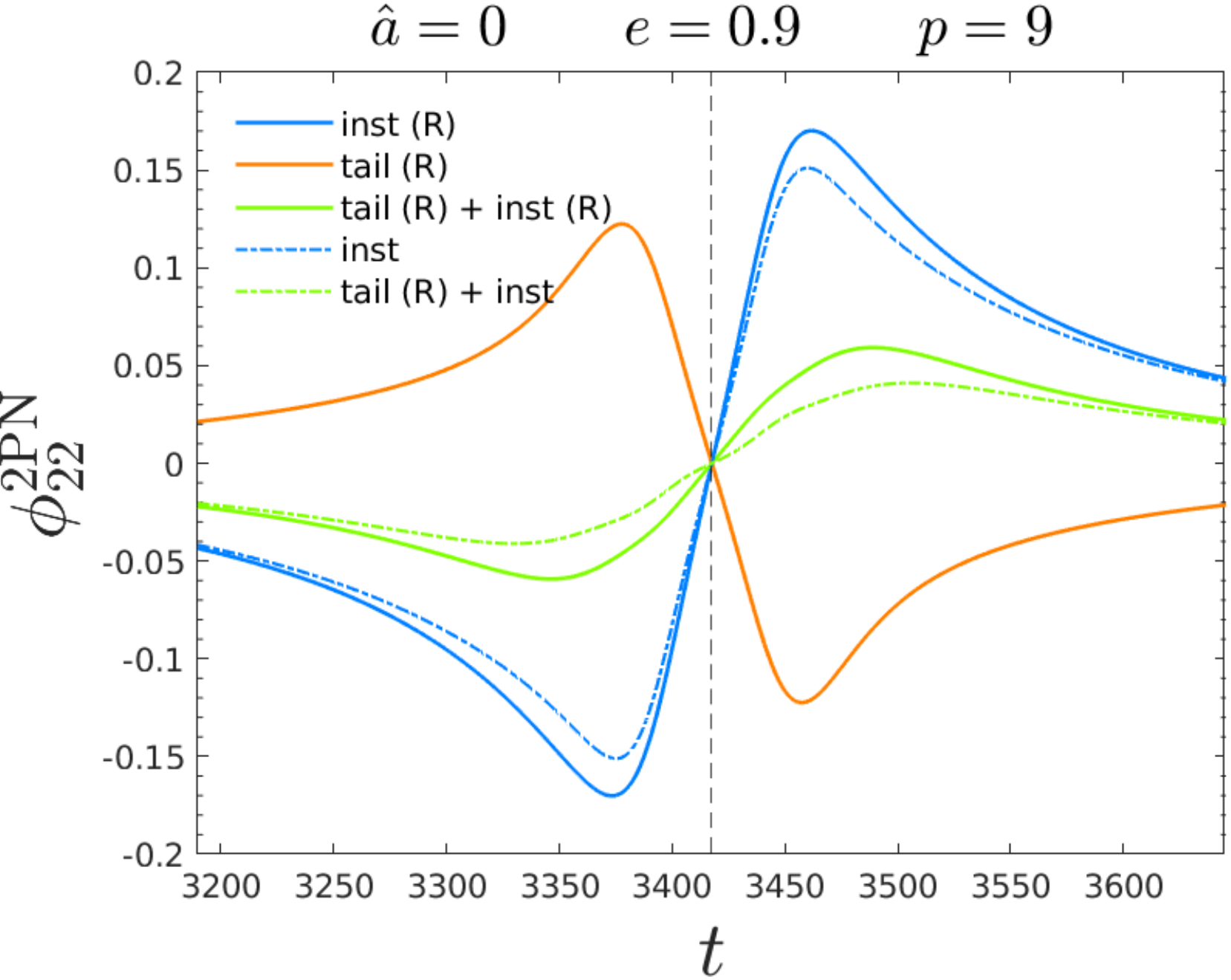}
		\includegraphics[width=0.22\textwidth]{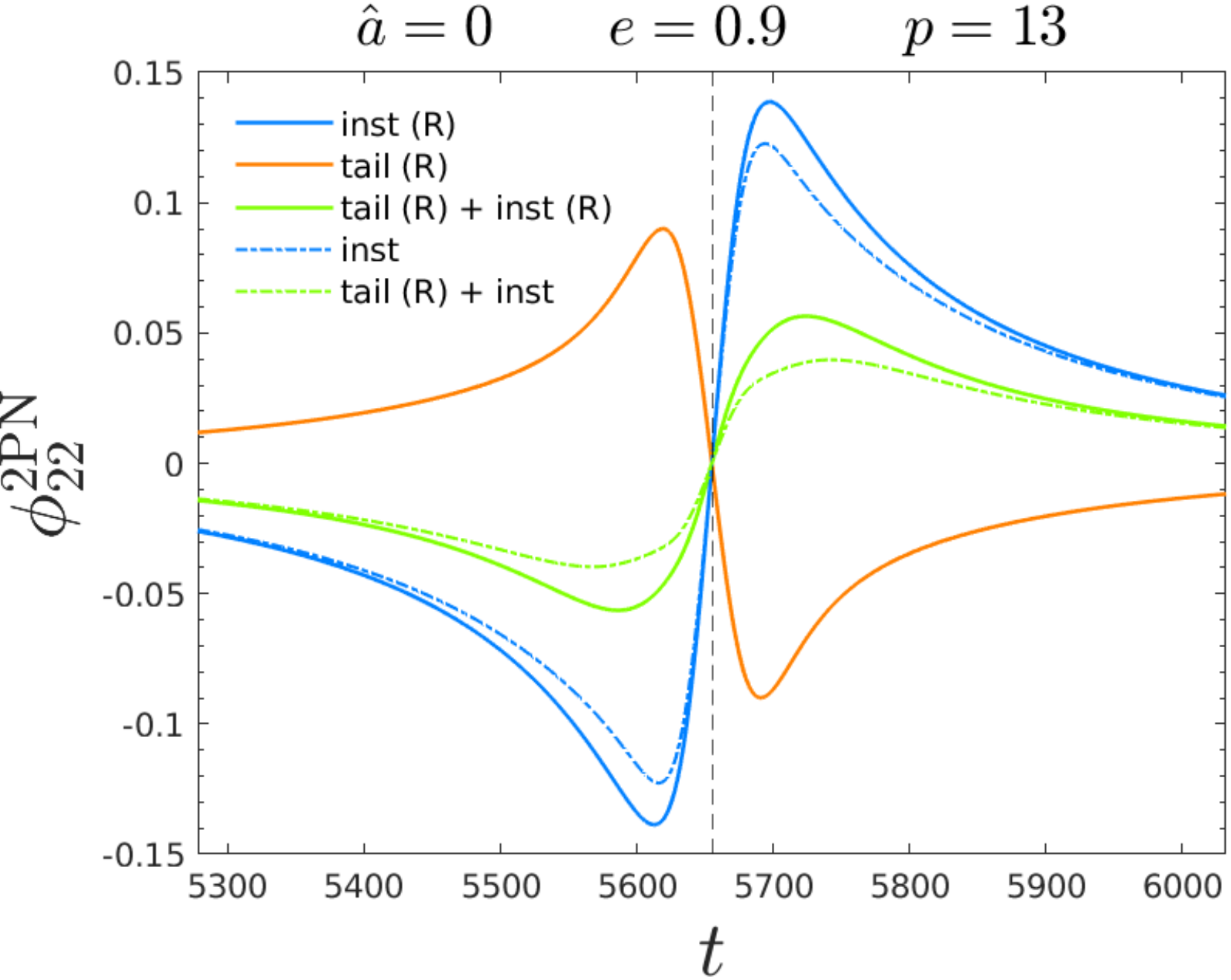}
		\caption{\label{fig:compensations_inst} Analogous to Fig.~\ref{fig:compensations_tail}, but
			here we focus on the relevance of the resummation for the instantaneous part. 
			The orange solid line is the phase contribution of the resummed eccentric tail, while the blue
			lines correspond to the instantaneous phase contributions: dash-dotted for the nonresummed 
			results, solid line for the resummed ones. 
			The corresponding sums between tail and instantaneous are shown in green with the same 
			style-scheme of the instantaneous terms. The vertical dashed line marks the periastron passage.
			For $e=0.9$ we do not show the whole radial period in order to highlight the periastron.}
	\end{figure}
	As an example, Fig.~\ref{fig:f_prstar_u} shows various truncations of $\hat{t}^{22}_{p_{r_*}}$.
	One sees that: (i) the various polynomial truncations become very large for
	values of $y$ of the order of those of the late inspiral and (ii) the sign alternation
	gives an oscillatory behavior that visually resembles the one that is typical of
	truncated PN expansions of the energy flux of a test-particle orbiting a Schwarzschild
	black hole on circular orbits (see e.g.~Ref.~\cite{Damour:1997ub}).
	On the basis of this analogy, and with the understanding that $\hat{t}^{22}_{p_{r_*}}$ is
	a suitable recasting of an expansion in the eccentricity (or in $p_{r_*}$ and $\dot{p}_r$),  we
	{\it interpret} the polynomial expression of $\hat{t}^{22}_{p_{r_*}}$ as the {\it truncated expansion}
	of an unknown function of $y$ that is expanded around $y =0$.
	As such, this function can be resummed, and we do it straightforwardly applying Pad\'e
	approximants. In Fig.~\ref{fig:f_prstar_u} we exhibit several (diagonal or nearly diagonal) 
	Pad\'e approximant that resum different truncation of the polynomials. The Pad\'e stabilizes
	the truncated series (e.g.~the results obtained resumming the truncation up to $y^6$ is
	equivalent to the Pad\'e of the full polynomial up to $y^7$) and considerably lowers the
	value of the function reached for $y\simeq 6$.
	Although we do not have a proof, the consistency between the $P^3_2$, $P^3_3$ and $P^4_3$ 
	approximants seems to suggest that the residual polynomial $\hat{t}^{22}_{p_{r_*}}$ is indeed the
	Taylor expansion of some unknown function and its resummation does make sense.
	A completely analogous behavior is found for the other three functions 
	$(\hat{t}^{22}_{p^3_{r_*}},\hat{t}^{22}_{p_{r_*}^2},\hat{t}^{22}_{p_{r_*}^4})$,
	which are thus also resummed. In practice, we replace the Taylor-expanded functions  
	with $(P^4_3[\hat{t}^{22}_{p_{r_*}}],P^4_4[\hat{t}^{22}_{p^3_{r_*}}],P^4_3[\hat{t}^{22}_{p_{r_*}^2}],P^4_4[\hat{t}^{22}_{p_{r_*}^4}])$.
	The quality of the resummed $\ell=m=2$ waveform is shown in Fig.~\ref{fig:testmass_inspl_resum_tail},
	which is the analogous of Fig.~\ref{fig:testmass_inspl_expanded_tail} where the Taylor-expanded 
	functions have been replaced by the Pad\'e resummed ones. The analytical/numerical phase 
	agreement not only improves (and largely) during the plunge and merger phase, but also during 
	the eccentric inspiral. In particular, it is now evident the improvement with respect to the
	simple Newtonian prefactor all over, notably without pathological behaviors towards merger.

	\subsection{Resummation of the instantaneous residual noncircular factor}
	\label{sec:resum_inst}
	A priori, the same resummation strategy should be implemented for the residual instantaneous 2PN
	corrections which exhibits an analogous structure with polynomials in $y$.
	We explore this on both the residual noncircular amplitude correction $f_{22}^{\rm nc_{inst}}$
	and phase $\delta_{22}^{\rm nc_{inst}}$. For the amplitude, we find that any 
	choice of Pad\'e approximant
	for the various residual polynomials in $y$ of Eq.~\eqref{APNnc22} 
	develops spurious poles, so that our resummation
	strategy cannot be pursued\footnote{This is the current situation with the 2PN-accurate 
		waveform. The procedure
		will have to be investigated again in the future using results at 3PN order.}. 
	This is not of great concern, since the generic Newtonian prefactor alone already
	gives an excellent approximation to the exact waveform.
	This can be clearly seen in Fig.~\ref{fig:testmass_inspl_resum_tail}, where the amplitudes 
	with and without 2PN noncircular corrections produce analytical/numerical 
	relative differences that are comparable.
	
	By contrast, for the instantaneous residual noncircular phase given 
	in Eq.~\eqref{PhPNnc22} the procedure is robust. 
	More precisely, we resum the $y$-polynomials $\hat{\delta}^{22}$ of Eq.~\eqref{PhPNnc22}
	using the Pad\'e approximants 
	$P^1_0[\hat{\delta}^{22}_{u^{\rm 1PN}}]$,
	$P^1_1[\hat{\delta}^{22}_{p_{r_*}^{\rm 1PN}}]$,
	$P^1_2[\hat{\delta}^{22}_{u^{\rm 2PN}}]$, and
	$P^1_1[\hat{\delta}^{22}_{p_{r_*}^{\rm 2PN}}]$.
	Note that the latter polynomial, written explicitly in Eq.~\eqref{eq:hatdelta22_4}, 
	is at fourth-order in $y$, 
	but we only use $O(y^2)$-terms since the $P^2_1$ approximant produces unphysical behaviors 
	for large $y$ and the other higher-order Pad\'e approximants have spurious poles in the equal-mass case.
	The improvements introduced by this resummation 
	are shown in Fig.~\ref{fig:testmass_inspl_resum_inst}, where we compare the analytical/numerical
	phase differences of the new obtained waveform with the phase differences of 
	the previous prescription, where the resummation was applied only to the
	eccentric hereditary terms.
	While a slight improvement in the phase accuracy can be seen in the reported 
	cases\footnote{The only exception is the $\ha = -0.4$ case, but bear in mind that
		we are not including spin terms in the noncircular corrections.}, the 
	resummation of the instantaneous phase correction is less relevant than the resummation of the 
	eccentric tail. Nonetheless, through this paper we will use the resummed instantaneous 
	phase as our default option for the 2PN noncircular corrections.
	
	\subsection{Discussion: Compensation between instantaneous and hereditary contributions}
	\label{sec:discussion}
	The results shown in the section above require some discussion. On the one hand, as noted in 
	previous works~\cite{Chiaramello:2020ehz,Albanesi:2021rby}, the Newtonian prefactor is quite
	effective in capturing the behavior of the correct waveform, both in amplitude and phase. 
	As a consequence, the missing analytical information is rather tiny and special resummation 
	procedures should be implemented to make the additional PN information really useful.
	By separately analyzing the cumulative action of the instantaneous and hereditary contributions
	to the waveform one finds that the good performance of our resummed waveform is  due to 
	{\it compensations} between the two. This eventually yields only a tiny correction to the Newtonian
	noncircular prefactor. More importantly, one notices that  the instantaneous 
	contributions alone
	tend to {\it overestimate} the analytical phase, eventually yielding phase differences with the 
	numerical waveform that are {\it larger} than those obtained with the simple Newtonian prefactor.
	This is very clear when inspecting Fig.~\ref{fig:compensations_tail} that illustrates this effect
	for four different geodesic configurations: $(e,p)=(0.3,9), (0.5,9), (0.9,9), (0.9,13)$. 
	Indeed at high eccentricity and relatively small 
	semilatus rectum, the resummation of the tail factor is a crucial 
	aspect in order to have a compensation between instantaneous and hereditary terms. 
	The benefits of the resummation can be seen even at milder eccentricities or larger semilatera
	recta, even if it is less crucial.
	In Fig.~\ref{fig:compensations_inst} we also show the effect of the resummation of the 
	instantaneous factor for the same configurations considered in Fig.~\ref{fig:compensations_tail}.
	While the effect of the resummation is clearly visible, it is also evident that the resummation
	of the instantaneous part is less relevant than the tail resummation.

	\subsection{Subdominant modes}
	\label{sec:HM}
	The factorization and resummation outlined in the previous section can be similarly applied to higher modes.
	The resulting factorized (though nonresummed) expressions are all reported in Appendix~\ref{App:HM}.
	We explicitly discuss analytical/numerical comparisons for the modes $(2,1)$, $(3,3)$, $(3,2)$ and $(4,4)$.
	While for the modes $(2,1)$ and $(3,3)$ the tail contribution is present, for the modes $(3,2)$ and $(4,4)$ it is absent,
	at 2PN order. This has implications on the waveform performance,
	as we will see below. We use several Pad\'e approximants. To make this clear, it is convenient
	to rewrite here explicitly the tail factor at 2PN order for the modes $(2,1)$, $(3,3)$
	
	\begin{widetext}
		\begin{align}
			\hat{h}_{21}^{\rm {nc}_{tail}}&=1 + \dfrac{1}{c^3}\pi\bigg[-i \left(\dfrac{3029}{1920} u p_{r_*} \, \hat{t}^{21}_{p_{r_*}} +\frac{619}{576} p_{r_*}^3 \, \hat{t}^{21}_{p_{r_*}^3} \right)+ \dfrac{635 }{768 } \dfrac{p_{r_*}^2}{p_\varphi}\, \hat{t}^{21}_{p_{r_*}^2} - \dfrac{61}{256} \frac{p_{r_*}^4}{p_\varphi u} \,  \hat{t}^{21}_{p_{r_*}^4} \bigg] ,\\
			\hat{h}_{33}^{\rm nc_{tail}} &= 1 +\dfrac{1}{c^3}\dfrac{\pi}{p_\varphi^2(7+2 p_\varphi^2 u)^2}\bigg[- i \left( \dfrac{4763}{384} p_{r_*}  \hat{t}^{33}_{p_{r_*}}-\dfrac{4763 }{24}\dfrac{p_\varphi^2 p_{r_*}^3}{p_\varphi^2 u^2 (7+2 p_\varphi^2 u)^2} \, \hat{t}^{33}_{p_{r_*}^3} \right)\cr
			&+\dfrac{4763}{96} \dfrac{p_{r_*}^2}{p_\varphi u (7+2 p_\varphi^2 u)}\, \hat{t}^{33}_{p_{r_*}^2}+ \dfrac{4763}{6} \dfrac{ p_\varphi^3 p_{r_*}^4}{p_\varphi^3 u^3 (7+2 p_\varphi^2 u)^3} \,  \hat{t}^{33}_{p_{r_*}^4}\bigg], 
		\end{align}
	\end{widetext}
	where
	%==========================================
	% Fig.8: insplunge: resummed delta-inst
	%==========================================
	\begin{figure*}[t]
		\center
		\includegraphics[width=0.22\textwidth]{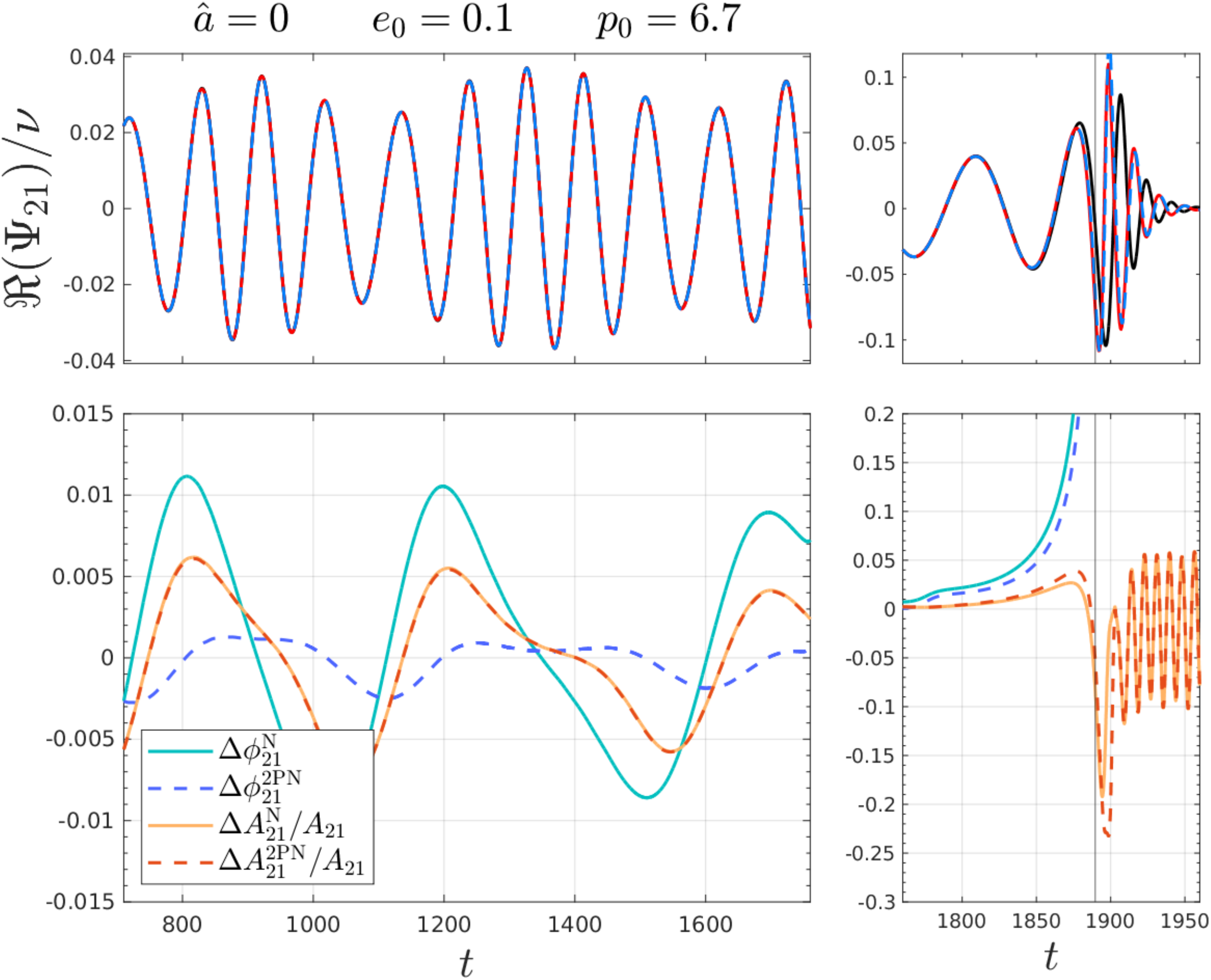}
		\hspace{0.2cm}
		\includegraphics[width=0.22\textwidth]{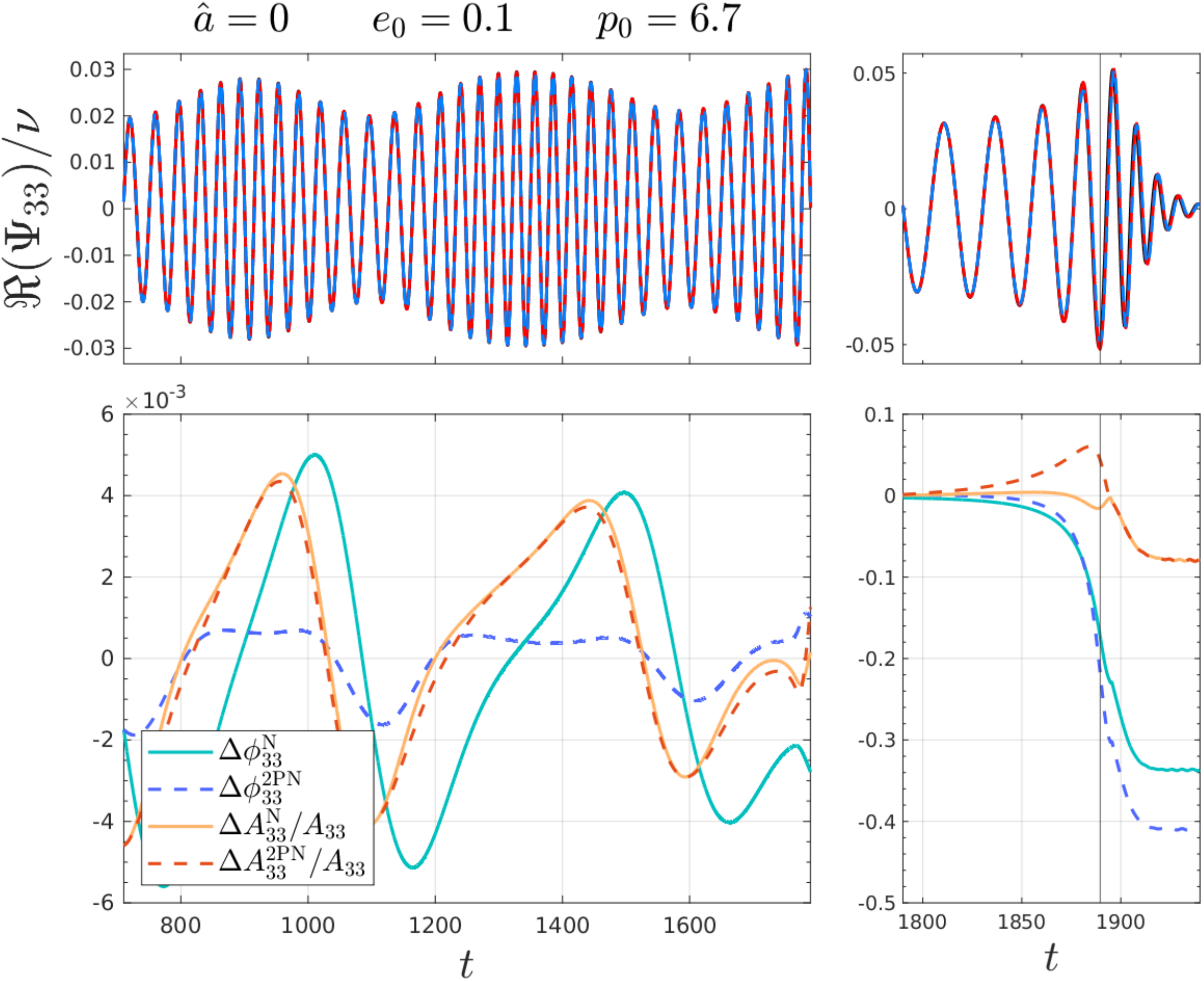}
		\hspace{0.2cm}
		\includegraphics[width=0.22\textwidth]{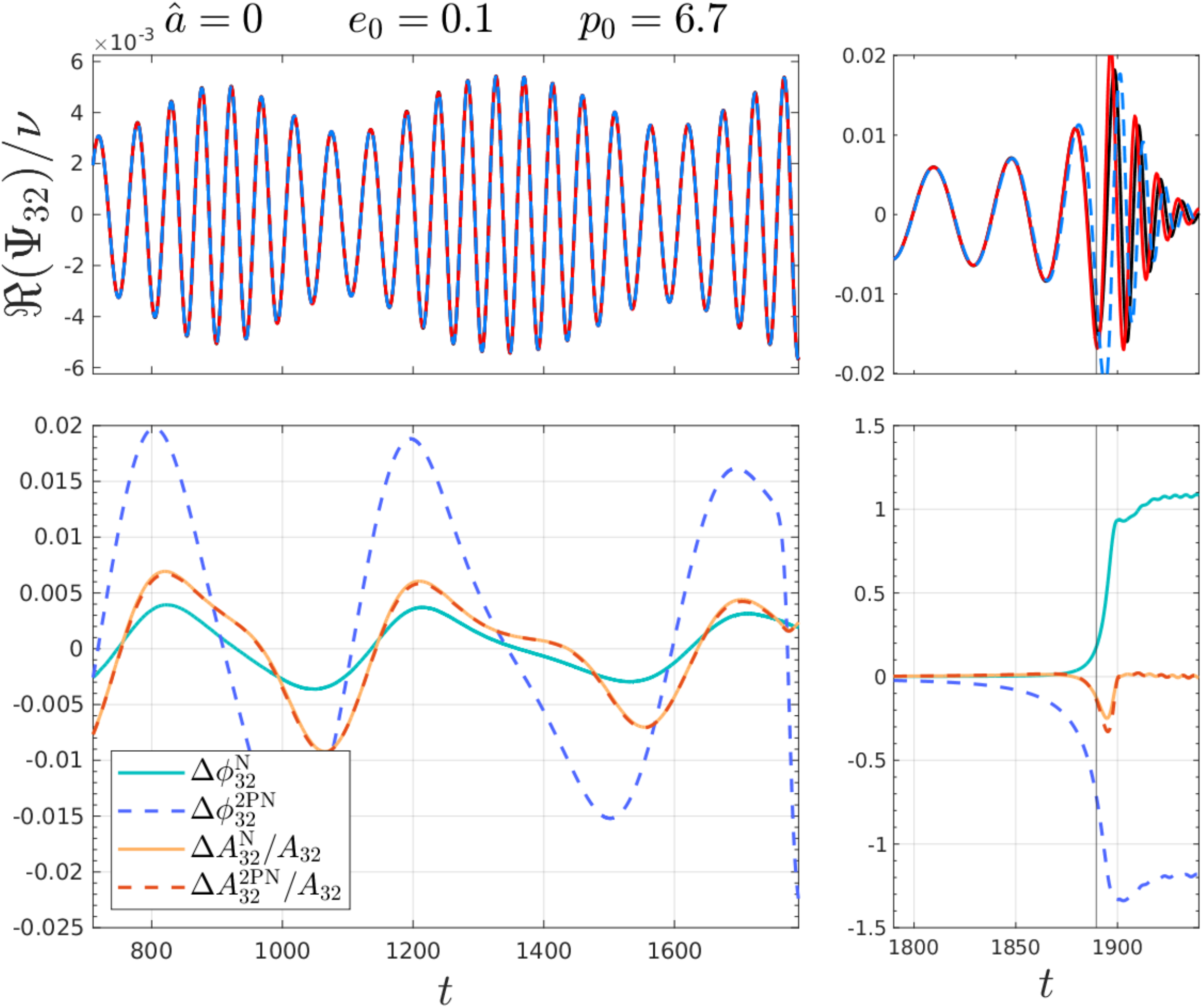}
		\hspace{0.2cm}
		\includegraphics[width=0.22\textwidth]{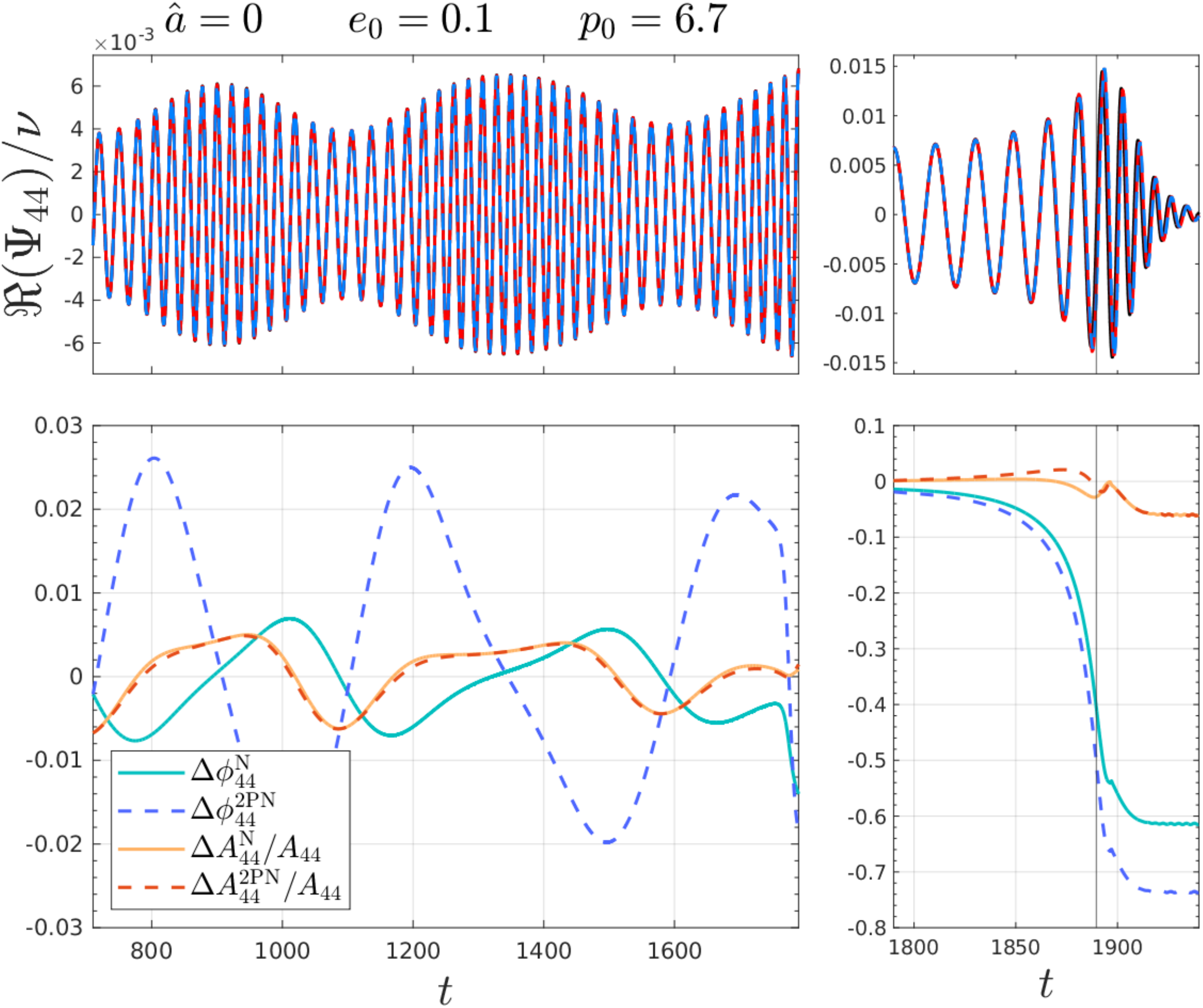}\\
		\vspace{0.3cm}
		\includegraphics[width=0.22\textwidth]{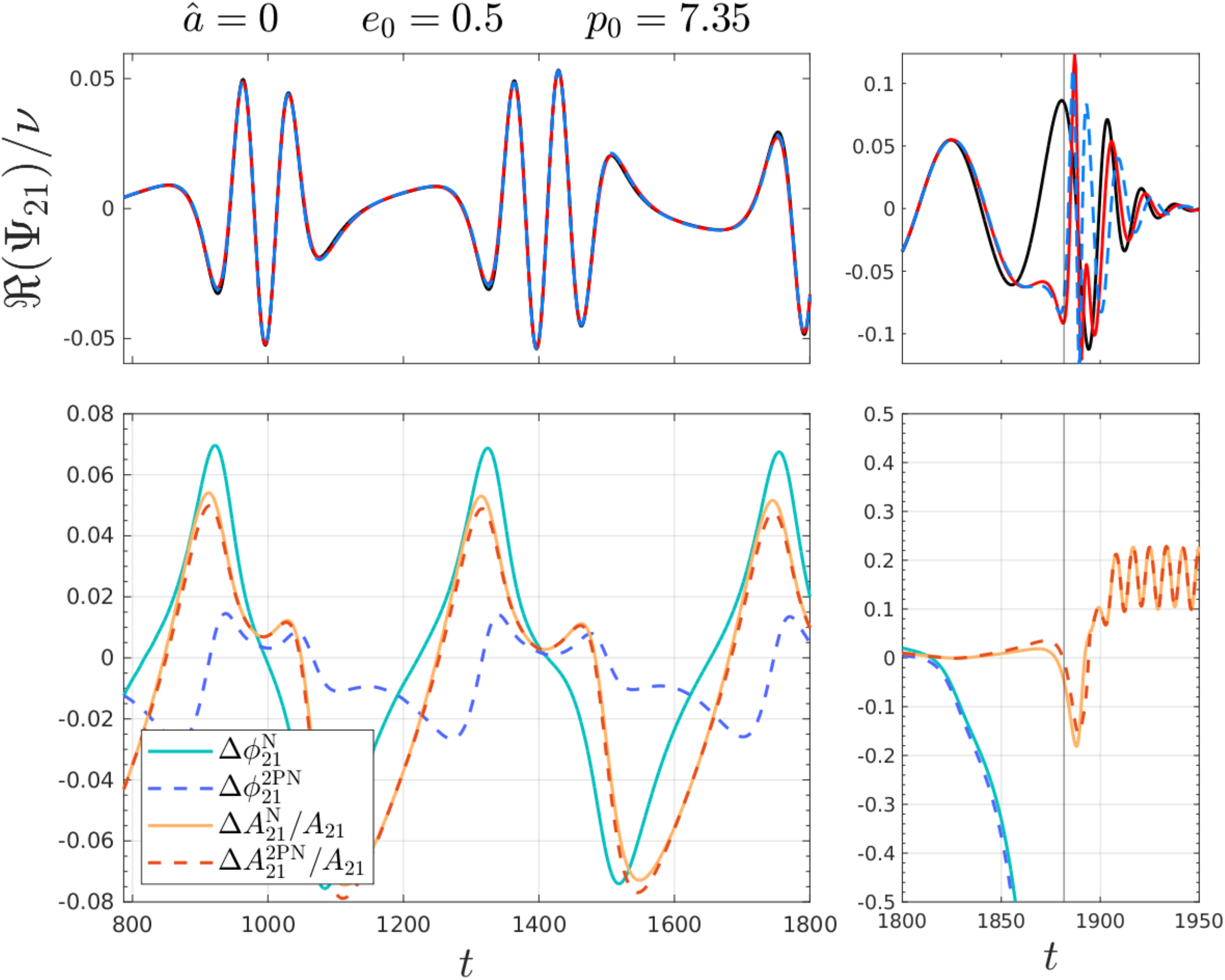}
		\hspace{0.2cm}
		\includegraphics[width=0.22\textwidth]{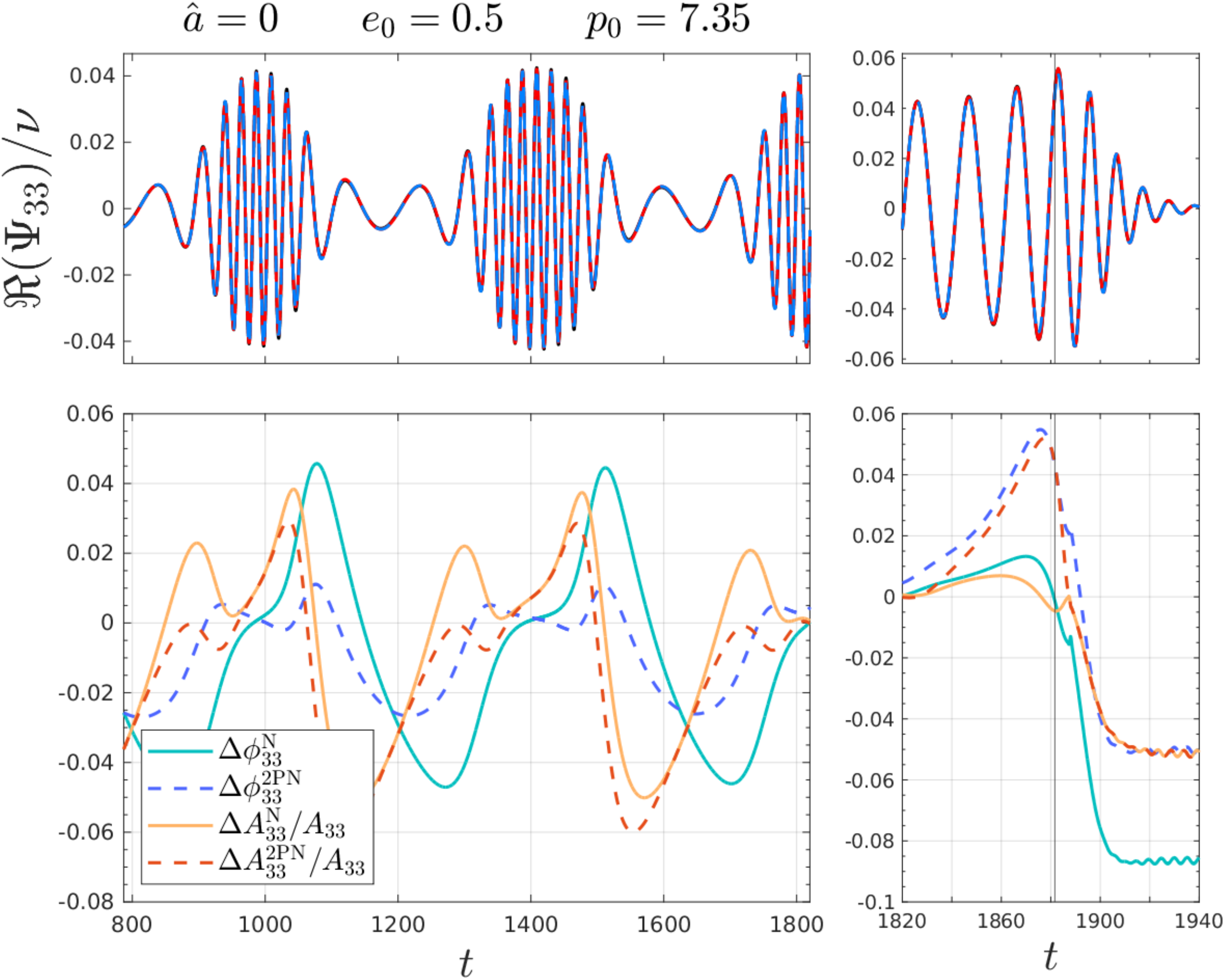}
		\hspace{0.2cm}
		\includegraphics[width=0.22\textwidth]{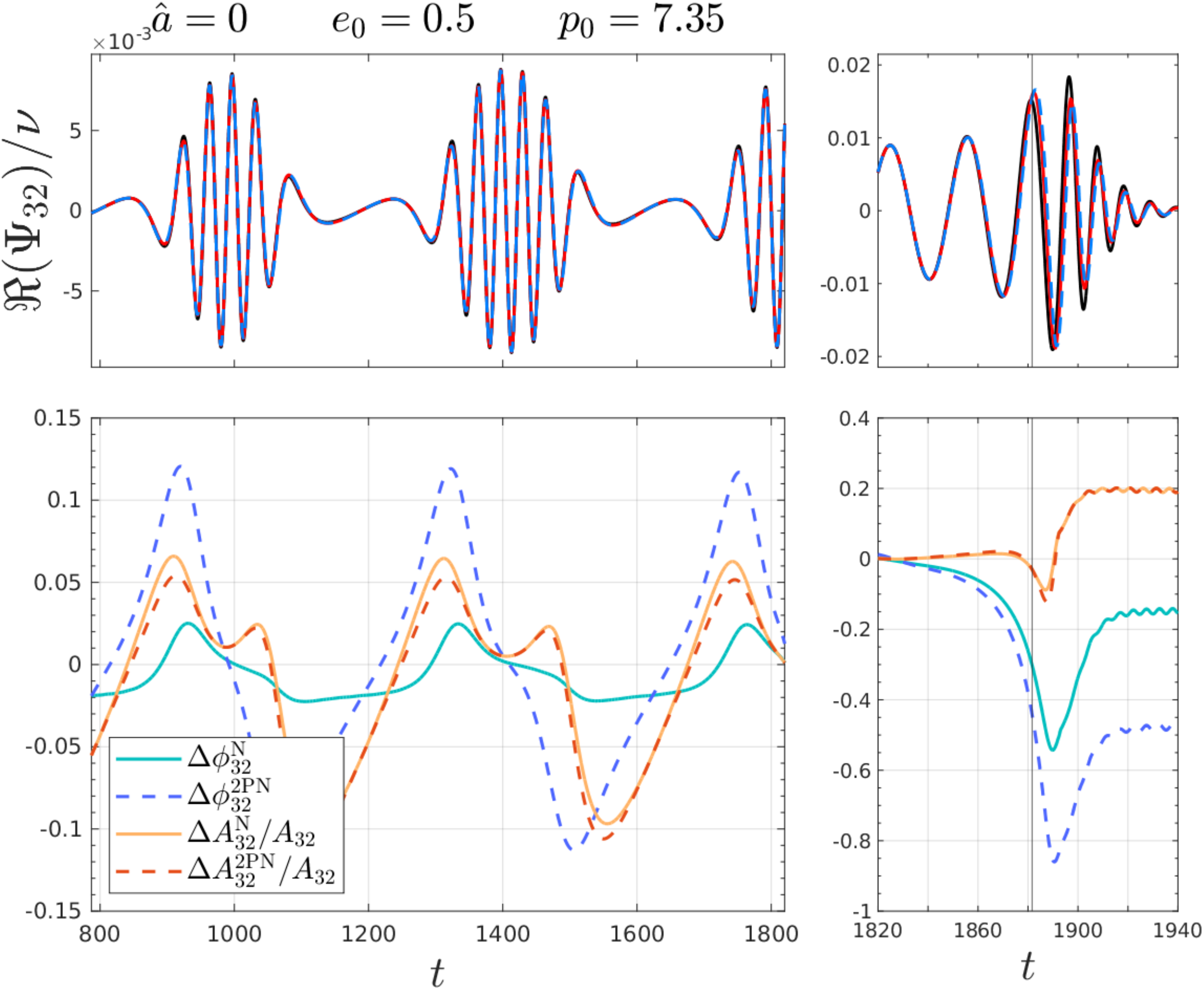}
		\hspace{0.2cm}
		\includegraphics[width=0.22\textwidth]{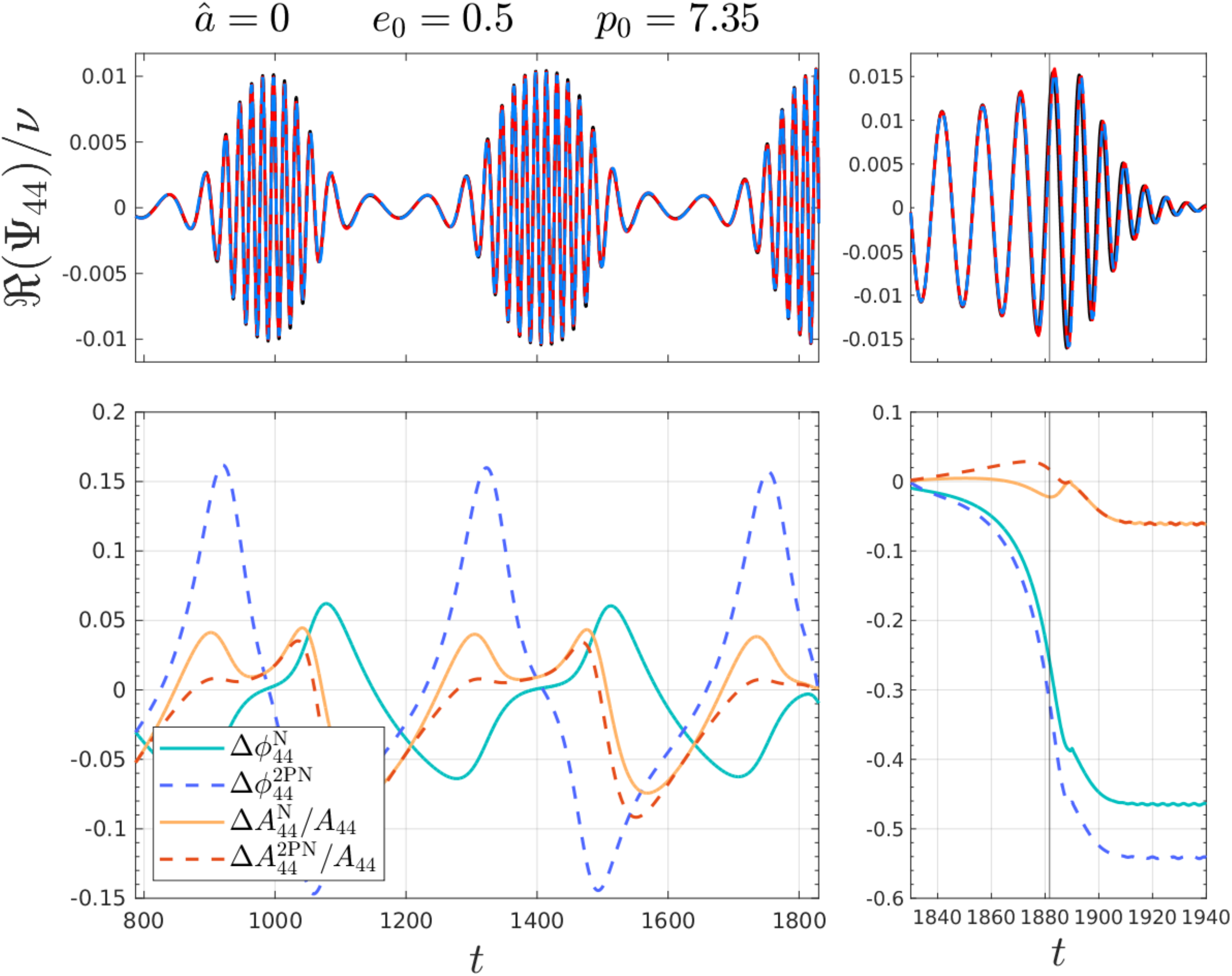}\\
		\vspace{0.3cm}
		\includegraphics[width=0.22\textwidth]{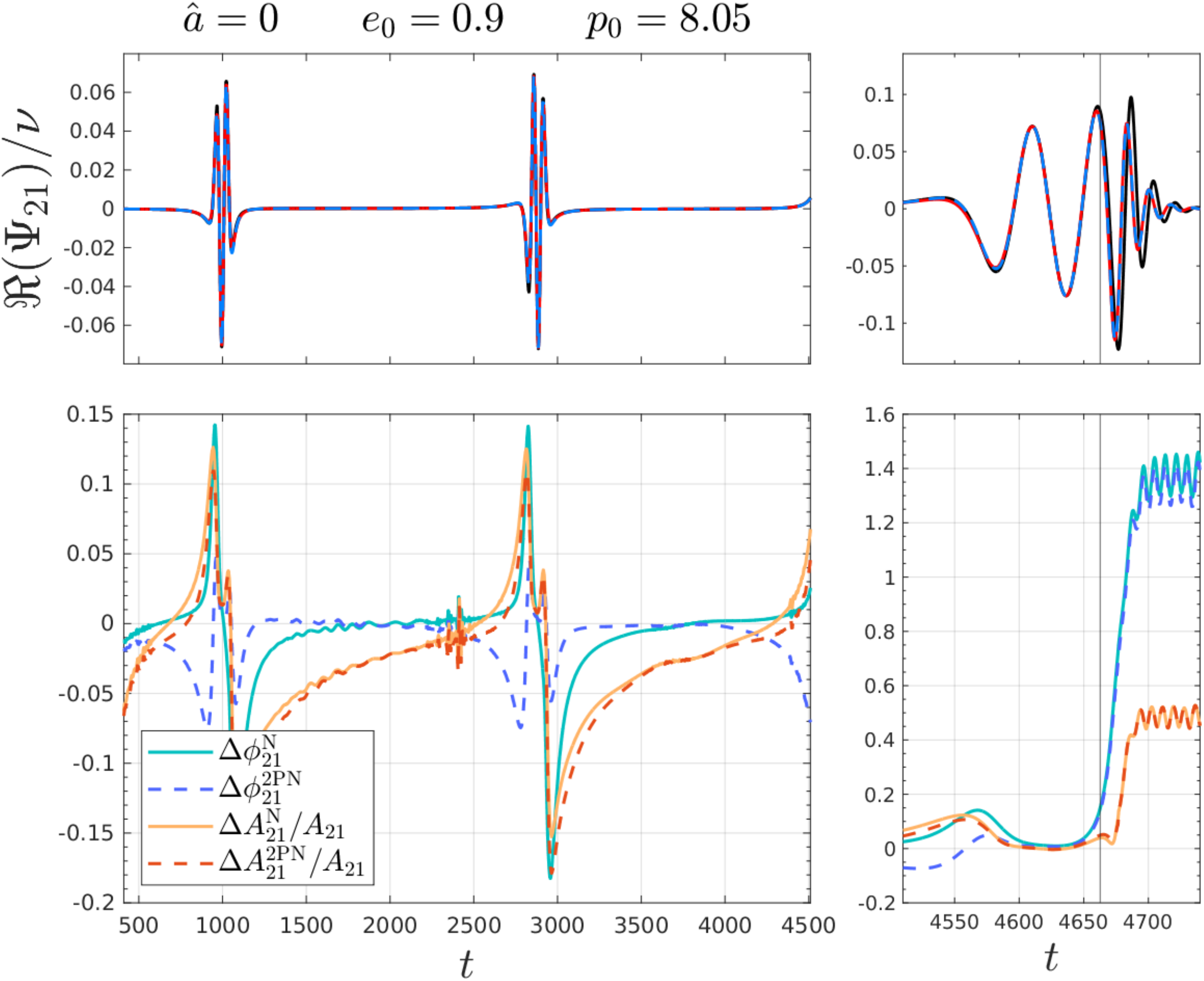}
		\hspace{0.2cm}
		\includegraphics[width=0.22\textwidth]{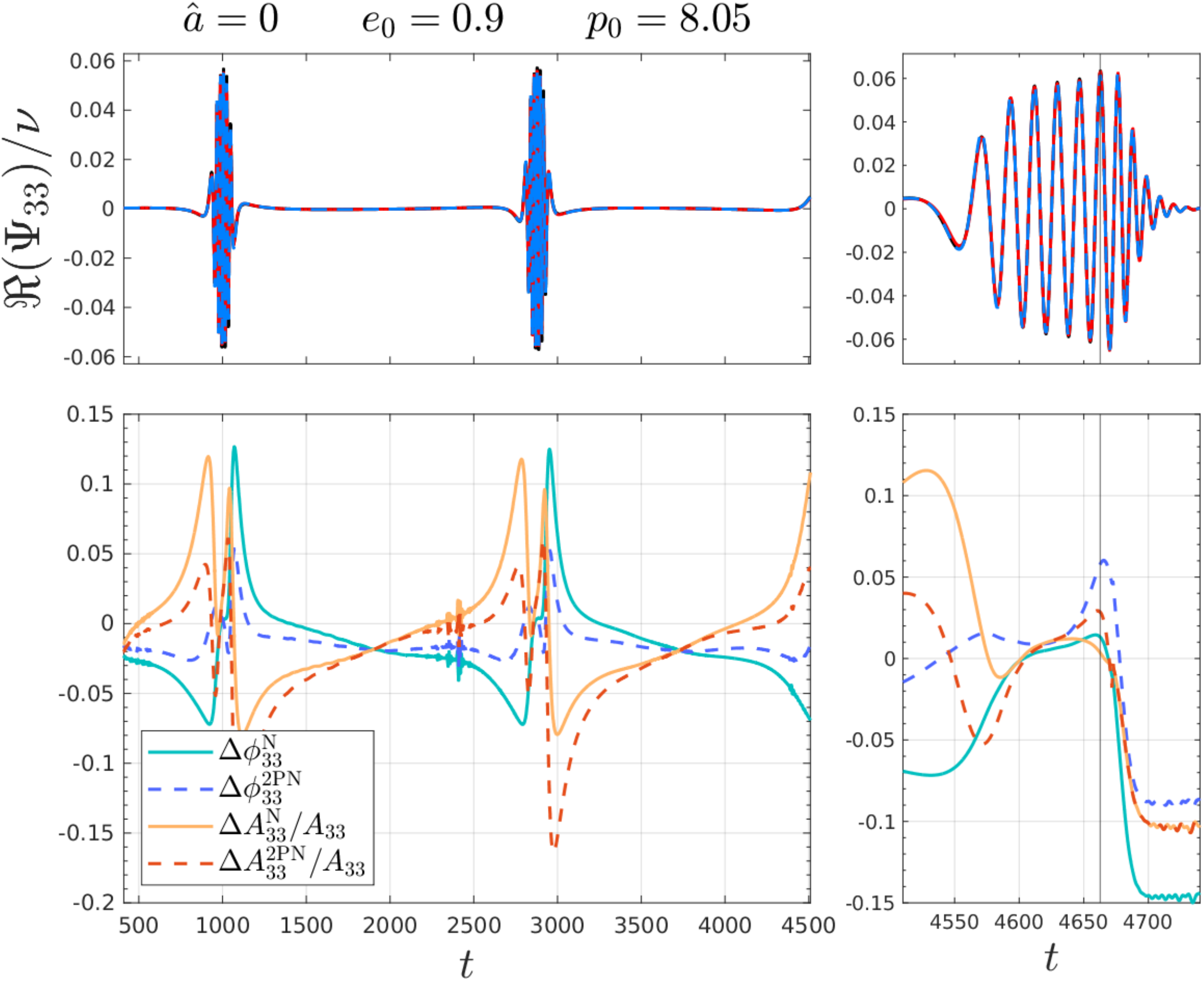}
		\hspace{0.2cm}
		\includegraphics[width=0.22\textwidth]{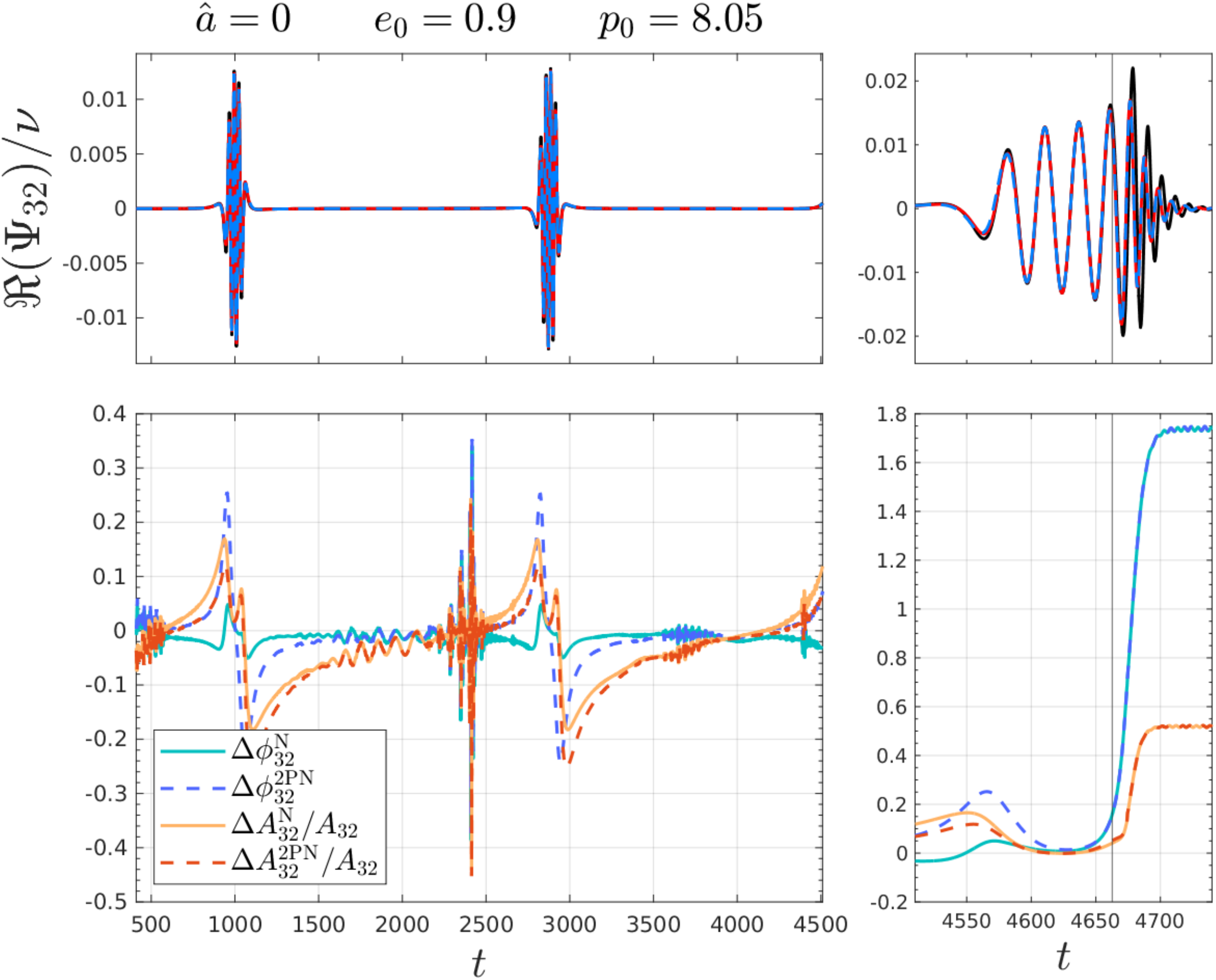}
		\hspace{0.2cm}
		\includegraphics[width=0.22\textwidth]{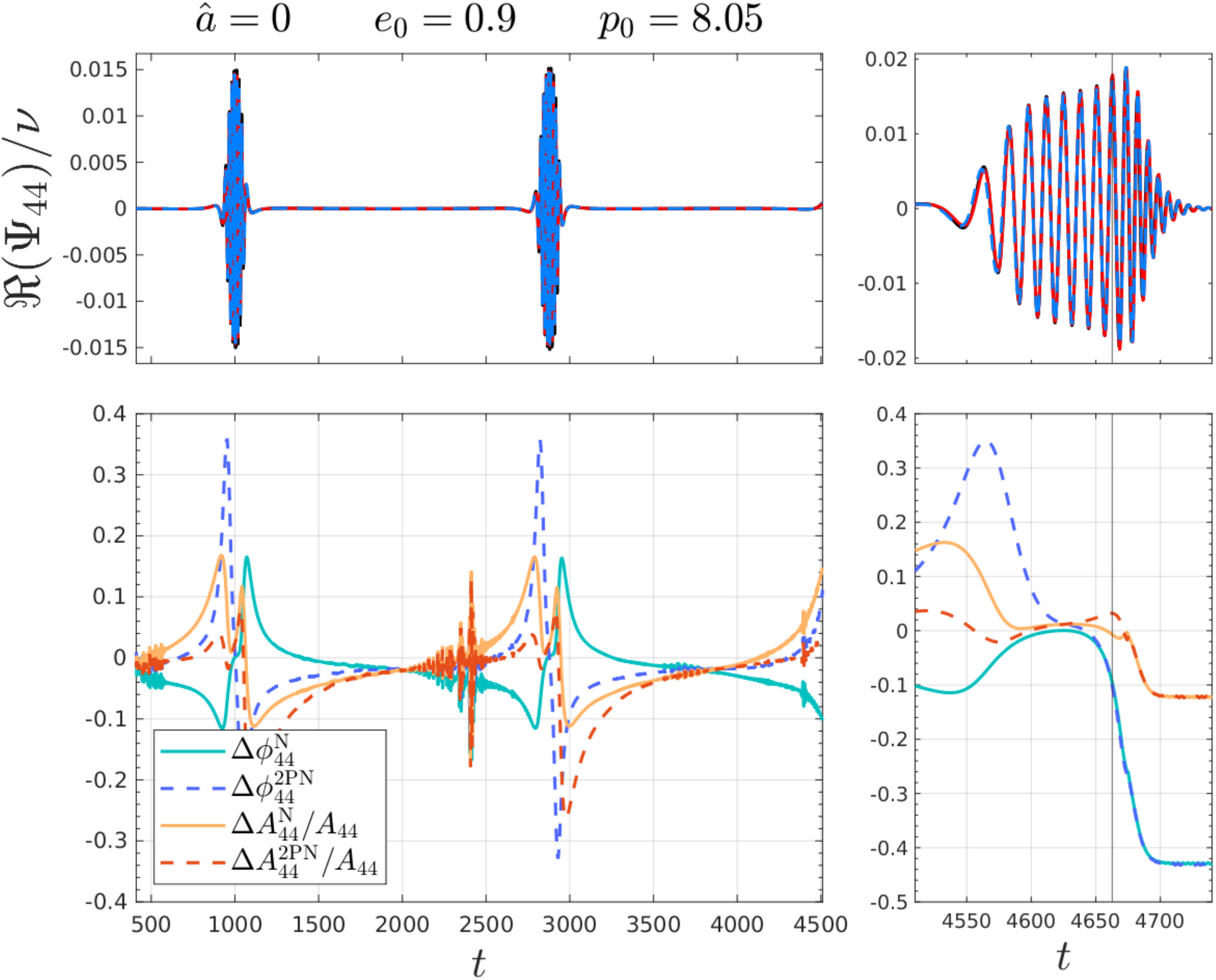}\\
		\vspace{0.3cm}
		\includegraphics[width=0.22\textwidth]{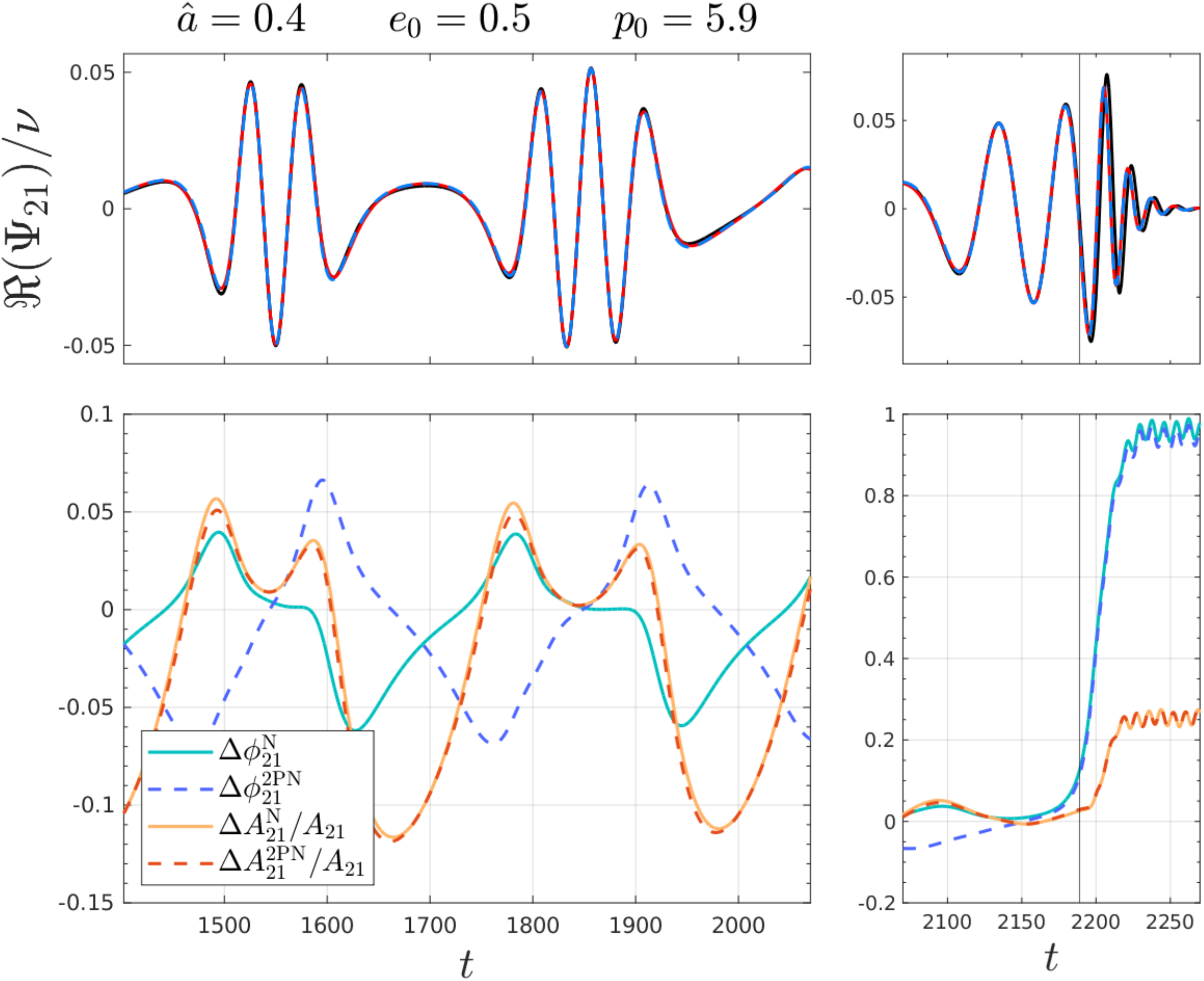}
		\hspace{0.2cm}
		\includegraphics[width=0.22\textwidth]{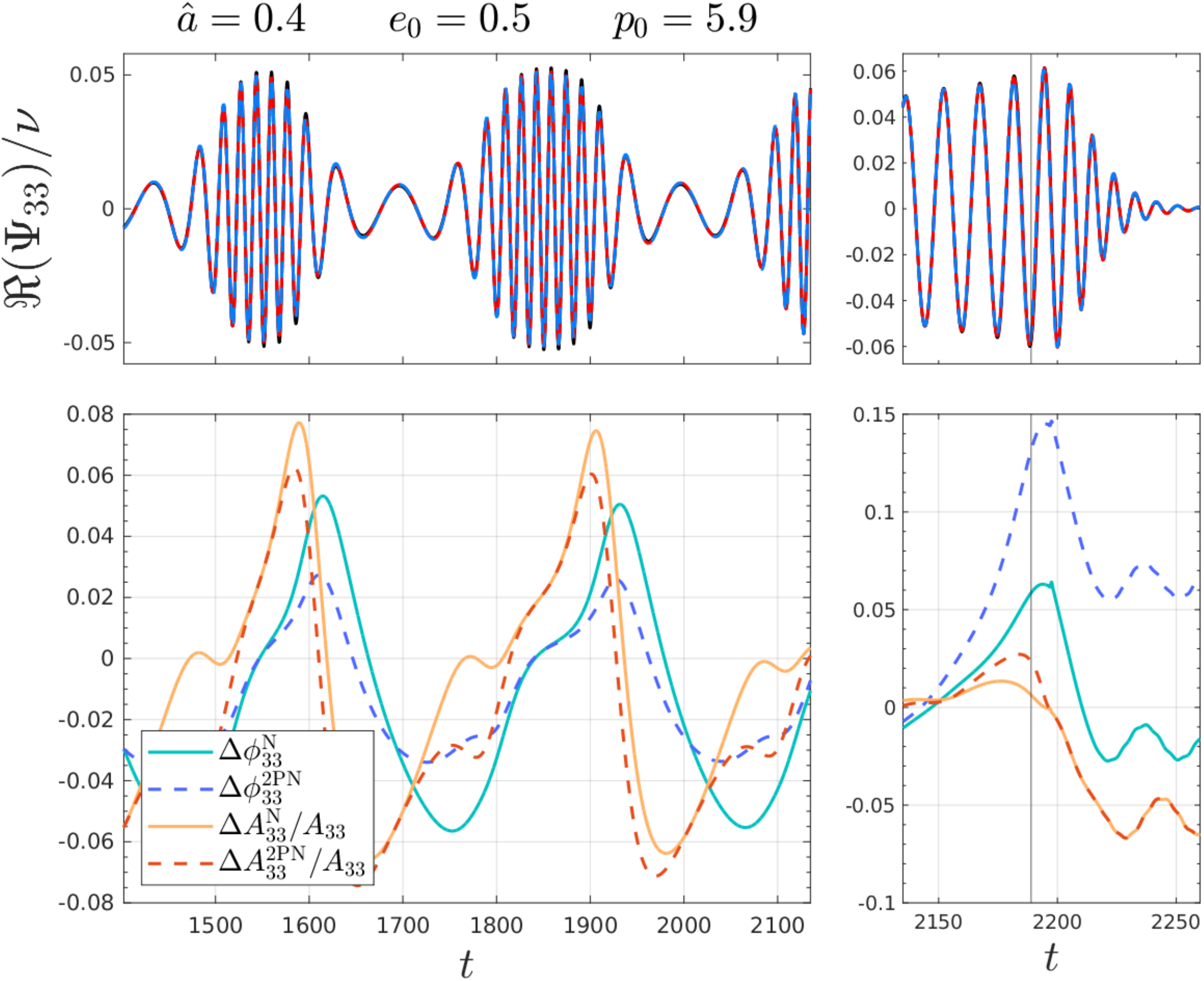}
		\hspace{0.2cm}
		\includegraphics[width=0.22\textwidth]{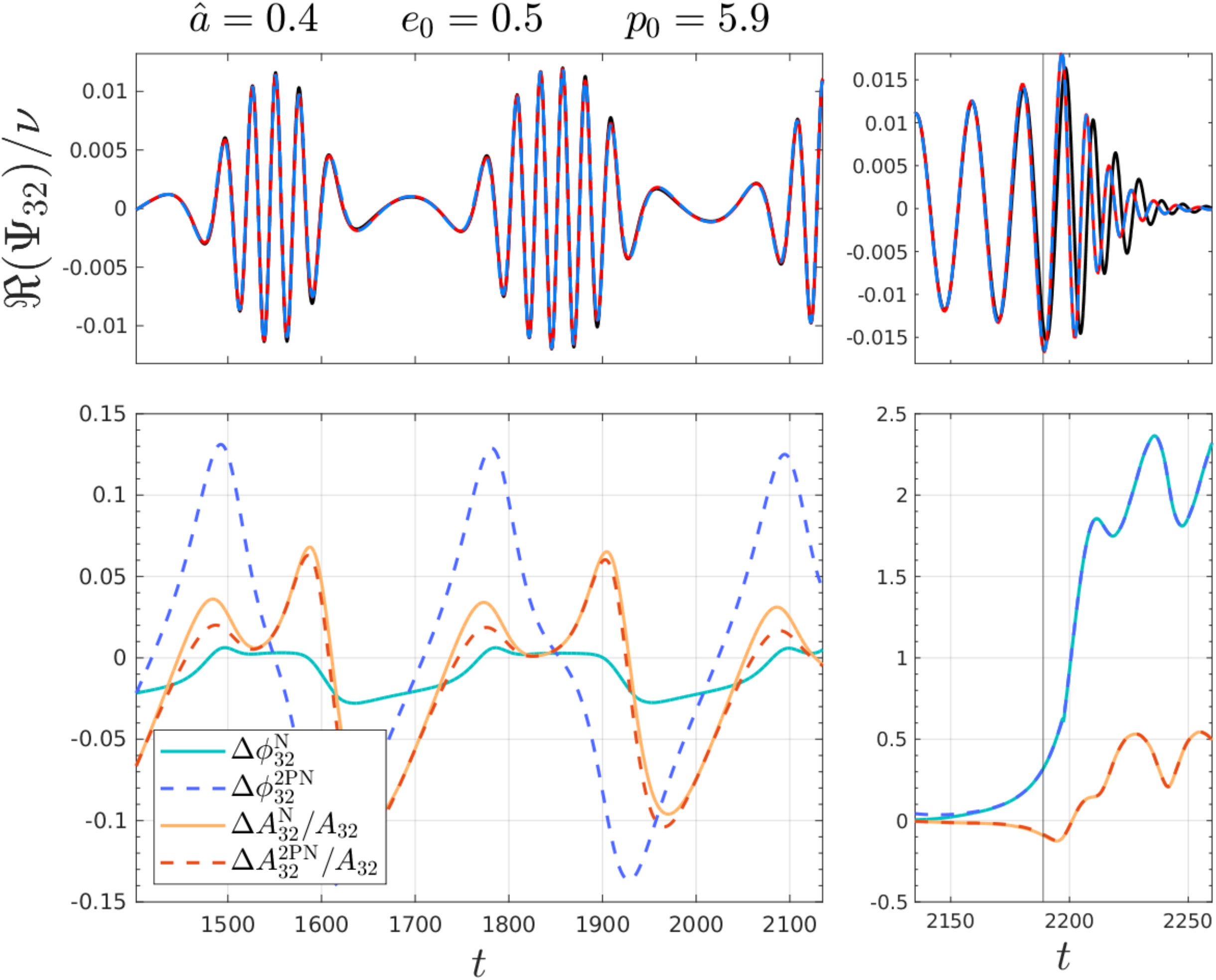}
		\hspace{0.2cm}
		\includegraphics[width=0.22\textwidth]{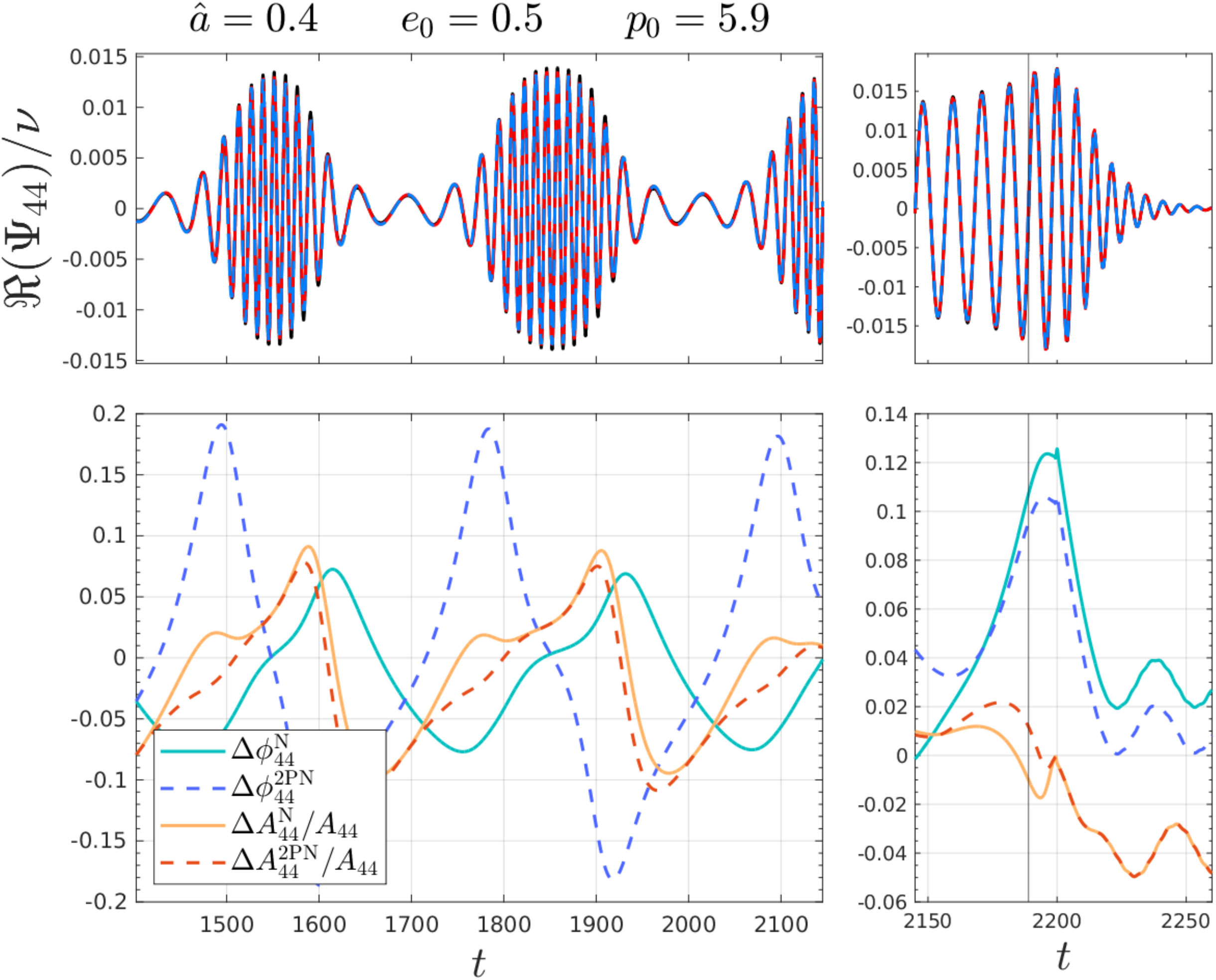}
		\caption{\label{fig:testmass_inspl_subdominant}
			Same color scheme of Fig.~\ref{fig:testmass_inspl_resum_tail}, but here in each row we 
			consider the subdominant modes (2,1), (3,3), (3,2), and (4,4) for the configurations
			$(e_0, \ha, p_0) = (0.1, 0, 6.7), (0.5, 0, 7.35), (0.9, 0, 8.05), (0.5, 0.4, 5.9)$.}
	\end{figure*}
	%%======
	\begin{align}
		\hat{t}^{21}_{p_{r_*}} = 1 &+\frac{6035 y}{3029}-\frac{10870 y^2}{3029}+\frac{8350 y^3}{3029}-\frac{3215 y^4}{3029} \cr
		&+\frac{511 y^5}{3029},\\
		\hat{t}^{21}_{p_{r_*}^2} = 1&-\frac{1388 y}{635}+\frac{666 y^2}{635}-\frac{92 y^3}{635}-\frac{13 y^4}{635},\\
		\hat{t}^{21}_{p_{r_*}^3}  = 1 &-\frac{981 y}{619}+\frac{573 y^2}{619}-\frac{115 y^3}{619},\\
		\hat{t}^{21}_{p_{r_*}^4} = 1 &-\frac{82 y}{183} - \frac{17 y^2}{183}, \\ \nonumber \\
		\hat{t}^{33}_{p_{r_*}} = 1 & +\frac{67183 y}{23815}+\frac{721737 y^2}{47630}-\frac{85973 y^3}{9526}\cr
		&+\frac{30812 y^4}{4763}-\frac{10722 y^5}{4763}+\frac{16769 y^6}{47630}\cr
		&-\frac{337 y^7}{47630},\\
		\hat{t}^{33}_{p_{r_*}^3} = 1 & + \frac{3125521 y}{762080} + \frac{5675333 y^2}{762080} -\frac{2623521 y^3}{69280} \nonumber \\ 
		&  +\frac{3513777 y^4}{95260} -\frac{2943211 y^5}{152416}+ \frac{7128059 y^6}{762080} 
		\nonumber \\
		&  -\frac{2725303 y^7}{762080} +\frac{35247 y^8}{95260} +  \frac{5233 y^9}{95260}, \\
		\hat{t}^{33}_{p_{r_*}^2} = 1 &+\frac{1407963 y}{381040}-\frac{343943 y^2}{47630}+\frac{2036583 y^3}{95260}\cr
		&-\frac{271775 y^4}{19052}+\frac{602219 y^5}{76208}-\frac{268357 y^6}{95260}\cr
		&+\frac{85037 y^7}{190520}+\frac{918 y^8}{23815},\\
		\hat{t}^{33}_{p_{r_*}^4}= 1 & +\frac{858779 y}{190520}+\frac{92901791y^2}{6096640} +\frac{66769 y^3}{7040} \nonumber \\
		& -\frac{94475723 y^4}{1219328}+\frac{9400891 y^5}{152416}-\frac{82481269y^6}{3048320} \nonumber \\
		&+\frac{17880961 y^7}{1524160}
		-\frac{6096411 y^8}{1524160}+\frac{2935 y^9}{13856}  \nonumber \\
		&+\frac{4009 y^{10}}{76208}.
	\end{align}
	
\begin{figure*}[t]
		\center
		\includegraphics[width=0.31\textwidth]{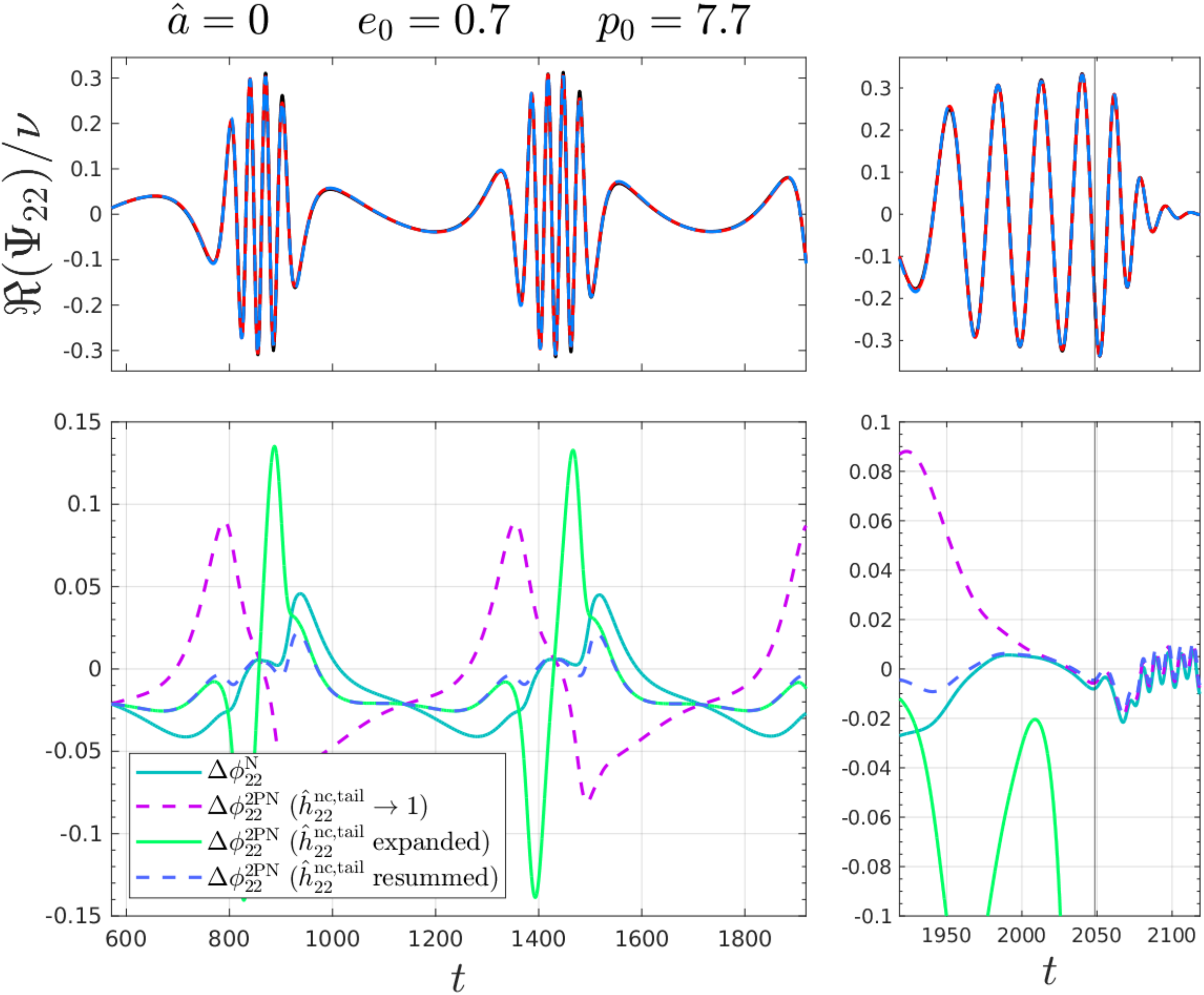}
		\hspace{0.2cm}
		\includegraphics[width=0.31\textwidth]{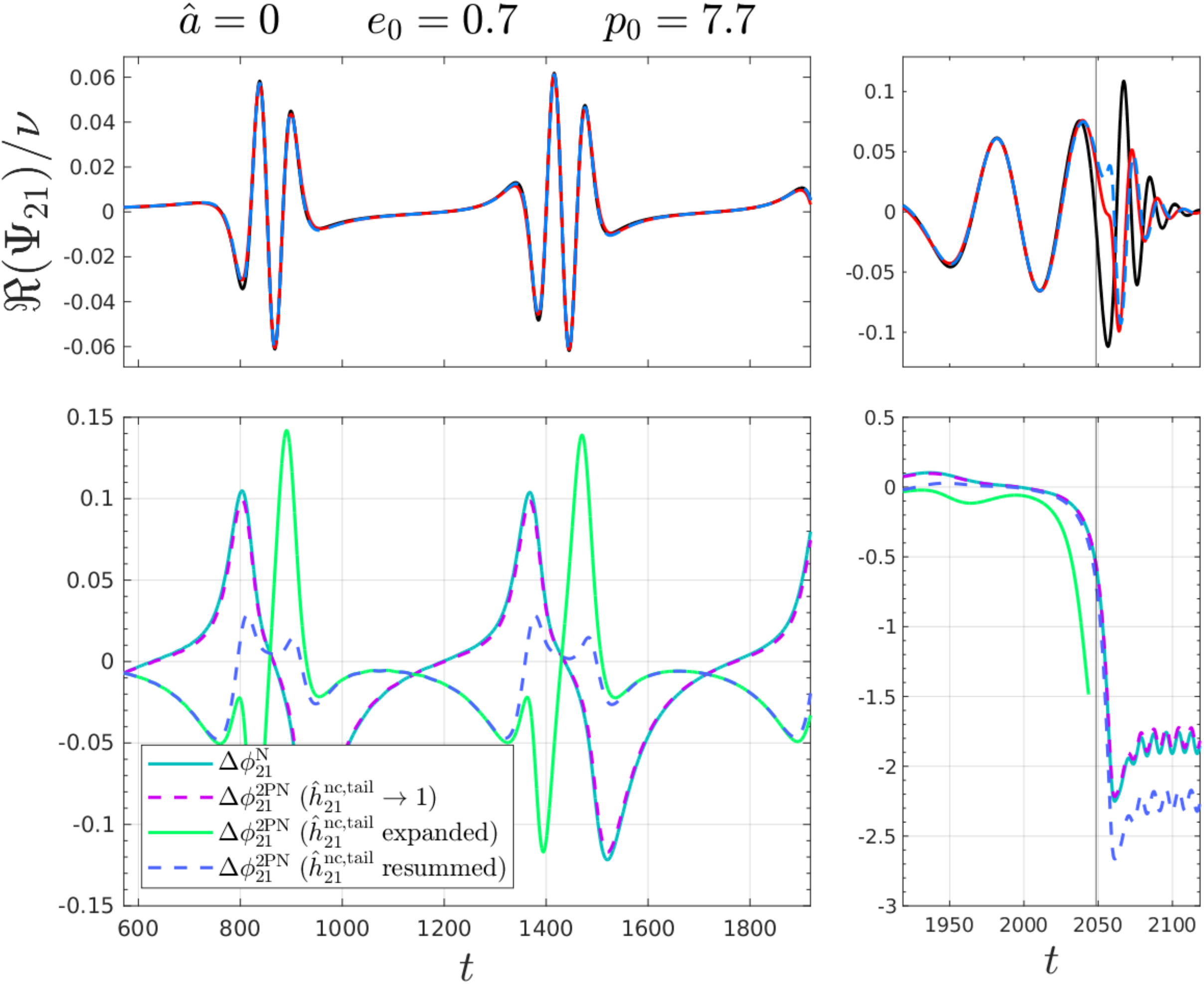}
		\hspace{0.2cm}
		\includegraphics[width=0.31\textwidth]{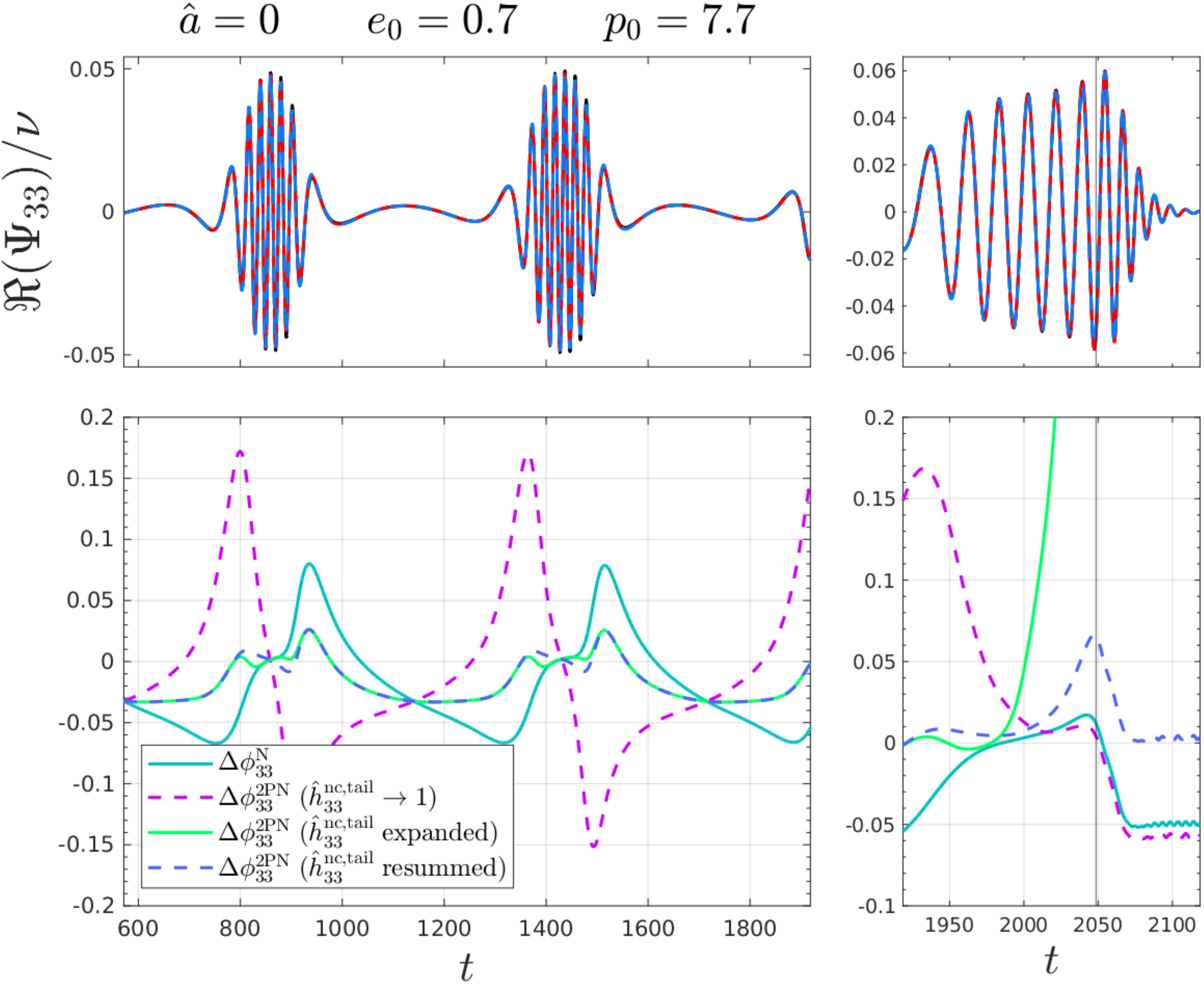}
		\caption{\label{fig:testmass_inspl_testnotail}
			Comparisons of the waveform modes (2,2), (2,1) and (3,3) for the nonspinning
			case $(e_0,p_0)=(0.7, 7.7)$. 
			Top panels: numerical waveforms (black), the waveforms with
			only Newtonian noncircular corrections (red online) 
			and the waveforms with 2PN noncircular corrections with
			the resummed eccentric tail factor (dashed blue).
			Bottom panels: analytical/numerical phase differences (in radians) for 
			different prescriptions: (i) 
			only Newtonian noncircular corrections (solid light blue) 
			(ii) only instantaneous noncircular corrections up to 2PN, i.e. without eccentric 
			tail (dashed purple) 
			(iii) waveform with 2PN noncircular corrections, 
			both instantaneous and hereditary in expanded form (solid aqua-green) 
			(iv) waveform with 2PN noncircular corrections, both instantaneous and hereditary, 
			with resummation applied to the latter (dashed blue).}
	\end{figure*}

	Each residual function is resummed using Pad\'e approximants. The choices we made are
	summarized in Table~\ref{tab:pade_submodes}. 
	\begin{table}[t]
		\caption{\label{tab:pade_submodes} Pad\'e used for the resummation of the 
			tail 2PN noncircular corrections for the modes  $(2,2)$, $(2,1)$ and $(3,3)$.
			Note that $\hat{t}^{33}_{p_{r_*}^3}$ has terms up to $y^8$, but we use the  
			$(3,2)$ Pad\'e. }
		\begin{center}
			\begin{ruledtabular}
				\begin{tabular}{ c | c c c c } 
					& \multicolumn{4}{c}{Pad\'e} \\
					\hline
					$(\ell, m)$ & $\hat{t}^{\lm}_{p_{r_*}}$ & $ \hat{t}^{\lm}
					_{p_{r_*}^2} $ & $\hat{t}^{\lm}_{p_{r_*}^3} $ & $ \hat{t}^{\lm}_{p_{r_*}^4}$ \\
					\hline
					$ (2,2)$ & $(4,3)$ & $(4,3)$ & $(4,4)$ & $(4,4)$ \\
					$ (2,1)$ & $(2,3)$ & $(1,3)$ & $(0,3)$ & $(2,0)$ \\
					$ (3,3) $ & $(4,3)$ & $(5,4)$ & $(3,2)$ & $(5,5)$ \\
				\end{tabular}
			\end{ruledtabular}
		\end{center}
	\end{table}
	The modes $(3,2)$ and $(4,4)$ are shown in Figure~\ref{fig:testmass_inspl_subdominant} 
	which compares analytical with numerical waveform for an illustrative, but significative, 
	set of configurations.
	We start by noticing that the  merger-ringdown waveform, especially for the mode  $(2,1)$, is more 
	accurate for high eccentricity than for small eccentricity. 
	This is related to the (yet unpublished) NQC/ringdown fit that we are using here
	and that will be presented in a future work. The behavior during merger/ringdown of this test-mass EOB
	model should be considered preliminary and will undergo further improvements.
	When analyzing the inspiral phase, a few comments are in order. 
	First, the phase and amplitude agreement during the inspiral phase for the modes $(2,1)$ and $(3,3)$ 
	is comparable to the $(2,2)$ mode, and the 2PN corrections are found to yield 
	a notable reduction of the analytical/numerical phase difference with respect to the simple 
	Newtonian prefactor. This is true for any value of the eccentricity considered.
	When moving to the modes $(3,2)$ and $(4,4)$ one faces instead the rather surprising 
	fact that the PN-corrected waveform performs {\it worse} than the leading-order one.
	We understand this result as due to the fact that these modes do not have a tail factor at 2PN order, in contrast to what happens for the modes $(2,2)$,  $(3,3)$ and $(2,1)$ where the aforementioned compensation between instantaneous and tail part can take place.  An illustration of this effect for initial
	eccentricity $e_0=0.7$ can be found in Fig.~\ref{fig:testmass_inspl_testnotail}. 
	Qualitatively,  with greater accuracy the same behavior should be found also for the modes $(3,2)$ and $(4,4)$. 
	Future work, that aims at incorporating all noncircular corrections up to 3PN in 
	our factorized and resummed waveform, will hopefully clarify these issues~\cite{placidi:2022}.
	%==========================================
	% Fig.10: test no tail
	%==========================================
	
	%==========================================
	% Fig.09: l2m0, geodesic
	%==========================================
	\begin{figure}[t]
		\center
		\includegraphics[width=0.23\textwidth]{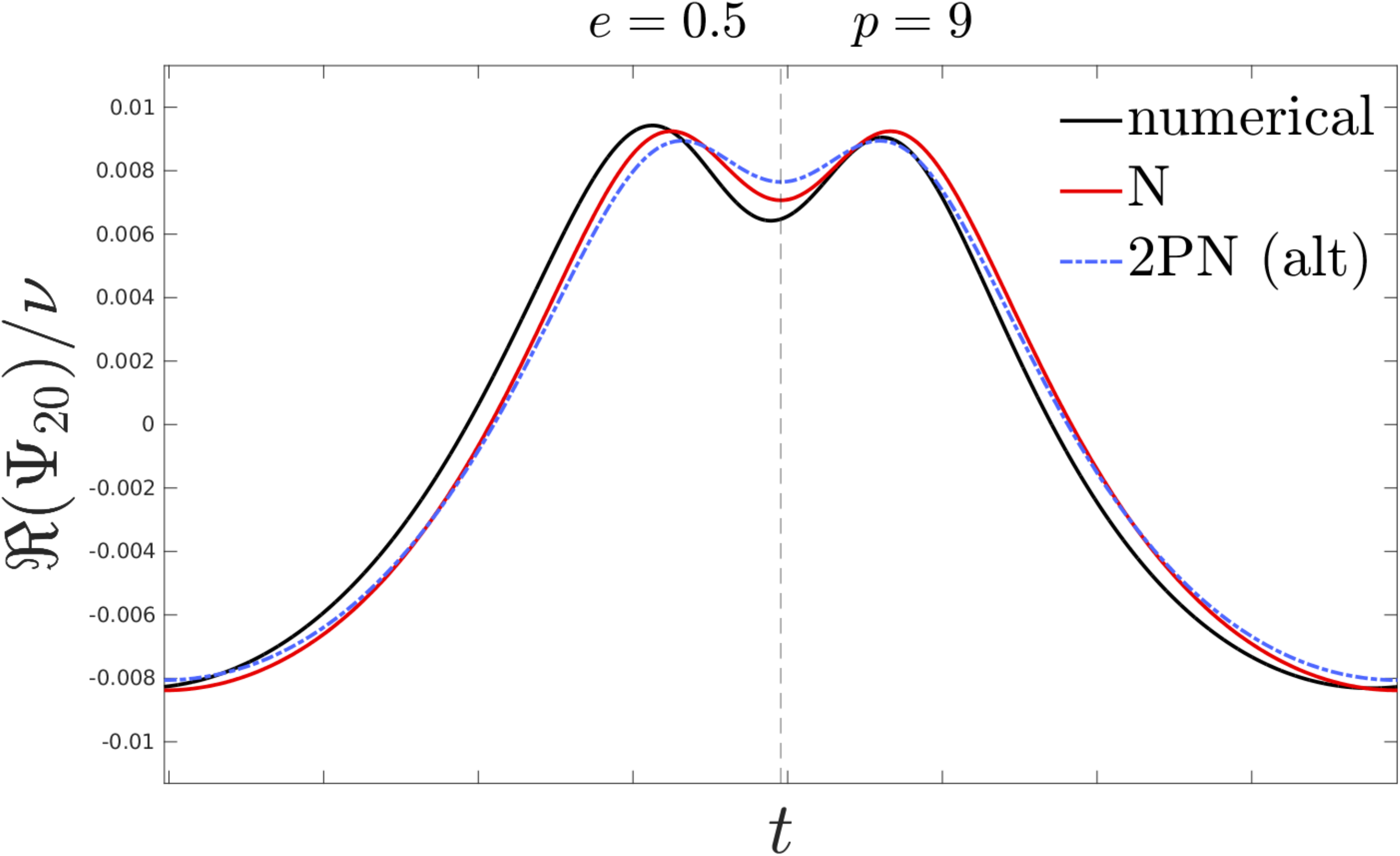}
		\includegraphics[width=0.23\textwidth]{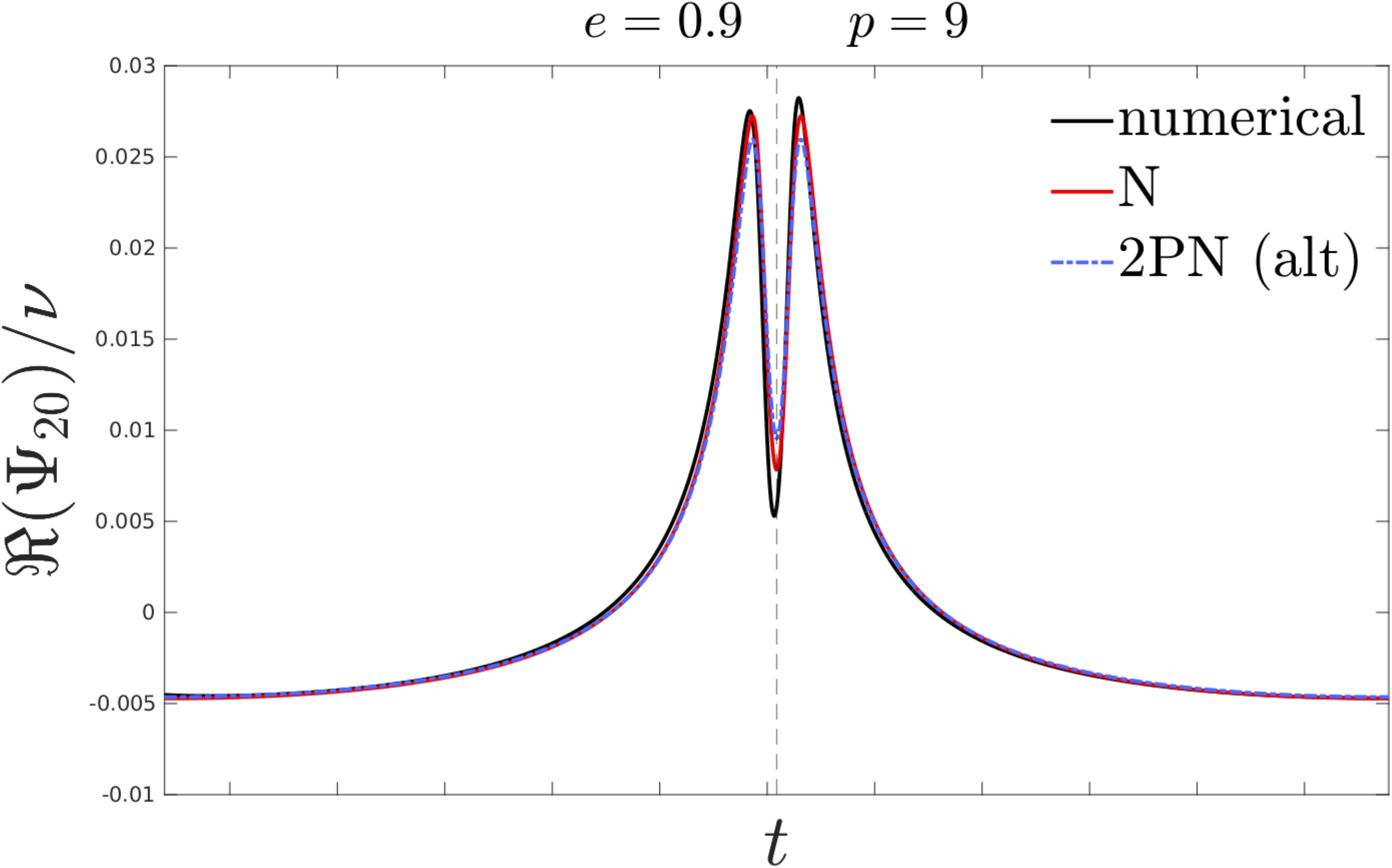}\\
		\vspace{0.2cm}
		\includegraphics[width=0.23\textwidth]{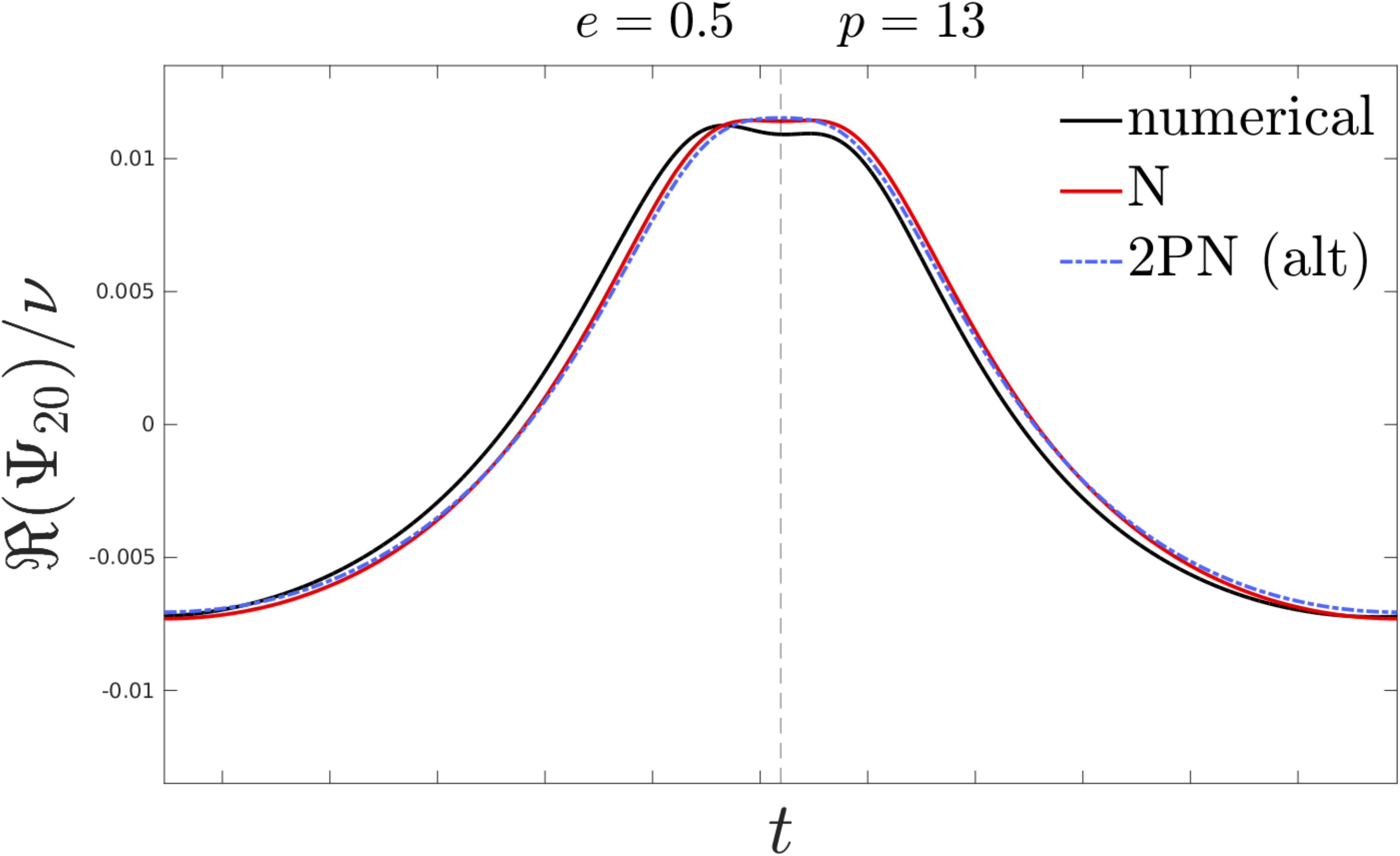}
		\includegraphics[width=0.23\textwidth]{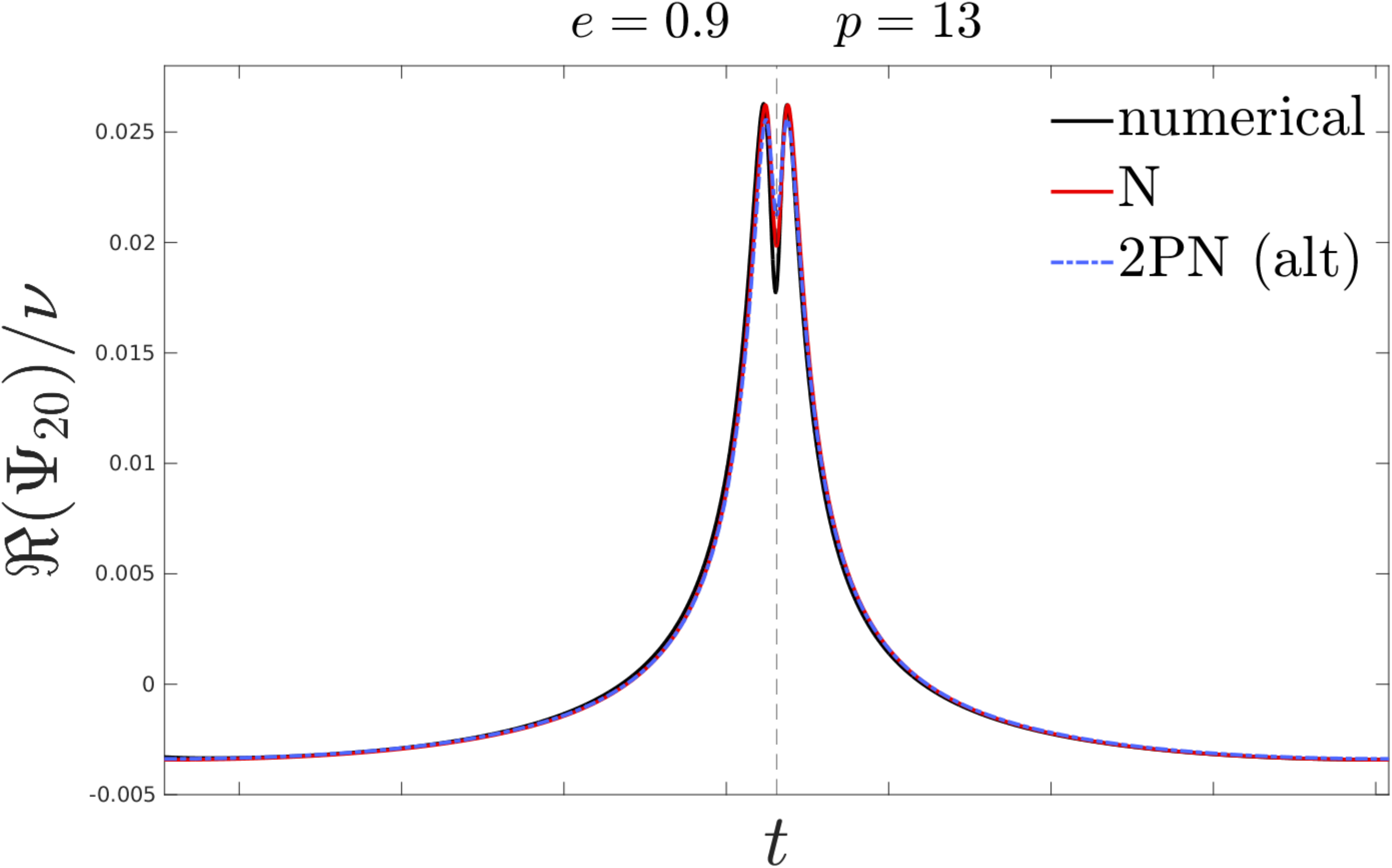}
		\caption{\label{fig:l2m0} Comparisons for the mode $(2,0)$ on nonspinning geodesic orbits 
			with $e=(0.5, 0.9)$ and $p=(9,13)$. We show the numerical waveform (black), the EOB
			waveform with noncircular corrections only at Newtonian level (red online) and with corrections
			at 2PN as discussed in Sec.~\ref{Sec:Alt_PN_corrections} (dashed blue).}
	\end{figure}

	\subsubsection{Multipole $\ell=2$, $m=0$}
	In Sec.~\ref{Sec:Alt_PN_corrections} we have already pointed out that we have to 
	use an alternative factorization for the modes with $m=0$. 
	In Fig.~\ref{fig:l2m0} we test the factorization proposed in Eq.~\eqref{eq:hlm_fact} 
	for different geodesic configurations in Schwarzschild. As it can be seen, the agreement between 
	numerical and analytical results is still qualitatively good, even if the other analytical
	modes discussed above are clearly more accurate 
	(both with only Newtonian and 2PN noncircular corrections). Here a source of disagreement
	is that the asymmetry of the $m=0$ numerical modes with respect to the apastron is not 
	negligible, even in the geodesic case. In any case, for the $m=0$ modes the 2PN corrections 
	do not seem to improve the analytical waveform with only the generic Newtonian prefactor.

	\section{Waveform validation: dynamical capture in the large mass ratio limit}
	\label{sec:testmass_hyp}
	
	\begin{table}
		\caption{\label{tab:Teukode_hyp} Hyperbolic capture configurations in the large mass ratio
			limit considered in this work. The symmetric mass ratio used to drive the dynamics
			is $\nu=10^{-2}$. 
			For each configuration we report the Kerr dimensionless 
			spin parameter $\ha$,  the initial energy $E_0$, the initial angular momentum $p_{\varphi,0}$, 
			the initial separation $r_0$, the number of peaks of the orbital frequency $N_{\Omega}^{\rm peaks}$, 
			and the merger time $t_{\rm mrg}$.  }
		\begin{center}
			\begin{ruledtabular}
				\begin{tabular}{c c c c c c} 
					$\ha$ & $E_0$ & $p_{\varphi,0}$ & $r_0$ & $N_{\Omega}^{\rm peaks}$ & $t_{\rm mrg}$  \\
					\hline
					\hline
					0 & 1.000711 & 4.01 & 120 & 2 & 2133\\
					0 & 1.000712 & 4.01 & 120 & 1 &  819 \\
					0 & 1.001240 & 4.01 & 120 & 1 &  731
				\end{tabular}
			\end{ruledtabular}
		\end{center}
		
	\end{table}
	
	%==========================================
	% Fig.11: dynamical captures
	%==========================================
	\begin{figure*}
		\center
		\hspace{0.4cm}
		\includegraphics[width=0.25\textwidth]{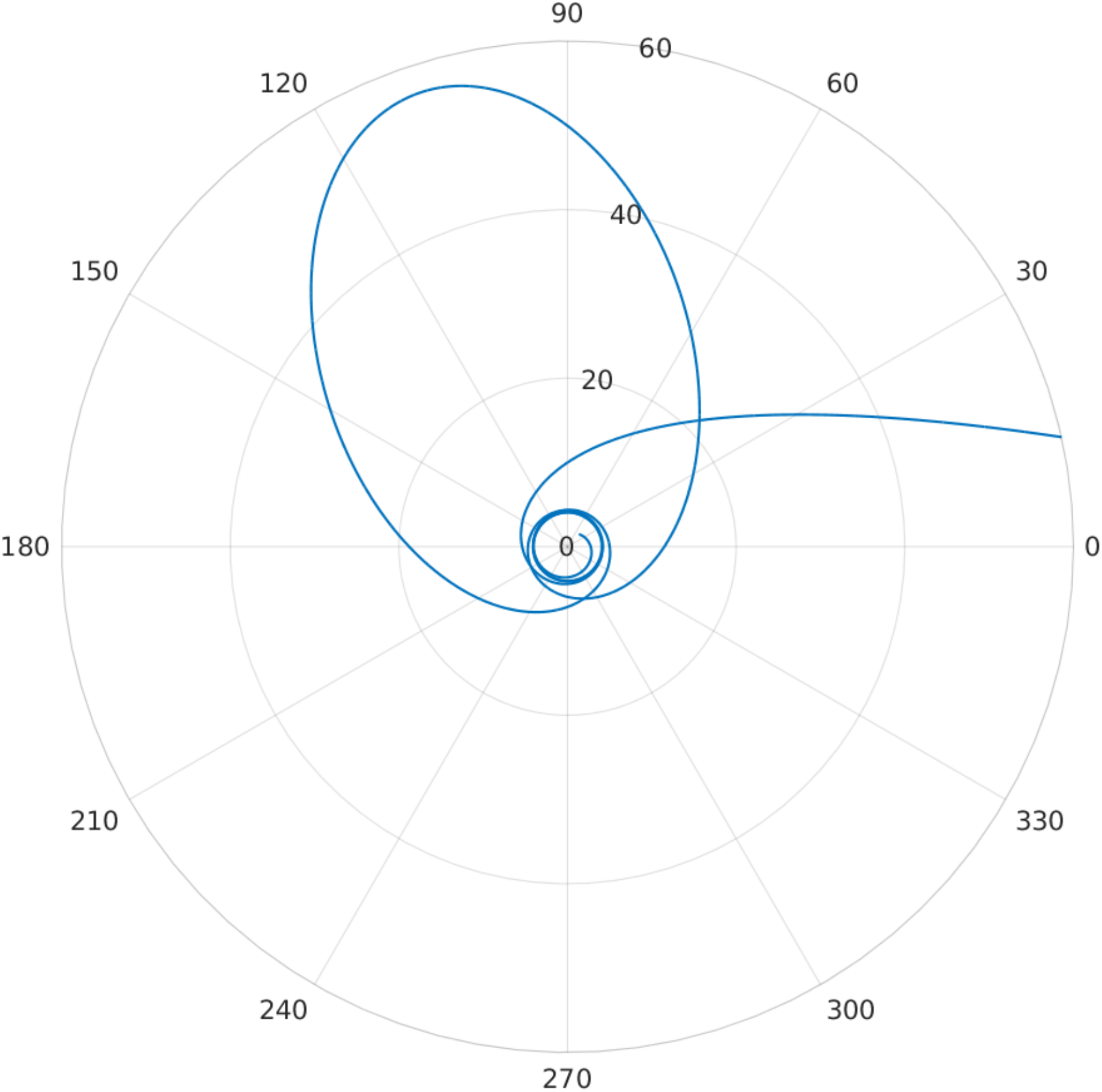}
		\hspace{1.3cm}
		\includegraphics[width=0.25\textwidth]{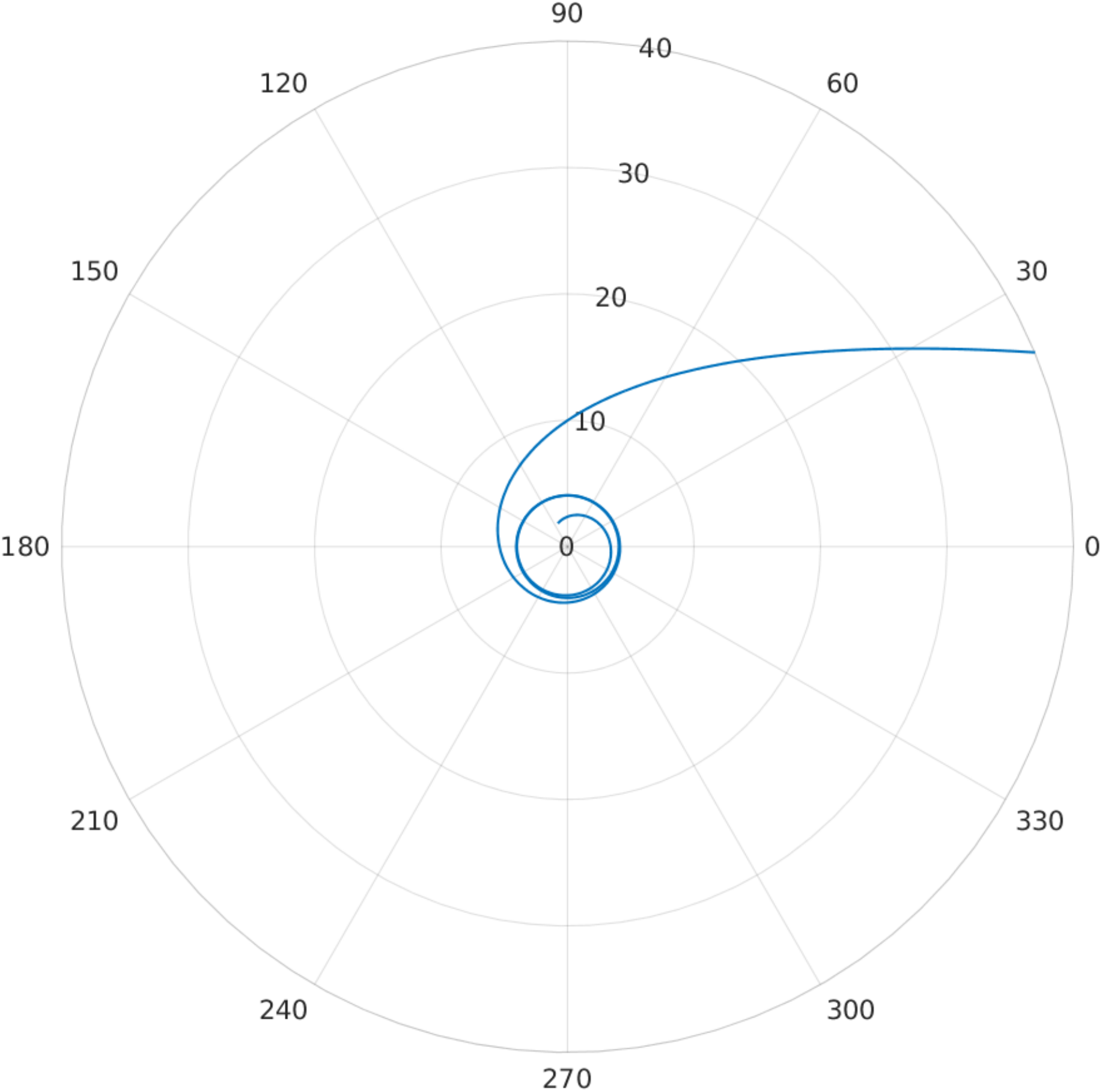}
		\hspace{1.3cm}
		\includegraphics[width=0.25\textwidth]{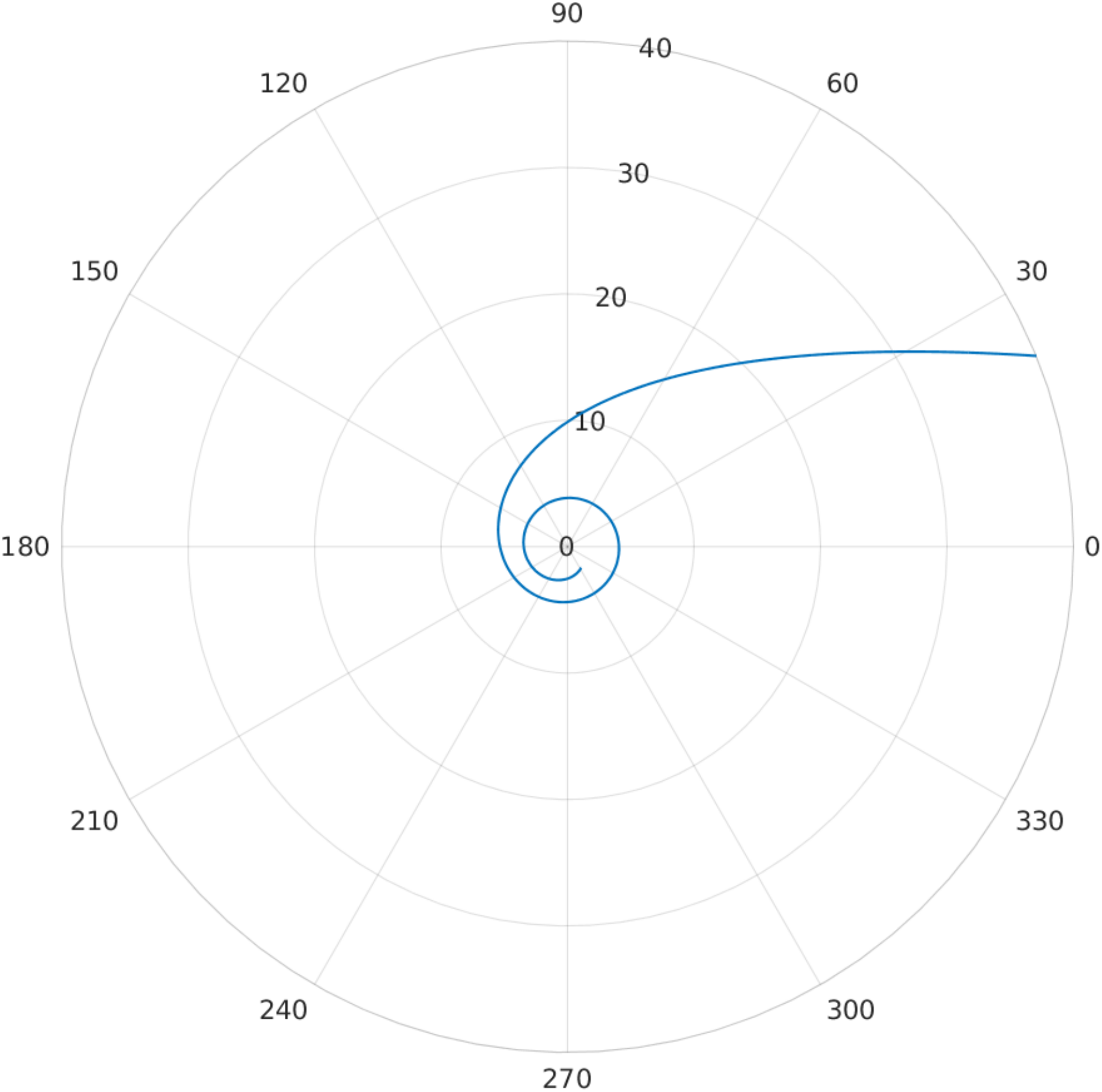}\\
		\vspace{0.3cm}
		\includegraphics[width=0.31\textwidth]{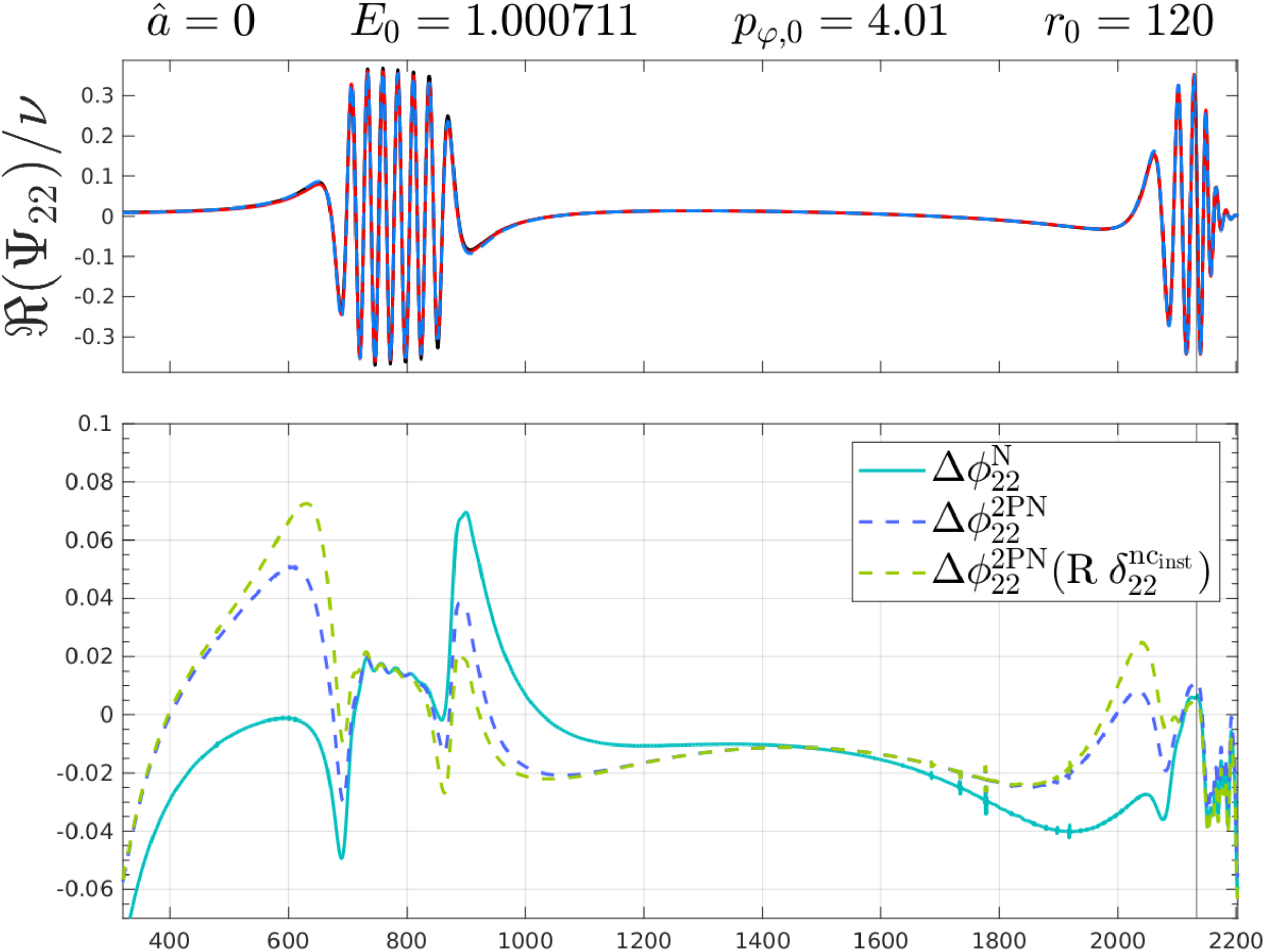}
		\hspace{0.2cm}
		\includegraphics[width=0.31\textwidth]{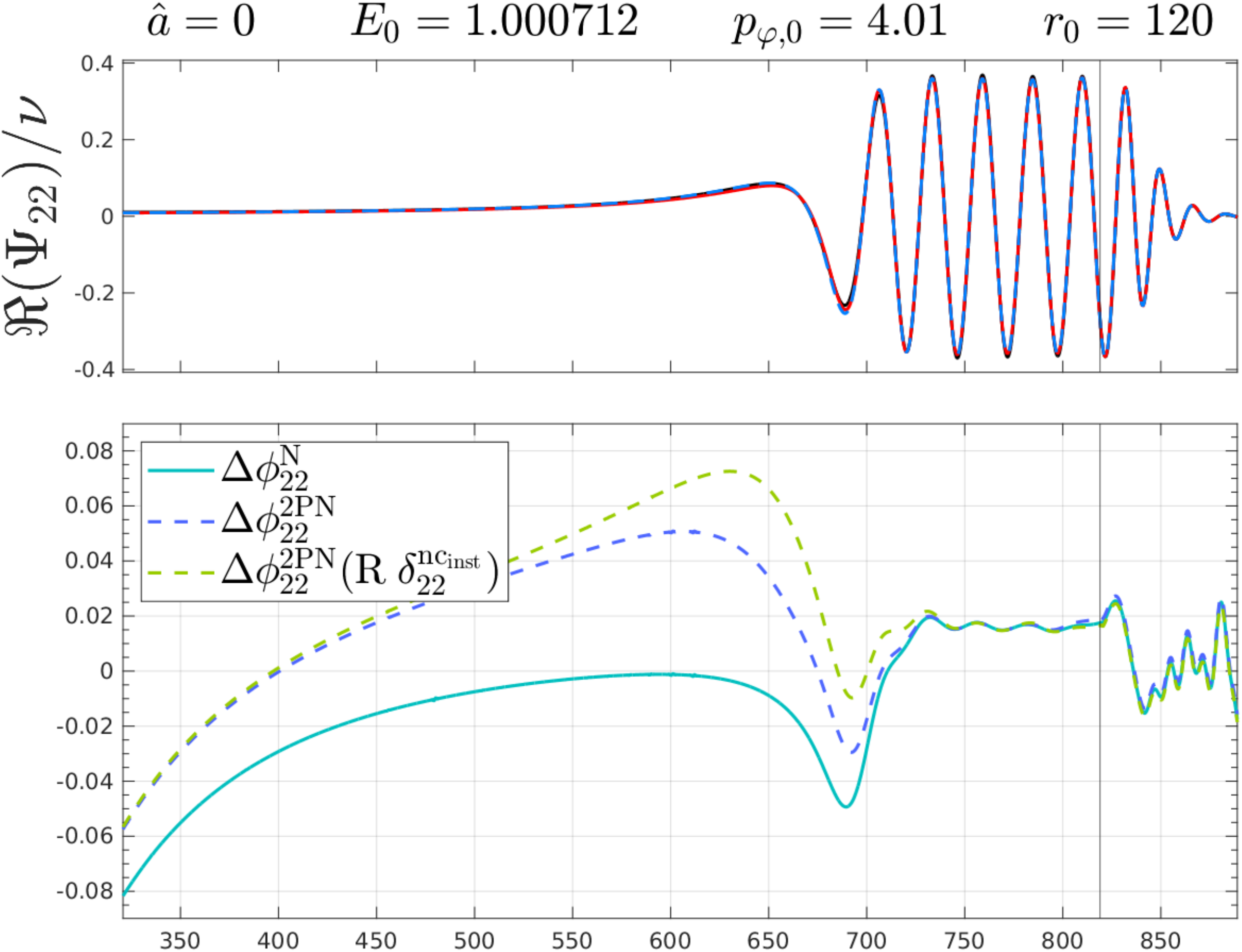}
		\hspace{0.2cm}
		\includegraphics[width=0.31\textwidth]{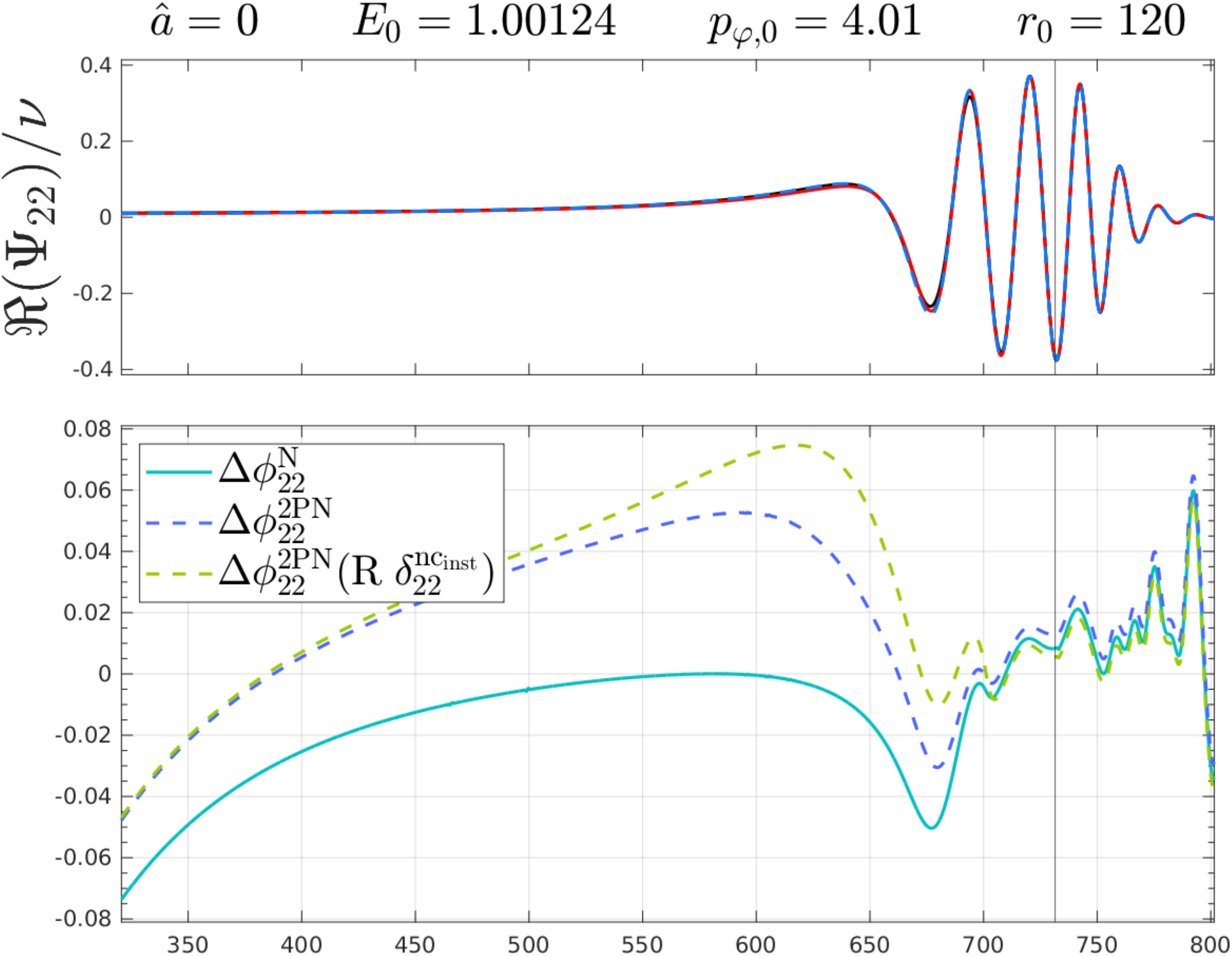}
		\caption{\label{fig:testmass_hyp} Dynamical capture of a particle on a 
			Schwarzschild black hole with $\nu=10^{-2}$, $p_{\varphi,0}=4.01$ and three 
			different initial energies $E_0=(1.000711, 1.000712, 1.00124)$. 
			Top panels: the trajectories, all starting from $r_0=120$,
			although the plot only focuses on the latest part.
			Middle panels: comparing the real part of the numerical waveform (black, barely
			indistinguishable) with two analytical waveforms: the one with only the Newtonian 
			noncircular corrections (red) and the one with the 2PN corrections where the tail 
			factor is resummed while the instantaneous factor is not (dashed blue).
			The corresponding phase differences are reported in the bottom panels 
			(solid clear blue and dashed blue, respectively). The same panels also show 
			the analytical/numerical phase difference obtained with the resummation
			applied to {\it both} the tail and the instantaneous 2PN noncircular 
			phase $\delta_{22}^{\rm nc_{inst}}$ (dashed green). The parameters of the 
			ringdown and of the NQC corrections, as well the merger time 
			(marked by the vertical line), are extracted from numerical data as in Ref.~\cite{Albanesi:2021rby}, see text.
			The closest analytical/numerical agreement is obtained resumming only the noncircular tail factor.}
	\end{figure*}
	
	We now turn our attention to dynamical captures in the large mass ratio limit. 
	In particular, we consider the configurations that were originally shown in 
	Fig. 14 of Ref.~\cite{Albanesi:2021rby}, whose parameters are also listed 
	in Table~\ref{tab:Teukode_hyp} for convenience.
	The analytical/numerical comparisons for the new prescriptions, compared with the original 
	Newtonian case, are shown in Figure~\ref{fig:testmass_hyp}. We report both the waveforms with 
	expanded and resummed instantaneous noncircular phase at 2PN. We also recall that in the hyperbolic case
	the parameters of the ringdown model, the NQC corrections and the merger time are extracted directly 
	from the numerical waveform. This is due to the fact that a fit over the parameter space of these 
	quantities is not currently available.
	See Sec.~V~C of Ref.~\cite{Albanesi:2021rby} for more details. For
	this reason, the last part of the waveform is artificially more accurate than the result that
	would be obtained with the same fitting-procedure followed for eccentric orbits.
	Nonetheless, this aspect is not very relevant for our discussion since in order to test the 
	reliability of the 2PN corrections we have to focus on the inspiral. 
	The phase differences of Figure~\ref{fig:testmass_hyp} show 
	that the 2PN noncircular corrections do not provide a better analytical/numerical agreement
	than the Newtonian wave in the hyperbolic scenario. Moreover, 
	the resummation of the instantaneous noncircular phase worsens the analytical/numerical
	agreement with respect to leaving the instantaneous phase terms in expanded form.
	This is an indication that the resummation of the instantaneous phase
	could be avoided. It is possible that employing terms beyond the 2PN order will
	clarify the procedure to follow for the instantaneous contribution. 
	
	\section{Waveform validation: the comparable mass case}
	\label{sec:comparable_masses}
	Let us move now to comparable-mass binaries. In this case, we incorporate our 2PN-improved
	eccentric waveform within the EOB eccentric model recently presented in Ref.~\cite{Nagar:2021xnh},
	that is currently the latest evolution of the scheme proposed  in Ref.~\cite{Chiaramello:2020ehz}.
	In doing so, we keep the same dynamics, informed by NR quasi-circular simulations, of Ref.~\cite{Nagar:2021xnh}.
	The 2PN noncircular corrections to the waveform have an essentially negligible impact on quasi-circular
	configurations and it is not worth to provide a new, optimized, determination of $(a_6^c,c_3)$ 
	using the 2PN resummed waveform. We thus explore here the performance of this new waveform 
	using both time-domain (phase-alignment) and frequency domain (unfaithfulness) comparisons. 
	%=====================
	% Table of eccentric datasets
	%=====================
	%=================================================
	% Final results at Dec, 1st 2021
	% To be inserted in table
	%=================================================
	\begin{table*}[t]
		\caption{\label{tab:SXS} SXS simulations with eccentricity analyzed in this work. From left to right: the 
			ID of the simulation; the mass ratio $q\equiv m_1/m_2\geq 1$ and the individual dimensionless 
			spins $(\chi_1,\chi_2)$; the time-domain NR phasing uncertainty at merger $\delta\phi^{\rm NR}_{\rm mrg}$; 
			the estimated NR eccentricity at first apastron $e_{\omega_a}^{\rm NR}$; 
			the NR frequency of first apastron $\omega_{a}^{\rm NR}$; 
			the initial EOB eccentricity $e^{\rm EOB}_{\omega_a}$ and apastron frequency $\omega_{a}^{\rm EOB}$ used to start the EOB evolution; 
			the maximal NR unfaithfulness uncertainty, $\bar{F}^{\rm max}_{\rm NR/NR}$, the initial frequency used in the EOB/NR unfaithfulness
			computation, $Mf_{\rm min}$, and the maximal EOB/NR unfaithfulness, $\bar{F}_{\rm EOB/NR}^{\rm max}$. }
		\begin{center}
			\begin{ruledtabular}
				\begin{tabular}{c| c  c c c c |l l| c|c c} 
					$\#$ & id & $(q,\chi_1,\chi_2)$ & $\delta\phi^{\rm NR}_{\rm mrg}$[rad]& $e^{\rm NR}_{\omega_a}$ & $\omega_a^{\rm NR}$ &$e^{\rm EOB}_{\omega_a}$ & $\omega_{a}^{\rm EOB}$ & $\bar{F}_{\rm NR/NR}^{\rm max}[\%]$  & $Mf_{\rm min}$ &$\bar{F}_{\rm EOB/NR}^{\rm max}[\%]$ \\
					\hline
					\hline
					1 & SXS:BBH:1355 & $(1,0, 0)$ & $+0.92$ & 0.0620  & 0.03278728 & 0.0888    & 0.02805750  & 0.012 & 0.0055 &0.13\\
					2 & SXS:BBH:1356 & $(1,0, 0)$& $+0.95$ & 0.1000 &  0.02482006  & 0.15038  & 0.019077  & 0.0077 & 0.0044 &0.24 \\
					3 & SXS:BBH:1358 & $(1,0, 0)$& $+0.25$ & 0.1023 & 0.03108936 & 0.18082   & 0.021238 & 0.016 &  0.0061 &0.22\\
					4 & SXS:BBH:1359 & $(1,0, 0)$& $+0.25$  & 0.1125 & 0.03708305 & 0.18240   & 0.021387 & 0.0024& 0.0065  &0.17\\
					5 & SXS:BBH:1357 & $(1,0, 0)$& $-0.44$  & 0.1096 & 0.03990101 & 0.19201   & 0.01960 & 0.028& 0.0061 &0.15\\
					6 & SXS:BBH:1361 & $(1,0, 0)$& +0.39    & 0.1634 & 0.03269520  & 0.23557   & 0.020991   & 0.057&0.0065  &0.35\\
					7 & SXS:BBH:1360 & $(1,0, 0)$& $-0.22$ & 0.1604 & 0.03138220 & 0.2440  & 0.019508   &0.0094  & 0.0065 &0.31\\
					8 & SXS:BBH:1362 & $(1,0, 0)$& $-0.09$ & 0.1999 & 0.05624375 & 0.3019     & 0.01914 & 0.0098 & 0.0065 &0.15\\
					9 & SXS:BBH:1363 & $(1,0, 0)$& $+0.58$ & 0.2048 & 0.05778104 &  0.30479    & 0.01908 & 0.07 & 0.006 &0.25 \\
					10 & SXS:BBH:1364 & $(2,0, 0)$& $-0.91$ & 0.0518 &  0.03265995   & 0.0844    & 0.025231   & 0.049  & 0.062 &0.15 \\
					11 & SXS:BBH:1365 & $(2,0, 0)$& $-0.90$ & 0.0650  &  0.03305974   & 0.110     & 0.023987 & 0.027&  0.062 &0.12\\
					12 & SXS:BBH:1366 & $(2,0, 0)$& $-6\times 10^{-4}$ & 0.1109 & 0.03089493 & 0.14989   & 0.02577 &  0.017  & 0.0052 &0.20  \\
					13 & SXS:BBH:1367 & $(2,0, 0)$& $+0.60$ & 0.1102 & 0.02975257 & 0.15095    & 0.0260  & 0.0076  & 0.0055 &0.15 \\
					14 & SXS:BBH:1368 & $(2,0, 0)$& $-0.71$ & 0.1043 & 0.02930360 &  0.14951  & 0.02512    & 0.026 & 0.0065 &0.13 \\
					15 & SXS:BBH:1369 & $(2,0, 0)$& $-0.06$ & 0.2053 & 0.04263738 & 0.3134     & 0.0173386  & 0.011& 0.0041  &0.25\\
					16 & SXS:BBH:1370 & $(2,0, 0)$& $+0.12$ & 0.1854 &  0.02422231 &  0.31708  & 0.016779  & 0.07& 0.006 &0.37 \\
					17 & SXS:BBH:1371 & $(3,0, 0)$& $+0.92$ & 0.0628 & 0.03263026  & 0.0912     & 0.029058   & 0.12  & 0.006  &0.19\\
					18 & SXS:BBH:1372 & $(3,0, 0)$& $+0.01$& 0.1035 & 0.03273944 & 0.14915      & 0.026070 & 0.06  & 0.006 &0.09 \\
					19 & SXS:BBH:1373 & $(3,0, 0)$& $-0.41$ & 0.1028 & 0.03666911 & 0.15035    & 0.02529 & 0.0034 &  0.0061 &0.13\\
					20 & SXS:BBH:1374 & $(3,0, 0)$& $+0.98$ & 0.1956  & 0.02702594 & 0.314   & 0.016938   & 0.067 & 0.0059 &0.1\\
					\hline
					21 & SXS:BBH:89   & $(1,-0.50, 0)$         &  $\dots$  & 0.0469  & 0.02516870 & 0.07194    & 0.01779  & $\dots$ & 0.0025 &0.18 \\
					22 & SXS:BBH:1136 & $(1,-0.75,-0.75)$   &  $-1.90$ & 0.0777  &0.04288969 &0.1209      & 0.02728 & 0.074 & 0.0058 &0.12 \\
					23 & SXS:BBH:321  & $(1.22,+0.33,-0.44)$& $+1.47$ & 0.0527  & 0.03239001 &0.07621     & 0.02694 & 0.015  & 0.0045  &0.27 \\
					24 & SXS:BBH:322  & $(1.22,+0.33,-0.44)$& $-2.02$  & 0.0658  &  0.03396319 &0.0984       & 0.026895 & 0.016 & 0.0061 &0.26 \\
					25 & SXS:BBH:323  & $(1.22,+0.33,-0.44)$& $-1.41$ & 0.1033  & 0.03498377 &0.1438      & 0.02584 & 0.019 & 0.0058 &0.17\\
					26 & SXS:BBH:324  & $(1.22,+0.33,-0.44)$& $-0.04$ & 0.2018  & 0.02464165 &0.29425        & 0.01894 & 0.098 & 0.0058 &0.19\\
					27 & SXS:BBH:1149 & $(3,+0.70,+0.60)$  &  $+3.00$  & 0.0371  &0.03535964 &$0.06237$   & $0.02664$ &0.025  & 0.005 &1.07\\
					28 & SXS:BBH:1169 & $(3,-0.70,-0.60)$    &  $+3.01$ & 0.0364  &0.02759632 &$0.04895$     & $0.024285$ & 0.033 & 0.004& 0.10     % 
					
				\end{tabular}
			\end{ruledtabular}
		\end{center}
	\end{table*}
	%================================================
	\subsection{Phase comparison in the time domain}
	\label{TD_phasing}
\begin{figure*}
		\center
		\includegraphics[width=0.19\textwidth]{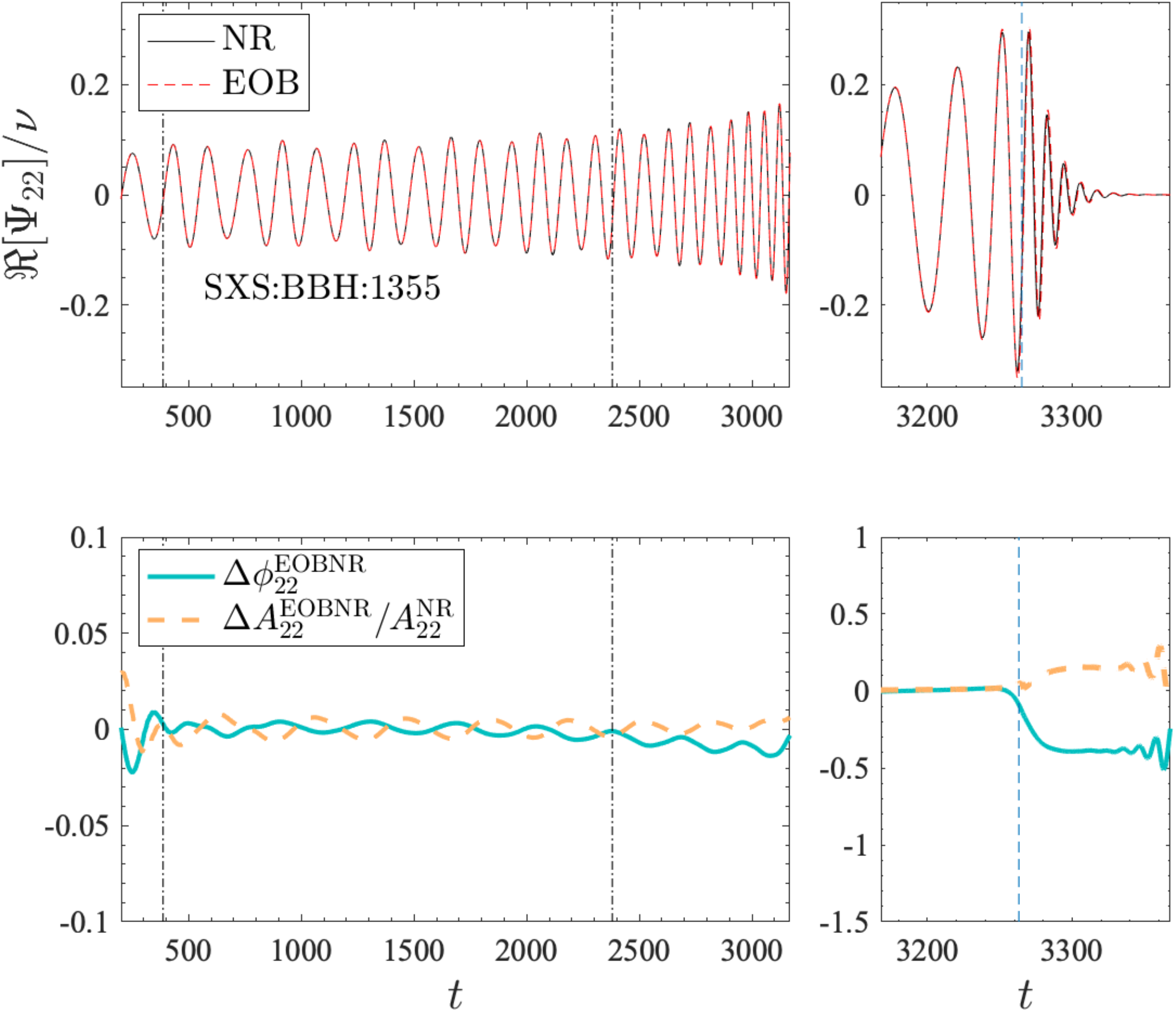}
		\includegraphics[width=0.19\textwidth]{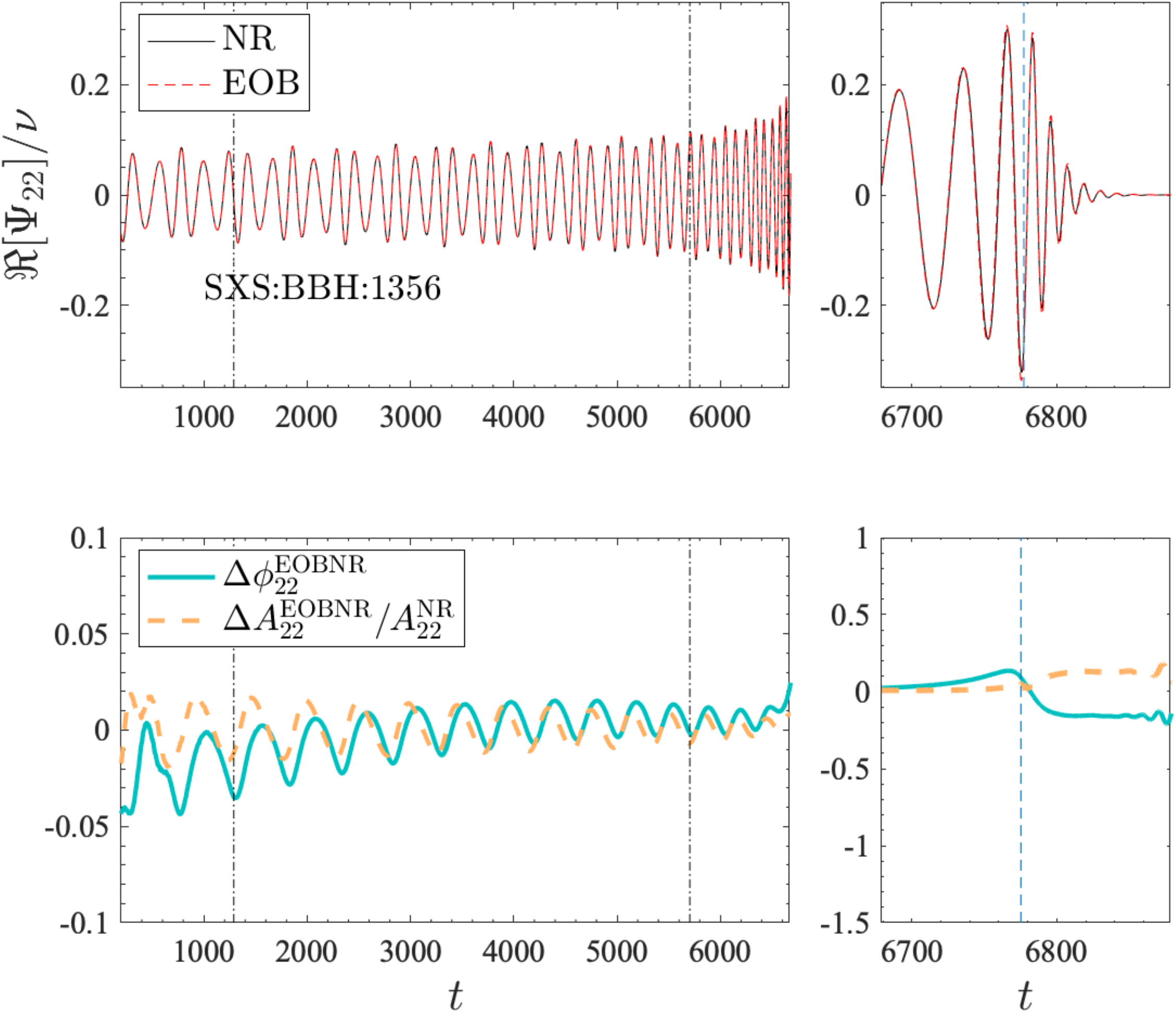}
		\includegraphics[width=0.19\textwidth]{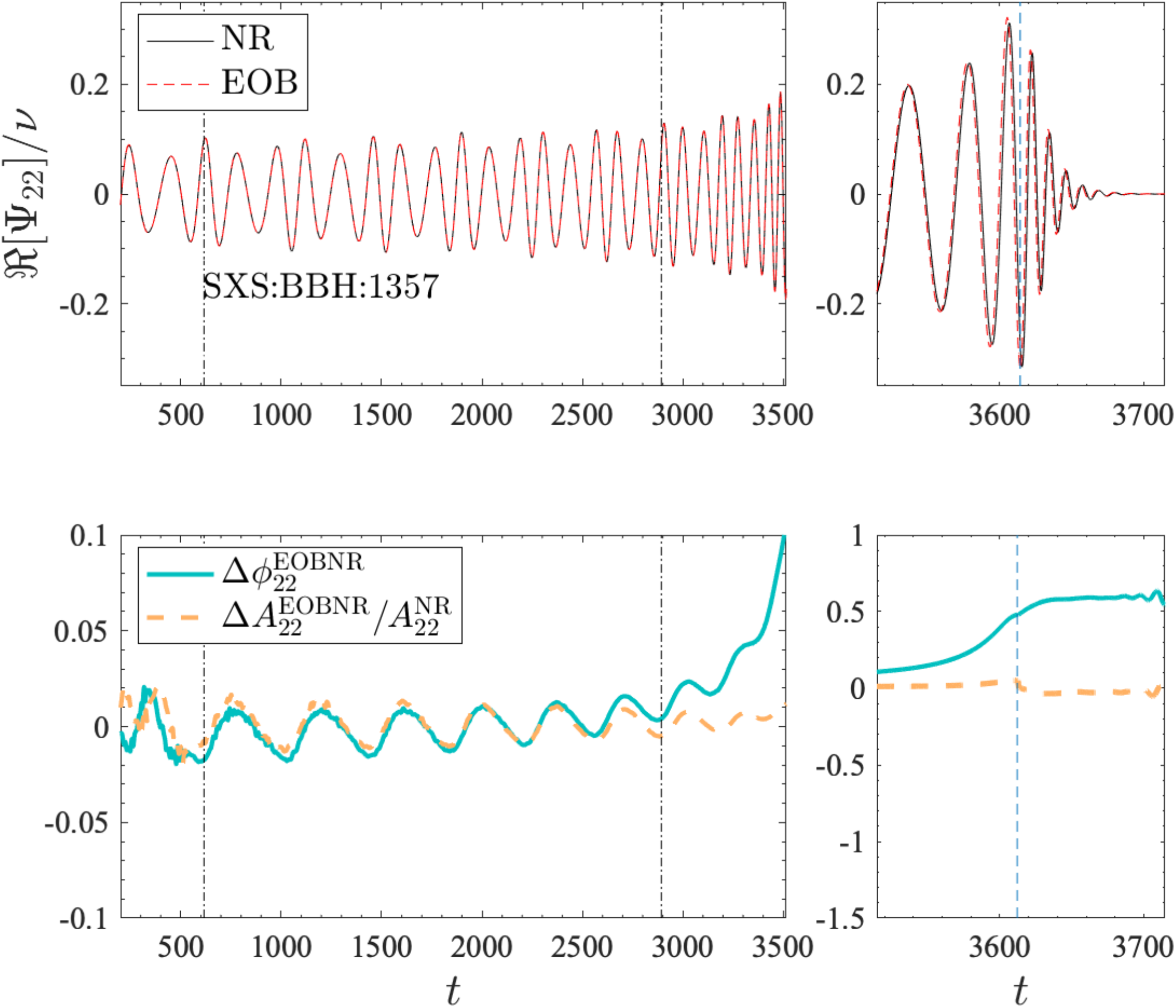}
		\includegraphics[width=0.19\textwidth]{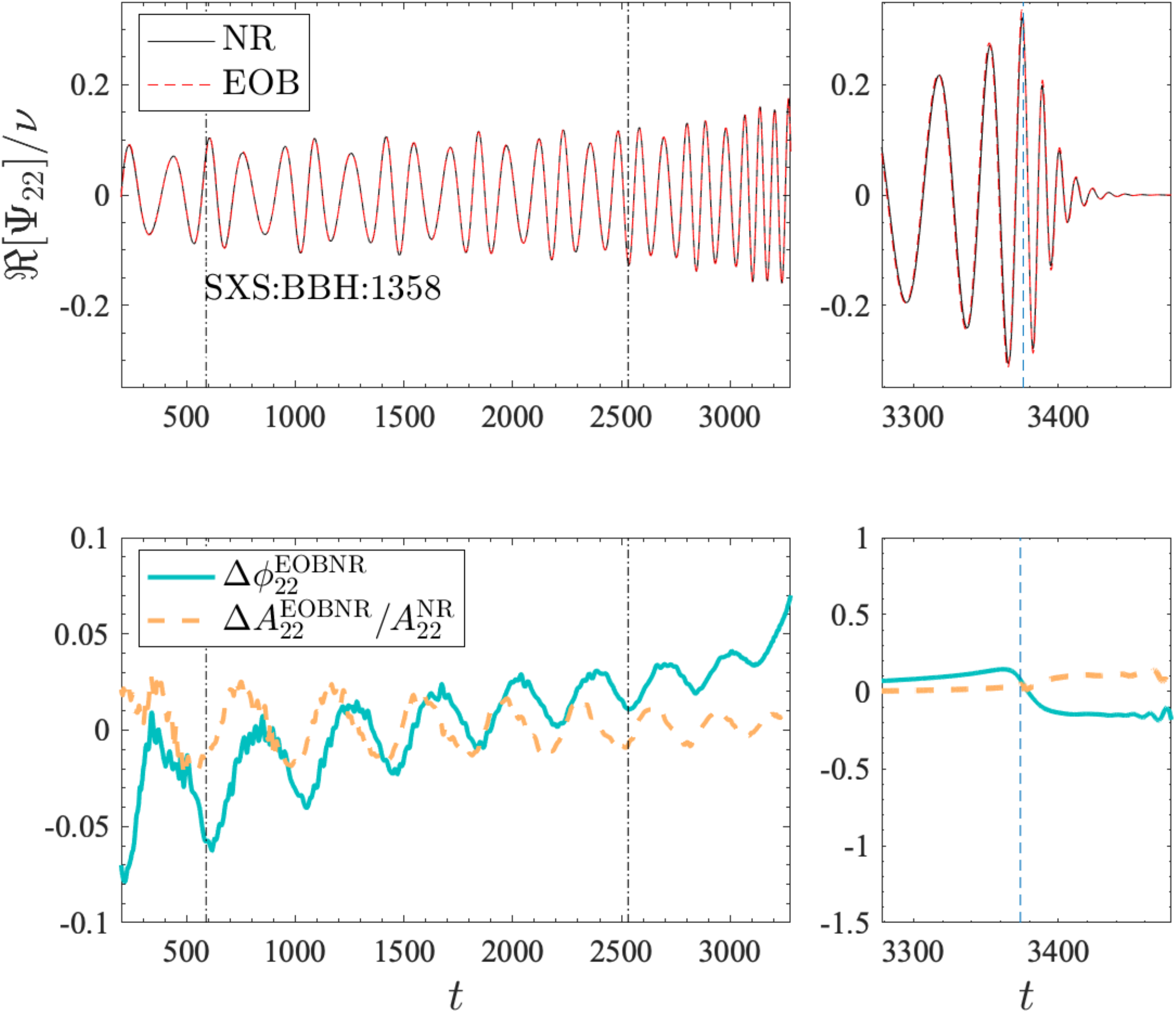}
		\includegraphics[width=0.19\textwidth]{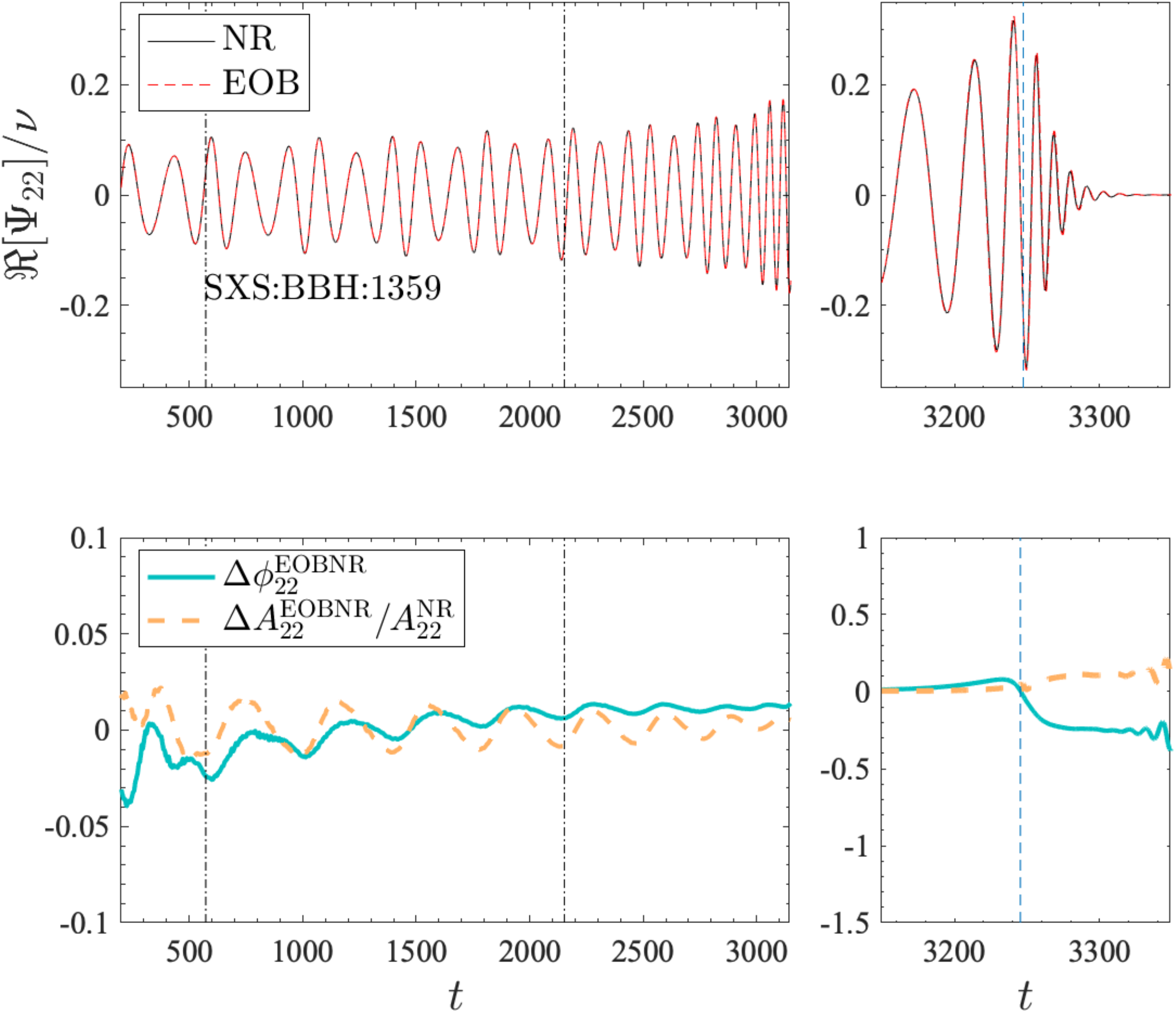}\\
		\vspace{4 mm}
		\includegraphics[width=0.19\textwidth]{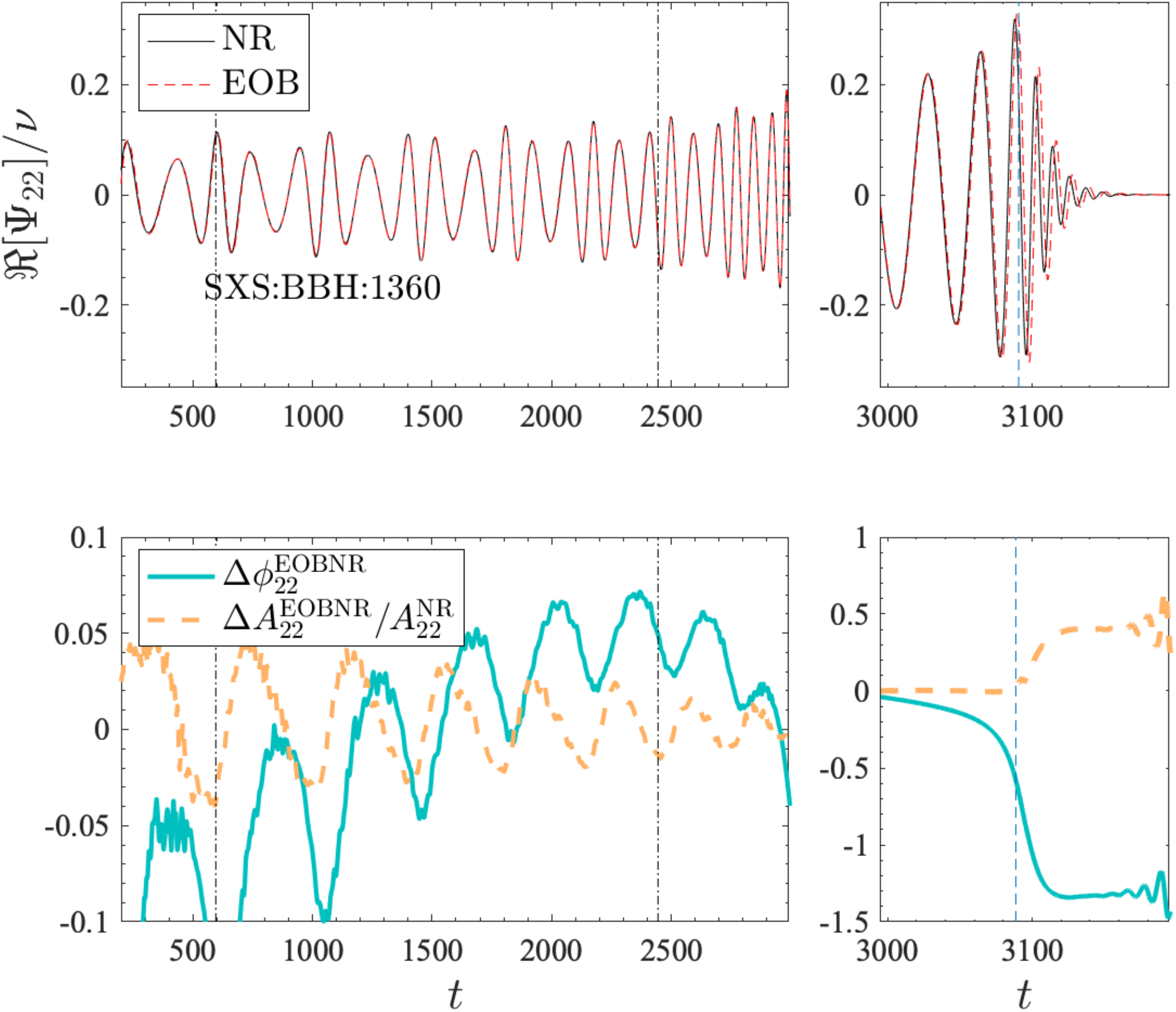}
		\includegraphics[width=0.19\textwidth]{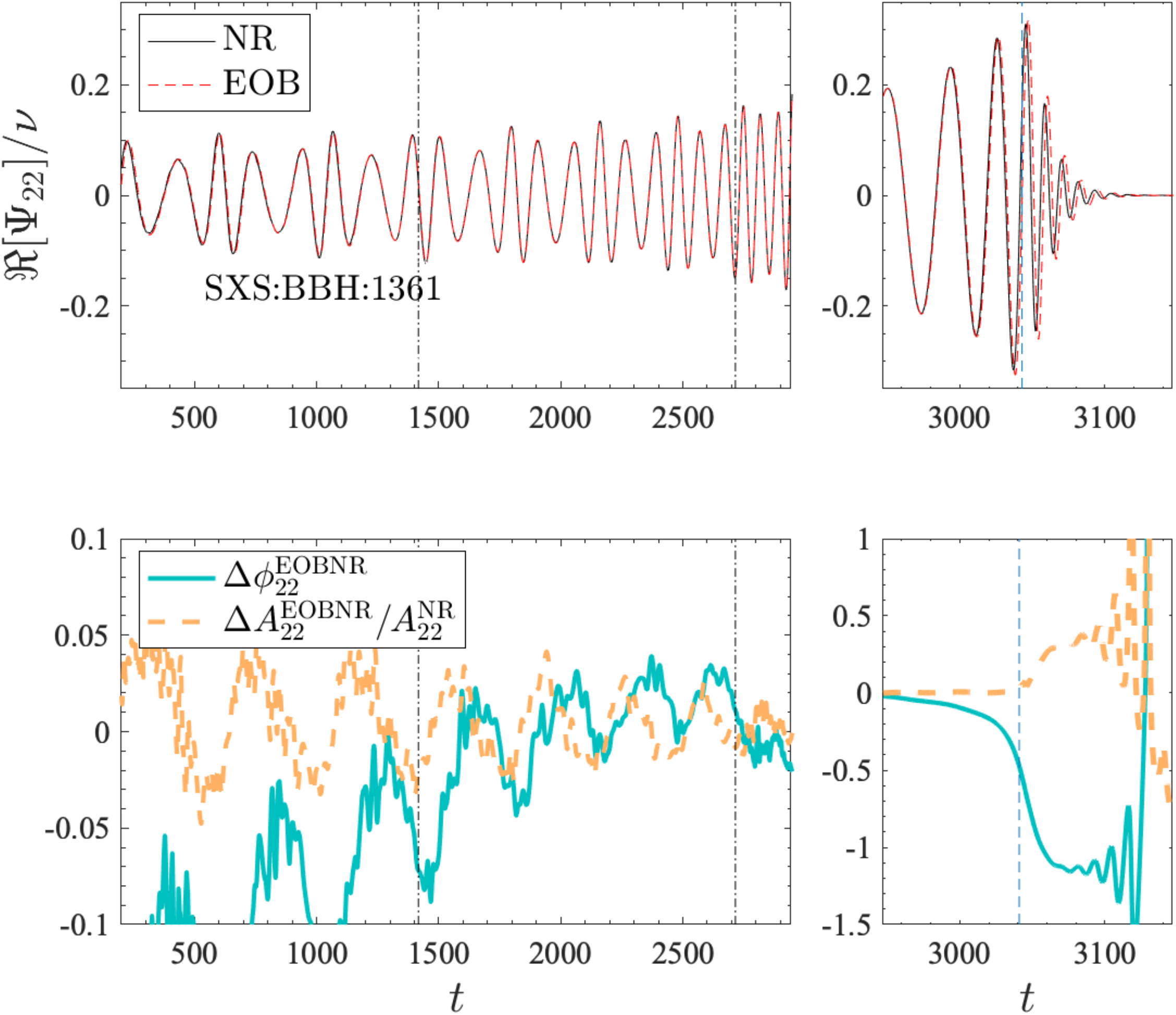}
		\includegraphics[width=0.19\textwidth]{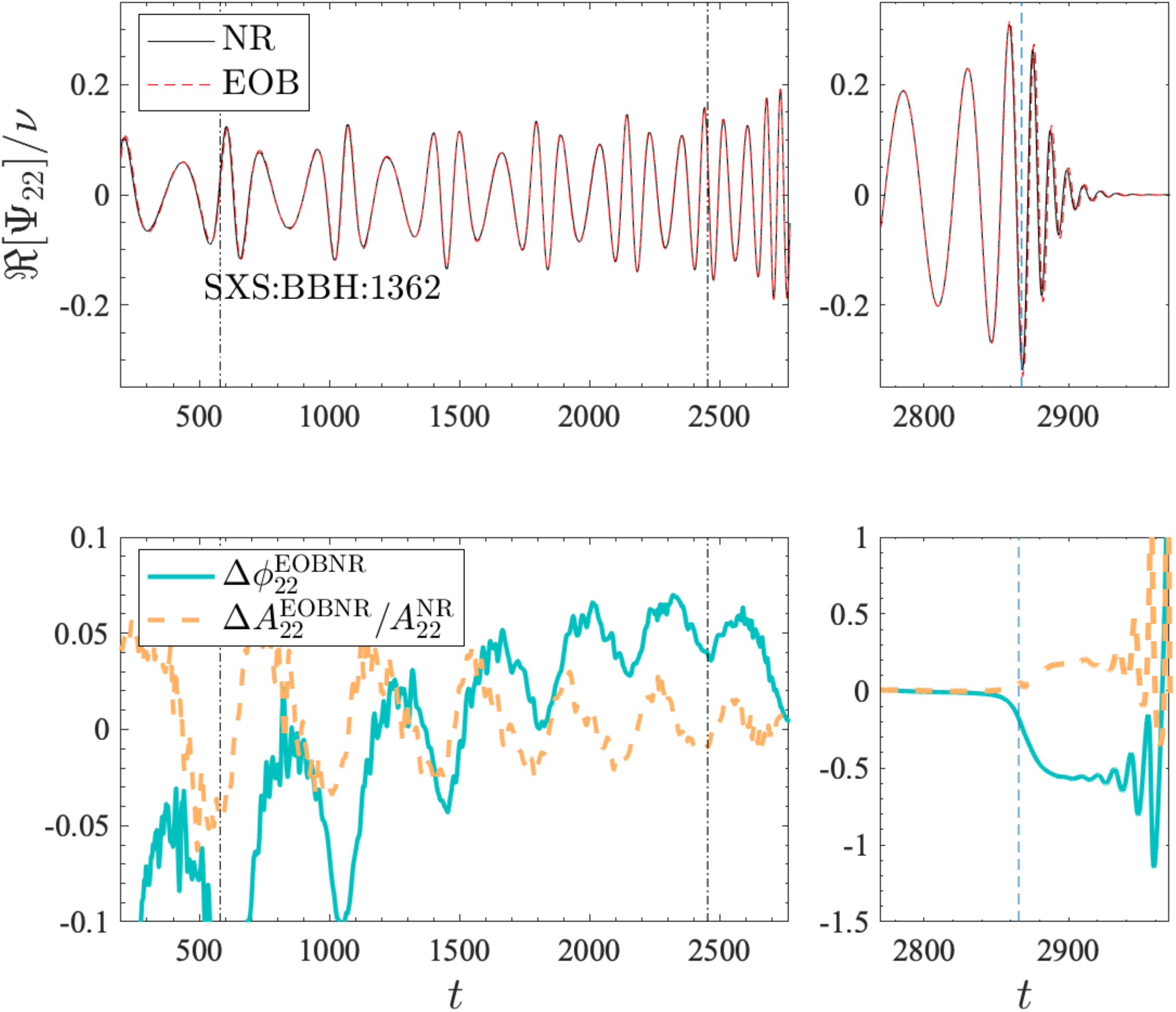}
		\includegraphics[width=0.19\textwidth]{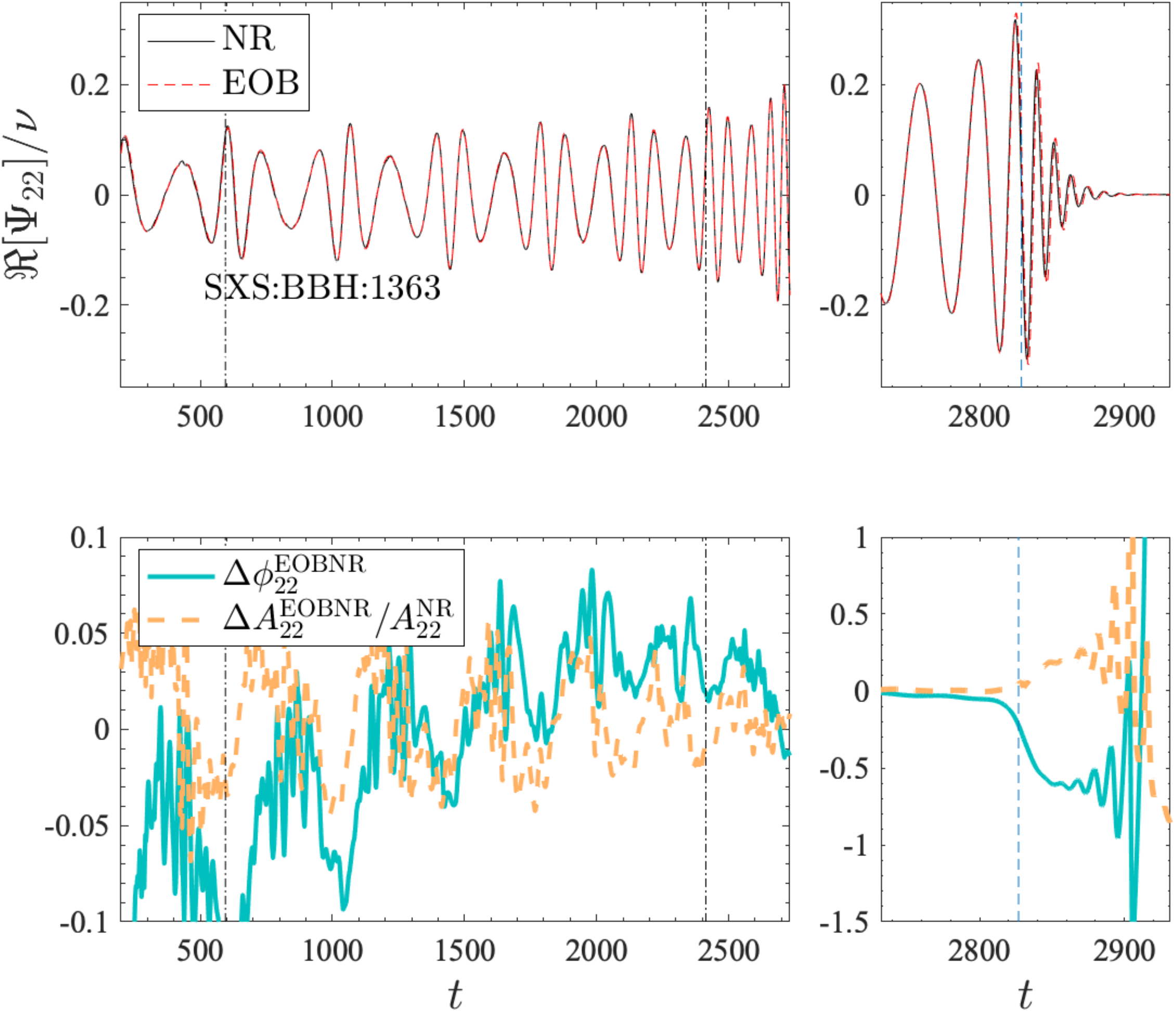}
		\includegraphics[width=0.19\textwidth]{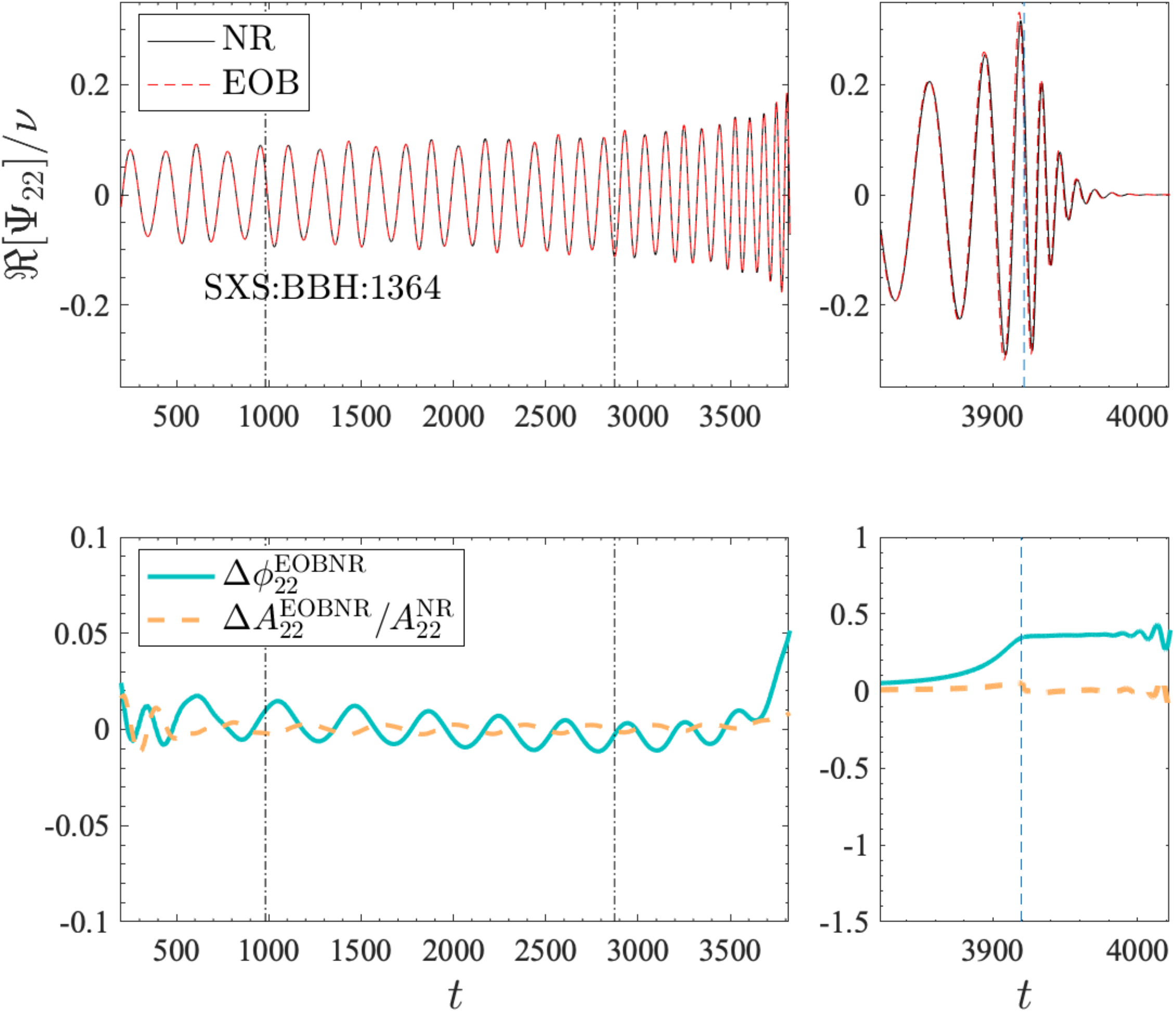}\\
		\vspace{4mm}
		\includegraphics[width=0.19\textwidth]{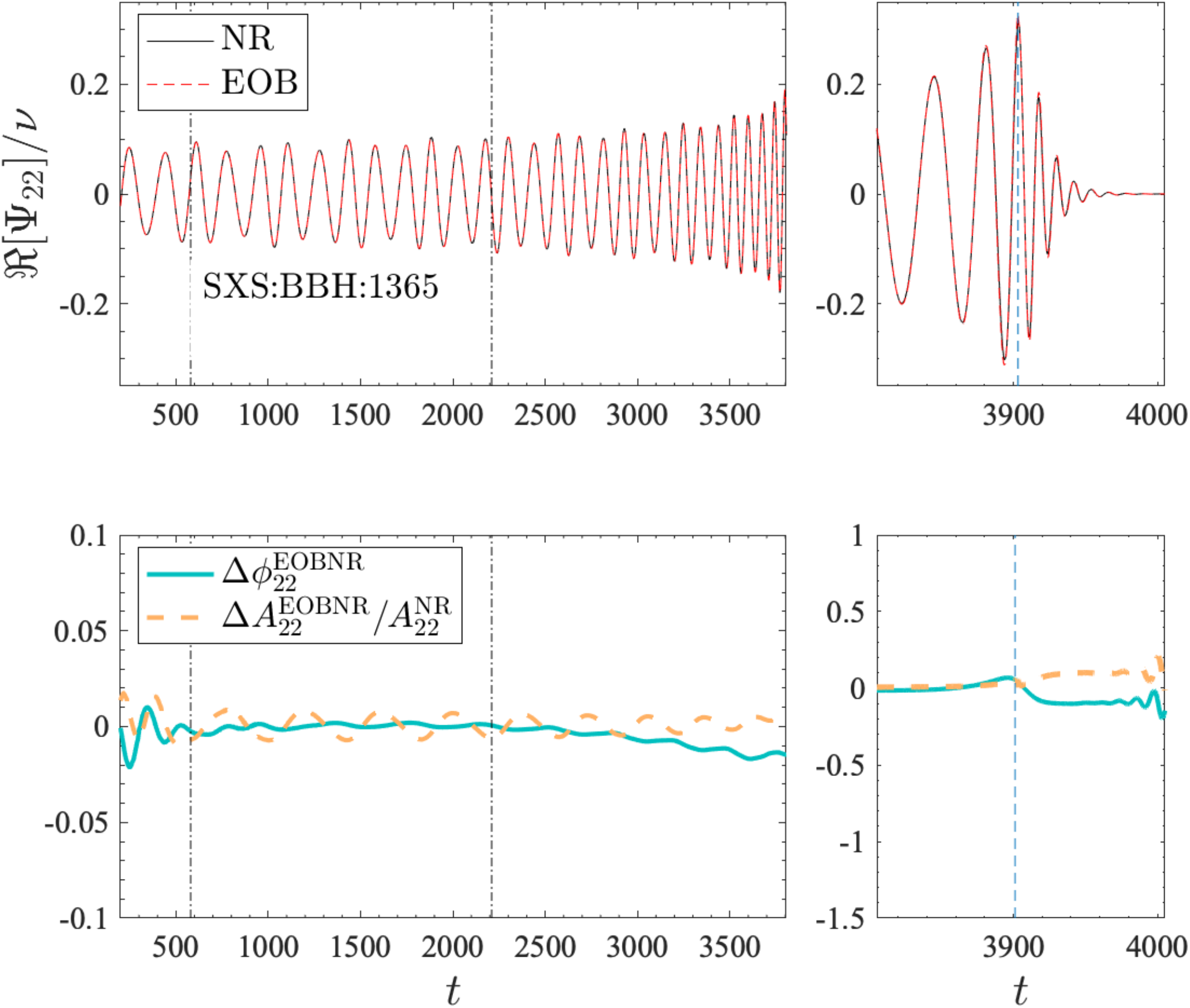}
		\includegraphics[width=0.19\textwidth]{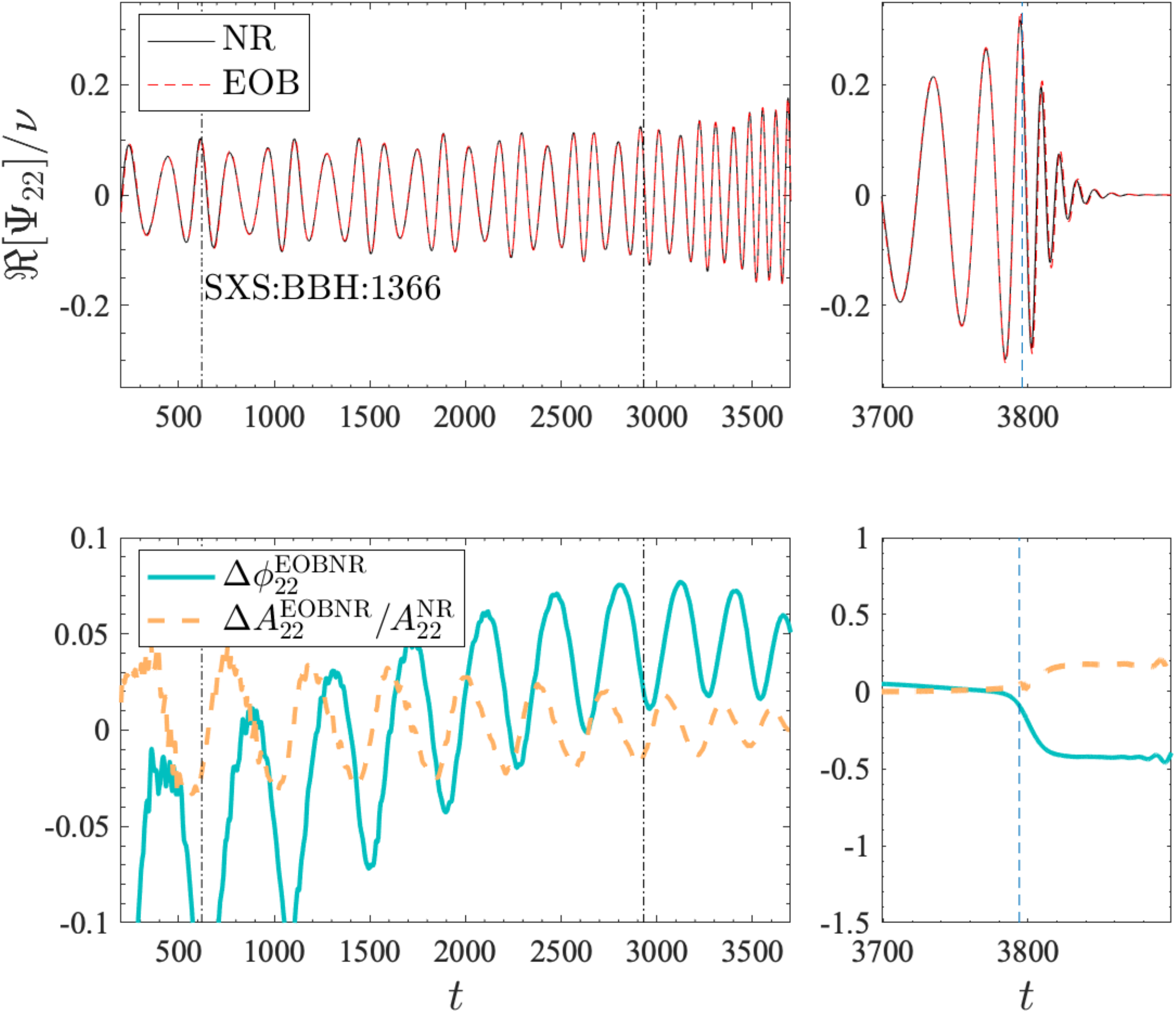}
		\includegraphics[width=0.19\textwidth]{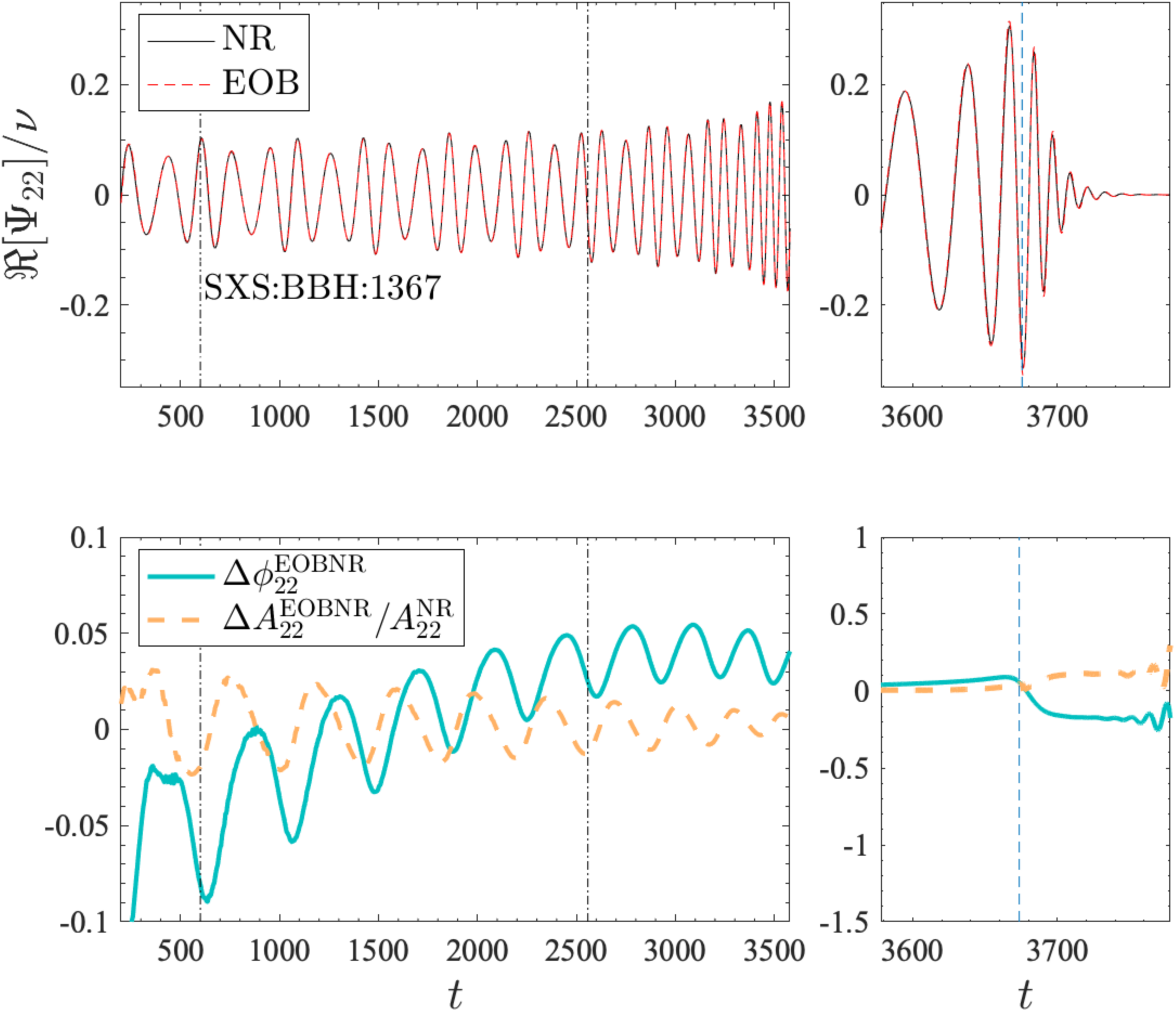}
		\includegraphics[width=0.19\textwidth]{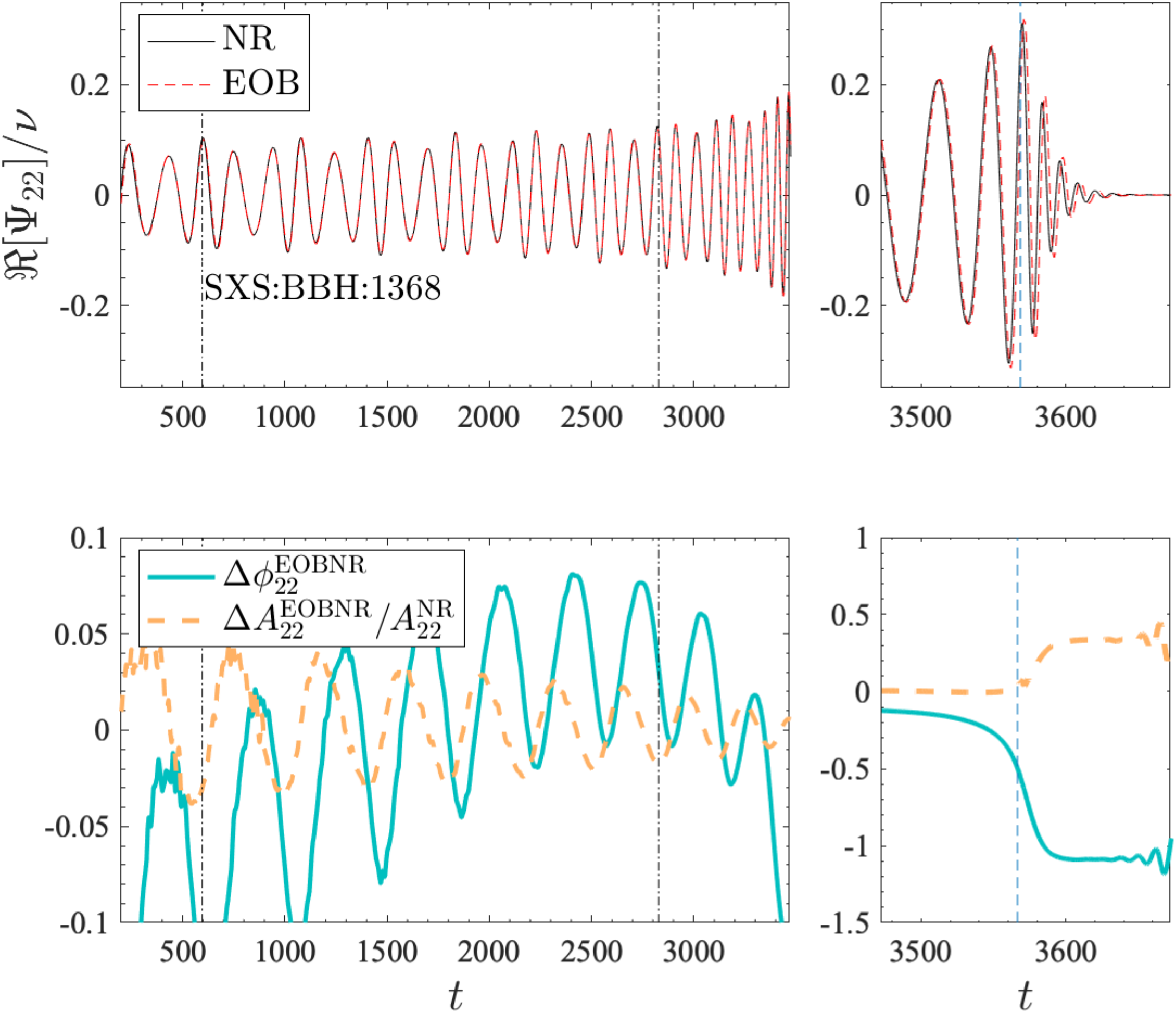}
		\includegraphics[width=0.19\textwidth]{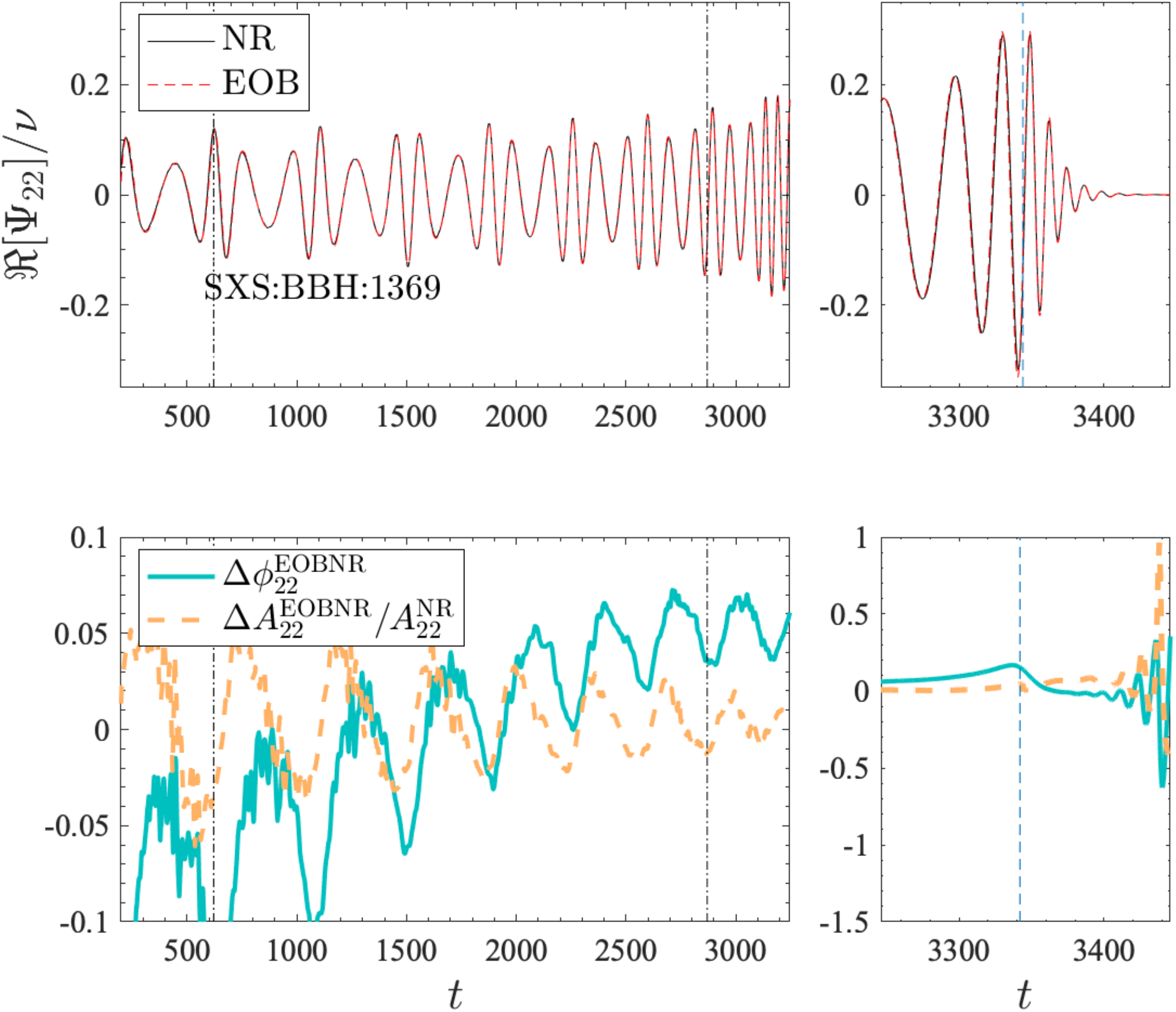}\\
		\vspace{4mm}
		\includegraphics[width=0.19\textwidth]{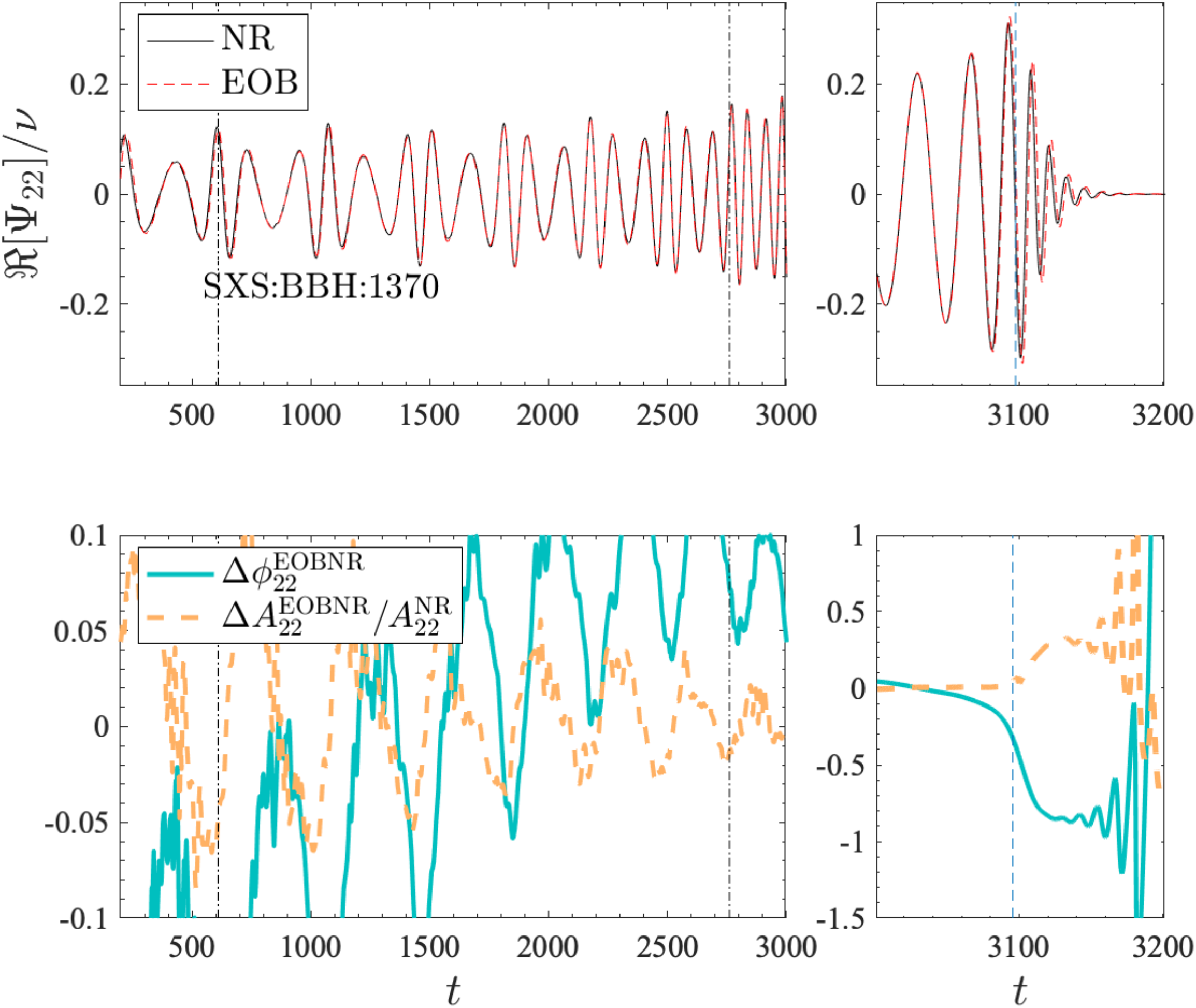}
		\includegraphics[width=0.19\textwidth]{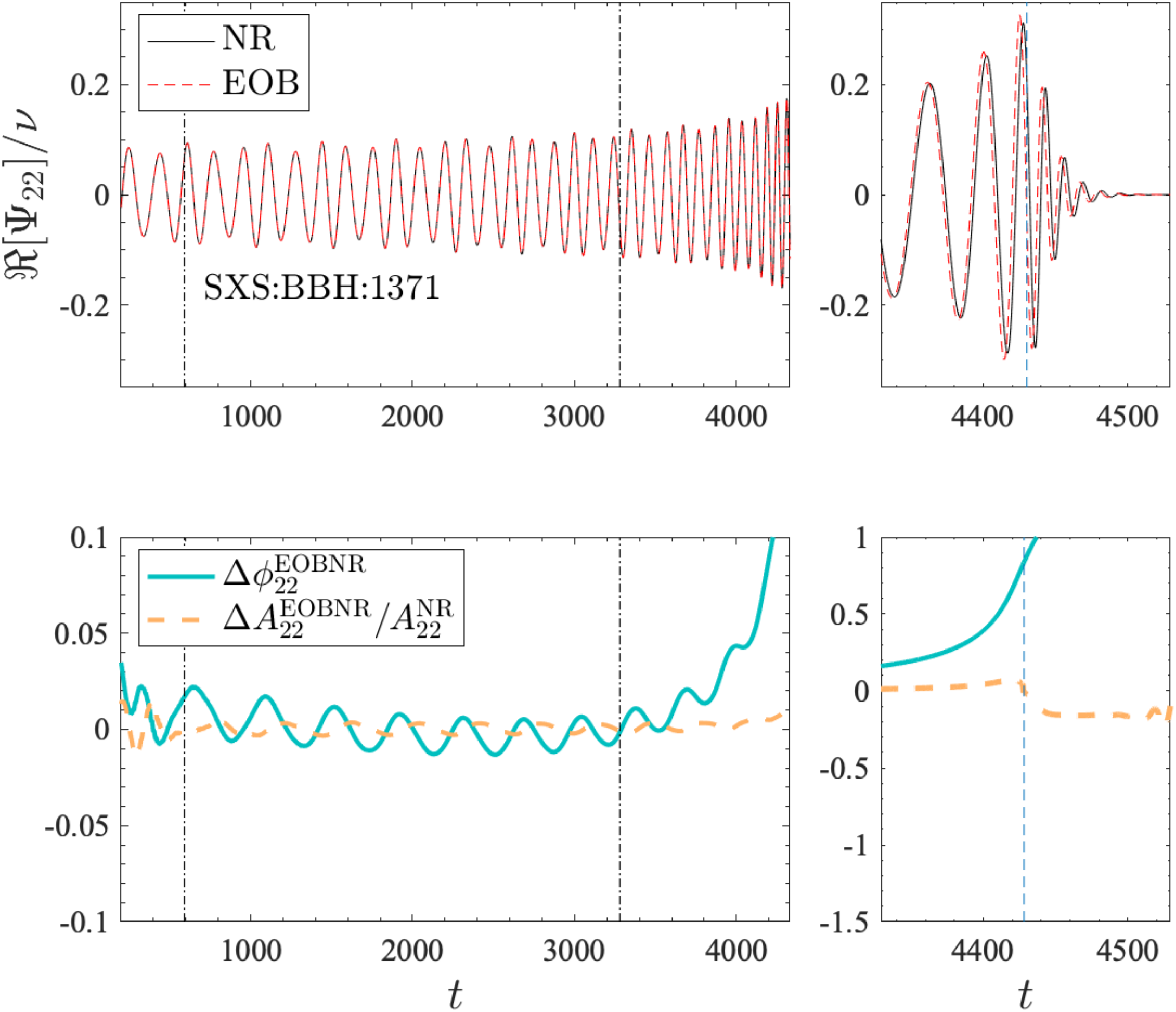}
		\includegraphics[width=0.19\textwidth]{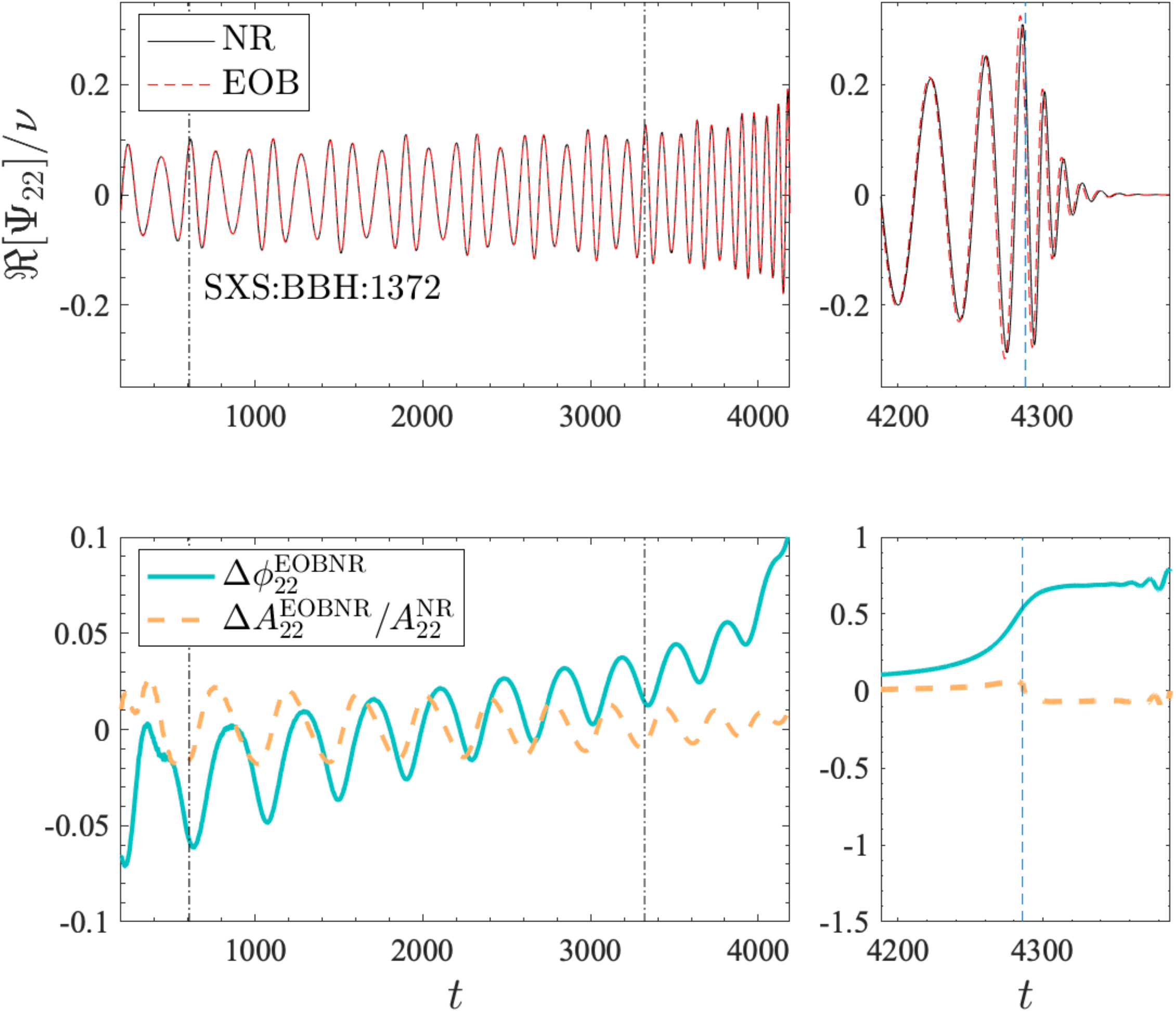}
		\includegraphics[width=0.19\textwidth]{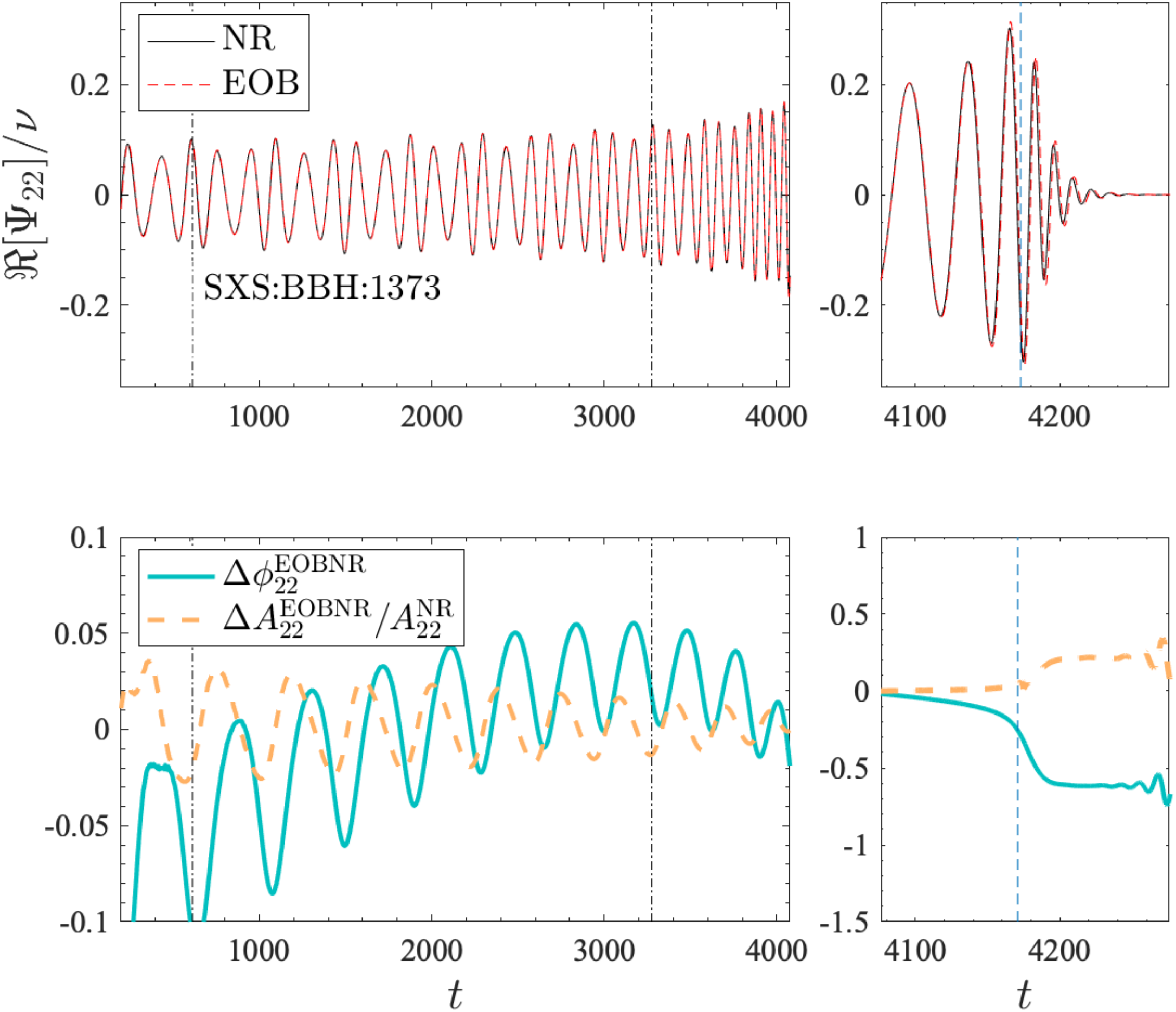}
		\includegraphics[width=0.19\textwidth]{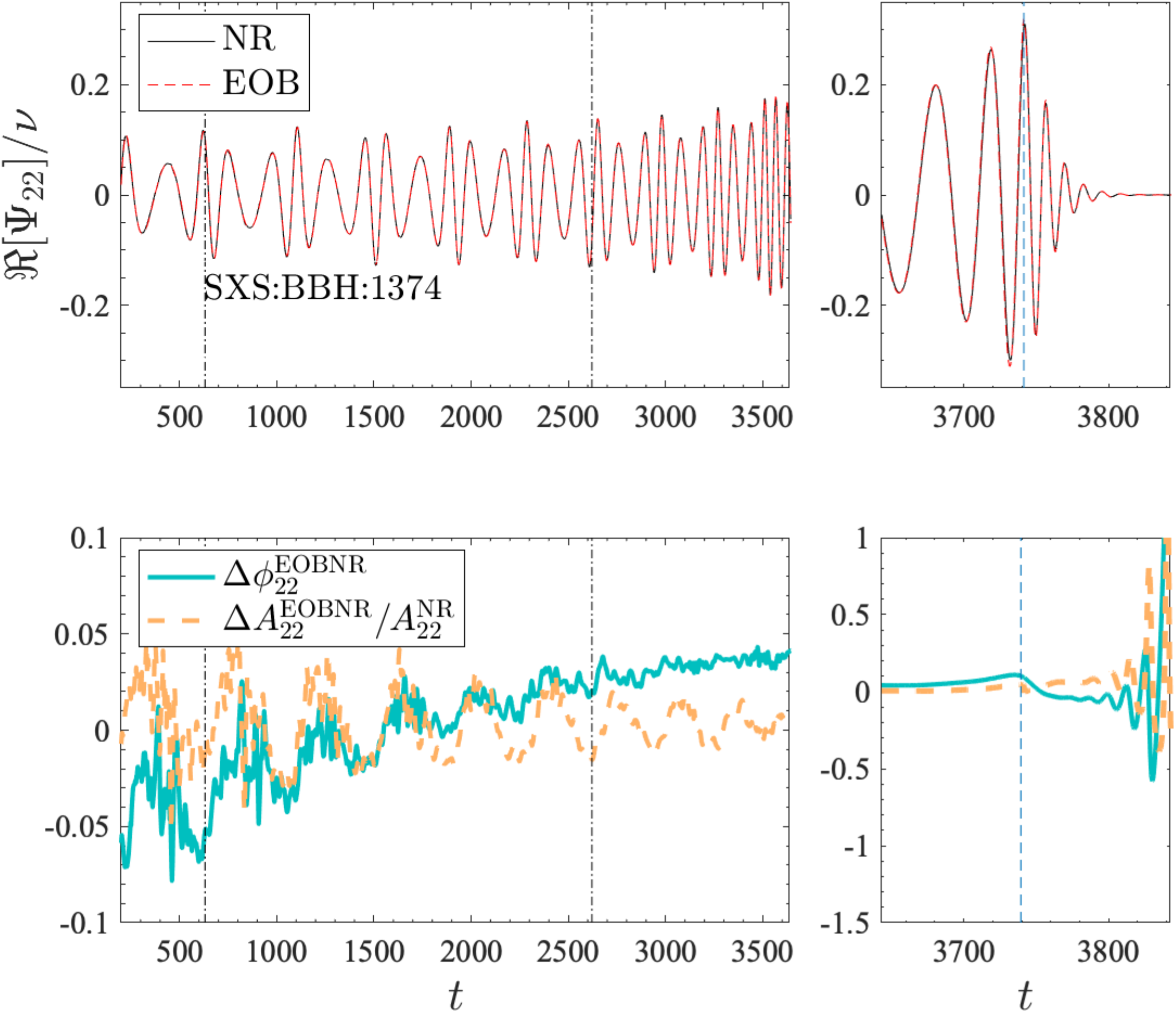}
		\caption{\label{fig:phasing_nospin}  EOB/NR time-domain phasing for all the nonspinning datasets considered. 
			The dashed vertical lines indicate the alignment window. For each configuration, we show together the NR (black) 
			and EOB (red) real part of the waveform (top panel), and the phase difference (blue online) and the relative amplitude 
			difference (orange online), bottom panel. }
		%\end{figure*}
	%-------------------------------
	% Time-domain: spinning
	%-------------------------------
	%\begin{figure*}
		%\center
		\vspace{1cm}
		\includegraphics[width=0.24\textwidth]{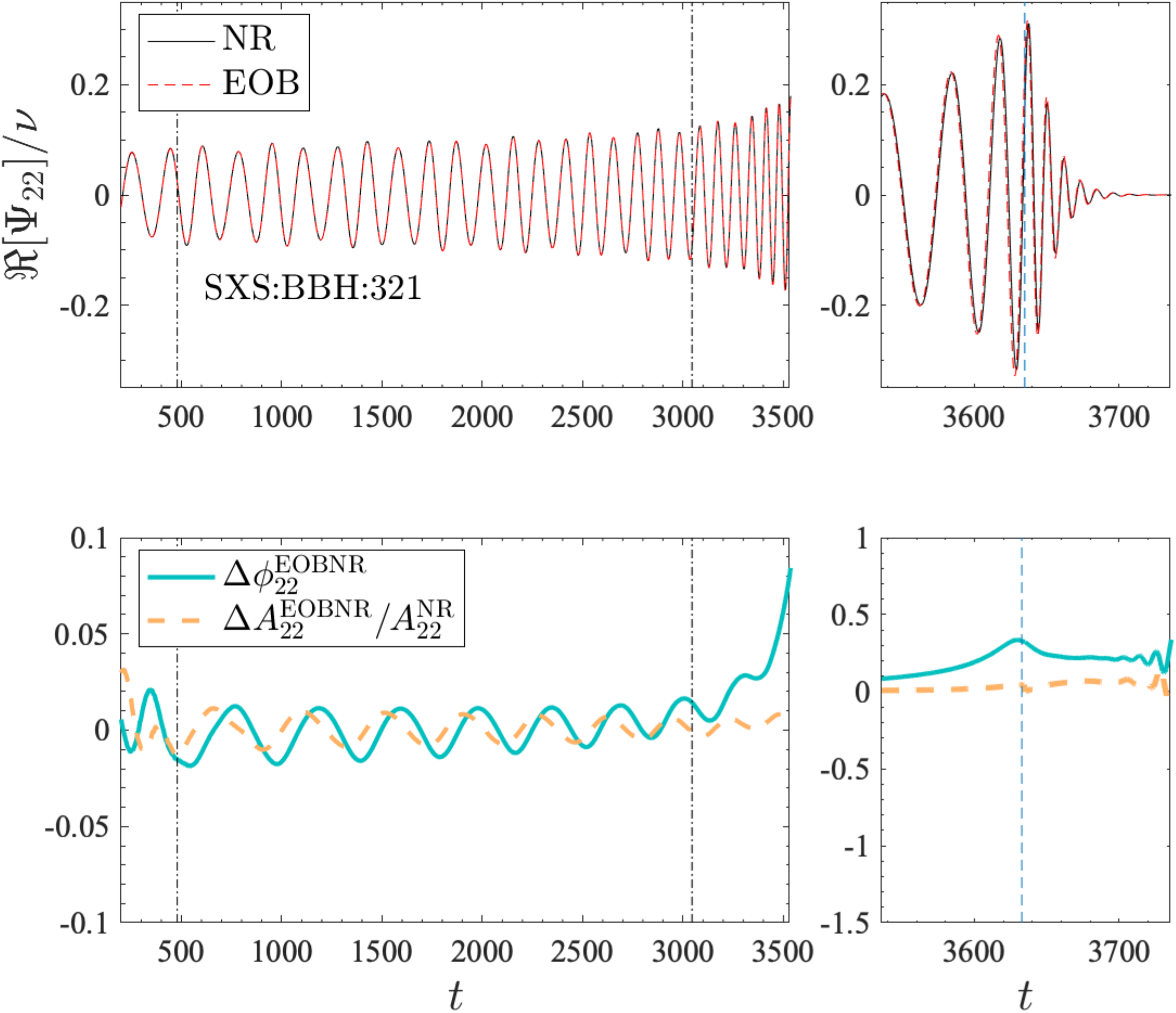}
		\includegraphics[width=0.24\textwidth]{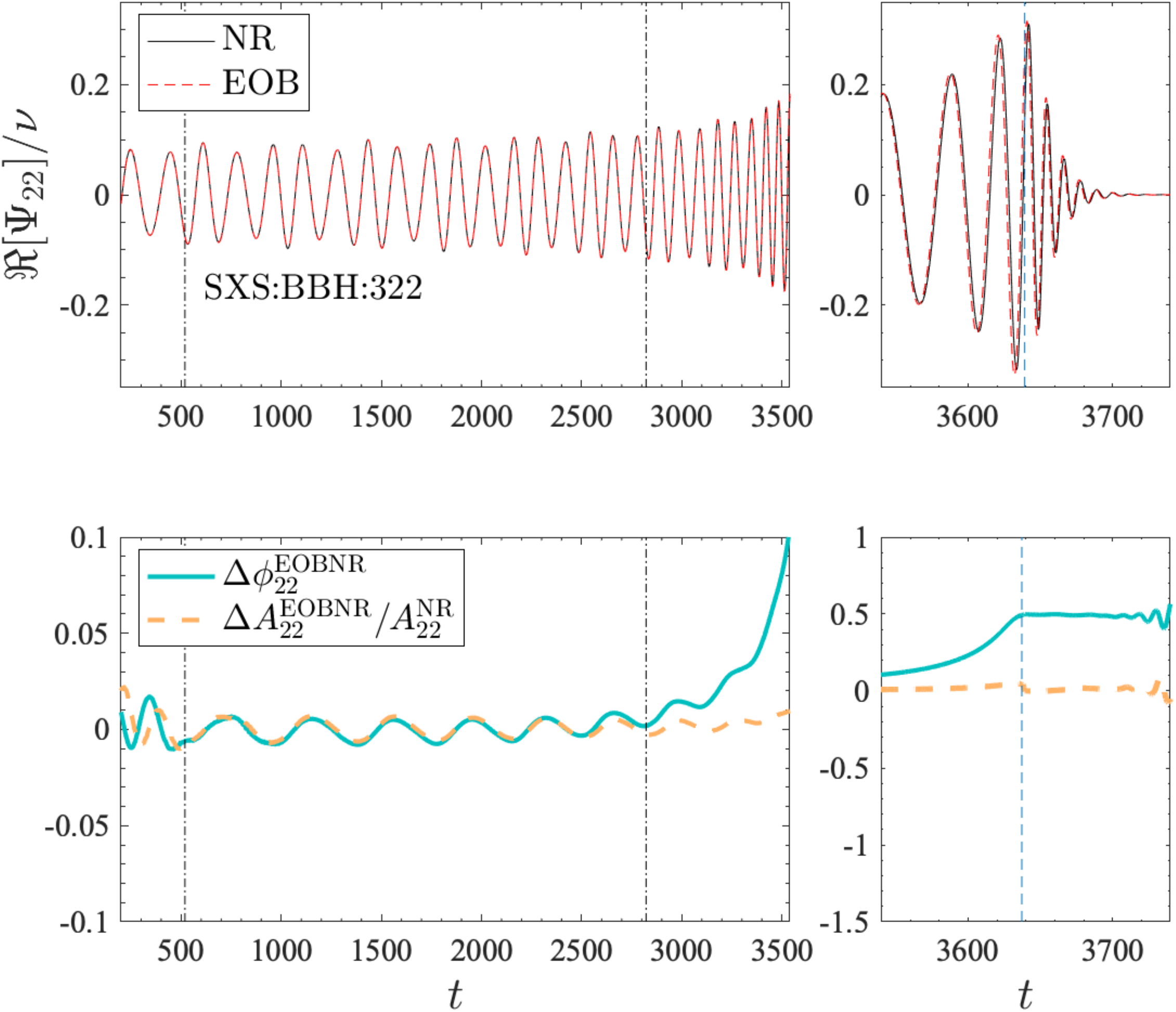}
		\includegraphics[width=0.24\textwidth]{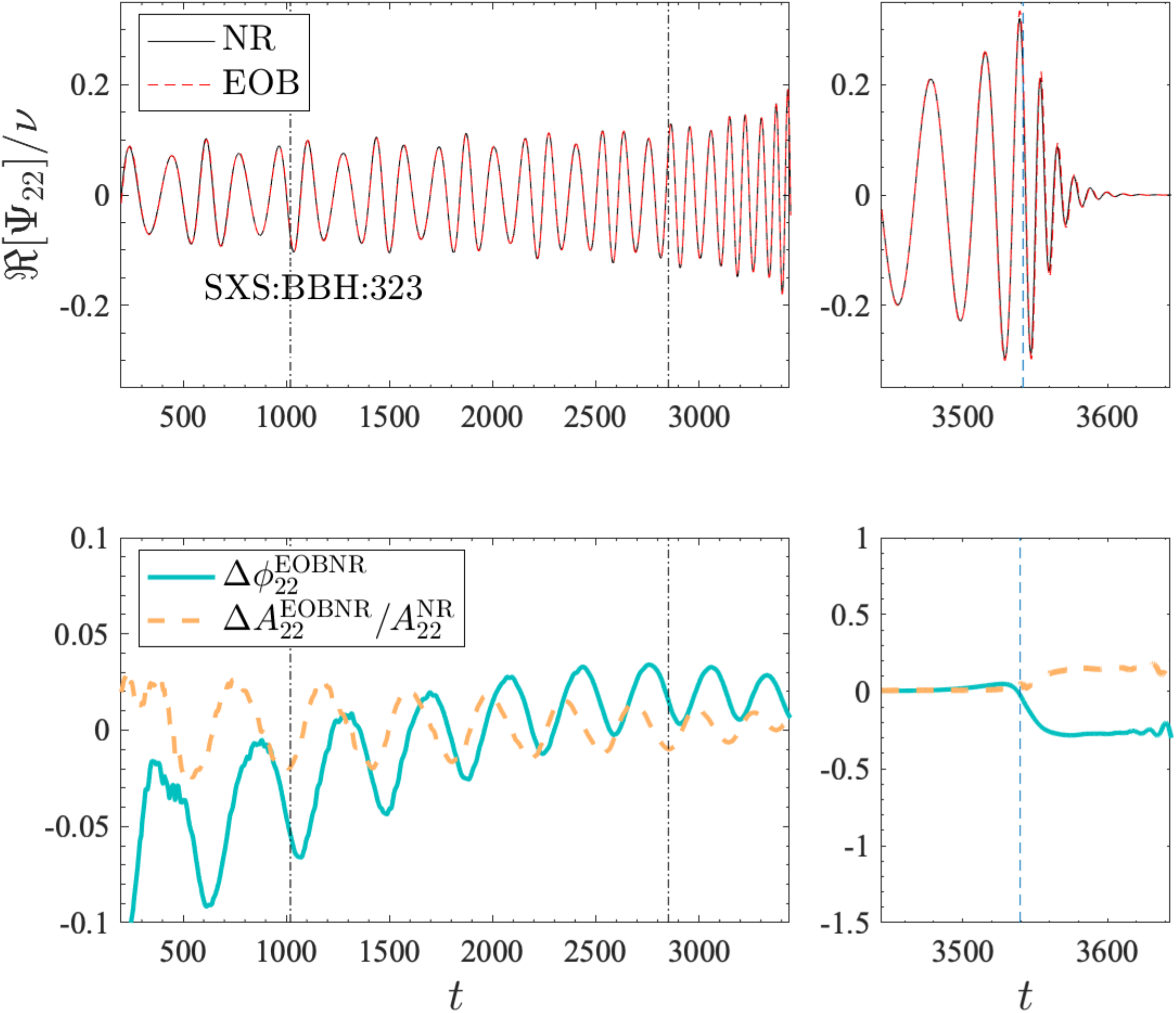}
		\includegraphics[width=0.24\textwidth]{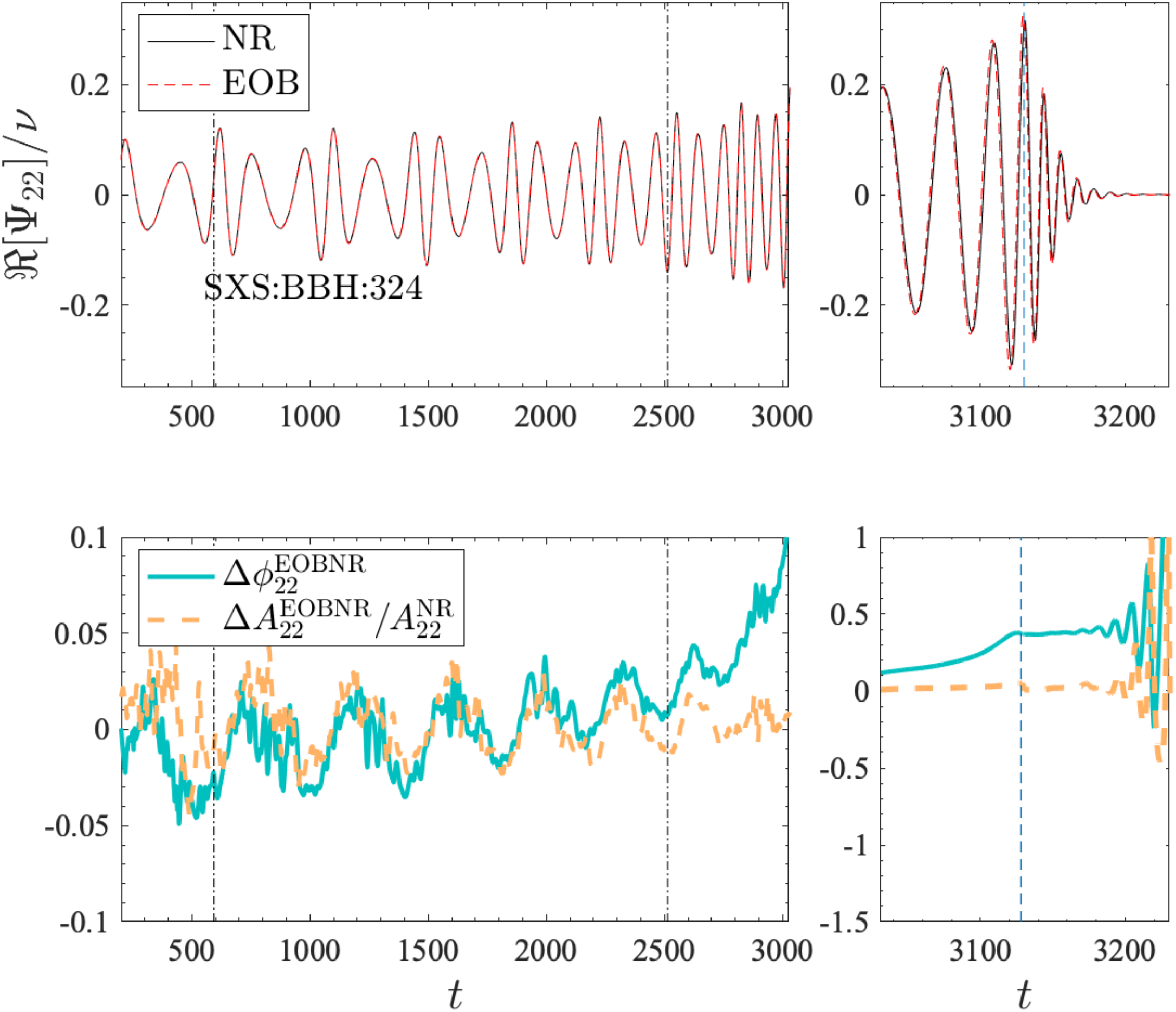} \\
		\vspace{4 mm}
		\includegraphics[width=0.24\textwidth]{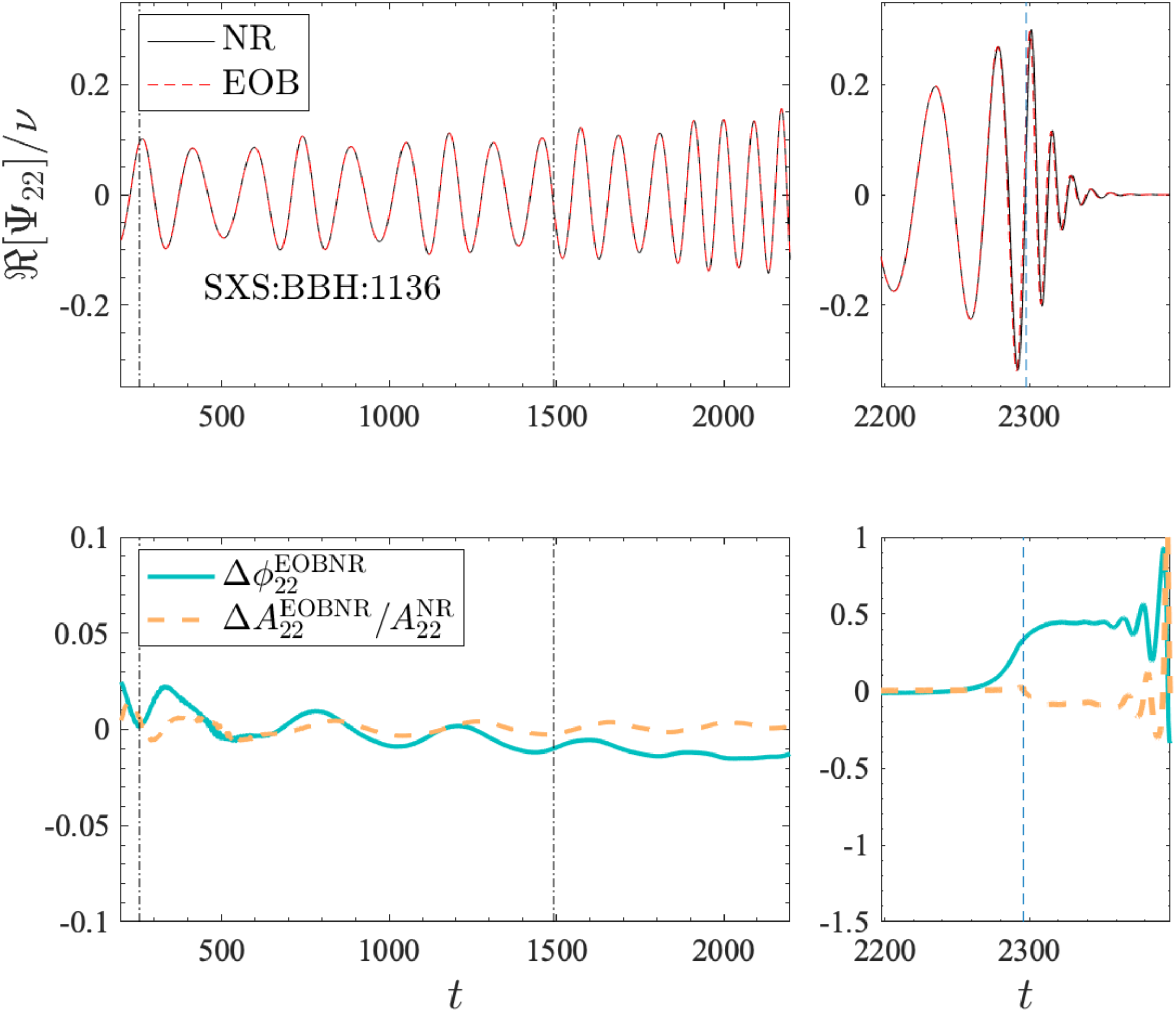}
		\includegraphics[width=0.24\textwidth]{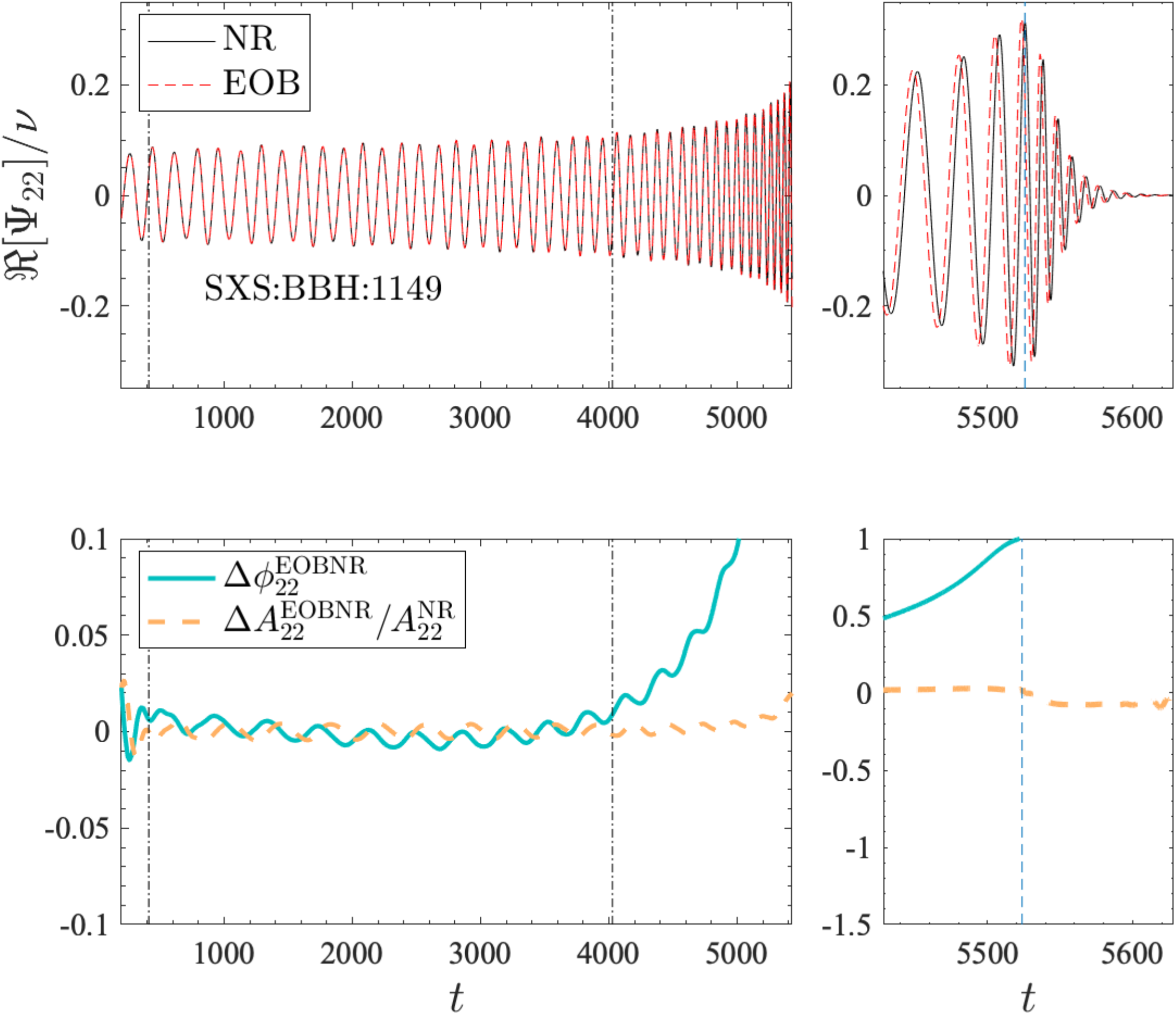}
		\includegraphics[width=0.24\textwidth]{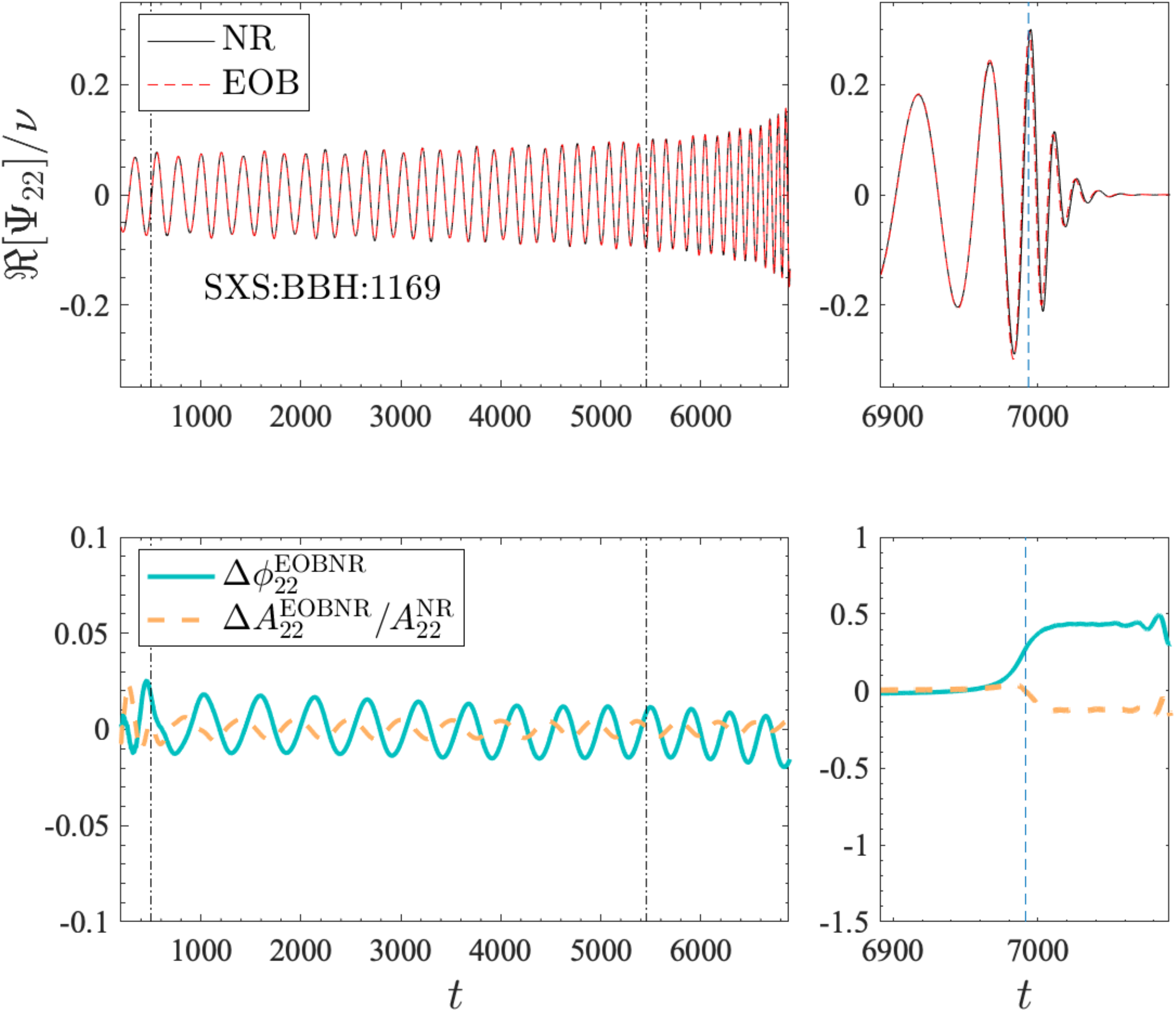}
		\includegraphics[width=0.24\textwidth]{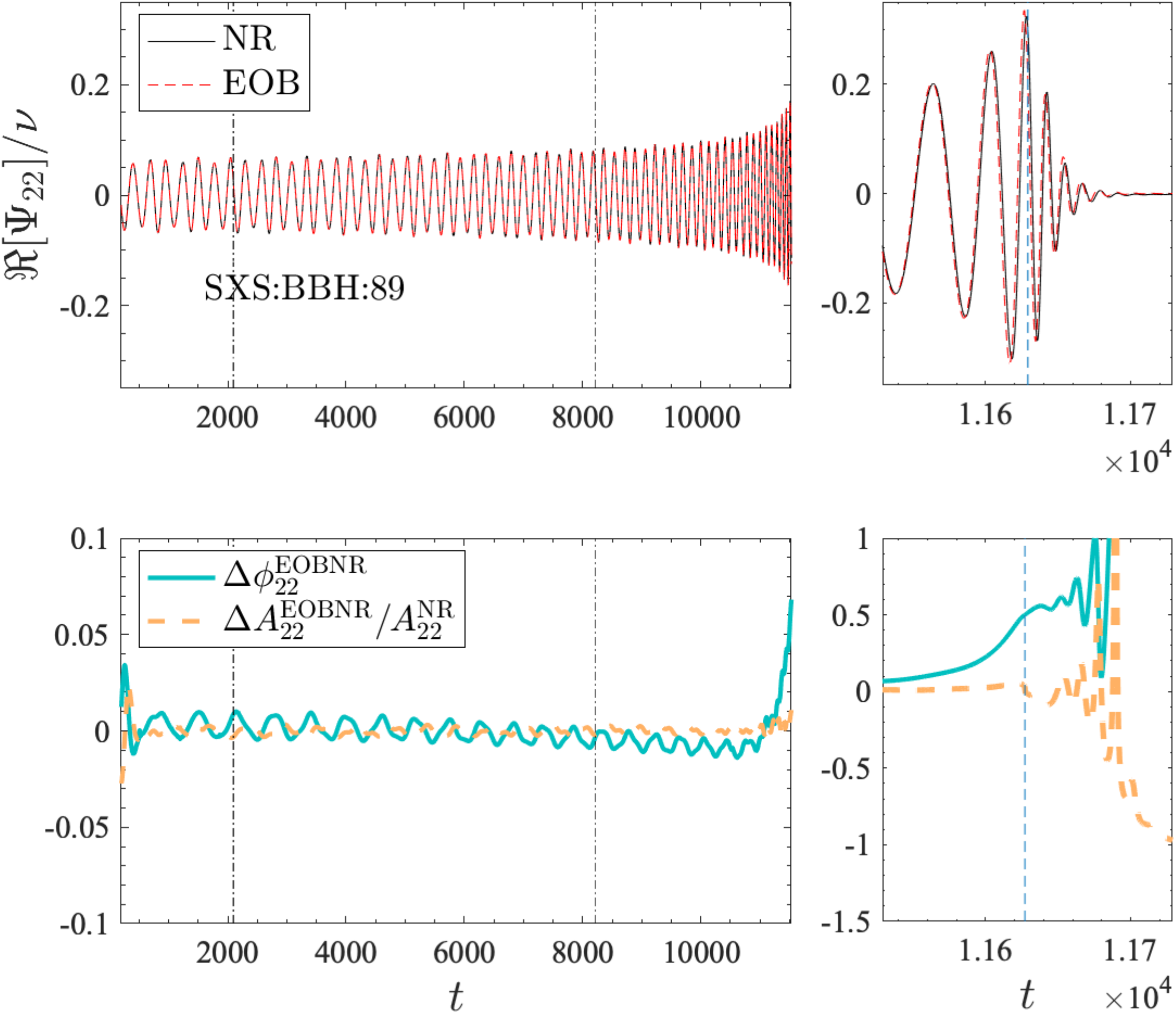}
		\caption{\label{fig:phasing_spin} EOB/NR time-domain phasing for all spinning datasets considered. The dashed vertical
			lines indicate the alignment window. For each configuration, we show together the NR (black) and EOB (red) real 
			part of the waveform (top panel), and the phase difference (blue online) and the relative amplitude difference (orange online),
			bottom panel. }
	\end{figure*}

	Let us consider first the time-domain phasing comparison with the 28 public eccentric datasets
	of the SXS catalog~\cite{Hinder:2017sxy}. We have 20 nonspinning datasets, with initial nominal
	eccentricities up to 0.3, and 8 spin-aligned datasets. In Ref.~\cite{Nagar:2021gss} we had performed specific 
	analyses of this data in order to complement the information available in previous work~\cite{Hinder:2017sxy},
	in particular: (i) computing a gauge-invariant estimate of the eccentricity during the evolution 
	and (ii) giving two different estimates on the NR uncertainty from the two highest resolutions available.
	For completeness, the datasets we consider are listed in Table~\ref{tab:SXS}. The Table reports the
	time-domain phase uncertainty at merger point $\delta\phi^{\rm NR}_{\rm mrg}$ as well as the analogous
	quantities for the unfaithfulness $\bar{F}^{\rm max}_{\rm NR/NR}$ on Advanced LIGO noise, as detailed
	in Ref.~\cite{Nagar:2021gss} according to the definitions that we will recall below.
	Table~\ref{tab:SXS} also reports, for each configuration, the parameters $(e^{\rm EOB}_{\omega_a},\omega_a^{\rm EOB})$
	used to initialize each EOB evolution at apastron (see Refs.~\cite{Chiaramello:2020ehz,Nagar:2021gss}).
	These values are updated with respect to previous work because they are determined by inspecting
	the EOB/NR phase difference in the time-domain and are tuned manually so to reduce as much as 
	possible the difference between the EOB and NR instantaneous GW frequencies~\cite{Nagar:2021gss}. 
	Let us note that our procedure for setting up initial data can be optimized. On the one hand, the
	manual procedure for determining $(e^{\rm EOB}_{\omega_a},\omega_a^{\rm EOB})$ could have
	been automatized. On the other hand, the initial conditions we use are the analogous of the 
	adiabatic initial conditions for circular orbits. As such, they do not reduce to the (iterated) post-adiabatic
	ones~\cite{Damour:2012ky,Nagar:2018gnk} in the quasi-circular limit, and some spurious eccentricity
	would be present in that case. These improvements are discussed in Ref.~\cite{RamosBuades:2021} 
	and will be taken into consideration in the future. It is understood that they can only improve the
	EOB/NR agreement, as pointed out in Ref.~\cite{RamosBuades:2021} for the version of the eccentric
	\TEOBResumS{} of Ref.~\cite{Nagar:2021gss}.
	To convey all available information, we find it useful to explicitly show the time-domain phasing
	comparisons in Figure~\ref{fig:phasing_nospin} (for the nonspinning datasets) and in 
	Figure~\ref{fig:phasing_spin} (for the spinning dataset).
	%------------------------------------
	% Time-domain: nonspinning
	%------------------------------------
	One appreciates that for several cases the careful choice of $(e^{\rm EOB}_{\omega_a},\omega_a^{\rm EOB})$ allows 
	one to obtain a rather flat EOB/NR phase difference, with residual oscillations of the order of 0.01~rad, with accumulated
	phase difference at merger compatible with the nominal NR uncertainty listed in Table~\ref{tab:SXS}. 
	However, for some datasets, notably those with larger initial eccentricities, the choice of the initial parameters 
	looks suboptimal, and the phase difference still shows a linear drift. 
	Typically, this effect is more prominent for dataset with larger initial eccentricity. 
	It might be related to either missing physics in the dynamics~\footnote{We remind the reader that the radiation 
		reaction we are using here only incorporates the noncircular Newtonian prefactor in the $\ell=m=2$ mode.}
	or to the need of further improving the initial data choice. Note however that this happens for NR simulations
	that are especially noisy in the frequency at early times, so this might prevent us from an optimal determination
	of the initial data that is based on the time-alignment procedure, which is in turn affected by the noise
	in the frequency. In any case, our choice of $(e^{\rm EOB}_{\omega_a},\omega_a^{\rm EOB})$ can
	be considered {\it conservative} for all the datasets considered and actually suggests that the analytical
	model can match the NR waveforms even better than what shown in Figs.~\ref{fig:phasing_nospin} 
	and~\ref{fig:phasing_spin}. The understanding that this is probably the case is motivated by the 
	observation that there are datasets with high eccentricity, e.g. SXS:BBH:1362 or SXS:BBH:1363,
	whose EOB/NR phase agreement is practically equivalent to that of less eccentric dataset (e.g. SXS:BBH:1358).
	
	%===========
	% Unfaithfulness
	%===========
	\begin{figure}[t]
		\center
		\includegraphics[width=0.45\textwidth]{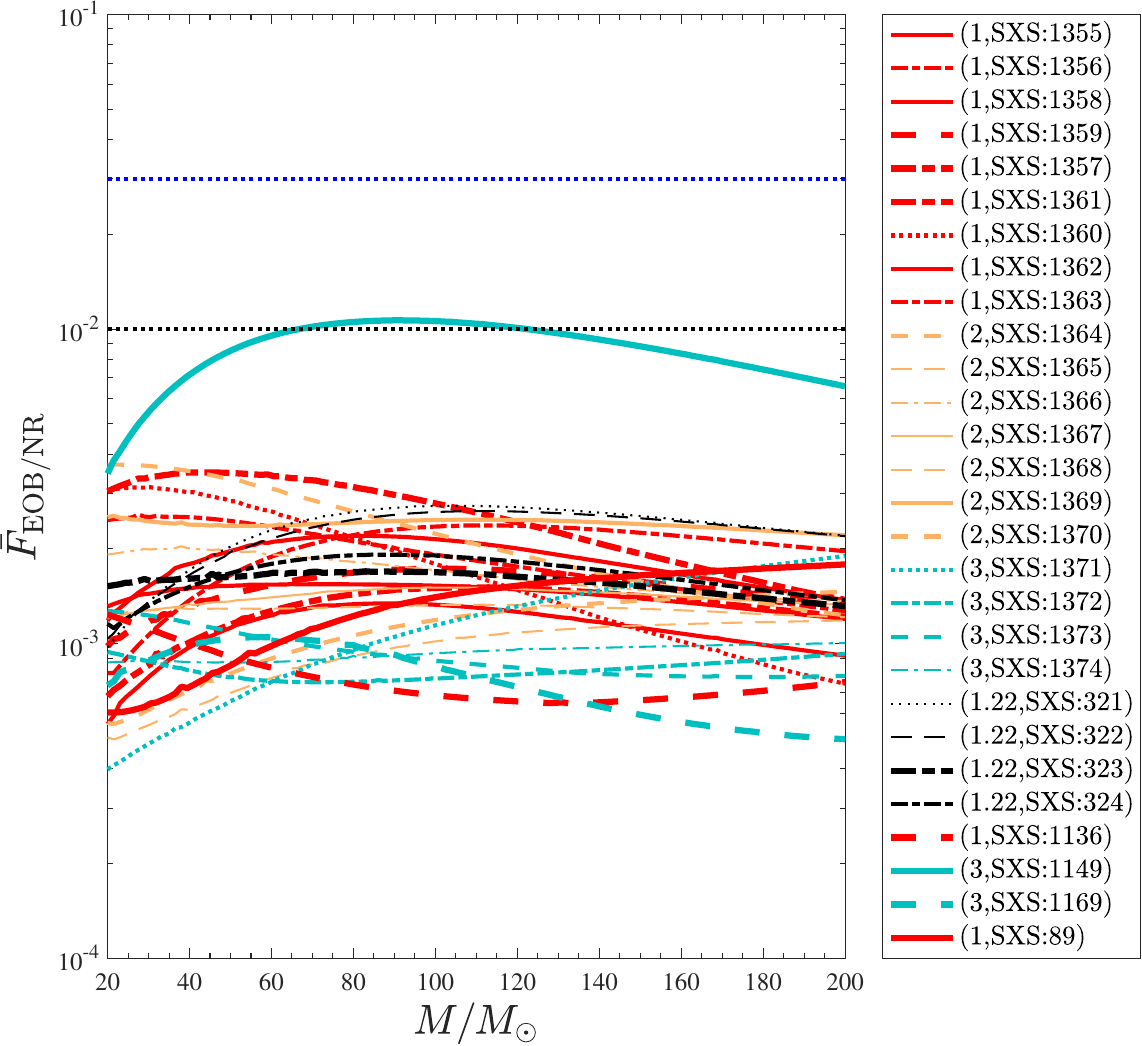}
		\caption{\label{fig:barF} EOB/NR unfaithfulness for the $\ell=m=2$ mode computed over the 
			eccentric SXS simulations publicly available, Table~\ref{tab:SXS}. The horizontal lines mark the 
			0.03 and 0.01 values. The value of $\bar{F}^{\rm max}_{\rm EOB/NR}$ does not exceed
			the $0.7\%$ except for the single outlier given by SXS:BBH:1149, corresponding to $(3,+0.70,+0.60)$ 
			with $e_{\omega_a}^{\rm NR}=0.037$, that is around $1\%$. This is consistent with the slight 
			degradation of the model performance for large positive spins already found in the quasi-circular limit,
			as pointed out in Ref.~\cite{Nagar:2021xnh}.}
	\end{figure}

	\subsection{EOB/NR unfaithfulness}
	\label{sec:barF}
	As done in previous works, as an additional figure of merit we evaluate the quality of the EOB waveform
	by computing the EOB/NR unfaithfulness weighted by the Advanced LIGO noise over all 
	the available configurations. Here we updated the analogous calculation done in Ref.~\cite{Nagar:2021xnh} that was only relying on the simple Newton-factorized waveform without the 2PN-accurate
	eccentric corrections. Considering two waveforms $(h_1,h_2)$, let us recall that the unfaithfulness 
	is a function of the total mass $M$ of the binary and is defined as
	\be
	\label{eq:barF}
	\bar{F}(M) \equiv 1-F=1 -\max_{t_0,\phi_0}\dfrac{\langle h_1,h_2\rangle}{||h_1||||h_2||},
	\ee
	where $(t_0,\phi_0)$ are the initial time and phase. We used $||h||\equiv \sqrt{\langle h,h\rangle}$,
	and the inner product between two waveforms is defined as 
	$\langle h_1,h_2\rangle\equiv 4\Re \int_{f_{\rm min}}^\infty \tilde{h}_1(f)\tilde{h}_2^*(f)/S_n(f)\, df$,
	where $\tilde{h}(f)$ denotes the Fourier transform of $h(t)$, $S_n(f)$ is the zero-detuned,
	high-power noise spectral density of Advanced LIGO~\cite{aLIGODesign_PSD} and
	$f_{\rm min}$ is the initial frequency approximately corresponding to the frequency of the 
	first apastron on each NR simulation, after the initial junk radiation has cleared.
	In practice, the integral is done up to a maximal frequency $f_{\rm end}$ that corresponds 
	to $|\tilde{h}(f_{\rm end})|\sim 10^{-2}$.
	Both EOB and NR waveforms are tapered\footnote{We use a hyperbolic tangent function function 
		with two tunable parameters, $(\alpha,\tau)$, of the form $w(t) = \left[1+\tanh(\alpha t - \tau)\right]/2$
		that multiplies both the NR and EOB waveforms.}
	in the time-domain so as to reduce high-frequency 
	oscillations in the corresponding Fourier transforms. In addition, as originally pointed out in
	Ref.~\cite{Hinder:2017sxy}, the accurate calculation of the Fourier transform of eccentric
	waveforms is a delicate matter and it may affect the calculation of the EOB/NR unfaithfulness,
	$\bar{F}_{\rm EOB/NR}$, if not  optimally chosen. These issues have been discussed to some extent 
	in Sec.~IV of Ref.~\cite{Nagar:2021xnh}, see in particular Fig.~15 and~16 therein. Here we only
	recall that the original waveform is padded with zeros in order to increase the frequency resolution
	and capture all the details of the Fourier transform. Similarly, we were careful to tune the tapering
	parameters so that the EOB and NR Fourier transform for each dataset visually agree likewise to
	the case shown in Fig.~16 of Ref.~\cite{Nagar:2021xnh}.
	The final outcome of the EOB/NR unfaithfulness computation versus $M$ is shown in Fig.~\ref{fig:barF}.
	The maximum values $\bar{F}^{\rm max_{\rm EOB/NR}}$ are also listed in the last column of Table~\ref{tab:SXS},
	together with the value of the initial frequency $M f_{\rm min}$ used in the integral.
	Figure~\ref{fig:barF}, complemented by Table~\ref{tab:SXS}, brings a minimal improvement with
	respect to Fig.~14 of Ref.~\cite{Nagar:2021xnh}, especially for low masses. Since we are using
	here a new choice of the parameters  $(e_{\omega_a}^{\rm EOB},\omega_a^{\rm EOB})$ (and consequently
	new tapering parameters) it is not really possible, within the context of equal-mass binaries, to precisely 
	state to which extent the small improvements found depend on these new choices or on the additional
	PN corrections in the waveforms. Globally, in view of the similarities between Fig.~14 of Ref.~\cite{Nagar:2021xnh}
	and our current Fig.~\ref{fig:barF}, we are prone to conservatively state that the factorized and resummed 
	2PN noncircular corrections to the waveform are not especially important on this specific corner of 
	the parameter space.

	\section{Testing the quasi-circular factorization of Khalil et al.~\cite{Khalil:2021txt}.}
	\label{sec:qc}
	%==========================================
	% Fig.05: insplunge: AEI waves
	%==========================================
	\begin{figure*}
		\center
		\includegraphics[width=0.3\textwidth]{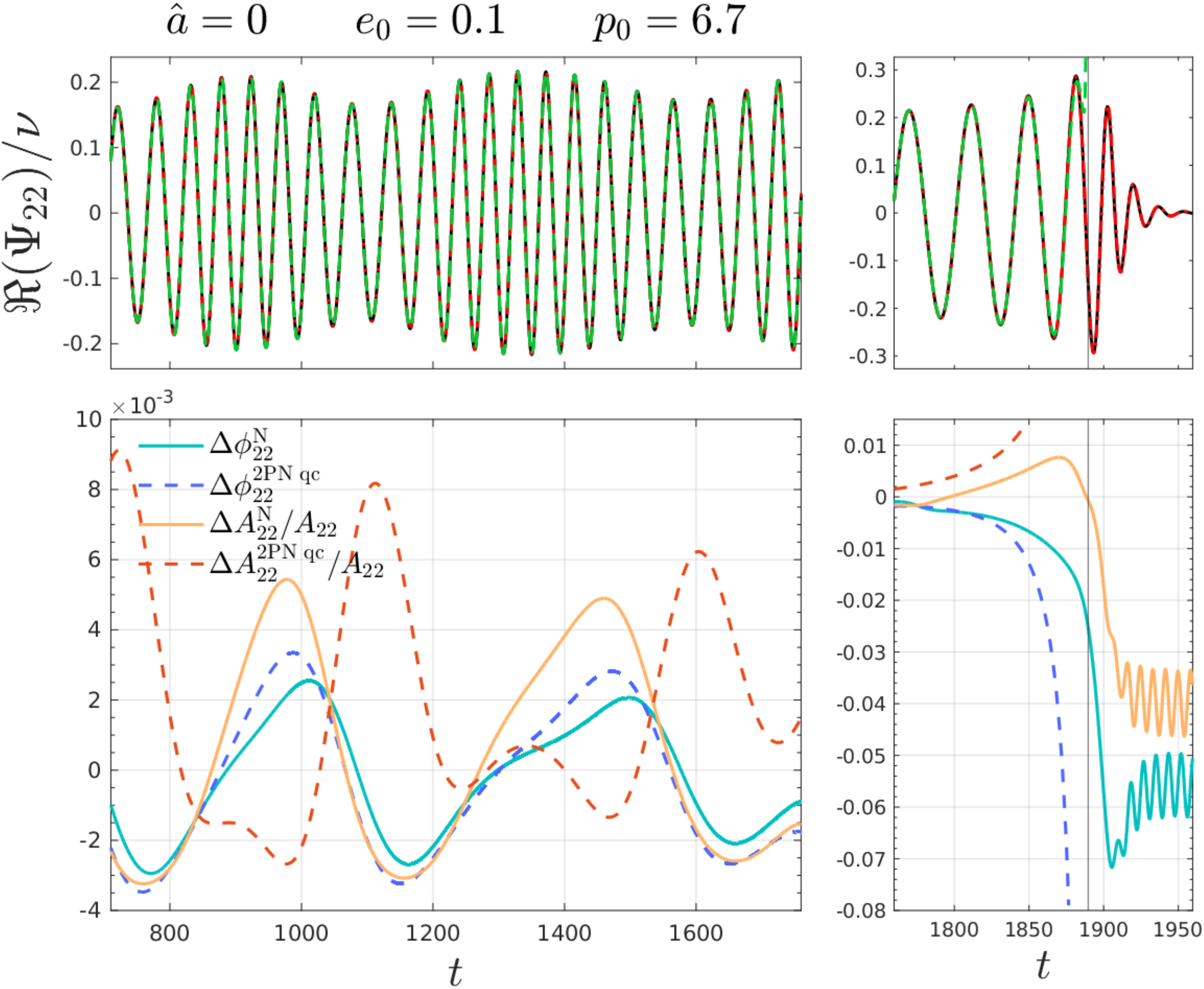}
		\hspace{0.3cm}
		\includegraphics[width=0.3\textwidth]{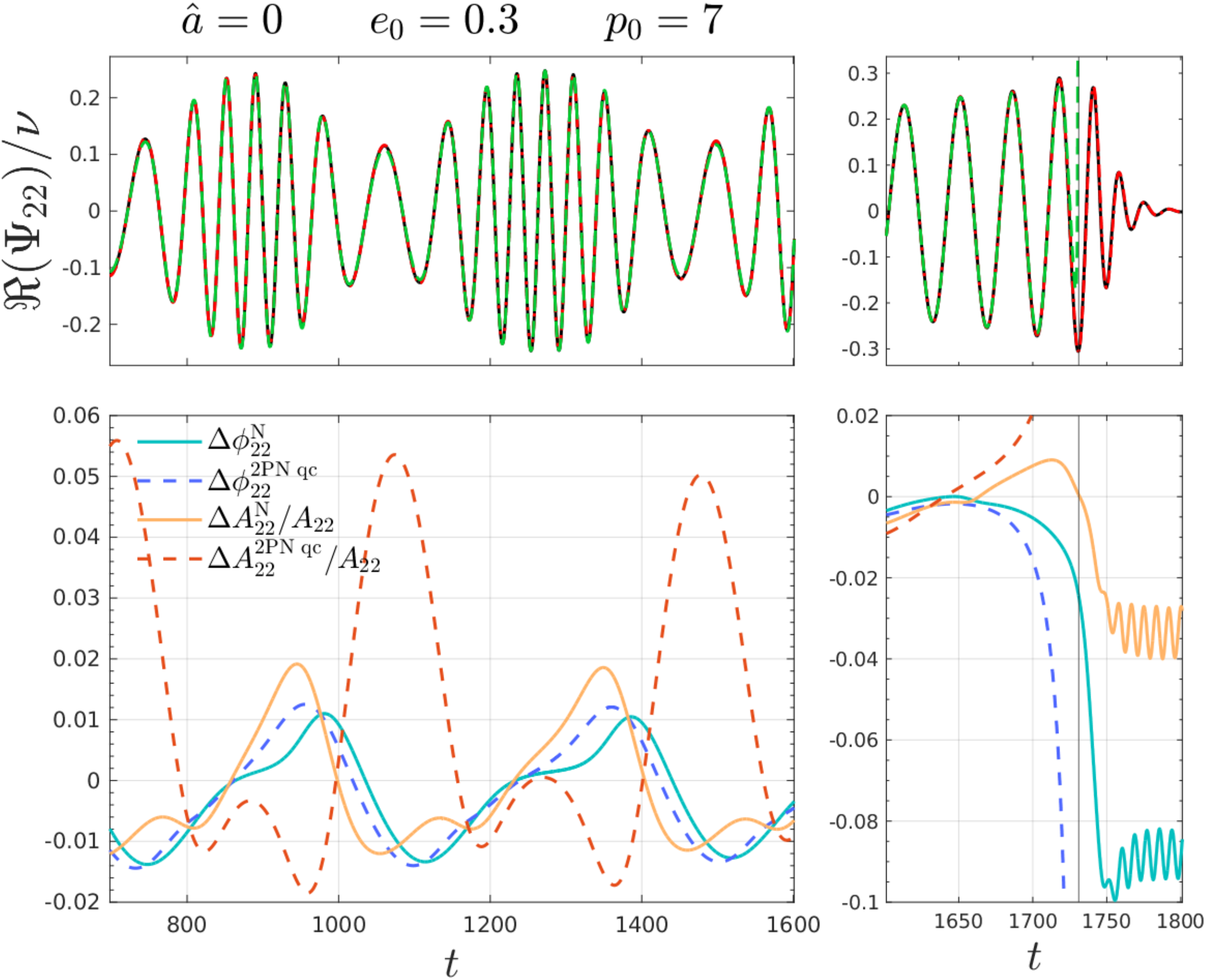}
		\hspace{0.3cm}
		\includegraphics[width=0.3\textwidth]{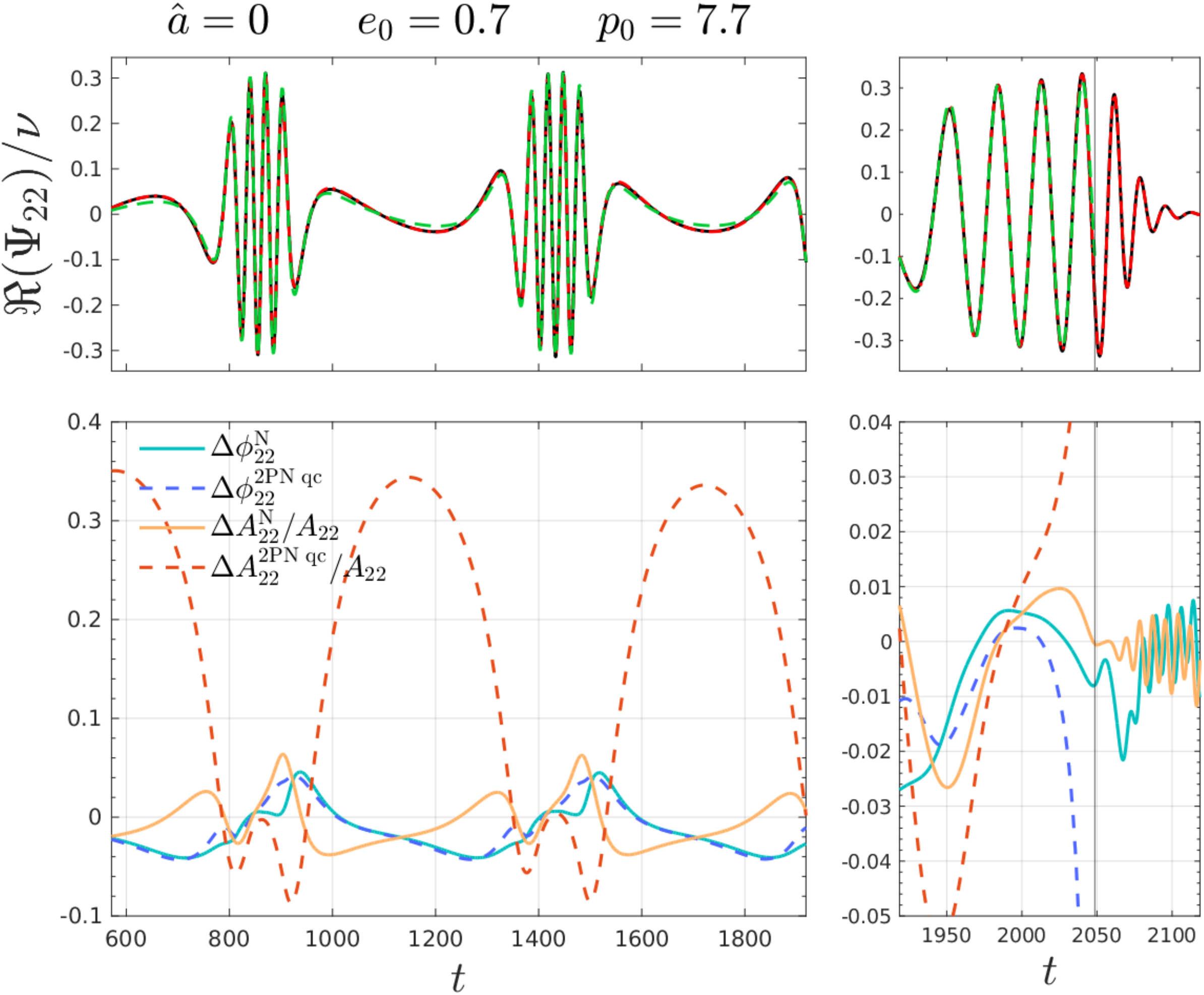}
		\caption{\label{fig:testmass_insplunge_AEI} 
			Testing the waveform factorization of Ref.~\cite{Khalil:2021txt},
			Eq.~\eqref{eq:AEIwave}: comparisons between the $\ell=m=2$ numerical and analytical waveforms
			emitted by the eccentric inspiral of a test particle on a Schwarzschild black hole.
			The initial eccentricities and semilatera recta are 
			$(e_0, p_0)$ =$(0.1, 6.7)$, $(0.3, 7)$, $(0.7, 7.7)$.
			Each top panel displays the numerical waveform (black, indistinguishable); the EOB waveform
			with the generic Newtonian prefactor (dash-dotted red, labeled N) and the 2PN accurate waveform 
			with the quasi-circular factorization of Eq.~\eqref{eq:AEIwave} (labeled ${\rm 2PN_{qc}}$).
			The corresponding phase differences and relative amplitude differences are shown in the 
			bottom panels.
			The vertical black line marks the merger-time, corresponding to the peak of the numerical 
			waveform amplitude.
			For simplicity, the ${\rm 2PN^{qc}}$ waveform has not been completed by NQC 
			corrections and ringdown.
			The analytical/numerical phase agreement is comparable for the two choices (blue lines);
			by contrast, the amplitude disagreement is always larger for the ${\rm 2PN_{qc}}$ prescription, and
			worsens up to $30\%$ when the eccentricity increases.}
	\end{figure*}
	%===========================
	During the development of the this work under the paradigm of the factorization
	of the general  Newtonian prefactor, Ref.~\cite{Khalil:2021txt} appeared. Besides
	providing some of the analytical expressions used here
	(e.g.~the explicit expression of the tail), Ref.~\cite{Khalil:2021txt} also 
	proposed a different waveform factorization where: (i) only the {\it quasi-circular} 
	Newtonian prefactor is factored out; (ii) all the noncircular effects are interpreted 
	as corrections to the quasi-circular baseline expression and (iii) the instantaneous 
	and hereditary contributions are included in additive form.
	
	Given our numerical waveforms in the test-mass limit, it is interesting to thoroughly 
	test also this analytical waveform proposal. We carefully follow Sec.~IIIB 
	of Ref.~\cite{Khalil:2021txt}
	and we report here all the equations needed to this aim. For $m>0$, Ref.~\cite{Khalil:2021txt} 
	proposes the following factorized expression
	\begin{equation}
		\label{eq:AEIwave}
		h^{\rm 2PN_{qc}}_\lm = h_\lm^{(N, \epsilon)_c} \hat{S}_{\rm eff} 
		\left( T_\lm + T_\lm^{\rm ecc} \right) e^{i \delta_\lm} 
		\left( f_\lm + f_\lm^{\rm ecc} \right)  ,
	\end{equation}
	where the eccentric terms $f_\lm^{\rm ecc}$ and $T_\lm^{\rm ecc}$ 
	are written as functions of ($r$, $p_r$, $\dot{p}_r$).
	For the leading-order quasi-circular hereditary term $T_\lm$ we use the standard prescription 
	introduced in Ref.~\cite{Damour:2008gu}, while for $\delta_\lm$, and $f_\lm$ 
	we follow Refs.~\cite{Nagar:2016ayt, Messina:2018ghh, Albanesi:2021rby}.
	We here focus explicitly on the $(2,2)$ mode only, precisely following
	the steps of Ref.~\cite{Khalil:2021txt}.
	More precisely, we use the full expression of $T_{22}^{\rm ecc}$ 
	presented in the supplemental material of Ref.~\cite{Khalil:2021txt}, that reads
	\begin{widetext}
		\begin{align}
			T_{22}^{\rm ecc} = 
			&  \eta ^3 \pi \bigg[ -\frac{3 i p_r}{2 r} -\dot{p}_r \sqrt{r}
			-\frac{p_r \left(2 p_r+i \dot{p}_r r^{3/2}\right)}{4 \sqrt{r}} + 
			\frac{1}{48} \left(-5 i \text{pr}^3-15 \text{pr}^2 
			\dot{p}_r r^{3/2}+9 i \text{pr} \dot{p}_r^2
			r^3-\dot{p}_r^3 r^{9/2}\right) \nonumber \\
			& +\frac{1}{32} \sqrt{r} \left(-10 p_r^4+15 i p_r^3 
			\dot{p}_r r^{3/2}  +8 p_r^2 \dot{p}_r^2 r^3-i p_r \dot{p}_r^3 r^{9/2} +2 
			\dot{p}_r^4 r^6\right)+\frac{r}{1920} \left(589 i
			p_r^5+1150 p_r^4 \dot{p}_r r^{3/2}\right. \nonumber \\
			& \left.-1060 i p_r^3 \dot{p}_r^2 r^3-505 p_r^2 \dot{p}_r^3
			r^{9/2}+55 i p_r \dot{p}_r^4 r^6-116 \dot{p}_r^5 r^{15/2}\right) +\frac{r^{3/2}}{11520} 
			\left(1974 p_r^6-4995
			i p_r^5 \dot{p}_r r^{3/2}\right. \nonumber \\
			& \left. -7470 p_r^4 \dot{p}_r^2 r^3+7240 i p_r^3 \dot{p}_r^3 r^{9/2}+3855
			p_r^2 \dot{p}_r^4 r^6-669 i p_r \dot{p}_r^5 r^{15/2}+560 \dot{p}_r^6 r^9\right)\bigg].
		\end{align}
	\end{widetext}
	For the instantaneous contribution, $f_{22}^{\rm ecc}$, we use Eq.~(122) of
	Ref.~\cite{Khalil:2021txt} that we rewrite here explicitly specified to the
	test-mass limit ($\nu = 0$),
	\begin{widetext}
		\begin{align}
			f_{22}^{\rm ecc} & =
			-\frac{p_r^2}{2 v_0^2}+i p_r r v_0+\frac{r^2 v_0^4}{2}+\frac{1}{2 r v_0^2}-1 +
			\eta ^2 \left(\frac{p_r^4}{4 r^3 v_0^8}+\frac{29 p_r^4}{84 v_0^2}+\frac{i p_r^3}{4 r^2
				v_0^5}-\frac{37}{84} i p_r^3 r v_0-\frac{p_r^2}{12 r^4 v_0^8}+\frac{1}{4} p_r^2 r^2
			v_0^4+\frac{31 p_r^2}{14 r v_0^2} \right. \nonumber \\
			& \left. -\frac{37}{84} i p_r r^3 v_0^7+\frac{i p_r}{6 r^3
				v_0^5}-\frac{209 i p_r v_0}{84}-\frac{1}{6 r^5 v_0^8}-\frac{2}{21} r^4 v_0^{10}-\frac{13}{12 r^2
				v_0^2}-\frac{59 r v_0^4}{84}+\frac{43 v_0^2}{21}\right)
			+
			\eta ^3 \left(-\frac{\ha p_r^2}{3 r^3 v_0^5}-\frac{4 i \ha p_r}{3 r^2
				v_0^2} \right. \nonumber \\
			& \left. +\frac{\ha}{3 r^4 v_0^5}-\frac{5 \ha v_0}{3 r}+\frac{4 \ha
				v_0^3}{3}\right)
			+
			\eta ^4 \left(-\frac{\ha^2 p_r^4}{6 r^4 v_0^8}-\frac{i \ha^2 p_r^3}{6 r^3
				v_0^5}+\frac{\ha^2 p_r^2}{6 r^5 v_0^8}-\frac{5 \ha^2 p_r^2}{6 r^2 v_0^2}+\frac{i
				\ha^2 p_r v_0}{r}+\frac{\ha^2}{2 r^3 v_0^2}-\frac{\ha^2
				v_0^4}{2}-\frac{p_r^6}{4 r^6 v_0^{14}} \right. \nonumber \\
			& \left. -\frac{31 p_r^6}{112 r^3 v_0^8}-\frac{277
				p_r^6}{1008 v_0^2}-\frac{5 i p_r^5}{32 r^5 v_0^{11}}-\frac{3 i p_r^5}{14 r^2
				v_0^5}+\frac{571 i p_r^5 r v_0}{2016}-\frac{p_r^4}{12 r^7 v_0^{14}}-\frac{179 p_r^4}{144
				r^4 v_0^8}-\frac{85}{252} p_r^4 r^2 v_0^4-\frac{1963 p_r^4}{1008 r v_0^2} \right. \nonumber \\
			& \left. -\frac{5 i
				p_r^3}{24 r^6 v_0^{11}}+\frac{349 i p_r^3 r^3 v_0^7}{1008}-\frac{953 i p_r^3}{1008 r^3
				v_0^5}+\frac{803}{504} i p_r^3 v_0+\frac{2 p_r^2}{9 r^8 v_0^{14}}+\frac{79 p_r^2}{168 r^5
				v_0^8}-\frac{71 p_r^2 r^4 v_0^{10}}{1008}+\frac{43 p_r^2}{42 r^3 v_0^4}-\frac{31
				p_r^2}{168 r^2 v_0^2} \right. \nonumber \\
			& \left. -\frac{31}{112} p_r^2 r v_0^4-\frac{43 p_r^2 v_0^2}{126}-\frac{5 i
				p_r}{72 r^7 v_0^{11}}+\frac{127 i p_r r^5 v_0^{13}}{2016}-\frac{83 i p_r}{504 r^4
				v_0^5}+\frac{5}{63} i p_r r^2 v_0^7-\frac{9221 i p_r v_0}{6048 r}+\frac{1}{9 r^9
				v_0^{14}}-\frac{1}{126} r^6 v_0^{16} \right. \nonumber \\
			& \left. +\frac{1}{6 r^6 v_0^8}+\frac{43}{63 r^4 v_0^4}+\frac{53 r^3
				v_0^{10}}{168}+\frac{13}{504 r^3 v_0^2}-\frac{43 r^2 v_0^8}{126}-\frac{43 v_0^2}{126 r}-\frac{11
				v_0^4}{18}\right),
		\end{align}
	\end{widetext}
	where 
	\begin{equation}
		v_0 = \frac{(1+\dot{p}_r r^2)^{1/6}}{\sqrt{r}}.
	\end{equation}
	Following the practice introduced in Ref.~\cite{Nagar:2006xv}, when considering the Hamiltonian
	formalism for a test particle around a Schwarzschild black hole it is useful
	to replace $p_r$ with $p_{r_*}$, the radial momentum conjugate to some
	Regge-Wheeler tortoise coordinate~\cite{Regge:1957td}. The scope of this is 
	to avoid the presence of spurious numerical singularities towards the horizon
	when the equations of motion are solved numerically. We thus rewrite the 
	above equations  replacing  ($p_r$, $\dot{p}_r$) with ($p_{r_*}$, $\dot{p}_{r_*}$).
	Using Hamilton's equations, the radial momentum and its derivative are
	written as (see Appendix E of Ref.~\cite{Khalil:2021txt})
	\begin{align}
		p_r & = \sqrt{\frac{B}{A}}p_{r_*}\equiv \xi^{-1}(r) p_{r_*}, \\
		\dot{p}_r & = - \bigg[   
		\left(\frac{\partial H}{\partial r}\right)_{p_{r_*}}  +
		\left(\frac{\partial H}{\partial p_{r_*}}\right)_{r} 
		\frac{p_{r_*}}{\xi(r)} \frac{d \xi(r)}{dr}   
		\bigg].
	\end{align}
	The explicit expressions for the Kerr metric functions $A$ and $B$ can be found 
	in Ref.~\cite{Damour:2014sva}. 
	Figure~\ref{fig:testmass_insplunge_AEI} shows the triple comparisons between:
	(i) the numerical waveform (black online, barely distinguishable in the top panels); 
	(ii) the waveform with the generic Newtonian prefactor~\cite{Chiaramello:2020ehz} (red online)
	and (iii) the 2PN-accurate waveform of above (dashed, green online). This is done 
	for three values of the initial eccentricity $e_0=(0.1, 0.3, 0.7)$. For each analytical waveform
	choice, the bottom panels of the figure reports the corresponding phase differences
	and fractional amplitude differences. For simplicity, the $h_\lm^{\rm 2PN_{ qc}}$
	is not completed through merger (via NQC corrections) and ringdown.
	As can be seen in the figure, the amplitude differences become relevant at 
	large eccentricities during the inspiral at apastra. Moreover, even at small eccentricity 
	the waveform with only Newtonian corrections seems to perform globally better 
	than the outcome of Eq.~\eqref{eq:AEIwave}. 
	%=====================
	% Understanding amplitude 1
	%=====================
	\begin{figure}
		\center
		\includegraphics[width=0.45\textwidth]{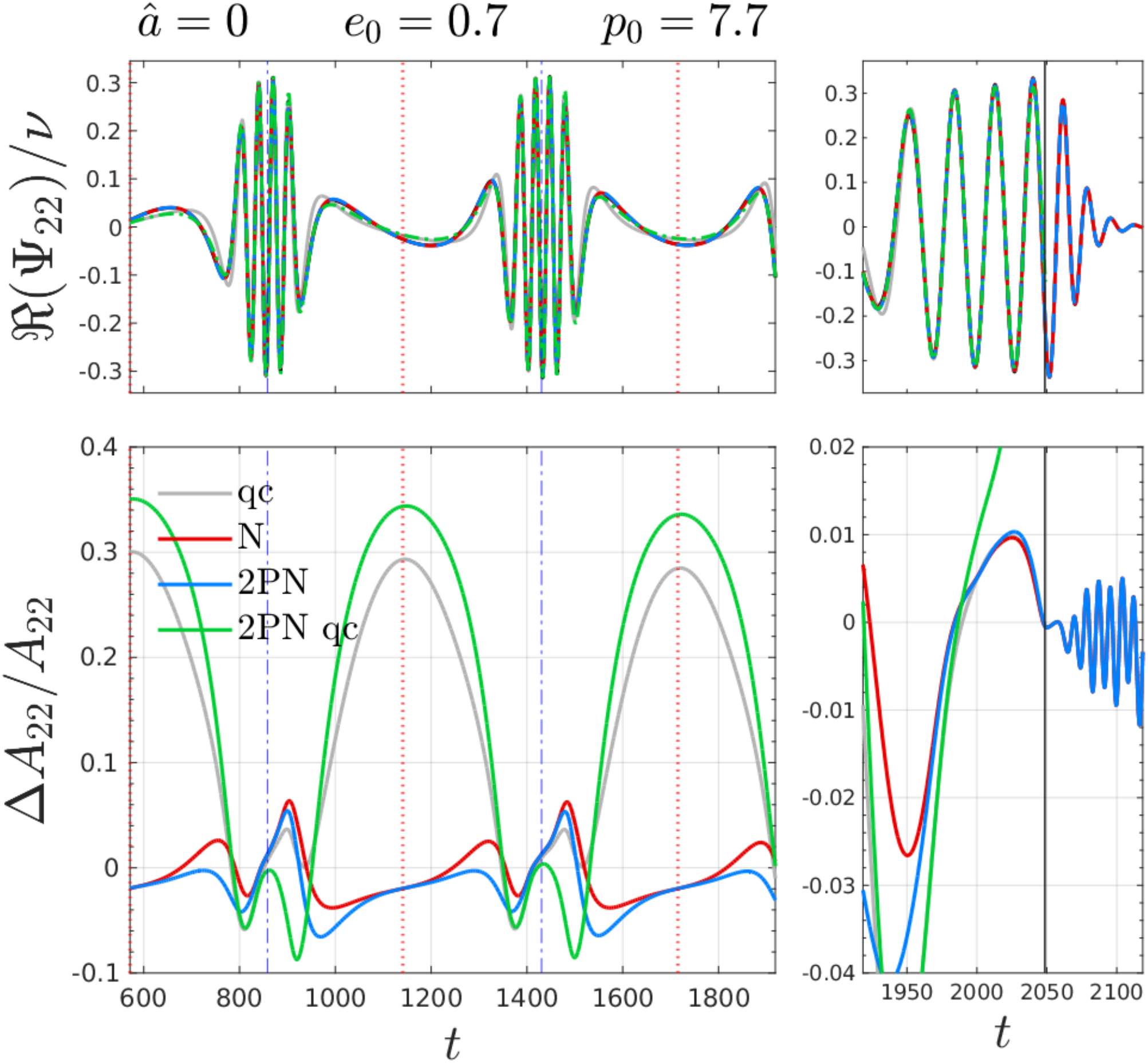}
		\caption{\label{fig:apastron_fig1} Comparing waveforms generated by a test-particle inspiralling and 
		plunging into a Schwarzschild black hole with $(e_0, p_0)=(0.7,7.7)$. Top panel:
		the numerical waveforms (black, indistinguishable); the quasi-circular EOB waveform (gray);
		the waveform with the Newtonian noncircular corrections (red); the waveform of Eq.~\eqref{eq:hlm_fact}
		that adds to the previous one the 2PN corrections (blue); and the
		waveform of Eq.~\eqref{eq:AEIwave} with the quasi-circular factorization and 2PN 
		noncircular corrections (green, without merger and ringdown).
		Bottom panel: relative amplitude differences with the numerical waveforms. Apastra are marked by dotted (red) vertical 
		lines, while periastra with dash-dotted (blue) lines. See text for discussion.}
	\end{figure}
	%
	%===========
	% Second check
	%===========
	\begin{figure}
		\center
		\includegraphics[width=0.45\textwidth]{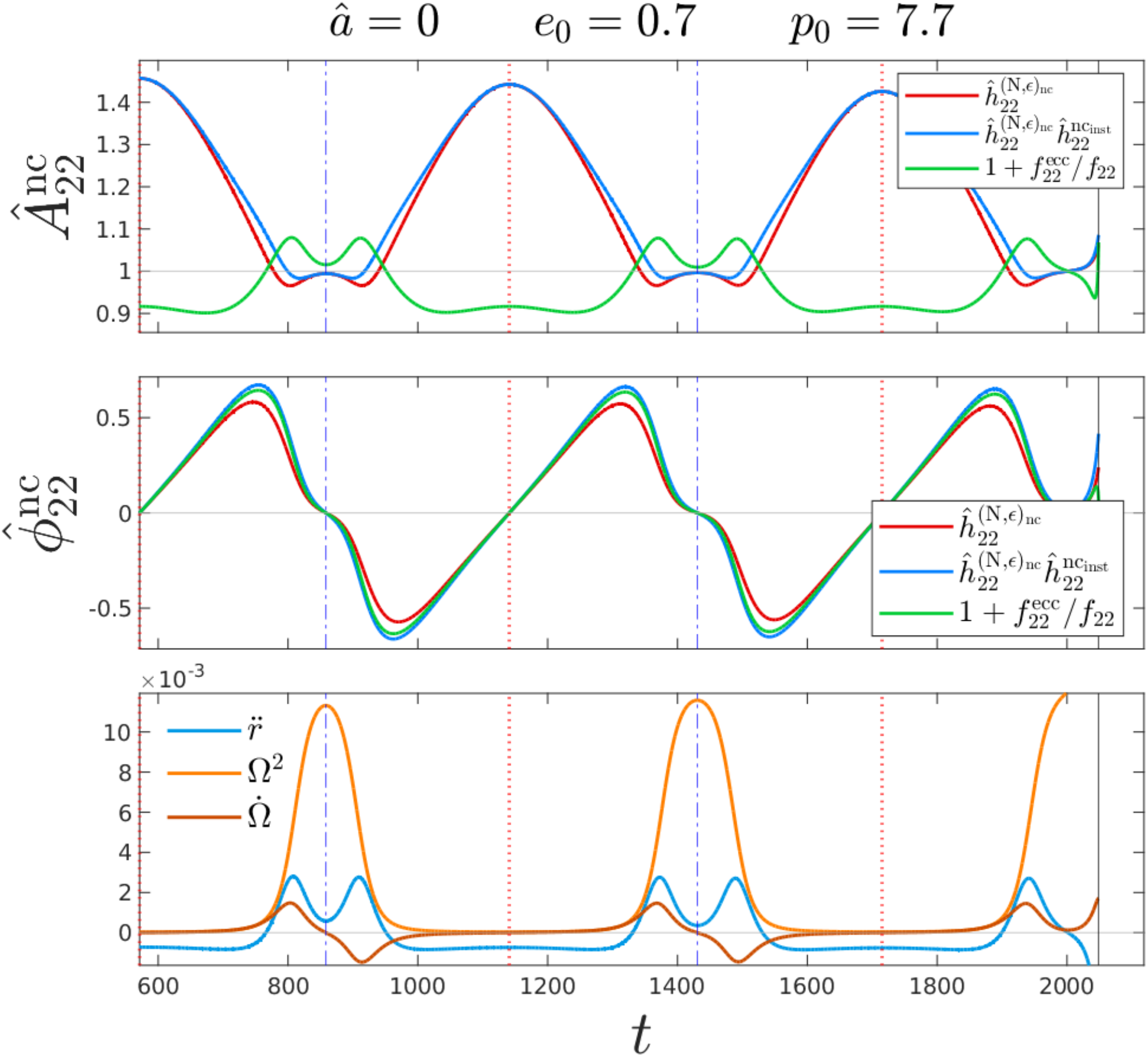}
		\caption{\label{fig:apastron_fig2} Contrasting different noncircular corrections for the
			same configuration of Fig.~\ref{fig:apastron_fig1}. Top and middle panels:
			the noncircular contributions to amplitude and phase. We consider the noncircular 
			Newtonian prefactor  $\hat{h}_\lm^{\rm (N,\epsilon)_{nc}}$ of Eq.~\eqref{eq:ncN22} (red),
			the Newtonian-factorized instantaneous corrections up to 2PN, 
			$\hat{h}_\lm^{\rm (N,\epsilon)_{nc}} \hat{h}_\lm^{\rm {nc}_{inst}}$, (blue), 
			and the 2PN noncircular corrections of Eq.~\eqref{eq:AEIwave} written as
			$1+f_{22}^{\rm ecc}/f_{22}$ (green). 
			The bottom panel shows $\ddot{r}$, $\Omega^2$ and $\dot{\Omega}$.
			The correction proportional to $\ddot{r}/(r\Omega^2)$ in Eq.~\eqref{eq:ncN22} yields
			larger values at apastron than $1+f_{22}^{\rm ecc}/f_{22}$. See text for discussion.}
	\end{figure}

	Figure~\ref{fig:testmass_insplunge_AEI}  also highlights an aspect that is a priori
	unexpected: the largest amplitude differences occur at {\it apastron} and not 
	at periastron. This might look puzzling because PN expansions are more accurate 
	in weak field than in strong field, while the plot seems to indicate the opposite.
	The reason of this behavior can be understood inspecting Figs.~\ref{fig:apastron_fig1} 
	and ~\ref{fig:apastron_fig2}. They refer to the same configuration ($e_0=0.7$) of
	the rightmost panel of Fig.~\ref{fig:testmass_insplunge_AEI}.
	In Fig.~\ref{fig:apastron_fig1} we compare different analytical quadrupolar waveforms 
	and their analytical/numerical relative amplitude differences. In particular:
	(i) the quasi-circular EOB waveform (gray online); (ii) the waveform with noncircular
	corrections included in the generic Newtonian prefactor (red online);
	(iii) the waveform of Eq.~\eqref{eq:hlm_fact}, where the 2PN noncircular effects
	are incorporated as a multiplicative correction to the Newtonian prefactor (blue online);
	and (iv) the waveform with quasi-circular factorization and 2PN noncircular corrections as written
	in Eq.~\eqref{eq:AEIwave} (green online). These plots are completed by Fig.~\ref{fig:apastron_fig2},
	that illustrates the noncircular {\it instantaneous} corrections to the amplitude and to the phase 
	for each analytical prescription.
	The instantaneous noncircular correction for the waveform of Eq.~\eqref{eq:AEIwave}
	is written as the multiplicative factor $1+f^{\rm ecc}_{22}/f_{22}$, for formal 
	consistency with the other analytical choices.
	All noncircular factors provide a relevant correction to the phase, as shown 
	in the middle panel of Fig.~\ref{fig:apastron_fig2}. The effect of these corrections is evident 
	in the top panel of Fig.~\ref{fig:apastron_fig1}, where the quasi-circular waveform (gray online) 
	is visibly dephased with respect to the other curves, This indicates that all noncircular 
	phase corrections discussed in this work eventually yield an improved numerical/analytical 
	phase agreement with respect to the quasi-circular EOB waveform.
	By contrast, the noncircular correction provided by Eq.~\eqref{eq:AEIwave}, at 2PN accuracy, 
	does not  provide a reliable amplitude at apastron, with differences that are rather close to
	those obtained using the standard circular waveform.
	
	To understand this aspect, let us focus for a moment on the Newtonian noncircular 
	prefactor of Eq.~\eqref{eq:ncN22}, whose time-evolution is shown for the case considered 
	in Fig.~\ref{fig:apastron_fig2} (red online). The figure shows that the contribution of the
	Newtonian prefactor is larger at apastron than at periastron. This follows from the fact 
	that in Eq.~\eqref{eq:ncN22} the orbital frequency $\Omega$ appears squared and 
	at the denominator of the noncircular correction, as a consequence of having factorized 
	the circular Newtonian contribution. This eventually amplifies the contribution of the whole function 
	in correspondence of the lowest values of $\Omega$, i.e.~at apastron. Note however 
	that the only nonvanishing contribution of Eq.~\eqref{eq:ncN22} at apastron 
        is the one proportional to $\ddot{r}$, that is thus the main reason behind the behavior seen in Fig.~\ref{fig:apastron_fig2}.
        The hierarchy between $\ddot{r}$ and $\Omega^2$ is clarified by the bottom panel of Fig.~\ref{fig:apastron_fig2}.
        By contrast, when considering Eq.~\eqref{eq:AEIwave}, without the crucial factorization of the Newtonian prefactor, 
        the amplitude correction remains substantially constant, and small, for the whole radial evolution, 
        see $1+f^{\rm ecc}_{22}/f_{22}$ in Fig.~\ref{fig:apastron_fig2} (green online). 
	This leads to the large analytical/numerical discrepancies for the amplitude, as  
	shown in Fig.~\ref{fig:testmass_insplunge_AEI} and Fig.~\ref{fig:apastron_fig1}.
	Loosely speaking, we can trace all this to the fact that Eq.~\eqref{eq:AEIwave}
	incorporates the PN expansion of Eq.~\eqref{eq:ncN22} through the replacement of
	$\Omega$ and $\ddot{r}$ via the 2PN equation of motion, and so the crucial 
	amplification related to the exact $\ddot{r}/\Omega^2$ contribution is lost.
	
\begin{figure}
		\center
		\includegraphics[width=0.5\textwidth]{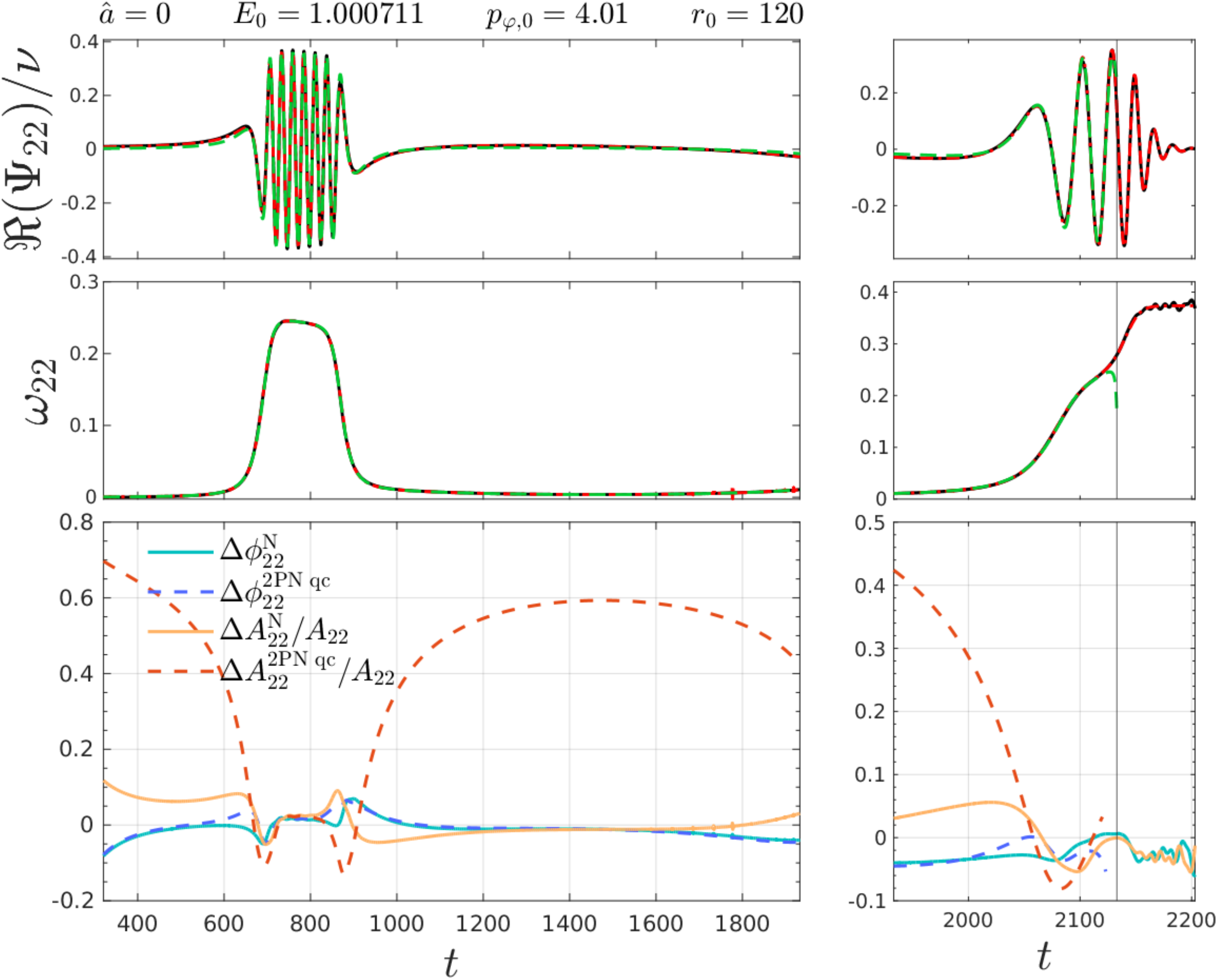}
		\caption{\label{fig:AEI_hyp}Same type of comparison of Fig.~\ref{fig:testmass_insplunge_AEI} 
			but considering the dynamical capture configuration in the left panels of Fig.~\ref{fig:testmass_hyp}.  
			Top panel: real part of the waveform. Middle panel: instantaneous frequency. Bottom panel: phase
			and fractional amplitude differences. The waveform of Ref.~\cite{Khalil:2021txt}, Eq.~\eqref{eq:AEIwave}, 
			accumulates rather large amplitude differences up to the apastron of the quasi-elliptic 
			orbit following the first encounter.}
	\end{figure}

	To conclude, we also mention that the problematic behavior of the waveform of 
	Eq.~\eqref{eq:AEIwave} is even more evident in the dynamical capture scenario. 
	Fig.~\ref{fig:AEI_hyp} refers to the leftmost configuration of Fig.~\ref{fig:testmass_hyp}. 
	The waveform of Ref.~\cite{Khalil:2021txt}, Eq.~\eqref{eq:AEIwave}, yields
	fractional amplitude differences $\sim 60\%$ at the apastron of the 
	quasi-elliptic orbit following the first encounter. 

	\section{Conclusions}
	\label{sec:end}
	We have presented a new factorized and resummed multipolar waveform for spin aligned binaries
	that is valid along noncircular orbits, i.e. either eccentric inspirals or dynamical capture scenarios.
	Our main findings can be summarized as follows:
	\begin{itemize}
		\item[(i)] We have exploited 2PN noncircular results within the paradigm of the factorization of
		the generic Newtonian prefactor. In this way, this 2PN information is simply recasted into correcting
		factors that can be directly used to improve the waveform of existing eccentric model based on
		the formalism of \TEOBResumS{}, notably those of Refs.~\cite{Chiaramello:2020ehz,Nagar:2021gss,Albanesi:2021rby,Nagar:2021xnh}.
		Differently from other works~\cite{Khalil:2021txt}, we choose to express the correcting 
		factors in terms of $p_\varphi^2 u$, without using powers of the derivative of the radial momentum $\dot{p}_r$. 
		This permits to obtain analytical expressions that are relatively simple, with a well organized analytical structure. 
		\item[(ii)] We thoroughly tested the performance of these analytical correcting factors  by performing 
		comparisons with numerical waveforms from eccentric inspirals (also through plunge and merger) in
		the test-mass limit. We showed that the analytical/numerical agreement through the plunge phase 
		(and for large eccentricity) can be largely improved by implementing straightforward resummations
		scheme (via Pad\'e approximants) of the residual polynomials in $p_\varphi^2 u$ entering the 
		noncircular tail factor. With this procedure, we can obtain analytical/numerical phase disagreements 
		of $\pm 0.04$ rad for $e=0.9$, and disagreements at most within $0.02$ rad for smaller 
		eccentricities (see Figs.~\ref{fig:testmass_inspl_resum_tail} and~\ref{fig:testmass_inspl_resum_inst}). 
		A similar behavior is also found for hyperbolic captures.
		\item[(iii)] The quadrupolar waveform can be further improved applying the same resummation procedure 
		adopted for the eccentric tail contribution also to the 2PN instantaneous noncircular contribution to the phase.
		Nonetheless, this step is not as crucial as the resummation of the hereditary contribution.
		\item[(iv)] When moving to the comparable-mass case, we applied the same 2PN-resummed noncircular
		correction to the EOB noncircular model of Ref.~\cite{Nagar:2021xnh} and we provided a new comparison
		with the 28 public NR simulations of eccentric inspirals from the SXS catalog. 
		For most of the configurations, the phase difference during the inspiral is mostly within the $\pm 0.05$~bandwidth. 
		The related EOB/NR unfaithfulness computations (using aLIGO noise for $20M_\odot \leq M \leq 200M_\odot$) 
		are below the $1\%$ threshold (at most $0.6\%$) except for the only outlier, SXS:BBH:1149 that grazes this limit 
		because of limitations inherited by the underlying quasi-circular model, as explained in Ref.~\cite{Nagar:2021xnh}.
		It should be noted that the new, factored and resummed, 2PN contributions discussed here seem to
		bring only marginal improvements to the model of Ref.~\cite{Nagar:2021xnh}.
		In this respect, the use of, well controlled, test-mass limit numerical data proves crucial to learn the actual
		importance of this additional analytical information.
		\item[(v)] In this respect, the availability of test-mass waveform data allowed us to thoroughly test 
		the waveform factorization proposed in Ref.~\cite{Khalil:2021txt}, whose main difference with our 
		prescription is that the noncircular part of the Newtonian prefactor is not factored out. 
		We found that the full Newtonian-factor factorization (even before the 
		presence of the 2PN noncircular corrections) looks more accurate and robust 
		(especially for the amplitude, with differences that can reach up to $60\%$ versus a $6\%$ at most) 
		all over the parameter space. The origin of the reliable behavior of the amplitude of the 
		Newton-factorized waveform (somehow overlooked in past works) traces to the crucial presence 
		of the the exact contribution $\ddot{r}/r\Omega^2$ in Eq.~\eqref{eq:ncN22}, as extensively
		explained in Sec.~\ref{sec:qc}.
	\end{itemize}
	In conclusion, our results indicate that the incorporation of high-PN noncircular 
	waveform term within EOB models is more effective if suitable factorizations and 
	resummation procedures are implemented. In particular, the use of the general Newtonian 
	prefactor, as originally proposed in Ref.~\cite{Chiaramello:2020ehz}, seems
	an essential element for constructing highly accurate analytical waveforms for 
	noncircular dynamics for either present or future gravitational wave detectors. 
	We think it is particularly important to emphasize that the good performance of 
	our waveform is not limited to mild values of eccentricities, but it remains 
	robust {\it also} for large eccentricities. In this respect, the waveform of 
	Ref.~\cite{Khalil:2021txt} implemented in the model of Ref.~\cite{RamosBuades:2021} 
	doesn't seem to be a robust choice when eccentricity is larger than $\sim 0.3$.
	Despite this issue, the model presented in~\cite{RamosBuades:2021} shows an 
	excellent EOB/NR performance over the 28 SXS datasets (with eccentricities up to $\sim 0.3$) 
	that is globally comparable to ours (and even better in the large-spin case).
	Although the focus of this paper is on the waveform, we finally note that from our
	resummed expressions it is possible to compute energy and angular momentum
	fluxes. These  can be used to construct more accurate representations of the analytical 
	radiation reaction (and especially the azimuthal part ${\cal F}_\varphi$) along general 
	(equatorial) orbits that include the 2PN information, thus going beyond the 
	Newtonian prefactor paradigm\footnote{We remind in passing that Ref~\cite{Chiaramello:2020ehz} 
		compared several choices of ${\cal F}_\varphi$ with the exact flux computed in the test-mass limit.
		Among these choices, there was a circular-factorized and resummed version of
		the results of~\cite{Bini:2012ji}, Eq.~(5) of Ref.~\cite{Chiaramello:2020ehz}, 
		analogous to the prescriptions suggested in Ref.~\cite{Khalil:2021txt}, modulo the additional Pad\'e resummation
		adopted in Ref.~\cite{Chiaramello:2020ehz} to enhance the robustness of the analytical expression
		in strong field. This study~\cite{Chiaramello:2020ehz,Chiaramello:2020Th} 
		indicated that the general-Newtonian prefactor factorization yielded a closer agreement of the fluxes 
		with test-mass numerical data, thus suggesting that another route is necessary to consistently 
		incorporate 2PN results in radiation reaction.} of Ref.~\cite{Chiaramello:2020ehz}.
	This additional analytical information in radiation reaction (as well as the Newton-factor dressing) 
	is currently omitted also in Ref.~\cite{RamosBuades:2021}.
	Our prescription for radiation reaction has been extensively tested in Ref.~\cite{Albanesi:2021rby},
	that clearly indicated that more analytical information is needed to obtain accurate
	analytical fluxes for large eccentricities. Exploiting our current results for radiation
	reaction purposes is postponed to future work.
	As a last observation, considering that this result is obtained using only 2PN information, our procedure
	seems to indicate a viable route to process a large amount of (high-order) PN
	information~\cite{Munna:2019fjz,Munna:2020iju,Munna:2020som} in order to construct 
	strong-field accurate, EOB-based, fully nonadiabatic, analytical waveform models for 
	eccentric extreme mass ratio inspirals in view of providing waveform templates 
	for LISA~\cite{Katz:2021yft, Hughes:2021exa}.

	\begin{acknowledgments}
		S.~B. acknowledges support by the EU H2020 under ERC Starting Grant, no.~BinGraSp-714626.  
		M.~O. and G.~G.~acknowledge support from the project ``Black holes, neutron stars and gravitational waves" 
		financed by Fondo Ricerca di Base 2018 of the University of Perugia. G.~G., M.~O. and A.~P.~thank 
		Niels Bohr Institute for hospitality. 
		S.~A. thanks Institute for Pure and Applied Mathematics (UCLA) for hospitality. 
		The numerical simulations using {\Teukode}~\cite{Harms:2013ib,Harms:2014dqa}
		were performed on the Virgo ``Tullio'' server in Torino, supported by INFN.
		We thank A.~Ramos-Buades, A.~Buonanno, M.~Khalil and S.~Ossokine for useful discussions
		and for sharing within the LVK collaboration the content of Ref.~\cite{RamosBuades:2021} before publication.
		We are particularly grateful to A.~Ramos-Buades for a question that allowed us to improve our understanding
		of the noncircular corrections to the amplitude.
	\end{acknowledgments}
	
	\appendix
	
	\section{Newton-factorized noncircular higher modes}
	\label{App:HM}
	
	In this appendix we present the resulting 2PN noncircular relativistic factors of Sec.~\ref{Sec:NewFactors} 
	for all the subdominant modes up to\footnote{For all the modes with $\ell>4$, at 2PN accuracy, one has at most just the Newtonian contribution, which means $\hat{h}_\lm=1$ and no relativistic factors.} $\l=m=4$. Note that the contributions that are not explicitly written, are equal\footnote{These are: $ \hat{h}^{{\text{tail-nc}}}_{30}, \hat{h}^{{\text{tail-nc}}}_{32}, \hat{h}^{{\text{tail-nc}}}_{40}, \hat{h}^{{\text{tail-nc}}}_{41}, \hat{h}^{{\text{tail-nc}}}_{42}, \hat{h}^{{\text{tail-nc}}}_{43}$  $\hat{h}^{{\text{tail-nc}}}_{44}$.} to 1.   
	\begin{widetext}
		
		\subsubsection{Tail noncircular factors}
		\begin{align}
			\label{ncTail20}
			\hat{h}^{{\text{tail-nc}}}_{20} & = 1 - \frac{\pi  p_\varphi }{320 c^3\left(p_\varphi^2 u-1\right)} \bigg[p_{r_*}^2 u \bigg(1364-2885 p_\varphi^2 u+3490 p_\varphi^4 u^2-2400 p_\varphi^6 u^3+890 p_\varphi^8 u^4-139 p_\varphi^{10} u^5\bigg)\cr
			& 
			-\frac{p_{r_*}^4}{3 \left(p_\varphi^2 u-1\right)}\bigg(4262-8790 p_\varphi^2 u+10390 p_\varphi^4 u^2-7155 p_\varphi^6 u^3+2670 p_\varphi^8 u^4-417 p_\varphi^{10} u^5\bigg)\bigg]. \\ \cr
			\label{ncTail21}
			\hat{h}^{{\text{tail-nc}}}_{21} & = 1 - \frac{\pi }{11520 c^3} \bigg[ 6i p_{r_*} u \bigg(3029+6035 p_\varphi^2 u-10870 p_\varphi^4 u^2+8350 p_\varphi^6
			u^3-3215 p_\varphi^8 u^4+511 p_\varphi^{10} u^5\bigg)\cr
			& -\frac{15 p_{r_*}^2}{p_\varphi}\bigg(635-1388 p_\varphi^2 u+666 p_\varphi^4 u^2-92 p_\varphi^6 u^3-13
			p_\varphi^8 u^4\bigg)+20 i p_{r_*}^3 \bigg(619-981 p_\varphi^2 u+573 p_\varphi^4 u^2\cr
			& -115 p_\varphi^6 u^3\bigg)+ \frac{15 p_{r_*}^4}{p_\varphi u} \bigg(183-82 p_\varphi^2 u-17 p_\varphi^4 u^2\bigg)\bigg] . \\ \cr
			%
			%
			%
			%\label{ncTail30}
			%\hat{h}^{{\text{tail-nc}}}_{30} & = 1.\\ \cr
			%
			%
			%
			\label{ncTail31}
			\hat{h}^{{\text{tail-nc}}}_{31} & = 1 - \frac{\pi}{1920 c^3p_\varphi^2 \left(7-6 p_\varphi^2 u\right)^2} \bigg[i p_{r_*} \bigg(88130-107366 p_\varphi^2 u+89843 p_\varphi^4 u^2-388835 p_\varphi^6
			u^3+588840 p_\varphi^8 u^4\cr
			&-397460 p_\varphi^{10} u^5+139731 p_\varphi^{12} u^6-20563 p_\varphi^{14} u^7\bigg)+\frac{p_{r_*}^2}{2 p_\varphi u \left(7-6 p_\varphi^2 u\right)} \bigg(2115120-3821769 p_\varphi^2 u\cr
			& +915328 p_\varphi^4 u^2-2431548 \text{p$\phi
				$}^6 u^3+9399380 p_\varphi^8 u^4-10528645 p_\varphi^{10} u^5+5655444
			p_\varphi^{12} u^6-1592318 p_\varphi^{14} u^7\cr
			& +186288 \text{p$\phi
				$}^{16} u^8\bigg) - \frac{ip_{r_*}^3}{3 p_\varphi^2 u^2 \left(7-6 p_\varphi^2 u\right)^2} \bigg(38072160-92454747 p_\varphi^2 u+79749569 p_\varphi^4 u^2-137703015
			p_\varphi^6 u^3\cr
			&+356358768 p_\varphi^8 u^4-482607515 p_\varphi^{10} u^5+363100527 p_\varphi^{12} u^6-161234979 p_\varphi^{14}
			u^7+40408200 p_\varphi^{16} u^8\cr
			&-4441608 p_\varphi^{18} u^9\bigg) - \frac{p_{r_*}^4}{2 p_\varphi^3 u^3 \left(7-6 p_\varphi^2 u\right)^3} \bigg(304577280-928941216 p_\varphi^2 u+1082387695 p_\varphi^4 u^2\cr
			&-1365588354
			p_\varphi^6 u^3+3139841017 p_\varphi^8 u^4-4991336104 p_\varphi^{10} u^5+4638849326 p_\varphi^{12} u^6-2632737900 p_\varphi^{14}
			u^7\cr
			&+916090404 p_\varphi^{16} u^8-181221912 p_\varphi^{18} u^9+15589584
			p_\varphi^{20} u^{10}\bigg)
			\bigg].\\ \cr
			%
			%
			%
			%\label{ncTail32}
			%\hat{h}^{{\text{tail-nc}}}_{32} & = 1.\\ \cr
			%
			%
			%
			\label{ncTail33}
			\hat{h}^{{\text{tail-nc}}}_{33} & = 1 - \frac{\pi }{3840 c^3 p_\varphi^2 \left(2 p_\varphi^2 u+7\right)^2} \bigg[i p_{r_*} \bigg(47630+134366 p_\varphi^2 u+721737 p_\varphi^4 u^2-429865 \text{p$\phi$}^6 u^3+308120 p_\varphi^8 u^4\cr
			&-107220 p_\varphi^{10} u^5+16769
			p_\varphi^{12} u^6-337 p_\varphi^{14} u^7\bigg) + \frac{ p_{r_*}^2}{2 p_\varphi u \left(2 p_\varphi^2 u+7\right)} \bigg(381040+1407963 p_\varphi^2 u\cr
			&-2751544 p_\varphi^4 u^2+8146332 p_\varphi^6 u^3-5435500 p_\varphi^8 u^4+3011095 p_\varphi^{10} u^5-1073428 p_\varphi^{12} u^6+170074 p_\varphi^{14} u^7+\cr
			&14688 p_\varphi^{16} u^8\bigg)-\frac{i p_{r_*}^3}{p_\varphi^2 u^2 \left(2 p_\varphi^2 u+7\right)^2}\bigg(762080+3125521 p_\varphi^2 u+5675333 p_\varphi^4 u^2\cr
			&-28858731 p_\varphi^6 u^3+28110216 p_\varphi^8 u^4-14716055 p_\varphi^{10} u^5+7128059 p_\varphi^{12} u^6-2725303 p_\varphi^{14} u^7\cr
			&+281976 p_\varphi^{16} u^8+41864 p_\varphi^{18} u^9 \bigg) - \frac{p_{r_*}^4}{2 p_\varphi^3 u^3 \left(2 p_\varphi^2 u+7\right)^3} \bigg( 6096640+27480928 p_\varphi^2 u+92901791 p_\varphi^4 u^2\cr
			&+57821954 p_\varphi^6 u^3-472378615 p_\varphi^8 u^4+376035640 p_\varphi^{10} u^5-164962538 p_\varphi^{12} u^6+71523844 p_\varphi^{14} u^7\cr
			&-24385644 p_\varphi^{16} u^8+1291400 p_\varphi^{18} u^9+320720 p_\varphi^{20} u^{10}\bigg)
			\bigg].
		\end{align}
		
		\subsubsection{Instantaneous noncircular factors}
		
		\begin{align}
			\label{APNnc20}
			f^{\text{inst-nc}}_{20} & = 1 +\frac{1}{c^2} \biggl\{\frac{p_{r_*}^2 }{-1+p_\varphi^2 u}\bigg[\left(\frac{1}{14}-\frac{31 \nu }{14}\right)+p_\varphi^2 u \left(\frac{9}{14}+\frac{15 \nu }{14}\right)\bigg]+\frac{p_{r_*}^4 }{u \left(-1+p_\varphi^2 u\right)^2} \left(-\frac{5}{7}+\frac{8 \nu }{7}\right) \biggr\}\cr
			& + \frac{1}{c^4}\biggl\{\frac{p_{r_*}^2}{\left(-1+p_\varphi^2 u\right)^2}\bigg[u \left(\frac{65}{252}+\frac{211 \nu }{126}+\frac{139 \nu ^2}{63}\right)+p_\varphi^2 u^2 \left(\frac{3683}{3528}-\frac{39371 \nu }{1764}-\frac{5099 \nu ^2}{882}\right)\cr
			& +p_\varphi^4 u^3 \left(\frac{1825}{1176}+\frac{5239 \nu }{294}+\frac{703 \nu ^2}{147}\right)+p_\varphi^6 u^4 \left(-\frac{209}{294}-\frac{355 \nu }{588}-\frac{355 \nu ^2}{294}\right)\bigg] \cr
			& +\frac{p_{r_*}^4}{\left(-1+p_\varphi^2 u\right)^3}\bigg[\left(\frac{1613}{504}-\frac{1567 \nu }{504}-\frac{71 \nu ^2}{72}\right)+p_\varphi^2 u \left(-\frac{16819}{1764}+\frac{92131 \nu }{3528}+\frac{7421 \nu ^2}{3528}\right)\cr
			&+p_\varphi^4 u^2 \left(\frac{323}{147}-\frac{19445 \nu }{1176}-\frac{1279 \nu ^2}{1176}\right)+p_\varphi^6 u^3 \left(-\frac{25}{168}+\frac{65 \nu }{168}-\frac{5 \nu ^2}{168}\right)\bigg]\biggr\}, \\
			\label{PhPNnc20}
			\delta^{\text{inst-nc}}_{20} &= 0. \\ \cr
			\label{APNnc21}
			f^{\text{inst-nc}}_{21} & = 1 + \frac{1}{c^2} p_{r_*}^2 \left( \frac{9}{14}+\frac{5 \nu }{7} \right) , \\
			\label{PhPNnc21}
			\delta^{\text{inst-nc}}_{21} &= \frac{1}{ c^2}  p_{r_*} p_\varphi \left(\frac{1}{14}+\frac{6 \nu }{7}\right)  .\\ \cr
			\label{APNnc30}
			f^{\text{inst-nc}}_{30} & = 1+\frac{1}{c^2}p_{r_*}^2 \frac{5-7 \nu -17 \nu ^2}{6-18 \nu } , \\
			\label{PhPNnc30}
			\delta^{\text{inst-nc}}_{30} &= 0.\\ \cr
			\label{APNnc31}
			f^{\text{inst-nc}}_{31} & = 1+\frac{1}{c^2}\biggl\{\frac{p_{r_*}^2}{p_\varphi^2 u \left(-7+6 p_\varphi^2 u\right)^3}\bigg[(-1076+1168 \nu )+p_\varphi^2 u \left(\frac{4783}{2}-4242 \nu \right)+p_\varphi^4 u^2 (-1520+4834 \nu )\cr
			&+p_\varphi^6 u^3 (6-2136 \nu )+p_\varphi^8 u^4 (180+288 \nu )\bigg]+ \frac{p_{r_*}^4}{p_\varphi^4 u^3 \left(-7+6 p_\varphi^2 u\right)^5} \bigg[ (154944-168192 \nu )\cr
			&+p_\varphi^2 u (-569222+740344 \nu )+p_\varphi^4 u^2 (840044-1241680 \nu )+p_\varphi^6 u^3 (-622914+1014636 \nu )\cr
			&+p_\varphi^8 u^4 (232812-408168 \nu )+p_\varphi^{10} u^5 (-35208+65232 \nu )\bigg]\biggr\}, \\
			\label{PhPNnc31}
			\delta^{\text{inst-nc}}_{31} &= \frac{1}{c^2} \biggl\{\frac{p_{r_*}}{p_\varphi \left(7-6 p_\varphi^2 u\right)^2}\bigg[\left(\frac{269}{3}-\frac{292 \nu }{3}\right)+p_\varphi^2 u (-146+222 \nu )+p_\varphi^4 u^2 (61-110 \nu )\bigg]\cr
			&+ \frac{p_{r_*}^3}{p_\varphi^3 u^2 \left(7-6 p_\varphi^2 u\right)^4} \bigg[(-12912+14016 \nu )+p_\varphi^2 u (39410-52984 \nu )+p_\varphi^4 u^2 (-45043+69758 \nu )\cr
			&+p_\varphi^6 u^3 (22704-39288 \nu )+p_\varphi^8 u^4 (-4248+8064 \nu )\bigg]\biggr\} .\\ \cr
			\label{APNnc32}
			f^{\text{inst-nc}}_{32} & = 1+\frac{1}{c^2}\biggl\{\frac{p_{r_*}^2}{p_\varphi^2 u (-1+3 \nu )} \bigg[\left(-\frac{187}{2880}+\frac{223 \nu }{576}-\frac{265 \nu ^2}{576}\right)+p_\varphi^2 u \left(-\frac{221}{192}+\frac{547 \nu }{192}+\frac{227 \nu ^2}{192}\right)\bigg]\cr
			&+ \frac{p_{r_*}^4}{p_\varphi^4 u^3 (-1+3 \nu )} \bigg[\left(\frac{187}{46080}-\frac{223 \nu }{9216}+\frac{265 \nu ^2}{9216}\right)+p_\varphi^2 u \left(\frac{61}{3072}-\frac{323 \nu }{3072}+\frac{317 \nu ^2}{3072}\right)\bigg]\biggr\}, \\
			\label{PhPNnc32}
			\delta^{\text{inst-nc}}_{32} &= \frac{1}{ c^2}\biggl\{\frac{p_{r_*}}{p_\varphi (-1+3 \nu )}\bigg[\left(-\frac{187}{720}+\frac{223 \nu }{144}-\frac{265 \nu ^2}{144}\right)+p_\varphi^2 u \left(-\frac{433}{240}+\frac{355 \nu }{48}-\frac{157 \nu ^2}{48}\right)\bigg]\cr
			&+\frac{p_{r_*}^3}{p_\varphi^3 u^2 (-1+3 \nu )}\bigg[\left(\frac{187}{11520}-\frac{223 \nu }{2304}+\frac{265 \nu ^2}{2304}\right)+p_\varphi^2 u \left(\frac{61}{768}-\frac{323 \nu }{768}+\frac{317 \nu ^2}{768}\right)\bigg]\biggr\}. \\ \cr
			\label{APNnc33}
			f^{\text{inst-nc}}_{33} & = 1+\frac{1}{c^2} \biggl\{\frac{p_{r_*}^2}{p_\varphi^2 u \left(7+2 p_\varphi^2 u\right)^3}\bigg[\left(\frac{1076}{9}-\frac{1168 \nu }{9}\right)+p_\varphi^2 u \left(\frac{9161}{18}+\frac{2626 \nu }{9}\right)+p_\varphi^4 u^2 \left(\frac{1292}{3}-\frac{214 \nu }{3}\right)\cr
			&+p_\varphi^6 u^3 \left(\frac{1430}{3}-\frac{424 \nu }{3}\right)+p_\varphi^8 u^4 \left(\frac{20}{3}+\frac{32 \nu }{3}\right)\bigg]+\frac{p_{r_*}^4}{p_\varphi^4 u^3 \left(7+2 p_\varphi^2 u\right)^5} \bigg[\left(-\frac{17216}{9}+\frac{18688 \nu }{9}\right)\cr
			&+p_\varphi^2 u \left(-\frac{55034}{9}+\frac{52744 \nu }{9}\right)+p_\varphi^4 u^2 \left(-\frac{23140}{3}+\frac{23984 \nu }{3}\right)+p_\varphi^6 u^3 \left(\frac{11510}{3}-\frac{44260 \nu }{3}\right)\cr
			&+p_\varphi^8 u^4 \left(\frac{163540}{9}-\frac{57560 \nu }{9}\right)+p_\varphi^{10} u^5 \left(-\frac{11912}{3}+\frac{7888 \nu }{3}\right)\bigg]\biggr\}, \\
			\label{PhPNnc33}
			\delta^{\text{inst-nc}}_{33} &= \frac{1}{c^2}\biggl\{\frac{p_{r_*}}{p_\varphi \left(7+2 p_\varphi^2 u\right)^2}\bigg[\left(\frac{269}{9}-\frac{292 \nu }{9}\right)+p_\varphi^2 u \left(\frac{970}{9}-\frac{1142 \nu }{9}\right)+p_\varphi^4 u^2 \left(\frac{313}{3}-\frac{182 \nu }{3}\right)\bigg]\cr
			&+\frac{p_{r_*}^3}{p_\varphi^3 u^2 \left(7+2 p_\varphi^2 u\right)^4}\bigg[\left(-\frac{4304}{9}+\frac{4672 \nu }{9}\right)+p_\varphi^2 u \left(-\frac{12010}{9}+\frac{11288 \nu }{9}\right)+p_\varphi^4 u^2 (431-1318 \nu )\cr
			&+p_\varphi^6 u^3 \left(\frac{5456}{3}+\frac{968 \nu }{3}\right)+p_\varphi^8 u^4 (-1336+864 \nu )\bigg]\biggr\}.\\ \cr
			\label{APNnc40}
			f^{\text{inst-nc}}_{40} & = 1+\frac{1}{c^2} \biggl\{\frac{p_{r_*}^2}{\left(7-6 p_\varphi^2 u\right)^2 \left(-1+p_\varphi^2 u\right) (-1+3 \nu )}\bigg[\left(-\frac{785}{11}+\frac{11205 \nu }{22}-\frac{17673 \nu ^2}{22}\right)\cr
			&+p_\varphi^2 u \left(57-\frac{1627 \nu }{2}+\frac{3479 \nu ^2}{2}\right)+p_\varphi^4 u^2 \left(45+276 \nu -1092 \nu ^2\right)+p_\varphi^6 u^3 \left(-\frac{414}{11}+\frac{486 \nu }{11}+\frac{1890 \nu ^2}{11}\right)\bigg]\cr
			&+\frac{p_{r_*}^4}{u \left(-7+6 p_\varphi^2 u\right)^3  \left(-1+p_\varphi^2 u\right)^2 (-1+3 \nu )} \bigg[\left(-\frac{20367}{11}+\frac{94608 \nu }{11}-\frac{94608 \nu ^2}{11}\right)\cr
			&+p_\varphi^2 u \left(\frac{8133}{2}-19992 \nu +22128 \nu ^2\right)+p_\varphi^4 u^2 \left(-\frac{65727}{22}+\frac{169416 \nu }{11}-\frac{201888 \nu ^2}{11}\right)\cr
			&+p_\varphi^6 u^3 \left(\frac{8037}{11}-\frac{43056 \nu }{11}+\frac{54144 \nu ^2}{11}\right)\bigg]\biggr\}, \\
			\label{PhPNnc40}
			\delta^{\text{inst-nc}}_{40} &= 0.\\ \cr
			\label{APNnc41}
			f^{\text{inst-nc}}_{41} & = 1, \\
			\label{PhPNnc41}
			\delta^{\text{inst-nc}}_{42} &= 0.\\ \cr
			\label{APNnc42}
			f^{\text{inst-nc}}_{42} & = 1+\frac{1}{c^2} \biggl\{\frac{p_{r_*}^2}{\left(-7-3 p_\varphi^2 u+6 p_\varphi^4 u^2\right)^3 (-1+3 \nu )}\bigg[\left(-\frac{5495}{11}+\frac{78435 \nu }{22}-\frac{123711 \nu ^2}{22}\right)\cr
			&+p_\varphi^2 u \left(\frac{22794}{55}-\frac{95055 \nu }{22}+\frac{237351 \nu ^2}{22}\right)+p_\varphi^4 u^2 \left(\frac{2308683}{220}-\frac{778917 \nu }{22}+\frac{83871 \nu ^2}{22}\right)\cr
			&+p_\varphi^6 u^3 \left(-\frac{4346379}{220}+\frac{1713879 \nu }{22}-\frac{991377 \nu ^2}{22}\right)+p_\varphi^8 u^4 \left(\frac{692577}{55}-\frac{603477 \nu }{11}+\frac{491211 \nu ^2}{11}\right)\cr
			&+p_\varphi^{10} u^5 \left(-\frac{127386}{55}+\frac{136890 \nu }{11}-\frac{160650 \nu ^2}{11}\right)+p_\varphi^{12} u^6 \left(-\frac{2484}{11}+\frac{2916 \nu }{11}+\frac{11340 \nu ^2}{11}\right)\bigg]\cr
			&+\frac{p_{r_*}^4}{u \left(-7-3 p_\varphi^2 u+6 p_\varphi^4 u^2\right)^5 (-1+3 \nu )}\bigg[\left(-\frac{997983}{11}+\frac{4635792 \nu }{11}-\frac{4635792 \nu ^2}{11}\right)\cr
			&+p_\varphi^2 u \left(\frac{43866123}{55}-\frac{47651520 \nu }{11}+\frac{63443142 \nu ^2}{11}\right)+p_\varphi^4 u^2 \left(\frac{35844849}{22}-\frac{16122492 \nu }{11}-\frac{135164952 \nu ^2}{11}\right)\cr
			&+p_\varphi^6 u^3 \left(-\frac{525077109}{44}+\frac{441545616 \nu }{11}-\frac{53867457 \nu ^2}{11}\right)+p_\varphi^8 u^4 \left(\frac{4684992993}{220}-\frac{914171508 \nu }{11}+\frac{489089961 \nu ^2}{11}\right)\cr
			&+p_\varphi^{10} u^5 \left(-\frac{2041998687}{110}+\frac{865358424 \nu }{11}-\frac{634963050 \nu ^2}{11}\right)+p_\varphi^{12} u^6 \left(\frac{96254487}{11}-\frac{433938060 \nu }{11}+\frac{377733780 \nu ^2}{11}\right)\cr
			&+p_\varphi^{14} u^7 \left(-\frac{23603562}{11}+\frac{112003560 \nu }{11}-\frac{109437480 \nu ^2}{11}\right)\cr
			&+p_\varphi^{16} u^8 \left(\frac{1077948}{5}-1069200 \nu +1139184 \nu ^2\right)\bigg]\biggr\}, \\
			\label{PhPNnc42}
			\delta^{\text{inst-nc}}_{42} &= \frac{1}{c^2}\biggl\{\frac{p_{r_*} p_\varphi u}{\left(7+3 p_\varphi^2 u-6 p_\varphi^4 u^2\right)^2 (-1+3 \nu )}\bigg[\left(-\frac{5513}{55}+\frac{5854 \nu }{11}-\frac{7548 \nu ^2}{11}\right)\cr
			&+p_\varphi^2 u \left(-\frac{86061}{220}+\frac{13686 \nu }{11}+\frac{273 \nu ^2}{11}\right)+p_\varphi^4 u^2 \left(\frac{139401}{220}-\frac{29826 \nu }{11}+\frac{22599 \nu ^2}{11}\right)\cr
			&+p_\varphi^6 u^3 \left(-\frac{2556}{11}+\frac{12186 \nu }{11}-\frac{11988 \nu ^2}{11}\right)\bigg]\cr
			&+ \frac{p_{r_*}^3 p_\varphi}{\left(7+3 p_\varphi^2 u-6 p_\varphi^4 u^2\right)^4 (-1+3 \nu )}\bigg[\left(-\frac{3434319}{110}+\frac{1677774 \nu }{11}-\frac{1868580 \nu ^2}{11}\right)\cr
			&+p_\varphi^2 u \left(-\frac{1623252}{55}-\frac{321240 \nu }{11}+\frac{4439970 \nu ^2}{11}\right)+p_\varphi^4 u^2 \left(\frac{80592003}{220}-\frac{13474980 \nu }{11}+\frac{1467963 \nu ^2}{11}\right)\cr
			&+p_\varphi^6 u^3 \left(-\frac{138830193}{220}+\frac{27487404 \nu }{11}-\frac{15732441 \nu ^2}{11}\right)+p_\varphi^8 u^4 \left(\frac{26404461}{55}-\frac{22944114 \nu }{11}+\frac{18134496 \nu ^2}{11}\right)\cr
			&+p_\varphi^{10} u^5 \left(-\frac{9559404}{55}+\frac{8902440 \nu }{11}-\frac{8355960 \nu ^2}{11}\right)+p_\varphi^{12} u^6 \left(\frac{1342332}{55}-\frac{1324512 \nu }{11}+\frac{1395792 \nu ^2}{11}\right)\bigg]\biggr\}. \\ \cr
			\label{APNnc43}
			f^{\text{inst-nc}}_{43} & = 1, \\
			\label{PhPNnc43}
			\delta^{\text{inst-nc}}_{43} &= 0. \\ \cr
			\label{APNnc44}
			f^{\text{inst-nc}}_{44} & = 1+\frac{1}{c^2}\biggl\{\frac{p_{r_*}^2}{\left(7+51 p_\varphi^2 u+6 p_\varphi^4 u^2\right)^3 (-1+3 \nu )}\bigg[\left(\frac{5495}{11}-\frac{78435 \nu }{22}+\frac{123711 \nu ^2}{22}\right)\cr
			&+p_\varphi^2 u \left(\frac{210159}{55}-\frac{667917 \nu }{22}+\frac{914625 \nu ^2}{22}\right)+p_\varphi^4 u^2 \left(-\frac{9429801}{55}+\frac{12229341 \nu }{22}-\frac{2912121 \nu ^2}{22}\right)\cr
			&+p_\varphi^6 u^3 \left(-\frac{19606626}{55}+\frac{22336029 \nu }{22}+\frac{2602071 \nu ^2}{22}\right)+p_\varphi^8 u^4 \left(-\frac{7261839}{55}+\frac{4772385 \nu }{11}-\frac{1459053 \nu ^2}{11}\right)\cr
			&+p_\varphi^{10} u^5 \left(-\frac{564948}{11}+\frac{1931958 \nu }{11}-\frac{475470 \nu ^2}{11}\right)+p_\varphi^{12} u^6 \left(-\frac{2484}{11}+\frac{2916 \nu }{11}+\frac{11340 \nu ^2}{11}\right)\bigg]\cr
			&+\frac{p_{r_*}^4}{u \left(7+51 p_\varphi^2 u+6 p_\varphi^4 u^2\right)^5 (-1+3 \nu )}\bigg[\left(\frac{997983}{11}-\frac{4635792 \nu }{11}+\frac{4635792 \nu ^2}{11}\right)\cr
			&+p_\varphi^2 u \left(-\frac{144030369}{110}+\frac{92715672 \nu }{11}-\frac{152428248 \nu ^2}{11}\right)+p_\varphi^4 u^2 \left(-\frac{737730837}{11}+\frac{3011803560 \nu }{11}-\frac{2051555040 \nu ^2}{11}\right)\cr
			&+p_\varphi^6 u^3 \left(-\frac{2431571553}{55}+\frac{1633697784 \nu }{11}-\frac{187083648 \nu ^2}{11}\right)+p_\varphi^8 u^4 \left(\frac{486563193}{5}-405446040 \nu +297670464 \nu ^2\right)\cr
			&+p_\varphi^{10} u^5 \left(\frac{3125458143}{110}-\frac{956714976 \nu }{11}-\frac{188690472 \nu ^2}{11}\right)\cr
			&+p_\varphi^{12} u^6 \left(-\frac{3895064577}{55}+\frac{3530464704 \nu }{11}-\frac{3326475600 \nu ^2}{11}\right)\cr
			&+p_\varphi^{14} u^7 \left(-\frac{673781166}{11}+\frac{2335443840 \nu }{11}-\frac{497702880 \nu ^2}{11}\right)\cr
			&+p_\varphi^{16} u^8 \left(\frac{305560836}{55}-\frac{229080960 \nu }{11}+\frac{102316608 \nu ^2}{11}\right)\bigg]\biggr\}, \\
			\label{PhPNnc44}
			\delta^{\text{inst-nc}}_{44} &= -  \frac{ 1}{ c^2} \biggl\{\frac{p_{r_*} p_\varphi u}{\left(7+51 p_\varphi^2 u+6 p_\varphi^4 u^2\right)^2 (-1+3 \nu )} \bigg[\left(\frac{11026 }{55}-\frac{11708   \nu }{11}+\frac{15096   \nu ^2}{11}\right)\cr
			&+p_\varphi^2 u \left(\frac{555681  }{110}-\frac{233274   \nu }{11}+\frac{175788   \nu ^2}{11}\right)+p_\varphi^4 u^2 \left(\frac{939051  }{110}-\frac{377406   \nu }{11}+\frac{241380   \nu ^2}{11}\right)\cr
			&+p_\varphi^6 u^3 \left(\frac{172872  }{55}-\frac{126540   \nu }{11}+\frac{47736   \nu ^2}{11}\right)\bigg]\cr
			&+ \frac{p_{r_*}^3 p_\varphi}{\left(7+51 p_\varphi^2 u+6 p_\varphi^4 u^2\right)^4 (-1+3 \nu )}\bigg[\left(\frac{3434319}{55}-\frac{3355548 \nu }{11}+\frac{3737160 \nu ^2}{11}\right)\cr
			&+p_\varphi^2 u \left(\frac{68284914}{55}-\frac{53840352 \nu }{11}+\frac{31634784 \nu ^2}{11}\right)+p_\varphi^4 u^2 \left(-\frac{69674769}{55}+\frac{66841884 \nu }{11}-\frac{71727336 \nu ^2}{11}\right)\cr
			&+p_\varphi^6 u^3 \left(-\frac{221951034}{55}+\frac{179751888 \nu }{11}-\frac{118566720 \nu ^2}{11}\right)+p_\varphi^8 u^4 \left(\frac{64377018}{55}-\frac{49429008 \nu }{11}+\frac{25417152 \nu ^2}{11}\right)\cr
			&+p_\varphi^{10} u^5 \left(\frac{74535768}{55}-\frac{42300576 \nu }{11}-\frac{20036160 \nu ^2}{11}\right)\cr
			&+p_\varphi^{12} u^6 \left(-\frac{29497608}{55}+\frac{22006080 \nu }{11}-\frac{9517824 \nu ^2}{11}\right)\bigg]\biggr\}.
		\end{align}
		
	\end{widetext}

	\section{Independent derivation of the angular radiation reaction force of Ref.~\cite{Bini:2012ji}}
	\label{App:angularRRcheck}
	In Ref.~\cite{Bini:2012ji} Bini and Damour recomputed in the EOB formalism the instantaneous part 
	of the radiation reaction force $\hat{\F}$ valid for general orbits, which had been originally 
	obtained in harmonic coordinates by Iyer, Will and collaborators 
	(see for instance Ref.~\cite{Gopakumar:1997ng}). In order to provide a meaningful check for 
	the canonical transformations~\eqref{rhTransf}-\eqref{dotphihTransf} we performed an 
	alternative computation of the angular component $\hat{\F}_{\varphi}$ by exploiting 
	directly our results for the instantaneous 2PN spherical modes $h^{\text{inst}}_\lm$ 
	discussed in Sec.~\ref{SubSec:inst_hlm}.
	
	The radiation-reaction force components  $\hat{\F}_{r,\varphi}$ that drive the non conservative 
	part of the EOB dynamics can be related via the equations of motion to the system loss of 
	energy and angular momentum, which in turn are connected to the energy and 
	angular momentum fluxes at infinity $\dot{E}$ and $\dot{J}$ by the balance equations.  Overall we can write \cite{Bini:2012ji}
	\begin{align}
		\label{Balanceq_r}
		\dot{r} \hat{\F}_r + \Omega \hat{\F}_\varphi + \dot{E}_{\rm Schott} + \dot{E} & = 0, \\
		\label{Balanceq_phi}
		\hat{\F}_\varphi + \dot{J} & = 0,
	\end{align}
	where $E_{\rm Schott}$ is the Schott contribution to the energy of the system, due to its interaction with the local field. In general Eq.~\eqref{Balanceq_phi} contains an additional Schott contribution $\dot{J}_{\rm Schott}$ but in Sec.~II of \cite{Bini:2012ji} it has been shown that such term can be always gauged away. Therefore thanks to this equation and the expression of the flux $\dot{J}$ in terms of the multipolar waveform we can determine the angular component of the radiation-reaction force simply as
	\begin{equation}
		\label{F_phi_with_multipoles}
		\hat{\F}_\varphi  = - \dot{J} = \frac{1}{16 \pi} \sum_{\ell=2}^{\ell_{\text{max}}} \sum_{m=-\ell}^{\ell} m \, \Im \bigg[ \dot{h}_\lm h_\lm^*  \bigg].
	\end{equation}
	This paves the way to recompute independently the 2PN instantaneous result for $ \F_\varphi$ given in Eq.~(3.70) of Ref.~\cite{Bini:2012ji}, as mentioned above. The instantaneous modes that bring a non-zero contribution to $ \hat{\F}_\varphi$ at 2PN are $h^{\text{inst}}_{21}$, $h^{\text{inst}}_{22}$, $h^{\text{inst}}_{31}$, $h^{\text{inst}}_{32}$, $h^{\text{inst}}_{33}$, $h^{\text{inst}}_{42}$ and $h^{\text{inst}}_{44}$. The angular radiation reaction force coming from Eq.~\eqref{F_phi_with_multipoles} has proved to be consistent with the result of Ref.~\cite{Bini:2012ji}.
	
	%%%%%%%%%%%%%%%%%%%%%%%%%%%%%%%%%%%%%%%%%%%%%%%%%%%%%%%%%%%%%%%%%%%%%%%%
	
	\section{Alternative corrections using \texorpdfstring{$\dot{p}_{r_*}$}{TEXT}}
	\label{App:2PN_prrdot}
	
\begin{figure*}
		\center
		\includegraphics[width=0.31\textwidth]{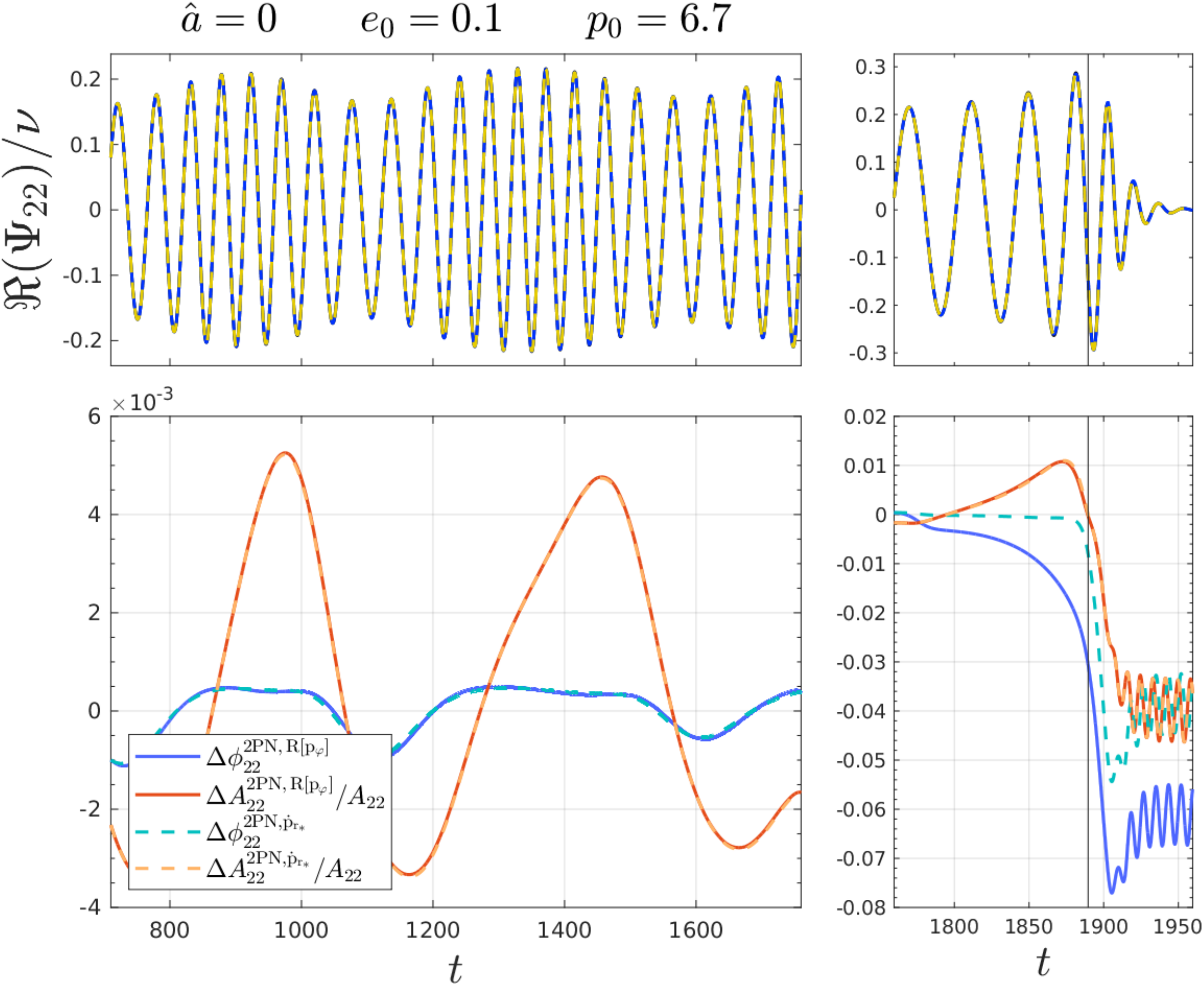}
		\includegraphics[width=0.31\textwidth]{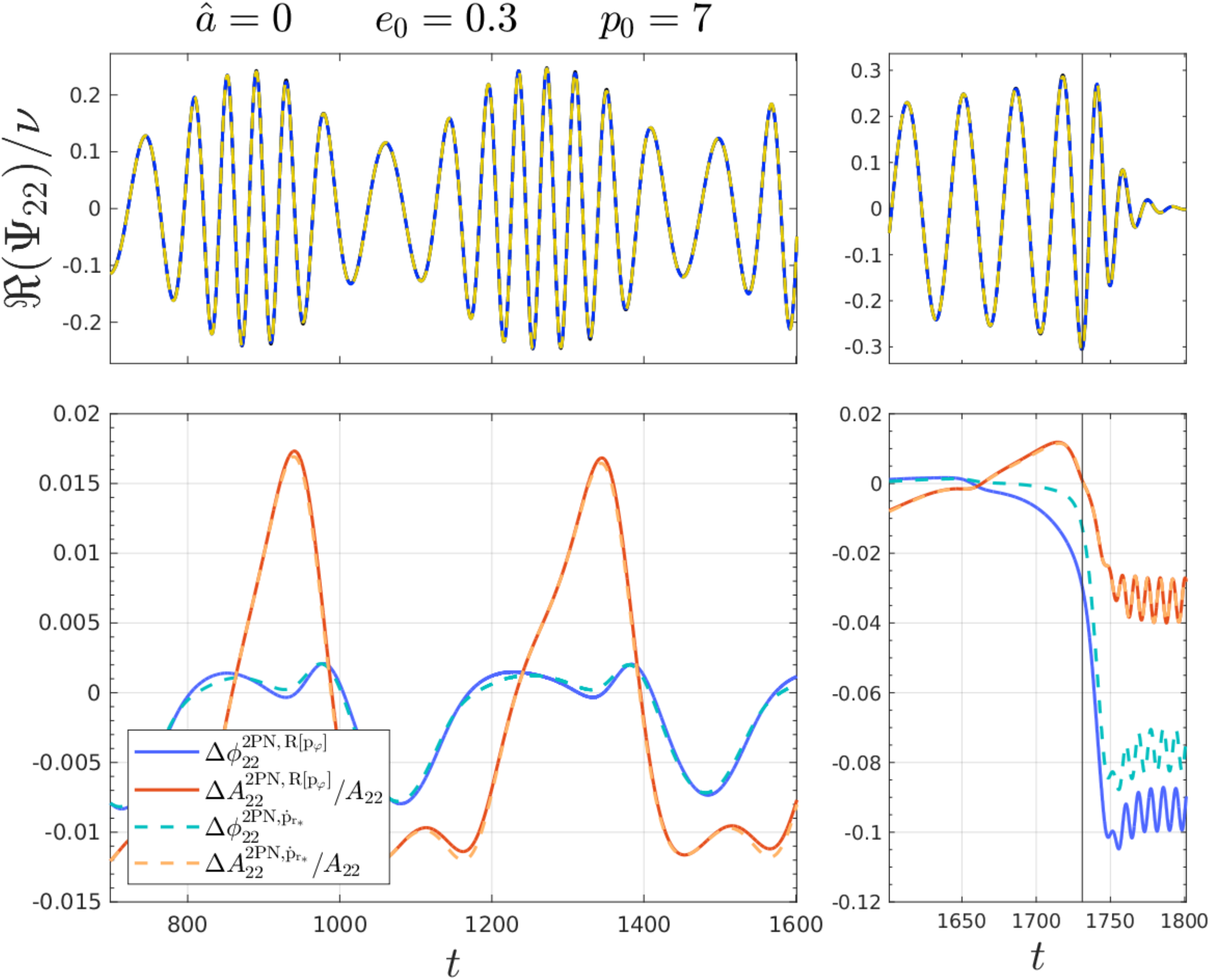}
		\includegraphics[width=0.31\textwidth]{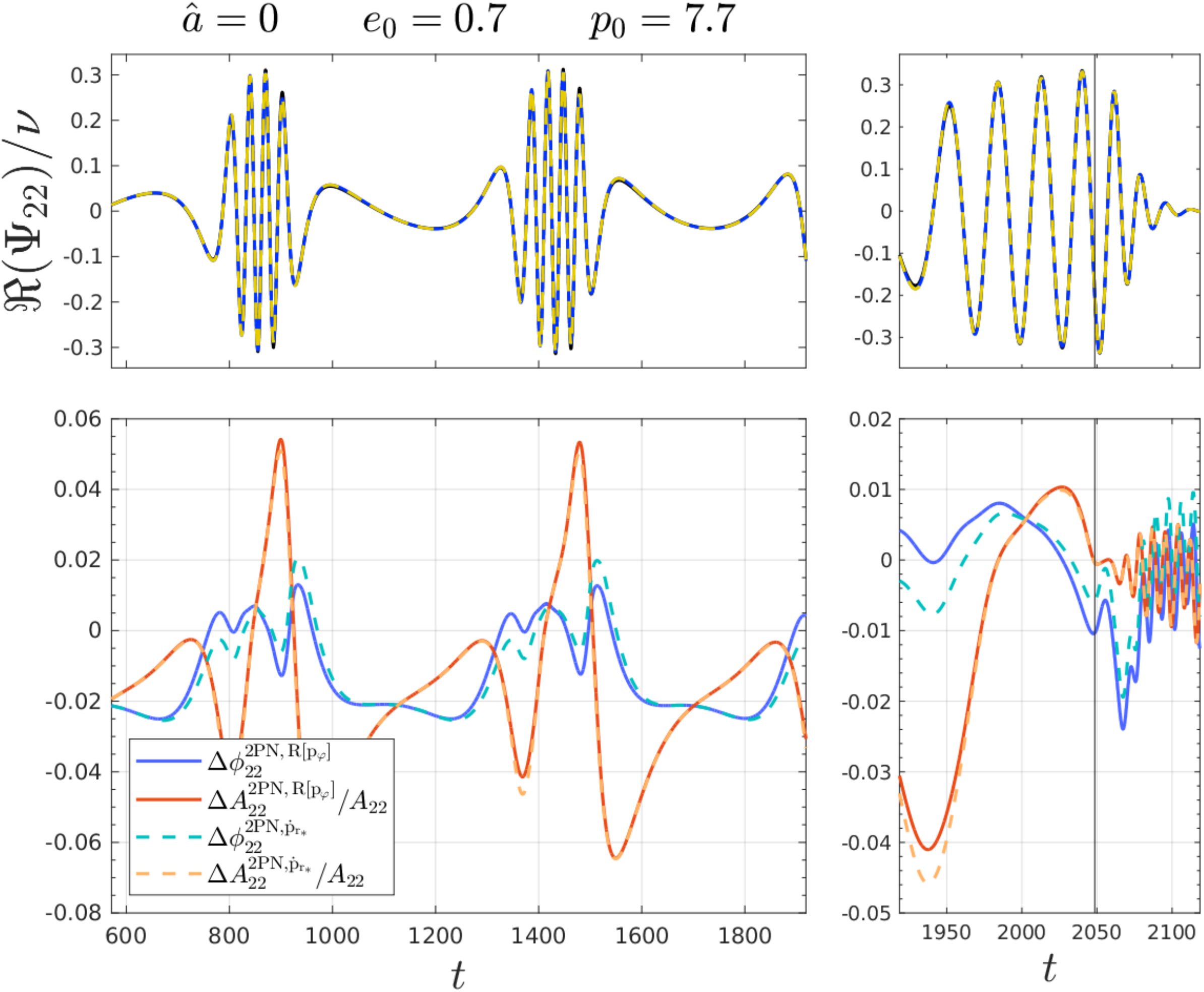}\\
		\includegraphics[width=0.31\textwidth]{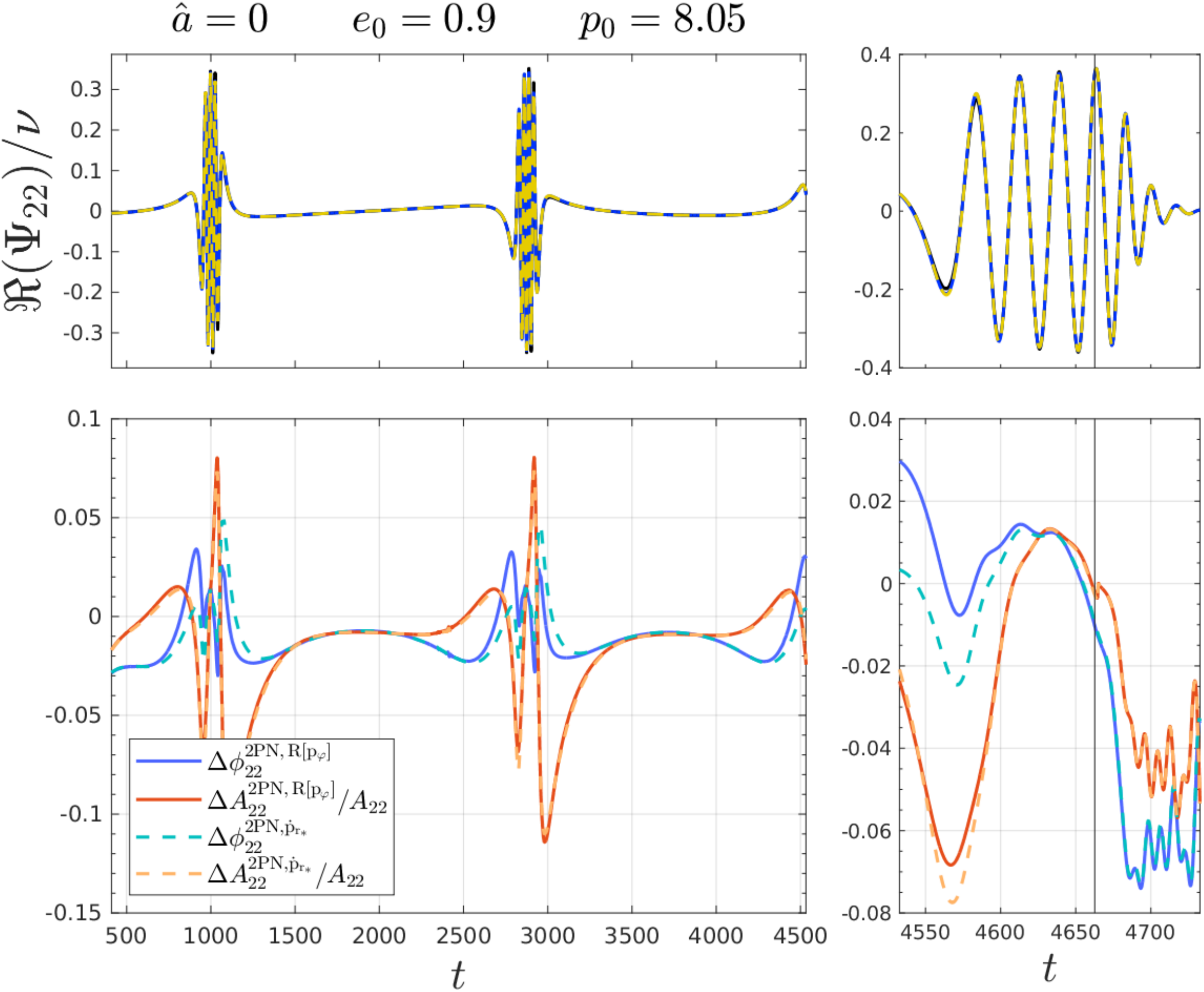}
		\includegraphics[width=0.31\textwidth]{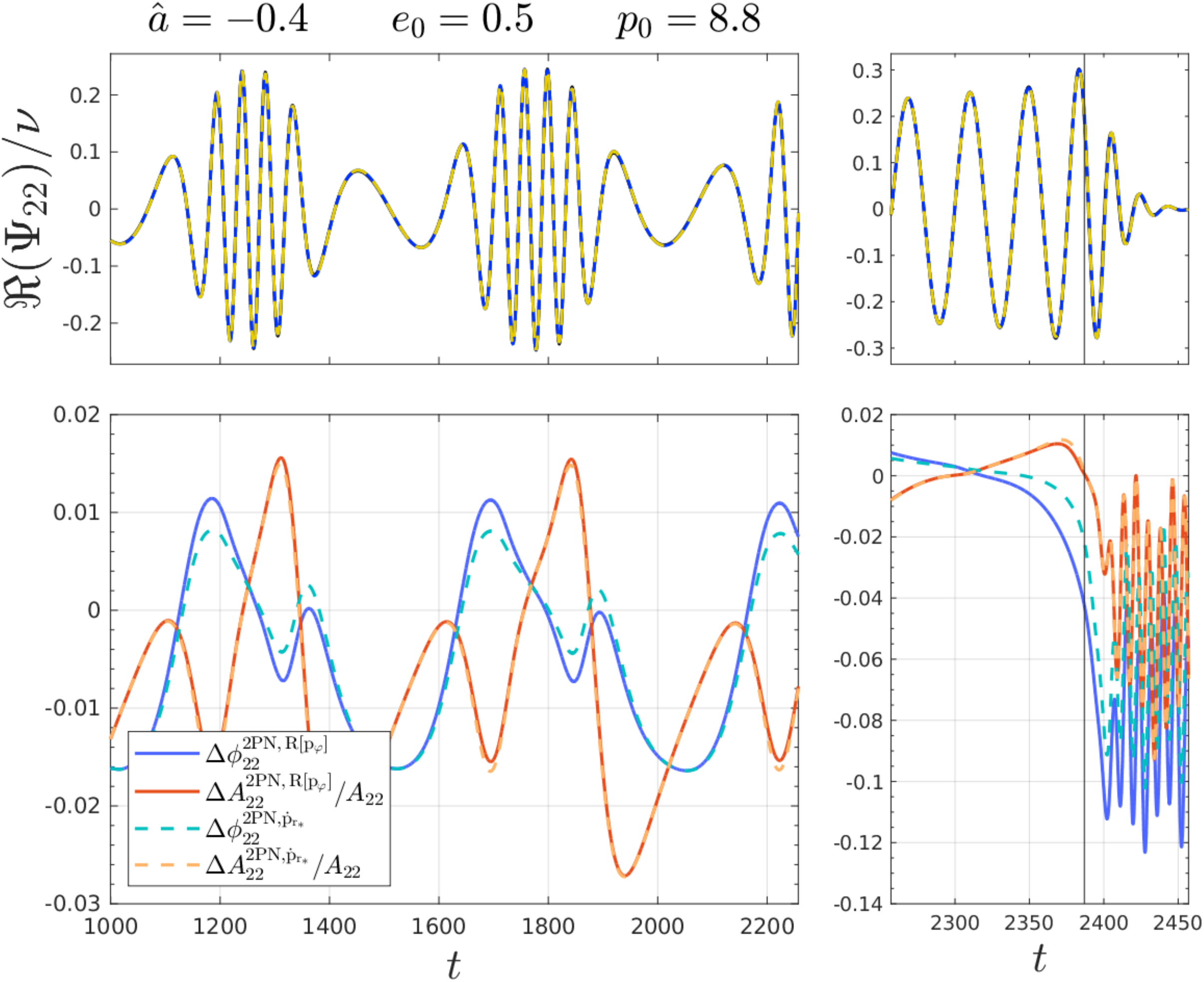}
		\includegraphics[width=0.31\textwidth]{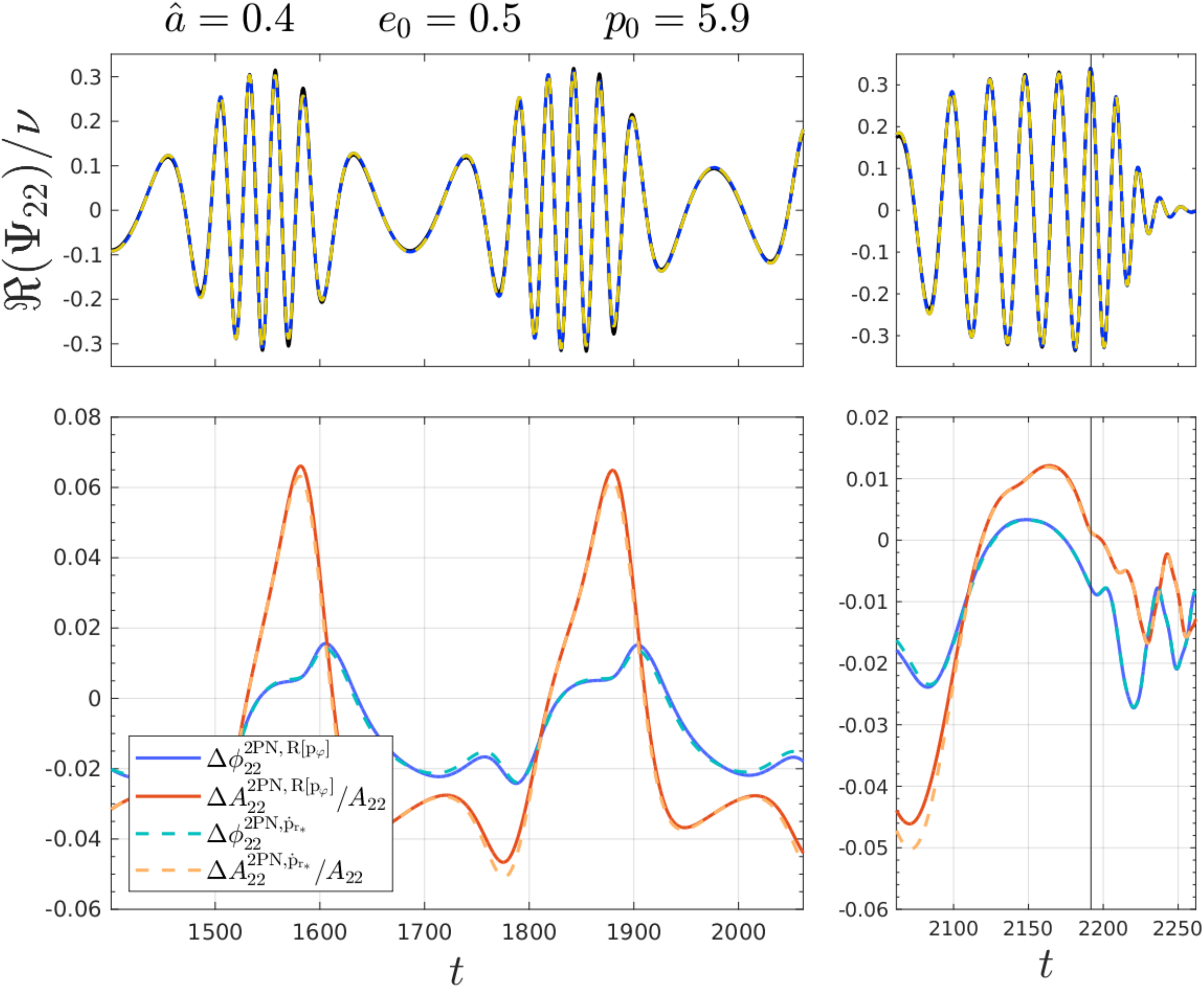}
		\caption{\label{fig:testmass_inspl_testprrdot} Analytical/numerical comparisons 
			of the $\l=m=2$ mode for
			different values of initial eccentricity and Kerr spin-parameter. 
			In the top panels we show the real part of the numerical waveform (black, almost 
			indistinguishable), 
			the analytical waveform with resummed 2PN noncircular corrections written in terms of 
			$(p_{r_*}, p_\varphi)$ (solid blue) and the analytical one with corrections written in terms of 
			$(p_{r_*}, \dot{p}_{r_*})$ (dashed yellow). The corresponding analytical/numerical 
			phase differences and amplitude relative  differences are shown in the bottom panels 
			(solid lines for resummed corrections with $p_\varphi$, dashed lines for 
			the corrections with $\dot{p}_{r_*}$). 
		}
	\end{figure*}

	As briefly discussed at the end of Sec.~\ref{Sec:Alt_PN_corrections} in this appendix we assay an alternative writing for the noncircular 2PN corrections for the modes with $m\neq0$. Starting from the results in terms of  $u$, $p_{r_*}$ and $p_\varphi$ we discussed in Sec.~\ref{Sec:NewFactors}, we exploit the 2PN equation of motion for $\dot{p}_{r_*}$, invert it with respect to  $p_\varphi$ and use the resulting relation to recast everything in terms of $u$, $p_{r_*}$ and $\dot{p}_{r_*}$.
	For the $(2,2)$ mode, again the primary target of our focus, we find
	\begin{widetext}
		\begin{align}
			\label{h22tail_prrdot}
			\left[	\hat{h}_{22}^{\rm nc_{tail}} \right]_{\dot{p}_{r_*}} & = 1 - \frac{\pi}{c^3} \bigg(\frac{3 }{2} i p_{r_*} u + \frac{i p_{r_*} \dot{p}_{r_*}}{4 u}+\frac{p_{r_*}^2 \sqrt{u}}{2} -\frac{3 i p_{r_*} \dot{p}_{r_*}^2}{16 u^3}+\frac{5 p_{r_*}^2 \dot{p}_{r_*}}{16 u^{3/2}}+\frac{5 i p_{r_*}^3}{48} +\frac{i p_{r_*} \dot{p}_{r_*}^3}{32 u^5}\cr
			&-\frac{p_{r_*}^2 \dot{p}_{r_*}^2}{4 u^{7/2}}-\frac{15 i p_{r_*}^3 \dot{p}_{r_*}}{32 u^2}+\frac{5 p_{r_*}^4}{16 \sqrt{u}} -\frac{11 i p_{r_*} \dot{p}_{r_*}^4}{384 u^7}+\frac{101 p_{r_*}^2 \dot{p}_{r_*}^3}{384 u^{11/2}}+\frac{53 i p_{r_*}^3 \dot{p}_{r_*}^2}{96 u^4}\cr
			&-\frac{115 p_{r_*}^4 \dot{p}_{r_*}}{192 u^{5/2}}-\frac{589 i p_{r_*}^5}{1920 u}-\frac{223 i p_{r_*} \dot{p}_{r_*}^5}{3840 u^9}+\frac{257 p_{r_*}^2 \dot{p}_{r_*}^4}{768 u^{15/2}}+\frac{181 i p_{r_*}^3 \dot{p}_{r_*}^3}{288 u^6}-\frac{83 p_{r_*}^4 \dot{p}_{r_*}^2}{128 u^{9/2}}\cr
			&-\frac{111 i p_{r_*}^5 \dot{p}_{r_*}}{256 u^3}+\frac{329 p_{r_*}^6}{1920 u^{3/2}} \bigg),\\
			\label{f22nc_prrdot}
			\left[	f^{\rm nc_{inst}}_{22} \right]_{\dot{p}_{r_*}} & = 1 + \frac{1}{c^2}\bigg[\frac{ (3 \nu +62)p_{r_*}^2 }{42} + \frac{(65-48 \nu ) p_{r_*}^2 \dot{p}_{r_*}}{84 u^2}+ \frac{(114 \nu -115) p_{r_*}^2 \dot{p}_{r_*}^2}{168 u^4}+\frac{(10-9 \nu ) p_{r_*}^4}{21 u}\cr
			&+\frac{(115-156 \nu ) p_{r_*}^2 \dot{p}_{r_*}^3}{336 u^6}+\frac{(144 \nu -125) p_{r_*}^4 \dot{p}_{r_*}}{168 u^3}+\frac{(174 \nu -65) p_{r_*}^2 \dot{p}_{r_*}^4}{672 u^8}+\frac{(55-144 \nu ) p_{r_*}^4 \dot{p}_{r_*}^2}{336 u^5}+\frac{(6 \nu -5) p_{r_*}^6}{24 u^2}\bigg]\cr
			& + \frac{1}{c^4} \bigg[\frac{(1685-1506 \nu -336 \nu ^2)p_{r_*}^2 u}{504} -\frac{(76039-79256 \nu +29480 \nu ^2) p_{r_*}^2 \dot{p}_{r_*}}{42336 u}\cr
			& -\frac{(40120-41507 \nu +6257 \nu ^2) p_{r_*}^2 \dot{p}_{r_*}^2}{10584 u^3}+\frac{(2609 +4538 \nu +382 \nu ^2 ) p_{r_*}^4}{3024}\cr
			&+\frac{(227327-281328 \nu +48696 \nu ^2) p_{r_*}^2 \dot{p}_{r_*}^3}{56448 u^5}-\frac{(198001-378512 \nu +96248 \nu ^2)p_{r_*}^4  \dot{p}_{r_*}}{84672 u^2}\cr
			&-\frac{11 (9290-14020 \nu + 3019 \nu ^2)p_{r_*}^2  \dot{p}_{r_*}^4}{42336 u^7}+\frac{(749981-1363564 \nu +247480 \nu ^2) p_{r_*}^4 \dot{p}_{r_*}^2}{169344 u^4}\cr
			&-\frac{(4687-20688 \nu +5916 \nu ^2) p_{r_*}^6}{14112 u}\bigg],\\
			\label{delta22nc_prrdot}
			\left[	\delta^{\rm nc_{inst}}_{22} \right]_{\dot{p}_{r_*}} & = \frac{1}{c^2} \bigg[\frac{(25-26 \nu ) p_{r_*} \sqrt{u}}{14} +\frac{(25-12 \nu ) p_{r_*} \dot{p}_{r_*}}{42 u^{3/2}}-\frac{(25-18 \nu) p_{r_*} \dot{p}_{r_*}^2}{48 u^{7/2}}+\frac{(6 \nu +5) p_{r_*}^3}{84 \sqrt{u}}\cr
			&+\frac{5 (10-9 \nu ) p_{r_*} \dot{p}_{r_*}^3}{168 u^{11/2}}-\frac{5 (31-30 \nu ) p_{r_*}^3 \dot{p}_{r_*}}{168 u^{5/2}}-\frac{(775-918 \nu ) p_{r_*} \dot{p}_{r_*}^4}{5376 u^{15/2}}+\frac{3 (45-58 \nu )p_{r_*}^3  \dot{p}_{r_*}^2}{224 u^{9/2}}\cr
			&-\frac{(25-26 \nu ) p_{r_*}^5}{56 u^{3/2}}+ \frac{(325-576 \nu)p_{r_*} \dot{p}_{r_*}^5}{5376 u^{19/2}}-\frac{(145-498 \nu) p_{r_*}^3 \dot{p}_{r_*}^3}{1344 u^{13/2}}+\frac{3  (5-8 \nu) p_{r_*}^5 \dot{p}_{r_*}}{56 u^{7/2}}\bigg]\cr
			&+\frac{1}{c^4}\bigg[\frac{(31259-10972 \nu -11360 \nu ^2)p_{r_*} u^{3/2}}{10584}-\frac{(30467-58216 \nu -26216 \nu ^2)p_{r_*} \dot{p}_{r_*}}{21168 \sqrt{u}}\cr
			&-\frac{(197641-182480 \nu +20600 \nu ^2)p_{r_*} \dot{p}_{r_*}^2}{84672 u^{5/2}}-\frac{(25213-81572 \nu -27568 \nu ^2) p_{r_*}^3 \sqrt{u}}{21168}\cr
			&+\frac{(430967-462772 \nu +66832 \nu ^2)p_{r_*} \dot{p}_{r_*}^3}{169344 u^{9/2}}-\frac{(15711-16846 \nu +11500 \nu ^2) p_{r_*}^3 \dot{p}_{r_*}}{14112 u^{3/2}}\cr
			&-\frac{(2456467-2958740 \nu +594704 \nu ^2) p_{r_*} \dot{p}_{r_*}^4}{1354752 u^{13/2}}+\frac{(866783-1234624 \nu +221608 \nu ^2) p_{r_*}^3 \dot{p}_{r_*}^2}{169344 u^{7/2}}\cr
			&-\frac{(38639-34420 \nu +21376 \nu ^2)p_{r_*}^5}{42336 \sqrt{u}}+\frac{(2807687-4018384 \nu +1077064 \nu ^2) p_{r_*} \dot{p}_{r_*}^5}{2709504 u^{17/2}}\cr
			&-\frac{(149307-248258 \nu +44612 \nu ^2) p_{r_*}^3  \dot{p}_{r_*}^3}{37632 u^{11/2}}+\frac{(324131-573220 \nu +81808 \nu ^2) p_{r_*}^5 \dot{p}_{r_*}}{84672 u^{5/2}}\bigg].
		\end{align}
	\end{widetext}
	In Fig.~\ref{fig:testmass_inspl_testprrdot} we look at the test-particle case
	and compare the above corrections with those of 
	Sec.~\ref{Sec:NewFactors}, specifically those with all the resummations.
	Without the need of a resummation scheme, the
	accuracy of the former turns out to be practically equivalent to the one of the latter,
	especially for low or moderate eccentricity. 
	Nevertheless, the evaluation of $\dot{p}_{r_*}$ 
	through the Hamilton's equation in the comparable 
	mass case is less efficient than using directly $p_\varphi$ and for this reason
	we deem preferable the resummed corrections containing $p_\varphi$ presented 
	in Sec.~\ref{Sec:NewFactors}.
	
	%==========================================================
	\bibliography{refs,local}
	%==========================================================

\end{document}